\documentclass[11pt]{article}
\usepackage{graphicx,amsmath}
\usepackage[margin=1.25in]{geometry}
\usepackage{authblk}
\usepackage{url}
\usepackage[colorlinks = true,
            linkcolor = blue,
            urlcolor  = blue,
            citecolor = blue,
            anchorcolor = blue]{hyperref}
            
\usepackage{caption}
\usepackage{subcaption}

\usepackage{graphicx}
\usepackage{amssymb}
\usepackage{slashed}
\usepackage{cancel}
\usepackage{xspace}
\usepackage[dvipsnames]{xcolor}

\usepackage{booktabs}
\usepackage{floatrow}
\newfloatcommand{capbtabbox}{table}[][\FBwidth]
\usepackage{multirow}
\usepackage{braket}
\usepackage[perpage]{footmisc}
\renewcommand{\thefootnote}{\fnsymbol{footnote}}

\usepackage[htt]{hyphenat}

\newenvironment{Abstract}{\begin{quotation} \begin{center}
                       ABSTRACT
     \end{center}\bigskip  }{\end{quotation}}

\def\beq{\begin{equation}}
\def\eeq#1{\label{#1}\end{equation}}
\def\eeqn{\end{equation}}

\newenvironment{Eqnarray}%
   {\arraycolsep 0.14em\begin{eqnarray}}{\end{eqnarray}}
\def\beqa{\begin{Eqnarray}}
\def\eeqa#1{\label{#1}\end{Eqnarray}}
\def\eeqan{\end{Eqnarray}}

\let\bar=\overbar

\def\lsim{\mathrel{\raise.3ex\hbox{$<$\kern-.75em\lower1ex\hbox{$\sim$}}}}
\def\gsim{\mathrel{\raise.3ex\hbox{$>$\kern-.75em\lower1ex\hbox{$\sim$}}}}

\def\del{\partial}
\def\Dslash{\not{\hbox{\kern-4pt $D$}}}
\def\dslash{\not{\hbox{\kern-2pt $\del$}}}
\def\pslash{\not{\hbox{\kern-2pt $p$}}}
\def\ETmiss{\not{\hbox{\kern-4pt $E$}}_T}

\def\Dlr{\mathrel{\raise1.5ex\hbox{$\leftrightarrow$\kern-1em\lower1.5ex\hbox{$D$}}}}

\def\MSB{{\bar{M \kern -2pt S}}}
\def\msb{{\bar{\scriptsize M \kern -1pt S}}}

\def\drb{{\bar{\scriptsize D \kern -1pt R}}}

\def\ifb{{\rm fb}^{-1}}

\def\MeV{{\rm MeV}}
\def\GeV{{\rm GeV}}
\def\TeV{{\rm TeV}}

\newcommand{\unit}[1]{\ensuremath{\text{\,#1}}\xspace}

\newcommand{\PYTHIA}{{\textsc{pythia}}\xspace}
\newcommand{\MADGRAPH}{{\textsc{MadGraph5}}\xspace}
\newcommand{\FEYNRULES}{{\textsc{FeynRules}}\xspace}
\newcommand{\Delphes}{{\textsc{Delphes}}\xspace}

\newcommand{\HT}{\ensuremath{H_{\mathrm{T}}}\xspace}
\newcommand{\RT}{\ensuremath{R_{\mathrm{T}}}\xspace}
\newcommand{\MT}{\ensuremath{M_{\mathrm{T}}}\xspace}
\newcommand{\met}{\ensuremath{\cancel{E}_{\mathrm{T}}}\xspace}
\newcommand{\pt}{\ensuremath{p_{\mathrm{T}}}\xspace}
\newcommand{\mj}{\ensuremath{m_{j}}\xspace}
\newcommand{\ptd}{$p_{\textrm{T}}D$}

\newcommand{\Pp}{\ensuremath{\mathrm{p}}\xspace}
\newcommand{\znnjets}{\ensuremath{{\PZ}{(\rightarrow\Pgn\Pagn)\text{+jets}}}\xspace}
\newcommand{\wlnjets}{\ensuremath{{\PW}{(\rightarrow\ell\Pgn)\text{+jets}}}\xspace}
\newcommand{\ttbar}{{\Pqt{}\Paqt}\xspace}
\newcommand{\Paqt}{\ensuremath{\overline{\mathrm{t}}}\xspace}
\newcommand{\Pge}{\ensuremath{\mathrm{e}}\xspace}
\newcommand{\Pgm}{\ensuremath{\mu}\xspace}
\newcommand{\Pgn}{\ensuremath{\nu}\xspace}
\newcommand{\Pagn}{\ensuremath{\overline{\nu}}\xspace}
\newcommand{\PW}{\ensuremath{\mathrm{W}}\xspace}
\newcommand{\PZ}{\ensuremath{\text{Z}}\xspace}
\newcommand{\Pg}{\ensuremath{\mathrm{g}}\xspace}
\newcommand{\Pq}{\ensuremath{\mathrm{q}}\xspace}
\newcommand{\mq}{\ensuremath{m_{\Pq}}\xspace}
\newcommand{\Paq}{\ensuremath{\mathrm{\bar{q}}}\xspace}
\newcommand{\Paqi}{\ensuremath{\overline{\mathrm{q}}_{i}}\xspace}
\newcommand{\Pqi}{\ensuremath{\mathrm{q}_{i}}\xspace}
\newcommand{\Pqj}{\ensuremath{\mathrm{q}_{j}}\xspace}
\newcommand{\mqi}{\ensuremath{m_{\Pqi}}\xspace}
\newcommand{\mqj}{\ensuremath{m_{\Pqj}}\xspace}
\newcommand{\Pqb}{\ensuremath{\mathrm{b}}\xspace}
\newcommand{\Pqu}{\ensuremath{\mathrm{u}}\xspace}
\newcommand{\Pqc}{\ensuremath{\mathrm{c}}\xspace}
\newcommand{\Pqt}{\ensuremath{\mathrm{t}}\xspace}
\newcommand{\Pqd}{\ensuremath{\mathrm{d}}\xspace}
\newcommand{\Pqs}{\ensuremath{\mathrm{s}}\xspace}
\newcommand{\mb}{\ensuremath{m_{\Pqb}}\xspace}

\newcommand{\rinv}{\ensuremath{r_{\text{inv}}}\xspace}
\newcommand{\Uonedark}{\ensuremath{U(1)^{\prime}}\xspace}

\newcommand{\Pqdark}{\ensuremath{\mathrm{q}_\mathrm{D}}\xspace}
\newcommand{\Paqdark}{\ensuremath{\overline{\mathrm{q}}_\mathrm{D}}\xspace}
\newcommand{\PqdarkO}{\ensuremath{\mathrm{q}_\mathrm{D 1}}\xspace}
\newcommand{\PaqdarkO}{\ensuremath{\overline{\mathrm{q}}_\mathrm{D 1}}\xspace}
\newcommand{\PqdarkT}{\ensuremath{\mathrm{q}_\mathrm{D 2}}\xspace}
\newcommand{\PaqdarkT}{\ensuremath{\overline{\mathrm{q}}_\mathrm{D 2}}\xspace}
\newcommand{\PqdarkTh}{\ensuremath{\mathrm{q}_\mathrm{D 3}}\xspace}
\newcommand{\PaqdarkTh}{\ensuremath{\overline{\mathrm{q}}_\mathrm{D 3}}\xspace}
\newcommand{\PqdarkA}{\ensuremath{\mathrm{q}_{\mathrm{D} \alpha}}\xspace}
\newcommand{\PaqdarkA}{\ensuremath{\overline{\mathrm{q}}_{\mathrm{D} \alpha}}\xspace}
\newcommand{\PqdarkB}{\ensuremath{\mathrm{q}_{\mathrm{D} \beta}}\xspace}

\newcommand{\PqdarkI}{\ensuremath{\mathrm{q}_{\mathrm{D} i}}\xspace}
\newcommand{\PpidarkIJ}{\ensuremath{\pi_{\mathrm{D} ij}}\xspace}
\newcommand{\PrhodarkIJ}{\ensuremath{\rho_{\mathrm{D} ij}}\xspace}

\newcommand{\Ppidark}{\ensuremath{\pi_{\mathrm{D}}}\xspace}
\newcommand{\PpidarkZ}{\ensuremath{\Ppidark^{\text{``}0\text{''}}}\xspace}
\newcommand{\PpidarkPM}{\ensuremath{\Ppidark^{\text{``}\pm\text{''}}}\xspace}
\newcommand{\Pkdark}{\ensuremath{\text{K}_\mathrm{D}}\xspace}
\newcommand{\PkdarkZ}{\ensuremath{\Pkdark^{\text{``}0\text{''}}}\xspace}
\newcommand{\PkdarkPM}{\ensuremath{\Pkdark^{\text{``}\pm\text{''}}}\xspace}
\newcommand{\Prhodark}{\ensuremath{\rho_\mathrm{D}}\xspace}
\newcommand{\Petadark}{\ensuremath{\eta_\mathrm{D}}\xspace}
\newcommand{\Petaprimedark}{\ensuremath{\eta'_\mathrm{D}}\xspace}
\newcommand{\Pomegadark}{\ensuremath{\omega_\mathrm{D}}\xspace}

\newcommand{\PdeltadarkI}{\ensuremath{\Delta_{\mathrm{D} i}}\xspace}
\newcommand{\Pphodark}{\ensuremath{\mathrm{A}^{\prime}}\xspace}

\newcommand{\mdark}{\ensuremath{m_\mathrm{D}}\xspace}
\newcommand{\mqdark}{\ensuremath{m_{\Pqdark}}\xspace}
\newcommand{\mpidark}{\ensuremath{m_{\Ppidark}}\xspace}
\newcommand{\mrhodark}{\ensuremath{m_{\Prhodark}}\xspace}
\newcommand{\momegadark}{\ensuremath{m_{\Pomegadark}}\xspace}
\newcommand{\metaprimedark}{\ensuremath{m_{\Petaprimedark}}\xspace}

\newcommand{\ctpidark}{\ensuremath{c\tau_{\Ppidark}}\xspace}
\newcommand{\fpidark}{\ensuremath{f_{\Ppidark}}\xspace}
\newcommand{\probrho}{\ensuremath{P_{\Prhodark}}\xspace}

\newcommand{\Ncdark}{\ensuremath{N_{c_\mathrm{D}}}\xspace}
\newcommand{\Nfdark}{\ensuremath{N_{f_\mathrm{D}}}\xspace}
\newcommand{\lamdark}{\ensuremath{\Lambda_\mathrm{D}}\xspace}
\newcommand{\tlamdark}{\ensuremath{\tilde{\Lambda}_\mathrm{D}}\xspace}
\newcommand{\adark}{\ensuremath{\alpha_\mathrm{D}}\xspace}
\newcommand{\adarkHigh}{\ensuremath{\adark^{\text{high}}}\xspace}
\newcommand{\adarkLow}{\ensuremath{\adark^{\text{low}}}\xspace}
\newcommand{\adarkPeak}{\ensuremath{\adark^{\text{peak}}}\xspace}
\newcommand{\lamdarkPeak}{\ensuremath{\lamdark^{\text{peak}}}\xspace}
\newcommand{\Qdark}{\ensuremath{Q_\mathrm{D}}\xspace}
\newcommand{\Tdark}{\ensuremath{T_\mathrm{D}}\xspace}

\newcommand{\PZprime}{\ensuremath{\mathrm{Z}^{\prime}}\xspace}
\newcommand{\mZprime}{\ensuremath{m_{\PZprime}}\xspace}
\newcommand{\gq}{\ensuremath{g_{\Pq}}\xspace}
\newcommand{\gqdark}{\ensuremath{g_{\Pqdark}}\xspace}

\newcommand{\Pbifun}{\ensuremath{\Phi}\xspace}
\newcommand{\cbifun}{\ensuremath{\kappa_{\alpha i }}\xspace}
\newcommand{\cbifunbeta}{\ensuremath{\kappa^{*}_{\beta j }}\xspace}
\newcommand{\mbifun}{\ensuremath{m_{\Pbifun}}\xspace}

\newcommand\snowmass{\begin{center}\rule[-0.2in]{\hsize}{0.01in}\\\rule{\hsize}{0.01in}\\
\vskip 0.1in Submitted to the  Proceedings of the US Community Study\\ 
on the Future of Particle Physics (Snowmass 2021)\\ 
\rule{\hsize}{0.01in}\\\rule[+0.2in]{\hsize}{0.01in} \end{center}}

\author[a]{Guillaume Albouy}
\author[h]{Jared Barron}
\author[b]{Hugues Beauchesne}
\author[c]{Elias Bernreuther}
\author[d]{Marcella Bona}
\author[e]{Cesare Cazzaniga}
\author[o]{Cari Cesarotti}
\author[f]{Timothy Cohen}
\author[e]{Annapaola de Cosa}
\author[h]{David Curtin}
\author[aa]{Zeynep Demiragli}
\author[v,y]{Caterina Doglioni}
\author[d]{Alison Elliot}
\author[c]{Karri Folan DiPetrillo}
\author[e]{Florian Eble}
\author[aa]{Carlos Erice}
\author[z]{Chad  Freer}
\author[g]{Aran Garcia-Bellido}
\author[h]{Caleb Gemmell}
\author[a,*]{Marie-Hélène Genest}
\author[i]{Giovanni Grilli di Cortona}
\author[j]{Giuliano Gustavino}
\author[v]{Nicoline Hemme}
\author[bb]{Tova Holmes}
\author[x]{Deepak Kar}
\author[k]{Simon Knapen}
\author[l,*]{Suchita Kulkarni}
\author[z]{Luca Lavezzo}
\author[r]{Steven Lowette}
\author[j]{Benedikt Maier}
\author[l]{Seán Mee}
\author[c]{Stephen Mrenna}
\author[u]{Harikrishnan Nair}
\author[e]{Jeremi Niedziela}
\author[cc]{Christos Papageorgakis}
\author[g]{Nukulsinh Parmar}
\author[z]{Christoph Paus}
\author[c]{Kevin Pedro}
\author[a]{Ana Peixoto}
\author[dd]{Alexx Perloff}
\author[w]{Tilman Plehn}
\author[m]{Christiane Scherb}
\author[m]{Pedro Schwaller}
\author[ee]{Jessie Shelton}
\author[u]{Akanksha Singh}
\author[x]{Sukanya Sinha}
\author[t]{Torbj\"orn Sj\"ostrand}
\author[h]{Aris G.B. Spourdalakis}
\author[n]{Daniel Stolarski}
\author[o]{Matthew J. Strassler}
\author[p]{Andrii Usachov}
\author[j]{Carlos Vázquez Sierra}
\author[q]{Christopher B. Verhaaren}
\author[cc]{Long Wang}

\affil[a]{Univ. Grenoble Alpes, CNRS, Grenoble INP, LPSC-IN2P3, 38000 Grenoble, France}
\affil[b]{Physics Division, National Center for Theoretical Sciences, Taipei 10617, Taiwan}
\affil[c]{Fermi National Accelerator Laboratory, Batavia, IL 60510 USA}
\affil[d]{Queen Mary University of London}
\affil[e]{ETH Zurich - Institute for Particle Physics and ~~~~~~~~~~~~~~~~~~~~~~~~~~~~~~~~~~~ Astrophysics (IPA), Zurich, Switzerland}
\affil[f]{Institute for Fundamental Science, University of Oregon, ~~~~~~~~~~~~~~~~~~~~~~~~~~~~~~~~~~~ Eugene, Oregon 97403, USA}
\affil[g]{Department of Physics and Astronomy, University of Rochester, ~~~~~~~~~~~~~~~~~~~~~~~~~~~~~~~~~~~ Rochester, NY 14627}
\affil[h]{Department of Physics, University of Toronto, Toronto, Ontario M5S 1A7, Canada}
\affil[i]{Istituto Nazionale di Fisica Nucleare, Laboratori Nazionali di Frascati, ~~~~~~~~~~~~~~~~~~~~~~~~~~~~~~~~~~~ C.P. 13, 00044 Frascati, Italy}
\affil[j]{European Organization for Nuclear Research (CERN), Geneva, Switzerland}
\affil[k]{Theoretical Physics Group, Lawrence Berkeley National Laboratory, Berkeley, CA 94720, USA and Berkeley Center for Theoretical Physics, Department of Physics, University of California, Berkeley, CA 94720, USA }
\affil[l]{Institute of Physics, NAWI Graz, University of Graz,Universit\"atsplatz 5, ~~~~~~~~~~~~~~~~~~~~~~~~~~~~~~~~~~~ A-8010 Graz, Austria}
\affil[m]{PRISMA$^+$ Cluster of Excellence \& Mainz Institute for Theoretical Physics, Johannes Gutenberg University, 55099 Mainz, Germany}
\affil[n]{Ottawa-Carleton Institute for Physics, Carleton University, 1125 Colonel By Drive, Ottawa, Ontario K1S 5B6, Canada}
\affil[o]{Department of Physics, Harvard University, Cambridge, MA, 02138}
\affil[p]{Nikhef National Institute for Subatomic Physics, Amsterdam, Netherlands}
\affil[q]{Department of Physics and Astronomy, Brigham Young University, ~~~~~~~~~~~~~~~~~~~~~~~~~~~~~~~~~~~ Provo, UT, 84602, USA}
\affil[r]{Vrije Universiteit Brussel, Brussels, Belgium}
\affil[s]{PRISMA+ Cluster of Excellence and Mainz Institute for Theoretical Physics, Johannes Gutenberg-Universitaet Mainz, 55099 Mainz, Germany}
\affil[t]{Theoretical Particle Physics, Department of Astronomy and Theoretical Physics, Lund University, S\"olvegatan 14A, 223 62 Lund, Sweden}
\affil[u]{University of Mumbai}
\affil[v]{Fysikum, Division of Particle Physics, Lund University, Lund, Sweden}
\affil[w]{Institut f\"ur Theoretische Physik, Universit\"at Heidelberg, Germany}
\affil[x]{University of Witwatersrand, South Africa}
\affil[y]{University of Manchester}
\affil[z]{Massachusetts Institute of Technology, Cambdrige, MA, USA}
\affil[aa]{Boston University}
\affil[bb]{University of Tennessee, Knoxville}
\affil[cc]{University of Maryland, College Park}
\affil[dd]{University of Colorado Boulder}
\affil[ee]{Department of Physics, University of Illinois at Urbana-Champaign\vspace{5pt} }
\affil[*]{ {\bf Corresponding authors: M-H.~Genest (genest@lpsc.in2p3.fr) and S.~Kulkarni (suchita.kulkarni@uni-graz.at)} }

\date{}

\begin{document}

\title{Theory, phenomenology, and experimental avenues for dark showers: a Snowmass 2021 report}

\maketitle

\vspace{-60pt}
\snowmass
\vspace{-30pt}
 \begin{Abstract}
{In this work, we consider the case of a strongly coupled dark/hidden sector, which extends the Standard Model (SM) by adding an additional non-Abelian gauge group. These extensions generally contain matter fields, much like the SM quarks, and gauge fields similar to the SM gluons. We focus on the exploration of such sectors where the dark particles are produced at the LHC through a portal and undergo rapid hadronization within the dark sector before decaying back, at least in part and potentially with sizeable lifetimes, to SM particles, giving a range of possibly spectacular signatures such as emerging or semi-visible jets. Other, non-QCD-like scenarios leading to soft unclustered energy patterns or glueballs are also discussed. After a review of the theory,  existing benchmarks and constraints, this work addresses how to build consistent benchmarks from the underlying physical parameters and present new developments for the \PYTHIA Hidden Valley module, along with jet substructure studies. Finally, a series of improved search strategies is presented in order to pave the way for a better exploration of the dark showers at the LHC.  }
\end{Abstract}

\tableofcontents

\def\thefootnote{\fnsymbol{footnote}}
\setcounter{footnote}{0}

\section{Introduction}

As the experimental program of the LHC searches for physics beyond the Standard Model (SM) is maturing, the community has started devoting significant effort to investigating alternative models and their associated phenomenology, especially those providing exotic signatures which would not have been directly addressed yet in the existing searches. Of interest here is the case of a strongly coupled dark sector or hidden sector, which extends the SM with an additional non-Abelian gauge group. Considering the non-trivial structure of the SM QCD, we should be open to the idea of a potentially complicated dark sector via non-Abelian gauge groups. These extensions generally contain matter fields, much like the SM quarks and gauge fields similar to the SM gluons. There are no a priory expectations on the gauge group dimension (number of colors), or that of matter fields (number of flavors) that the theory may have. 

When the dark sector confines below some confinement scale (\lamdark) dark hadrons are formed, and depending on the symmetries of the theory, some of them could be stable leading to dark matter candidates. To allow for the production of dark states at the LHC, which could be either dark quarks or hadrons, the dark sector is coupled to the SM via a portal. 
The realisation of associated LHC phenomenology of such dark/hidden sector is however very much dependent on the details of the model. Nevertheless, some generic expectations can be set. For example, at the LHC, cases where dark quark masses (\mqdark) and corresponding confinement scale are much smaller than the collider centre-of-mass energy ($\mqdark \lesssim \lamdark \ll \sqrt{s}$) lead to spectacular signatures in terms of emerging or semi-visible jets~\cite{Schwaller:2015gea,Cohen:2015toa,Daci:2015hca}. Increasing \lamdark implies heavier bound states, which in turn decreases the final state multiplicities for a given $\sqrt{s}$, as the allowed phase space decreases. This means that as the limit $\lamdark \sim \sqrt{s}$ is approached, depending on the relevant production mechanisms, $2\to 2$ SM initial state to dark meson final state processes become prevalent and resonance-like searches for dark bound states may prove useful~\cite{Kribs:2018ilo,Hochberg:2015vrg,Kribs:2016cew,Butterworth:2021jto}. Finally, cases where $\mqdark \gg \lamdark,  \mqdark \lesssim \sqrt{s}$ lead to unusual signals known as quirks~\cite{Kang:2008ea, Knapen:2016hky, Harnik:2008ax}. If the strongly-interacting sector is non-QCD like, other signatures such as Soft Unclustered Energy Patterns are also possible~\cite{Knapen:2016hky, Harnik:2008ax}. For a review pertaining to this discussion see~\cite{Kribs:2016cew}.

In this work, we focus on LHC exploration of such dark/hidden sectors where $\lamdark \ll \sqrt{s}$. In such cases, dark quarks could be produced at the LHC through a portal and undergo rapid hadronization within the dark sector before decaying back, at least in part and potentially with sizeable lifetimes, to SM particles. Models with such hidden dark sectors have been discussed e.g. in the context of twin Higgs models \cite{Chacko:2005pe,Craig:2015pha,Barbieri:2015lqa,Blennow:2010qp}, composite and/or asymmetric dark matter scenarios \cite{Hur:2007uz, Hur:2011sv, Frandsen:2011kt,Buckley:2012ky,Hambye:2013dgv,Bai:2013iqa,Bai:2013xga,Cline:2013zca,Boddy:2014yra,Hochberg:2014kqa,Antipin:2015xia,GarciaGarcia:2015fol,Hochberg:2015vrg,Dienes:2016vei,Lonsdale:2017mzg,Davoudiasl:2017zws, Choi:2018iit, Cacciapaglia:2020kgq, Cline:2021itd}, and string theory \cite{Berlin:2018tvf,Halverson:2016nfq}. Some example studies focusing on hidden valley phenomenology can be found in \cite{Acharya:2017szw,Strassler:2006im,Strassler:2006ri,Alexander:2016aln,Batell:2011tc,Agrawal:2011ze,Calibbi:2015sfa,Harnik:2008ax, Han:2007ae,Juknevich:2009gg,Strassler:2008fv,Juknevich:2009ji,Carloni:2010tw,Carloni:2011kk,Kribs:2016cew,Beauchesne:2019ato} with additional examples referred to elsewhere in this write-up. The above references show a rather large activity throughout the last four decades. 

These hidden valley models differ from most other Beyond the Standard Model scenarios because the infrared (IR) parameters of the theory can not be computed from ultraviolet (UV) definitions using perturbative techniques. This is in contrast to many other models e.g.~MSSM, which have a well-defined relationships between UV and IR, even if their parameter space is high dimensional. The other well known approach to characterise new physics is the use of simplified models which do not rely on such top-down priors but are defined by their minimality; these may not be effective for hidden valley scenarios either,  due to the inherent dependence on UV parameters in strongly interacting theories. Therefore, neither principles used otherwise to analyse new physics scenarios apply to hidden valley models. 

While on the one hand these considerations motivate avenues for model-building with applications to shortcomings of the SM such as e.g.~dark matter, LHC searches for hidden valleys are primarily motivated by the exotic phenomenology as stated above. Concretely, this means that we do not have a strong theory prior on e.g.~the number of colors and flavors, the mass hierarchies amongst the matter fields, or the possible patterns of flavor breaking. Out of the multiple choices at hand, however, it is nevertheless possible to try and build internally coherent models and develop tools to predict their phenomenology and guide the searches.

Despite the complexity and the challenges in analysing such non-Abelian new sectors, there has been an increased activity in the recent years that has focused on  understanding the signature parameter space of such models, both on the theory and the experimental sides. This report presents some of these developments, particularly concentrating on jet-like signatures, and puts in perspective the efforts necessary to make systematic progress in understanding, classifying and searching for such non-Abelian scenarios. Throughout this report we will consider dark sector scenarios where the dark quarks are uncharged under any SM group and the dark sector communicates to the SM via an additional mediator.

The report is organised as follows. QCD-like dark-sector scenarios are first reviewed in Section 2, addressing the theories in the s- and t-channels, and the existing benchmarks and limits for such models. Section 3 will address two possibilities of dark sector beyond the QCD-like scenarios: soft unclustered energy patterns (SUEP) and glueballs. Section 4 will be devoted to simulation tool limitations and how to build consistent benchmarks from the underlying physical parameters for semi-visible jets. After a discussion of consistent parameter setting, some improvements to the \PYTHIA Hidden Valley module and their validation will be presented, followed by some phenomenological studies on the jet substructure effects of varying the physical parameters. Finally, a series of improved search strategies will be discussed in Section 5, based on event-level variables, deep neural networks, autoencoder-based anomaly detection or better triggering algorithms.
\section{QCD-like scenarios of dark sector}
\label{existing_studies}
In SM QCD, the strong coupling constant $\alpha_s$ becomes weaker as the energy increases. This is known as asymptotic freedom. New non-Abelian sectors which display such asymptotic freedom fall into the category of QCD-like scenarios. 

This section outlines possible exotic signatures that QCD-like dark sector scenarios may exhibit, including a discussion of benchmark models that have been or are currently being employed by the community. We also discuss existing limits in the context of these benchmarks. These signatures and results in turn provide a strong motivation behind detailed studies of such scenarios, motivating further theory effort, as will be discussed in sec.~\ref{sec:model_building}.

\subsection{Theories of dark QCD}
\emph{Contributors: Timothy Cohen and Christiane Scherb}

As an organizing principle, we will assume that the dark sector communicates with the Standard Model via a so-called portal.  The Standard Model admits three renormalizable portals, in that there are three Standard Model gauge singlet operators with mass dimension less than four: the dark sector can couple to the field strength of the Hypercharge gauge boson $B_{\mu\nu}$~\cite{Holdom:1985ag}, to the Higgs bi-linear $|H|^2$~\cite{Patt:2006fw}, or to the neutrino via $H L$~\cite{Falkowski:2009yz}.  In this paper, we will focus on introducing a new mediator particle that serves as the portal.  This could be a \PZprime which would mediate $s$-channel production~\cite{Frandsen:2012rk,Buchmueller:2013dya,Dreiner:2013vla,Buchmueller:2014yoa,Hamaguchi:2014pja,Harris:2014hga,Jacques:2015zha,Liew:2016oon,Englert:2016joy,Bernreuther:2019pfb,Cheng:2019yai,Cohen:2017pzm}, as was proposed in the original hidden valley paper~\cite{Strassler:2006im}, or it could be a new scalar bi-fundamental which would mediate $t$-channel production~\cite{An:2013xka,Bai:2013xga,Chang:2013oia,Bai:2013iqa,Papucci:2014iwa,Garny:2014waa}.  In both of these cases, in the limit that the mediator mass is large, it may also be appropriate to integrate it out which would induce a contact operator.  There are of course many options beyond these two examples, but to keep our scope finite we will only discuss these $s$- and $t$-channel production models.

Dark sector particles can then be produced at hadron colliders via the portal. Similar to SM quarks, dark quarks shower and hadronize and form dark jets. The properties of dark jets are determined by the dynamics of the dark sectors, namely the coupling strength, the ratio of unstable to stable dark hadrons inside the dark jets, and the mass scale of the dark hadrons. Production of dark sector particles at hadron colliders lead to a broad class of exotic signatures: Depending on the lifetime of the dark hadrons final states can contain semi-visible jets, lepton jets, emerging jets, soft bombs, quirks, etc \cite{Knapen:2016hky,Costantino:2020msc,Schwaller:2015gea,ATLAS:2015itk,Buschmann:2015awa,ATLAS:2019tkk,Cohen:2015toa,Cohen:2017pzm,Park:2017rfb,Bernreuther:2020vhm,Kar:2020bws,Curtin:2022tou}.

Our focus will be on characterizing exotic signatures that could result at the LHC.  The space of possible dark sector models is vast, and furthermore many models can yield essentially the same LHC phenomenology.  For this reason, we work with a simplified model-like parameterization of the dark sector.  The phenomenology can be largely determined by specifying the dynamics of the dark sector shower (the number of dark colors, dark quark flavors, and the dark confinement scale), the mass spectrum, and decay patters of the dark mesons. We will largely frame the phenomenological implications of having a strongly coupled dark sector in terms of these variables.

\subsubsection{s-channel\label{schannel}}

As discussed above, we take as an organizing principle that the dark sector communicates with the visible sector via a portal. If the portal is heavy, then one can describe it by integrating out the mediator to obtain a so called contact operator \cite{Goodman:2010yf, Beltran:2010ww, Fox:2011pm}, for example 
\begin{align}
    \mathcal{L} \supset \frac{c_{ij\alpha\beta}}{\Lambda}\left(\bar{q}_i\gamma^\mu  q_j\right)\left(\PaqdarkA\gamma_\mu\PqdarkB\right),
\label{eq:contactoperator}
\end{align}
where $q$ are SM fermions, \Pqdark are dark sector quarks, $c_{ij\alpha\beta}$ are $\mathcal{O}(1)$ couplings encoding a possible flavor structure, and $\Lambda$ is the scale of the operator. Generally we use Roman indices as SM flavor indices and Greek indices for the dark sector flavor indices. 

We are assuming that the portal couples the dark sector to the Standard Model quarks.  Therefore, the observables of interest will be jets (which are expected to have non-QCD-like features) and missing energy that is likely to be aligned with the jets.  One way to organize thinking about the possible signature space is in terms of the average fraction of invisible particles that are contained within a final state jet
\begin{align}
r_\text{inv} \equiv \Braket{ \frac{\# \rm{stable\, dark\, hadrons}}{\# \rm{dark\, hadrons}}}.
\label{eq:rinv}\end{align}
  We will present results in terms of this variable in what follows.

\begin{figure}
    \centering
    \includegraphics[width=0.49\linewidth]{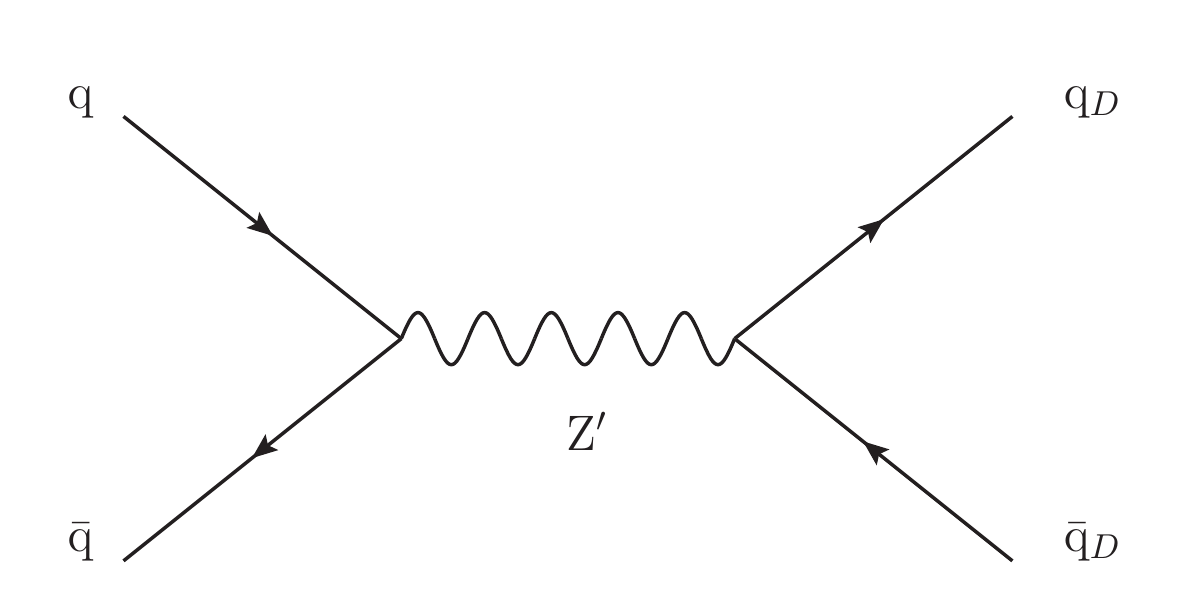}
    \caption{Diagram for the pair production of dark quarks through a \PZprime portal.}
    \label{fig:s-channel_production}
\end{figure}

One option for UV completing Eq.~\ref{eq:contactoperator} is to introduce a so-called $s$-channel mediator.
In such models, pairs of dark quarks can be produced via a heavy resonance \PZprime, that also couples to SM quarks via \cite{Frandsen:2012rk,Buchmueller:2013dya,Dreiner:2013vla,Buchmueller:2014yoa,Hamaguchi:2014pja,Harris:2014hga,Jacques:2015zha,Liew:2016oon,Englert:2016joy,Bernreuther:2019pfb,Cheng:2019yai,Cohen:2017pzm}
\begin{align}
    \mathcal{L} \supset -Z_\mu^\prime\left(\gq\bar{q}_i\gamma^\mu q_i + \gqdark \PaqdarkA\gamma^\mu \PqdarkA\right),
\label{eq:Zprime}\end{align}
where 
$g_{q,\Pqdark}$ are the respective coupling constants. In general, \PZprime can also couple to other SM particles, which would lead to many possibilities in the final state.  For concreteness here, we will focus on dark showers that result in SM jets + missing energy signatures.  Therefore, we limit ourselves here to coupling the \PZprime to quarks, see Figure \ref{fig:s-channel_production}.  We will also simply give the \PZprime a mass, and will not worry about the associated Higgs mechanism or related effects. We will not discuss the additional particle content needed to cancel anomalies, nor the $Z-\PZprime$ mixing structure needed for $\gqdark \ne \gq$ of models with a heavy \PZprime here (c.f. e.g. \cite{Ismail:2017ulg}), but will simply focus on the phenomenology.

Heavy resonances \PZprime are produced at a hadron collider in Drell Yan processes and will have a non-trivial branching ratio to decay to two dark quarks, which shower and hadronize in the dark sector.  Then some of the dark sector hadrons are assumed to decay back to Standard Model quarks, which subsequently shower and hadronize as usual. Consequently, the phenomenology of s-channel models is governed by the following parameters: the \PZprime mass $m_{Z'}$, its couplings to visible and dark quarks \gq and \gqdark, the dark sector shower (governed by the number of dark colors, dark flavors, and the scale of dark sector confinement \lamdark), the characteristic scale of the dark hadrons \mdark, and the average fraction of stable hadrons that are aligned with the visible jet \rinv. While the coupling to SM quarks determines the \PZprime production cross section, the other parameters determine the final state. The details of the shower and mass scale of the dark hadrons determine how many dark sector particles are produced, and \rinv determines the amount of missing energy in a dark jet. Depending on these parameters several interesting signatures and search methods can be defined, e.g semi-visible jets and searches for dark matter in the jet substructure. 

\begin{figure}
\begin{center}
\includegraphics[width=0.555\textwidth]{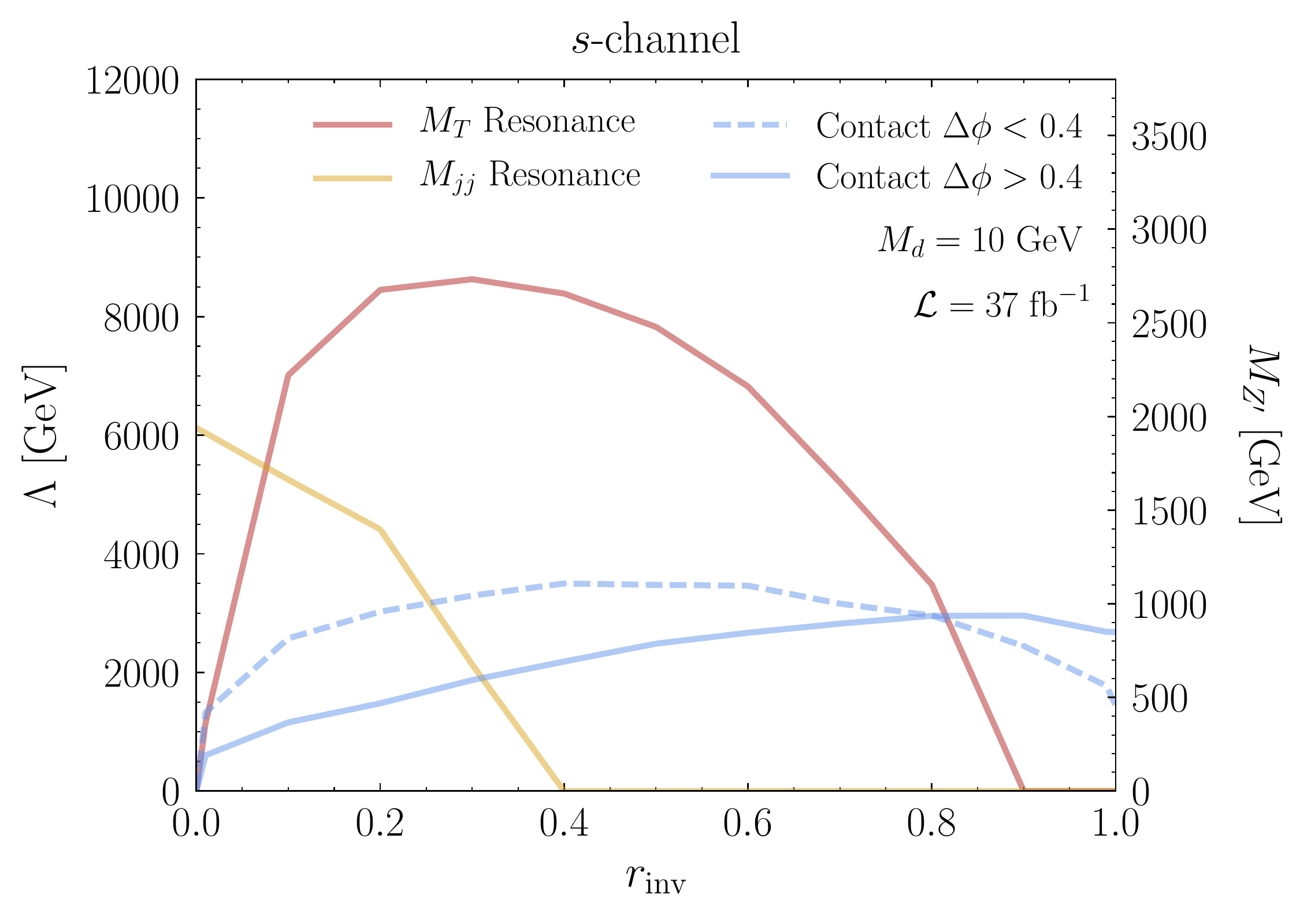}
\caption{Estimates of the projected limits for the $s$-channel model.  The `contact' limit uses a mono-jet search strategy, where $\Delta \phi$ is the angle between the missing transverse energy and the closest jet.  Figure taken from \cite{Cohen:2017pzm}.\label{fig:schannel}}
\end{center}
\end{figure}

There are two dominant strategies to search for the signatures of these models, that largely depend on the choice of \rinv.  When \rinv is small, most of the final state associated with the resonance is visible, and so a normal bump hunt strategy can be employed.  Then as \rinv gets larger, it becomes advantageous to perform a bump hunt using the standard transverse mass variable $M_T^2 = M_{jj}^2 + 2(\sqrt{M_{jj}^2 + p_{Tjj}^2} \cancel{E}_{T} - \vec{p}_{Tjj} \cdot \vec{\cancel{E}}_{T})$.  Then, once \rinv approaches unity, the final state is essentially dominated by missing energy, and so a ``mono-jet'' style strategy is most sensitive.  This is illustrated in Fig.~\ref{fig:schannel}~\cite{Cohen:2017pzm}, where we show the projected limits on the $s$-channel model for these different strategies.  This approach relies on very simple criteria, and so there is clearly much room for improvements that rely on additional characteristics of these models. Existing limits will be discussed in Section \ref{constraints}, while strategies for improvements will be addressed in Section \ref{sec:improved_strategies}. 

\subsubsection{t-channel}\label{tchannel}

Another simple option for UV completing the contact operator is to introduce a so-called $t$-channel mediator.
The $t$-channel UV completion is determined by the following interaction \cite{An:2013xka,Bai:2013xga,Chang:2013oia,Bai:2013iqa,Papucci:2014iwa,Garny:2014waa} 
\begin{align}
    \mathcal{L}\supset -\left(\cbifun\PqdarkA \Pbifun \Bar{q}_{R i}\right) + h.c.,
    \label{eq:t-channel}
\end{align}
which can be realized in either Minimal Flavor Violating models, where in addition to the SM flavor symmetry $U_q(3)\times U_u(3)\times U_d(3)$ the dark flavor symmetry $U_D(3)$ of the dark quarks \Pqdark is introduced \cite{Agrawal:2014aoa,Jubb:2017rhm,Blanke:2017fum}, or by enhancing the SM gauge group by a dark flavor symmetry $SU_D(\Ncdark)$ and introducing \Nfdark dark quarks \Pqdark \cite{Bai:2013xga,Schwaller:2015gea,Renner:2018fhh}.     
Then, the visible and dark sector communicate via a scalar bi-fundamental mediator \Pbifun charged under both the SM and the dark flavor symmetry  
and $q_R$ represent right-handed up-type and down-type quarks.  We consider the case where \Pbifun is an $SU(2)$-singlet, and so it will not have any couplings to the left-handed quark doublets.  (We note that generally $SU(2)$-doublet mediators coupling to left-handed SM quark doublets are also possible.) Depending on the hypercharge of \Pbifun, the dark sector communicates with either the up- ($Y_\Pbifun = 1/3$) or the down-type quarks ($Y_\Pbifun = -2/3$). In the following we will always use $\Ncdark = \Nfdark = 3$. In addition, we assume $\mqdark < \lamdark$, so that the pseudo-Nambu-Goldstone bosons (we will denote them dark pions in the following) of the spontaneously broken dark chiral symmetry are parametrically lighter than other dark hadrons. Consequently, heavier dark sector states, e.g. heavier dark hadrons or glueballs, will decay into dark pions and the dark pions will govern the phenomenology of such models. It is also worth pointing out that for $\Nfdark>3$ an unbroken $SU(\Nfdark-3)$ symmetry leads to one or more stable dark pion \cite{Renner:2018fhh}.

A particularly interesting feature of such models is the fact that the dark sector inherits the SM flavor structure via $\kappa_{\alpha\beta i j}$. The coupling $\kappa_{\alpha i}$ can generally be expressed as 
\begin{align}
    \kappa = VDU
\end{align}
with $D$ a diagonal $3\times3$ matrix of the form \cite{Agrawal:2014aoa}
\begin{align}
    D=diag\left(\kappa_0+\kappa_1,\kappa_0+\kappa_2,\kappa_0-(\kappa_1+\kappa_2)\right)
\end{align}
and $V$ and $U$ Hermitian $3\times3$ matrices. For $m_{Q_{\alpha\beta}}=\delta_{\alpha\beta}m_{Q_{\alpha\beta}}$ the resulting dark flavor symmetry $U_d(3)$ can be used to rotate $V$ away. Finally, $U$ can be decomposed into
\begin{align}
    U = U_{23}U_{13}U_{12},
\end{align}
with $U_{ij}$ the rotational matrices for $ij$. 

The phenomenology of $t$-channel dark sectors and dark mesons has been studied e.g. in \cite{Bai:2013xga,Schwaller:2015gea,Renner:2018fhh,Cohen:2017pzm,An:2013xka,Chang:2013oia,Bai:2013iqa,Papucci:2014iwa,Garny:2014waa,Kribs:2018ilo,Kribs:2018oad,Beauchesne:2018myj,Jubb:2017rhm,Agrawal:2014aoa,Blanke:2017fum,Carmona:2021seb,Mies:2020mzw}. At colliders, the particle content of $t$-channel dark sector models can be produced in various channels such as from mediator pair production ($gg/qq \to \Pbifun\Pbifun$) or associated mediator production ($gq \to \Pbifun \Pqdark$), as well as  the direct production of dark quarks.  
\begin{figure}
    \centering
    \includegraphics[width=0.49\linewidth]{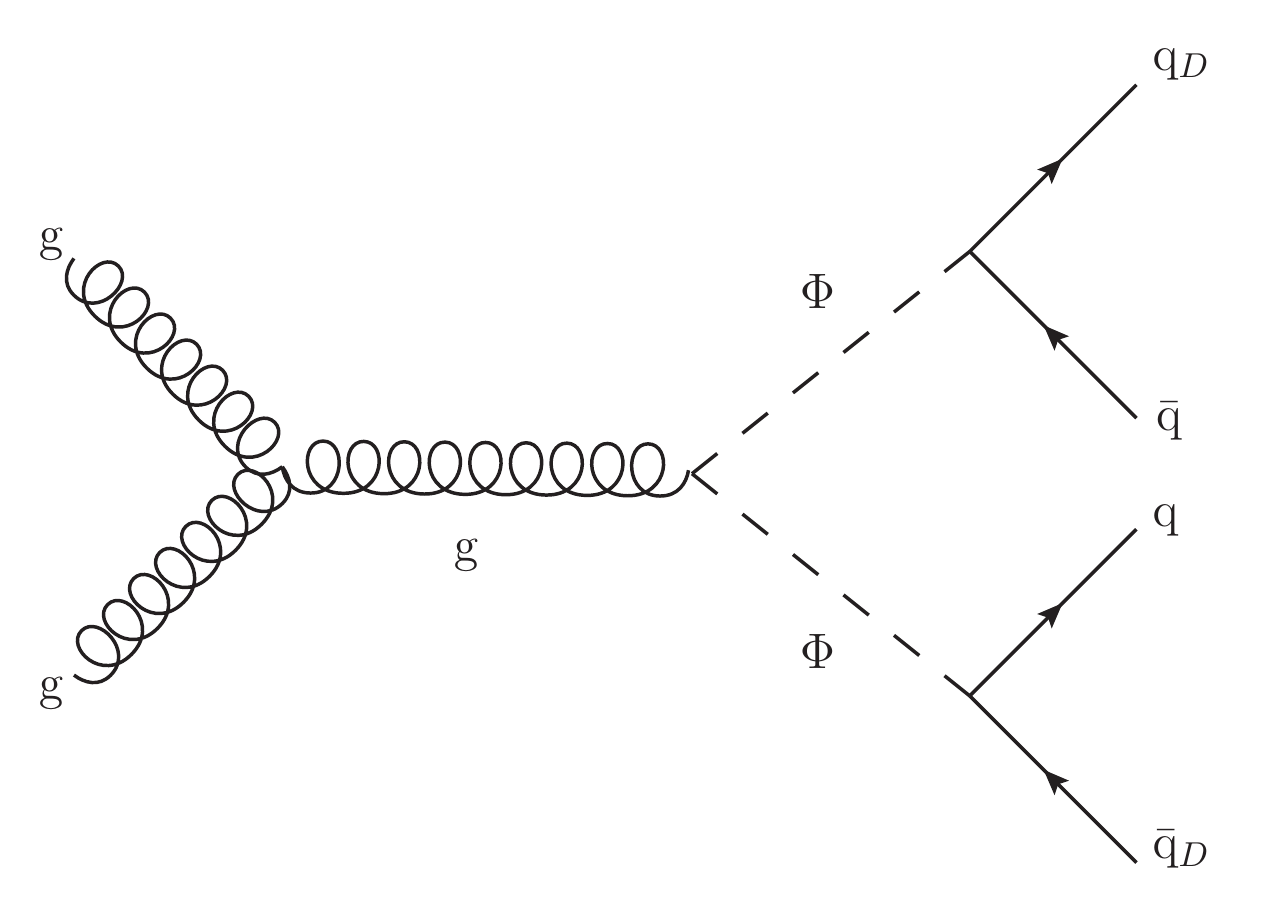}
    \includegraphics[width=0.49\linewidth]{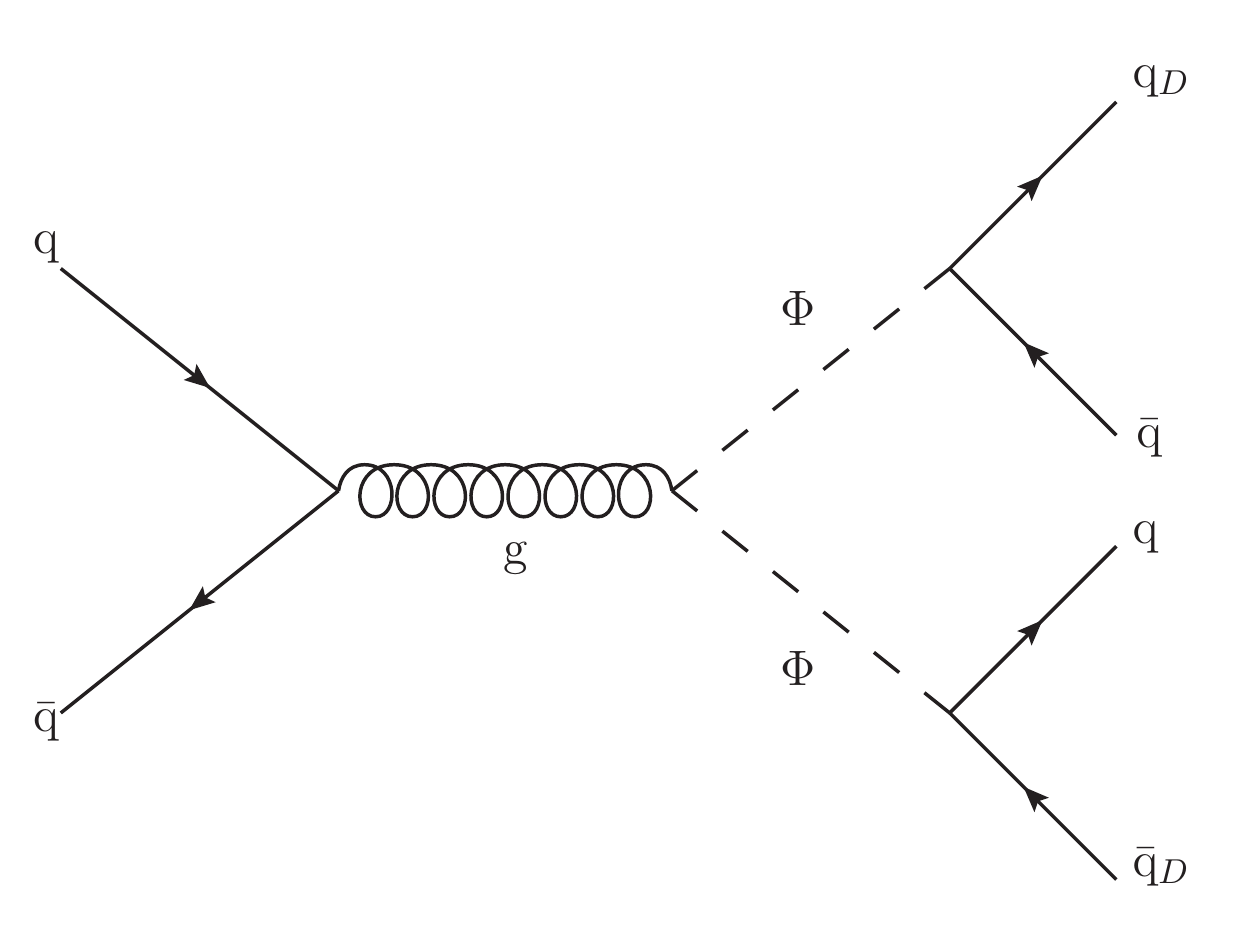}
    \caption{Diagram for the pair production of mediators and the subsequent decay to dark and SM quarks.}
    \label{fig:t-channel_production}
\end{figure}

Here, we will focus on mediator pair production. The diagrams for this process are shown in Fig.~\ref{fig:t-channel_production}. Both mediators decay subsequently to a visible and a dark quark, which undergo showering and hadronization, forming a SM and a dark jet.  Heavier dark hadrons decay promptly into the lightest dark hadron, the dark pions. Dark pions decay back into SM particles. Depending on the lifetime of the dark pions three different final states are possible:
\begin{itemize}
    \item the dark pions decay promptly and the final states consists of four prompt jets,
    \item the dark pions have intermediate lifetimes ($c\tau\sim 0.001-1$~m) and form emerging jets,
    \item the dark pions are stable on collider scales and are recorded as missing energy.
\end{itemize}
\begin{figure}[h]
    \centering
    \includegraphics[width=0.6\linewidth]{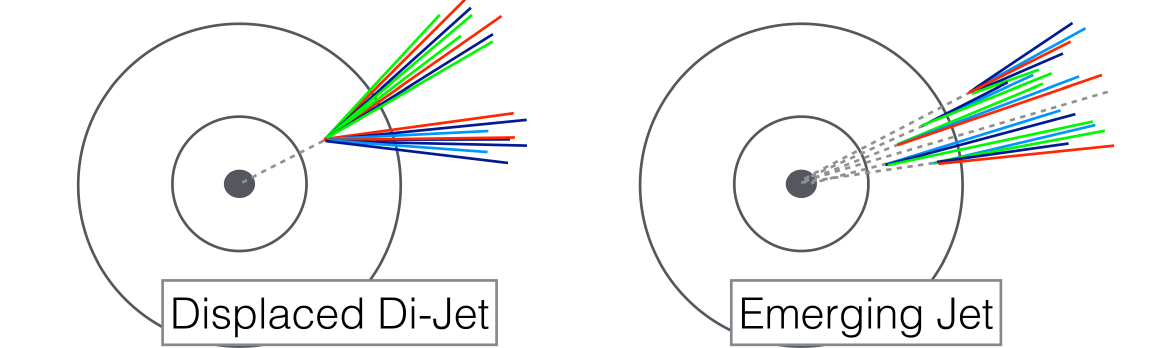}
    \caption{Emerging jet signature for t-channel dark QCD models in comparison to the displaced dijet signature. Figure taken from \cite{Schwaller:2015gea}.}
    \label{fig:emerging_jet}
\end{figure}
While in the first and last case the final states consists of typical SM objects, emerging jets provide a very distinct signature: Each pion of a dark jet will decay at a different length due to the boost of each dark pion depending on its individual momentum and the exponential distribution of the actual decay points for a given lifetime. Therefore, from a radial perspective, a dark jet deposits very little energy at the interaction point and then emerges with every dark pion decay into visible particles. The topology of an emerging jet is shown in Fig~\ref{fig:emerging_jet} in comparison to the displaced dijet signature.   

The emerging jet signature was first studied in \cite{Schwaller:2015gea} for a dark sector coupling to right-handed down-type quarks and a first search for this signature was performed with CMS \cite{CMS:2018bvr}. In \cite{Mies:2020mzw} this search has been combined with recasts of four jet and two jet plus missing energy (For more details on the recast c.f. Section \ref{sec:existing_constraints_t_channel}.). It was found that the dark pion mass does not change these bounds in a significant way. The obtained limits can be shown in the usual dark matter mass-mediator mass frame. To do so assumptions about the ratio of the dark pion and dark matter candidate, here taken as the dark proton ($p_D$), must been made.
\begin{figure}
    \centering
    \includegraphics[width=0.47\linewidth]{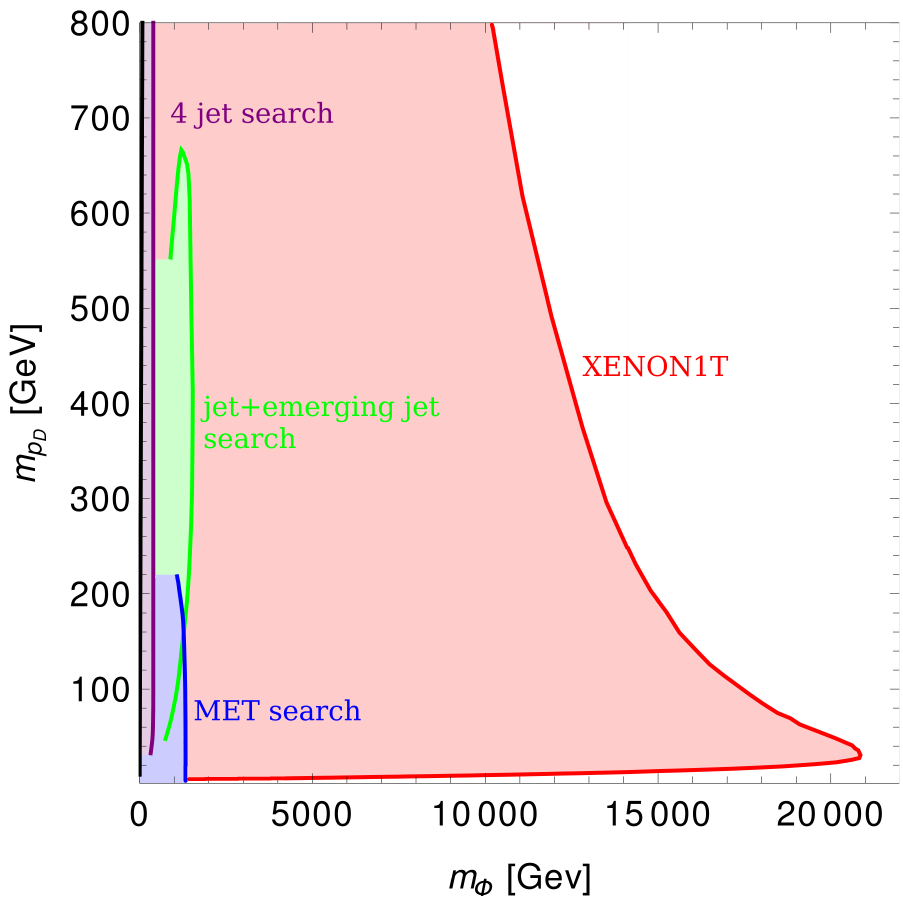}
    \quad
    \includegraphics[width=0.47\linewidth]{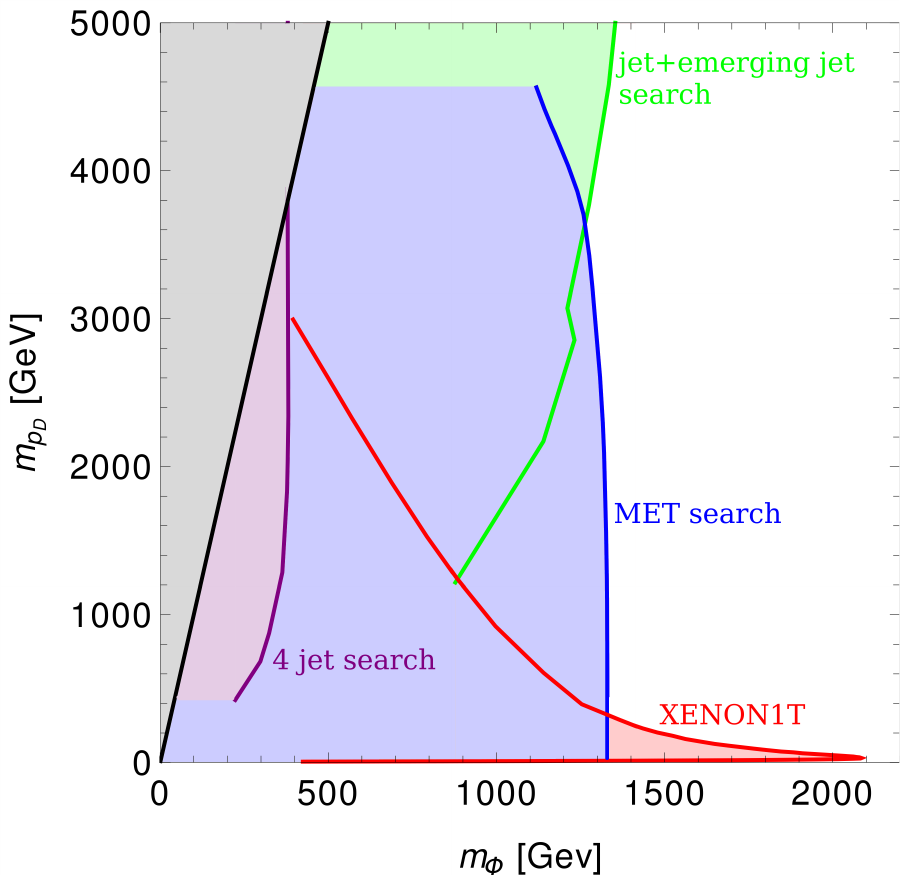}
    \caption{Bounds from direct detection and collider searches for $m_{p_D} = 10\mpidark$. Left: $\kappa = \text{diag}(1,1,1)$, right: $\kappa = \text{diag}\left(0.1,1,1\right)$. In the gray shaded region the dark matter is larger than the mediator mass. Figure taken from \cite{Mies:2020mzw}.}
    \label{fig:t-channel1}
\end{figure}
In Fig.~\ref{fig:t-channel1} the exclusion limits, combined with the constraints from direct detection experiments are shown as a function of the dark matter mass and the mediator mass for $m_{p_D} = 10\mpidark$ for two different choices of couplings: The left panel corresponds to $\kappa = \text{diag}\left(1,1,1\right)$, right to $\kappa = \text{diag}\left(0.1,1,1\right)$. This shows clearly how the coupling structure will influence the best search method.  

Due to the connection to the SM flavor structure such models also contribute to flavor processes such as neutral meson mixing and flavor violating Kaon $B$ and $D$ decays. The impact on flavor physics for dark sectors coupled to the down-type quarks has been studied in \cite{Renner:2018fhh,Agrawal:2014aoa,Blanke:2017fum}, and for couplings to up-type quarks in \cite{Blanke:2017fum,Jubb:2017rhm}, as well as models with CP violation \cite{Bensalem:2021qtj}, and for a simplified model in \cite{Carmona:2021seb}. In both cases the parameter space is largely unconstrained for dark pion masses above a few GeV. For the case of couplings to the up-type quarks this region could for example be probed via emerging jets from flavor violating top decays (c.f. \cite{Carmona:2022jid}).      

\subsection{Existing benchmarks}\label{benchmarks}
\emph{Contributors:  Elias Bernreuther, Florian Eble, Alison Elliot, Giuliano Gustavino, Simon Knapen, Benedikt Maier, Kevin Pedro, Jessie Shelton, Daniel Stolarski}
\label{sec:benchmarks}

Several attempts have been done in order to parametrize QCD-like dark sector theories in the literature. These are partly motivated by observations of parametric relationships between the confinement scale and rho and pion masses in the SM, and partly by inferred relationships between e.g. quark masses and meson masses within the \PYTHIA8 HV module. We list below some of the efforts in this context. It is important to note that these do not necessarily imply consistent UV and IR parameters of the underlying non-Abelian dynamics itself, however they have been useful in providing first phenomenological insight in the behaviour of such theories to guide the experiments. 

\subsubsection{CMS emerging jet search}\label{sec:cms-emj-benchmark}

The CMS emerging jet search~\cite{CMS:2018bvr} follows the class of models introduced in Ref.~\cite{Schwaller:2015gea}.
The specific process investigated is $\Pp\Pp\rightarrow\Pbifun\overline{\Pbifun}$, $\Pbifun\rightarrow\Pq\Paqdark$, depicted in Fig.~\ref{fig:t-channel_production}, where \Pbifun is a bifundamental scalar mediator with Yukawa couplings \cbifun between dark quarks $\PqdarkA$ and SM quarks $\Pq_i$, as shown in Eq.~\ref{eq:t-channel}.

The parameters of these models are briefly summarized here:
\begin{itemize}
\item $\Ncdark = 3$
\item $\Nfdark = 7$
\item $\lamdark = \mqdark$
\item $\mqdark = 2\mpidark$
\item $\mrhodark = 4\mpidark$
\item $\mbifun = 400\text{--}2000\,\GeV$
\item $\mpidark = 1\text{--}10\,\GeV$
\item $\ctpidark = 1\text{--}1000\unit{mm}$
\end{itemize}
The first two parameters are inspired by the dark matter model from~\cite{Bai:2013xga}. These two along with the next two are \PYTHIA parameters that control the shower but are not directly observable. The mas of \Prhodark is set such that $\Prhodark\rightarrow \Ppidark\Ppidark$ is kinematically allowed, and that decay will be dominant. The last three parameters in this list are treated as free parameters with their values varied as indicated.

The production cross section is controlled by $\mbifun$. The distinct emerging jets phenomenology is controlled by the dark pion decay length \ctpidark, treated as a free parameter encapsulating variations of both the dark pion decay constant \fpidark and the Yukawa coupling \cbifun:
\begin{equation}
\ctpidark \approx 80\unit{mm}
\left( \frac{1}{\cbifun}\right)^4
\left( \frac{2\,\GeV}{\fpidark} \right)^2
\left( \frac{100\,\MeV}{\mq} \right)^2
\left( \frac{2\,\GeV}{\mpidark} \right)
\left( \frac{\mbifun}{1\,\TeV} \right)^4,
\label{eqn:bssw}
\end{equation}
Because only pair production of \Pbifun is considered, \cbifun does not influence the production cross section.

The event generation and hadronization are done with the \PYTHIA 8.212 Hidden Valley module, modified to allow running of the dark coupling constant \adark.
Only the Yukawa couplings to down quarks are non-zero.
The dark rho mesons decay promptly to pairs of dark pions with branching fraction 0.999 or directly to pairs of down quarks with branching fraction 0.001, while the dark pions decay exclusively to pairs of down quarks with decay length \ctpidark.
The parameter \probrho, the probability of producing a vector rather than pseudoscalar meson during hadronization, is set to its default value of 0.75.
Dark baryons are expected to be stable and can act as dark matter candidates as in the model of~\cite{Bai:2013xga}, but are expected to be produced rarely compared to dark mesons (${\sim}10\%$ for SM QCD~\cite{Schwaller:2015gea}), so their presence is not simulated.

\subsubsection{Flavored emerging jet model}\label{sec:cms-emj-flavored}

The class of models described in Section~\ref{sec:cms-emj-benchmark} was extended to the case where the dark sector has a flavor structure related to SM QCD~\cite{Renner:2018fhh}.
The result is multiple scenarios in which different dark mesons have different lifetimes, varying over a wide range of values.
In particular, the following parameters are considered:
\begin{itemize}
\item $\Ncdark = 3$
\item $\Nfdark = 3$
\item $\lamdark > \mqdark$
\end{itemize}
The last condition implies that heavier dark mesons decay promptly to lighter dark mesons, so only the latter influence the final state kinematic behavior.
The spectrum of lighter dark mesons can be understood in analogy to SM pions and kaons: \PpidarkZ ($\PqdarkA\PaqdarkA$), \PpidarkPM ($\PqdarkO\PaqdarkT$, $\PqdarkT\PaqdarkO$), \PkdarkZ ($\PqdarkT\PaqdarkTh$, $\PqdarkTh\PaqdarkT$), \PkdarkPM ($\PqdarkO\PaqdarkTh$, $\PqdarkTh\PaqdarkO$); the quotation marks indicate that the superscripts do not represent actual charges.
The mass splittings between these different dark meson species are taken to be negligible.

In the simplest version of this flavored model, called the ``aligned'' scenario, there is no mixing of neutral dark mesons.
This scenario can be generated using a custom modification of \PYTHIA 8.230\footnote{\url{https://github.com/kpedro88/pythia8/tree/emg/230}} that gives the correct proportions of the dark meson species listed above.
The dark pion and kaon decay widths can be calculated as follows, for dark quark content $\PaqdarkA\PqdarkB$ and SM quark products $\Paqi\Pqj$:
\begin{equation}
\begin{split}
\Gamma_{\alpha\beta ij}=&\frac{\Ncdark \mpidark \fpidark^{2}}{8\pi \mbifun^{4}}\left\lvert\cbifun\cbifunbeta\right\rvert^2 \\
&\left(\mqi^{2}+\mqj^{2}\right) \sqrt{\left(1-\frac{(\mqi+\mqj)^2}{\mpidark^{2}}\right)\left(1-\frac{(\mqi-\mqj)^2}{\mpidark^{2}}\right)}
\label{eq:ctau_aligned}
\end{split}
\end{equation}

\subsubsection{CMS semi-visible jet search}\label{sec:cms-svj-benchmark}

The CMS search for semi-visible jets~\cite{CMS:2021dzg} is based on the class of models introduced in Refs.~\cite{Cohen:2015toa,Cohen:2017pzm}.
The specific process investigated is $\Pp\Pp\rightarrow\PZprime\rightarrow\Pqdark\Paqdark$,
where \PZprime is a leptophobic vector mediator with couplings to SM quarks \gq and couplings to dark quarks \gqdark, as shown in Eq.~\ref{eq:Zprime}.

The parameters of the models used in the CMS search are summarized below:

\begin{itemize}
\item $\Ncdark = 2$
\item $\Nfdark = 2$
\item $\mqdark = \mpidark/2$
\item $\mZprime = 1500\text{--}5100\,\GeV$
\item $\mpidark = \mrhodark = 1\text{--}100\,\GeV$
\item $\rinv = 0.0\text{--}1.0$ (see Eq.~\ref{eq:rinv})
\item $\adark = \adarkLow\text{--}\adarkHigh$
\end{itemize}
The last four parameters in this list are treated as free parameters with their values varied as indicated.

\begin{sloppypar} The unstable dark pions, as pseudoscalars, decay via a mass insertion preferentially to the most massive allowed SM quark species (bottom quarks unless ${\mpidark < 2\mb}$).
The branching fractions for the mass insertion decays are calculated with quark mass running included.
The unstable dark rho mesons, as vectors, decay democratically to pairs of any allowed SM quark.
All decays to SM quarks are assumed to be prompt.\end{sloppypar}

\begin{figure*}[htb!]
\centering
\includegraphics[width=0.49\linewidth]{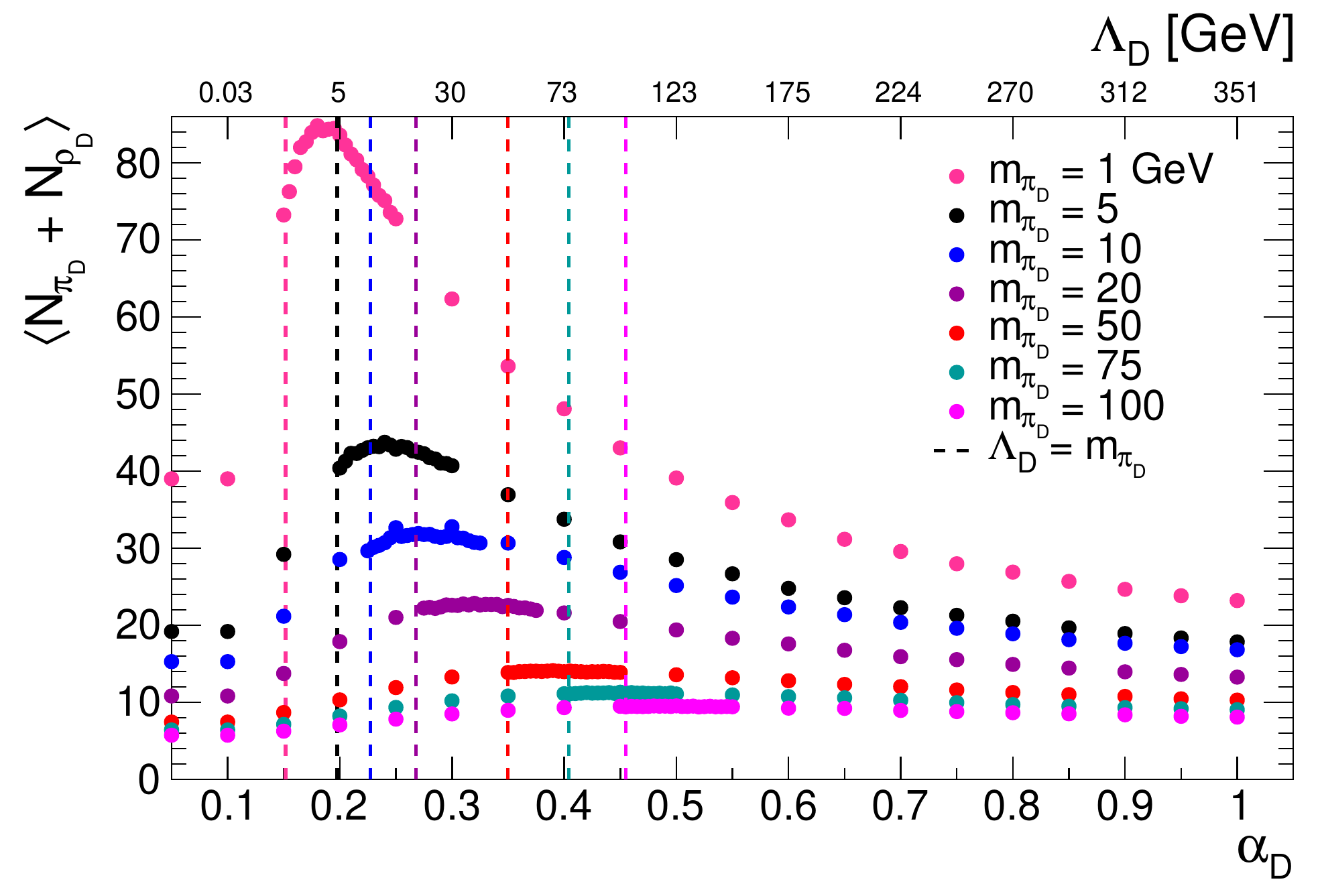}
\includegraphics[width=0.49\linewidth]{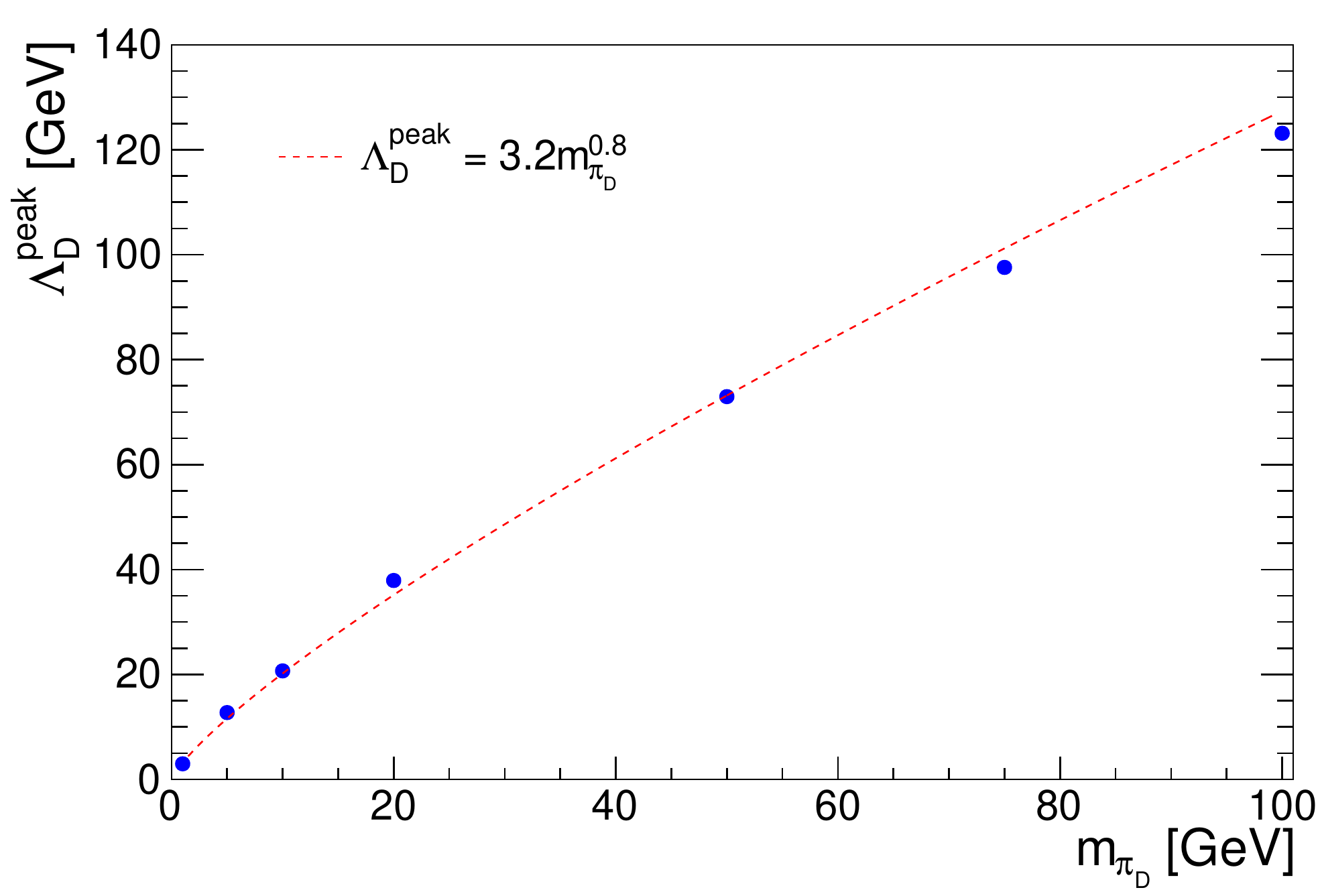}\\
\includegraphics[width=0.49\linewidth]{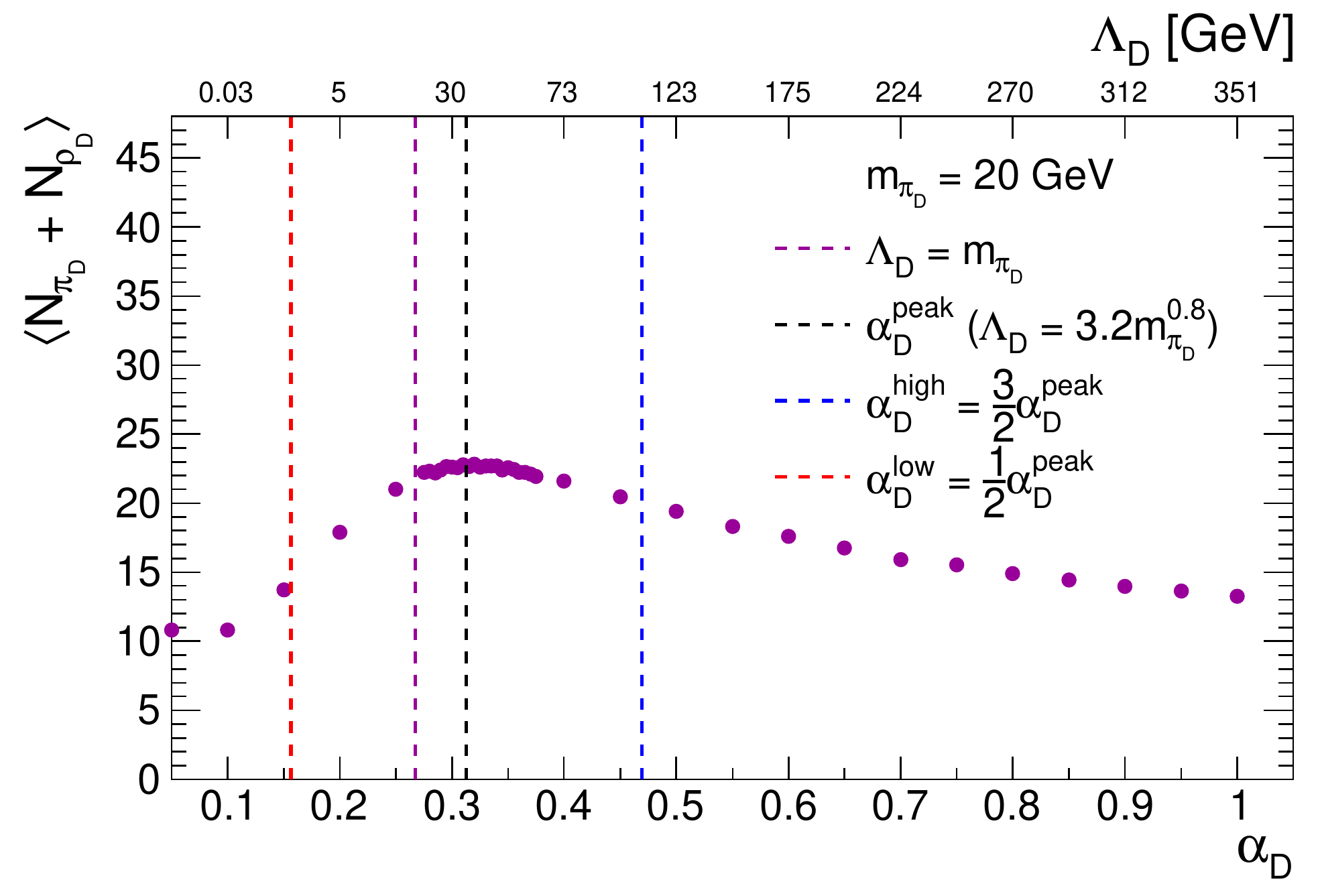}
\caption{Top left: the average number of dark hadrons, both stable and unstable, created per \PZprime event by \PYTHIA 8.230 for different \mpidark and \lamdark or \adark values.
The behavior for ${\lamdark < \mqdark}$ may not be physically accurate.
Top right: For each \mpidark value, the corresponding \lamdark value that maximizes the number of dark hadrons in the shower, \lamdarkPeak, is identified from the top left plot.
The relationship between \mpidark and \lamdarkPeak is represented with a power law.
Bottom: A comparison of different \adark variations for a representative dark hadron mass $\mpidark = 20\,\GeV$.
}
\label{fig:adarkpeak}
\end{figure*}

The treatment of the dark coupling constant \adark, or equivalently the dark scale \lamdark, follows a relationship that is derived based on the behavior shown in Fig.~\ref{fig:adarkpeak}.
The former can be expressed in terms of the latter: $\adark(\lamdark) = \pi/\left(b_{0}\log\left(\Qdark/\lamdark\right)\right)$,
with the factor $b_{0} = (11\Ncdark - 2\Nfdark)/6$ and $\Qdark = 1\,\TeV$.
The benchmark value for the scale is chosen according to the empirical fit ${\lamdarkPeak = 3.2(\mpidark)^{0.8}}$.
This maximizes the production of dark hadrons in the dark showers, because the effect of \lamdark depends on the dark hadron mass \mpidark.
The corresponding value \adarkPeak is then varied by ${\pm}50\%$ to give \adarkLow, \adarkHigh.
It is important to note that the results generated from \PYTHIA for ${\lamdark < \mqdark}$ may not be physically accurate. This is discussed further in Section~\ref{parameters}.

The event generation and hadronization are done with the \PYTHIA 8.230 Hidden Valley module.
The invisible fraction \rinv is implemented by reducing the branching fractions of dark hadrons to SM quarks, so the ``decay'' of a dark hadron to invisible particles represents a stable dark hadron.
A $\mathbb{Z}_{2}$ symmetry filter is enforced to reject events with an odd number of stable dark hadrons, as they would always be produced in pairs in a complete model.
The exact \PYTHIA settings used in Ref.~\cite{CMS:2018bvr} can be found on HEPData~\cite{hepdata:2021dzg}.

\subsubsection{Further semi-visible jet models under study by CMS}


The published CMS searches described in Sections~\ref{sec:cms-emj-benchmark},~\ref{sec:cms-svj-benchmark} use \PYTHIA for both event generation and hadronization. The processes that the \PYTHIA Hidden Valley module can generate are those shown in Figs.~\ref{fig:s-channel_production} and~\ref{fig:t-channel_production}.

\begin{figure*}[htb!]
\centering
\includegraphics[width=0.31\linewidth]{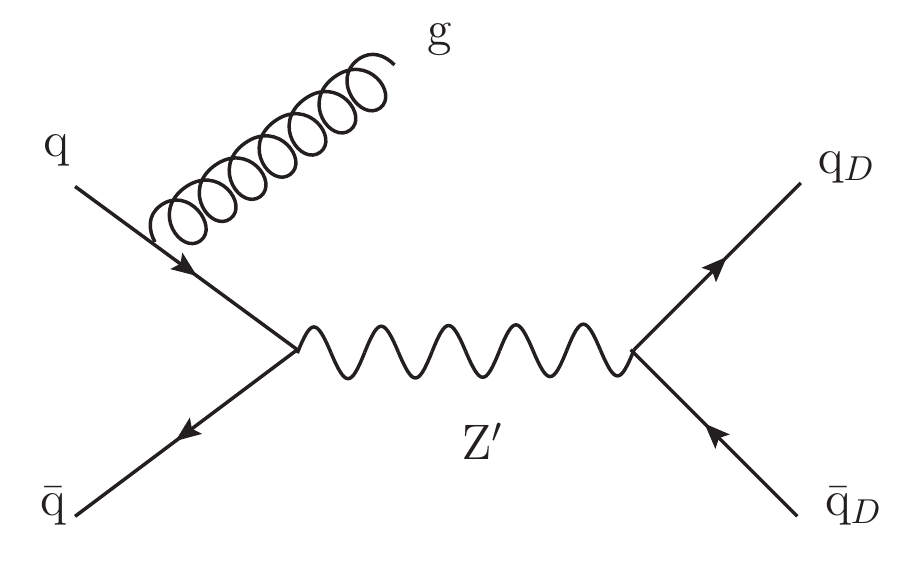}
\includegraphics[width=0.35\linewidth]{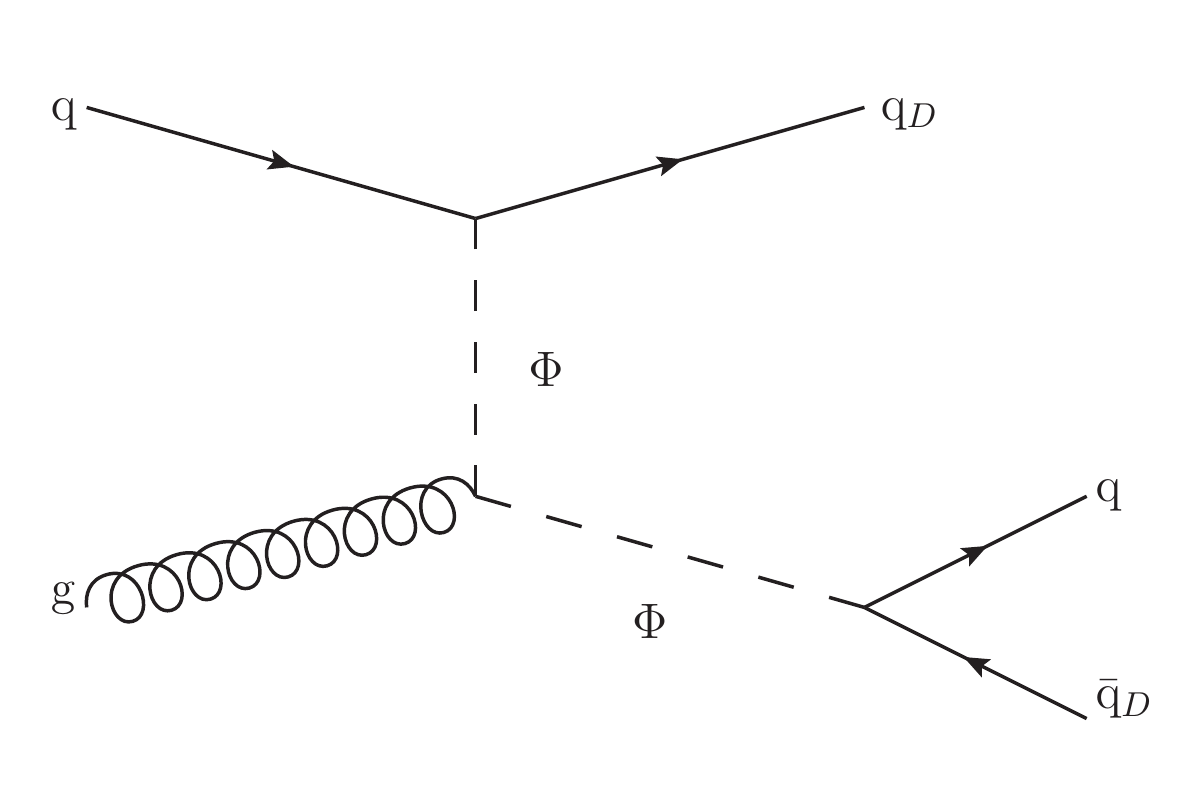}
\includegraphics[width=0.32\linewidth]{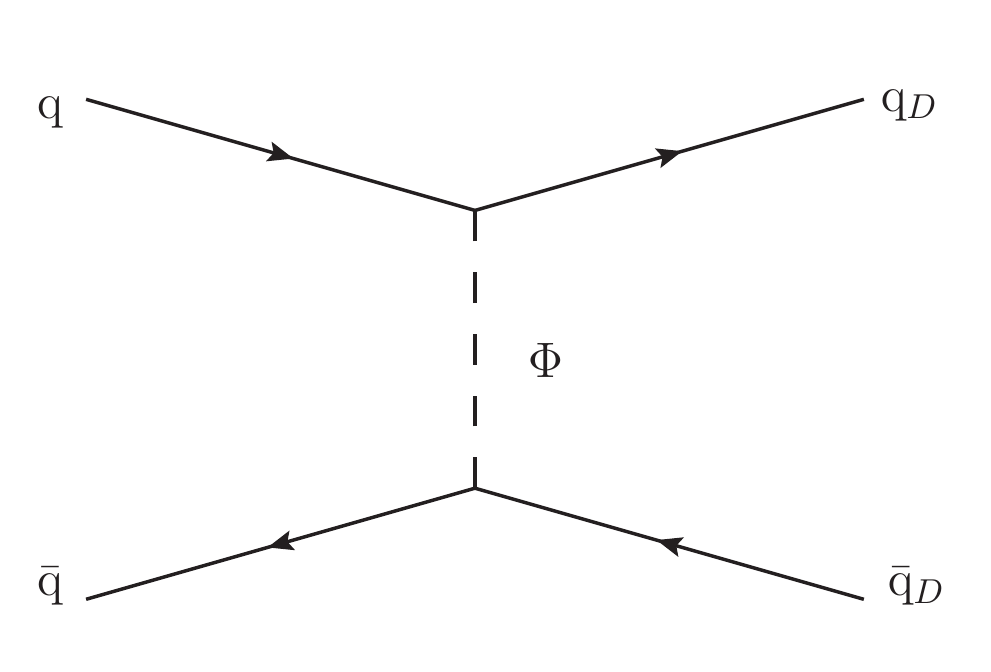}
\caption{Processes available via \FEYNRULES in \MADGRAPH: $\Pp\Pp\rightarrow\Pg\PZprime\rightarrow\Pg\Pqdark\Paqdark$ (boosted), $\Pq\Pg\rightarrow\Pqdark\Pbifun\rightarrow\Pqdark\Pq\Paqdark$, $\Pq\Paq\rightarrow\Pqdark\Paqdark$ (non-resonant).
}
\label{fig:madgraph_processes}
\end{figure*}

Studies are in progress to extend the CMS semi-visible jet program to low-mass boosted resonances and $t$-channel production via the bifundamental scalar mediator \Pbifun.
To generate these processes, depicted in Fig.~\ref{fig:madgraph_processes}, \MADGRAPH 2.6.5 is used.
(\MADGRAPH can also generate the processes in Figs.~\ref{fig:s-channel_production} and ~\ref{fig:t-channel_production}.)
The \FEYNRULES definitions are obtained from Ref.~\cite{smsharma:SVJ}, associated with Ref.~\cite{Cohen:2017pzm}.
The \PYTHIA Hidden Valley module is still used for hadronization.
To obtain accurate results, several additional steps are needed:
\begin{itemize}
\item Ensure that mediator decay widths are properly computed (\texttt{set param\_card decay [id] auto})
\item Increase the number of events when making gridpacks from 2000 to 10000 (to overcome instability in $t$-channel phase space integration)
\item convert PDG IDs to \PYTHIA conventions in LHE output
\end{itemize}

Both \MADGRAPH and \PYTHIA have some limitations for handling dark shower generation and hadronization.
For \PYTHIA, some useful future additions would be:
\begin{itemize}
\item Add processes like $\Pq\Pq\rightarrow\Pqdark\Pbifun\rightarrow\Pqdark\Pq\Paqdark$, $\Pg\Pg\rightarrow\Pqdark\Paqdark$ (non-resonant) that are currently only available via \FEYNRULES, to facilitate generator-level studies.
\item Include more theory uncertainties as event weights: PDF variations, renormalization and factorization scale variations, and Hidden Valley-specific parameters such as those that control the hadronization models.
\item Allow user control over the dark hadron spectrum for studies of flavored models (Section~\ref{sec:cms-emj-flavored}, Ref.~\cite{Renner:2018fhh}).
\item Add dark baryons for completeness (currently only dark mesons are explicitly produced).
\end{itemize}
Updates to the \PYTHIA Hidden Valley module made in the context of this Snowmass project and described in Section~\ref{newpythia}, are intended to address the last two points.
For \MADGRAPH, some useful updates and improvements would include:
\begin{itemize}
\item Better fixes for the items mentioned in the previous paragraph.
\item Central support for common processes such as $t$-channel production, to reduce the need for manual \FEYNRULES implementations.
\item Ability to add new $SU(N)$ gauge groups in a complete and consistent way, in order to model Hidden Valley radiation explicitly. A Hidden Valley jet matching procedure would also be needed to avoid overlap with radiation from \PYTHIA during the hadronization process.
\end{itemize}

\subsubsection{Semi-visible jet models under study by ATLAS} \label{sec:atlas-svj-benchmark}

The semi-visible jet models under study are based on those introduced in Refs.~\cite{Cohen:2015toa,Cohen:2017pzm}; the fully visible jet models described in \cite{Park:2017rfb} are also considered. The \PZprime and \Pbifun portals introduced in sections \ref{schannel} and \ref{tchannel} are considered, with the following parameter settings:

\begin{itemize}
\item $\Ncdark = 2$
\item $\Nfdark = 2$
\item $\mqdark = \mpidark/2$
\item $\mpidark = \mrhodark = 20\,\GeV$
\item $\mZprime / \mbifun = 750\text{--}5000\,\GeV$
\item $\rinv = 0.0\text{--}1.0$
\end{itemize}

The $\rinv$ and $\mZprime$ parameters are varied through the values as displayed in the above list. The $\adark$ coupling is chosen to be a running constant as it is in SM QCD.

The generated models differ between $s$-channel and $t$-channel in terms of the jet matching applied. For the $t$-channel, an MLM matching scheme is used, while for the $s$-channel, a CKKM-L matching has been used. 

The semi-visible jet models are generated in \MADGRAPH, and \PYTHIA~is used for showering. Fo the $s$-channel, \MADGRAPH~2.9.3 and \PYTHIA~8.245 are used. For the $t$-channel, \MADGRAPH~2.8.1 and \PYTHIA~8.244 are used. The fully visible jet models are entirely generated with \PYTHIA~8.

\subsubsection{Aachen model\label{aachen}}

This model was introduced in Ref.~\cite{Bernreuther:2019pfb} and subsequently used in Refs.~\cite{Bernreuther:2020vhm,Buss:2022lxw}.
It is designed to satisfy cosmological constraints and reproduce the observed dark matter relic abundance.
The dark sector is connected to the SM by a heavy vector mediator \PZprime arising from a new \Uonedark gauge group.

The baseline parameters of this model are summarized below:
\begin{itemize}
\item $\Ncdark = 3$
\item $\Nfdark = 2$
\item $\lamdark = \mrhodark$
\item $\mqdark = 0.5\,\GeV$
\item $\mZprime = 1\,\TeV$
\item $\mpidark = 4\,\GeV$
\item $\mrhodark = 5\,\GeV$
\item $\rinv = 0.75$
\end{itemize}
The particle masses $\mqdark$, $\mZprime$, $\mpidark$, and $\mrhodark$ are treated as free parameters that are set to the benchmark values indicated above.
In the benchmark model, only the $\Prhodark^{0}$ mesons decay (where 0 indicates the \Uonedark charge), while all other dark mesons are stable on the scale of the detector.
Dark baryons and other dark bound states, such as dark eta and dark omega mesons, are assumed to be too heavy to be produced frequently.
The study in Ref.~\cite{Bernreuther:2020vhm} includes variations $\mpidark = \mrhodark = 5\text{--}20\,\GeV$ and $\rinv = 0.1\text{--}0.9$.

\MADGRAPH 2.6.4 with \FEYNRULES 2.3.13 is used to generate the studied events, with \PYTHIA 8.240 used for hadronization.
For the benchmark case with $\rinv = 0.75$, the \PYTHIA Hidden Valley default settings are modified to $\probrho = 0.5$ in order to obtain the correct proportion of unstable dark mesons.
When \rinv is varied, it is implemented in the same manner described in Section~\ref{sec:cms-svj-benchmark}.
In this case, all unstable dark hadrons are assumed to decay democratically to pairs of \Pqu, \Pqd, \Pqs, or \Pqc quarks.
(The assumption that only $\Prhodark^{0}$ mesons decay is relaxed.)

\subsubsection{Decay portals}

Hidden valley models differ from most other models in two crucial aspects. The first is the lack of a very well-defined theory prior on what the model should look like, as already mentioned earlier in this document. 
The second complication is that, even if we somehow had a strongly preferred model, it would still be very challenging to extract accurate predictions for all observables due to the non-perturbative dynamics in the dark sector.

Both these challenges indicate that we may be best served by a suite of searches that is as model-independent as possible. This is easier said than done however, and some theory priors are always needed to design an experimental analysis, especially in the initial phases of this program. Ref.~\cite{Knapen:2021eip} has advocated to inject these theory priors in the decay portals that allow the dark sector to decay back to the Standard Model. This approach has the following advantages:
\begin{enumerate}
    \item The number of plausible options is relatively limited once one restricts to the set of decay portals that do not introduce dangerous flavor-changing neutral currents. A systematic survey is therefore very feasible. 
    \item We have good theoretical control over these decay portals, in terms of both the dark particle's decay length and its allowed branching ratios. This is in contrast to the process of dark sector hadronization, where we must resort to parameterizing our ignorance as best we can.
    \item If one moreover insists on a relatively minimal UV completion and/or no more than moderate fine tuning, one can moreover derive an approximate lower bound on the lifetime of the dark mesons. This gives the models a little bit more predictive power. The resulting lifetimes and branching ratios of the visibly-decaying dark particle are crucial for the design of experimental search strategies and a systematic survey of models featuring a minimal suite of decay portals can therefore be a good starting point for a comprehensive experimental search program. 
\end{enumerate}

\begin{figure}
\begin{floatrow}

\capbtabbox{%
  \begin{tabular}{p{2.5cm}p{4.cm}}
Decay portal&decay operator   \\\hline\hline
gluon &$\Petadark G^{\mu\nu}\tilde G_{\mu\nu}$ \\
photon &$\Petadark F^{\mu\nu}\tilde F_{\mu\nu}$ \\
vector & $\Pomegadark^{\mu\nu}F_{\mu\nu}$ \\
Higgs &$\Petadark H^\dagger H$  \\
dark photon &$\Petadark F^{\prime\mu\nu}\tilde F^{\prime}_{\mu\nu} + \epsilon F^{\prime\mu\nu} F_{\mu\nu}$   \\
\end{tabular}
}{%
  \caption{Overview of the decay portals considered in~\cite{Knapen:2021eip}. The \Petadark and \Pomegadark represent respectively the lightest spin-zero and spin-one meson in the dark sector, and $F'_{\mu\nu}$ is the field strength for an elementary dark photon  \Pphodark. The decay portal column indicates the operator(s) that allow the unstable dark meson to decay, defining the model. \label{tab:portals}}%
}

\ffigbox{%
  \includegraphics[width=0.50\textwidth]{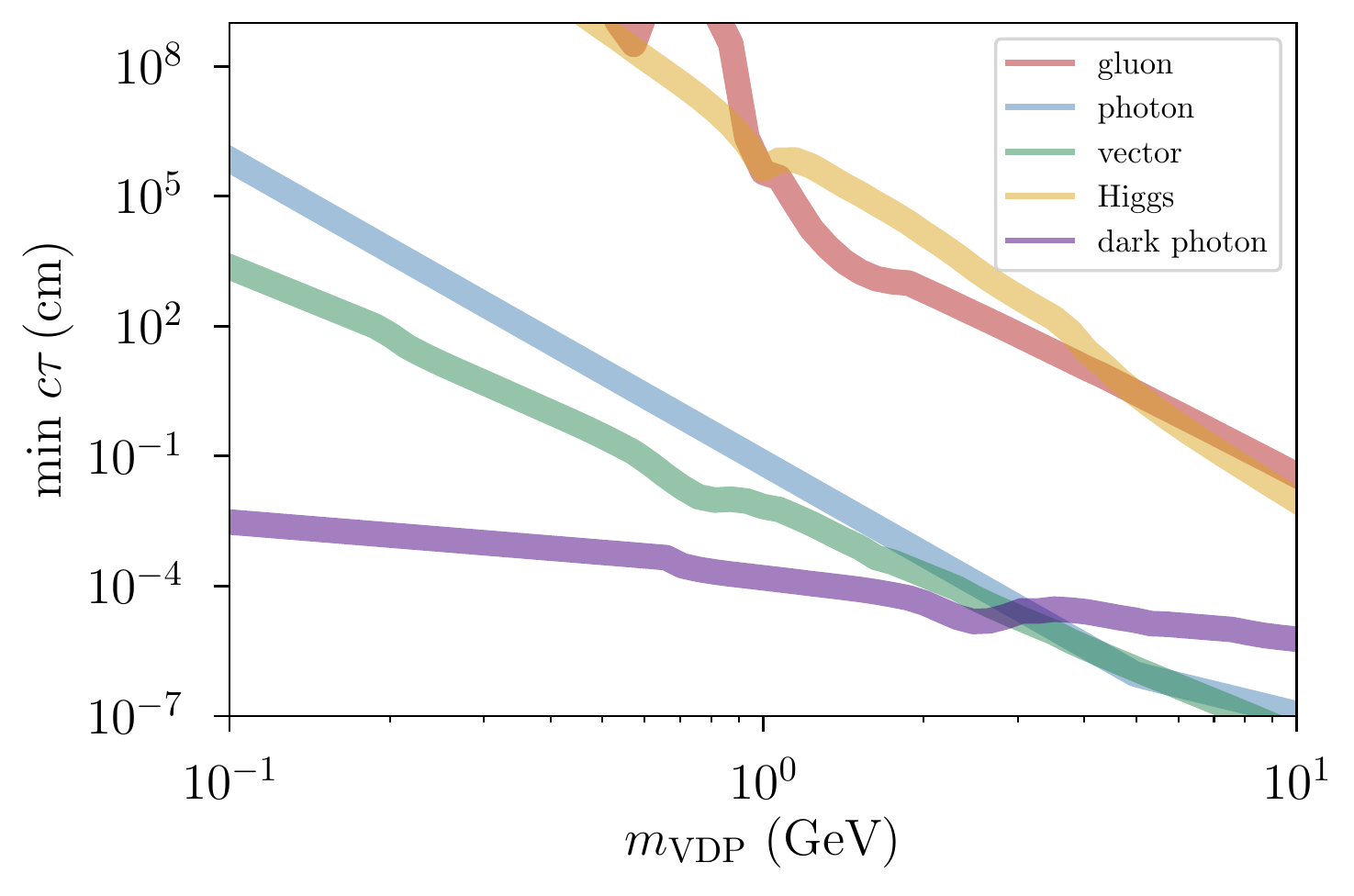}%
}{%
  \caption{Approximate lower bound on the proper lifetime of the visibly decaying particle (VDP) in the dark sector, for the decay portals in Table~\ref{tab:portals}. \label{fig:lifetime}}%
}
\end{floatrow}
\end{figure}

The portals considered in Ref.~\cite{Knapen:2021eip} are summarized in Table~\ref{tab:portals}. To maintain compatibility with the \PYTHIA Hidden Valley module at the time, a simplified dark sector was considered with a single (pseudo)scalar \Petadark and vector \Pomegadark meson. For the dark photon portal an additional elementary dark photon \Pphodark was added. For each of these portals the branching ratios and lower bound on the lifetime of the visibly decaying particle were computed (see Fig.~\ref{fig:lifetime}).

For the \textbf{gluon portal} one assumes that the dark sector pseudoscalar meson (\Petadark) decays through a dimension 5 coupling to the Standard Model gluons, leading to very \textbf{hadron-rich} final states. This coupling requires a low-scale UV completion, especially because \Petadark itself is a composite state and the $\Petadark G\tilde G$ coupling should be suppressed by dimensional transmutation in models with perturbative parton showers. Such a UV completion moreover requires the presence of new, colored particles, which could be produced at the LHC. The bound in Fig.~\ref{fig:lifetime} was obtained by assuming that any such states must have a mass $\gtrsim 2\,\TeV$. This assumption can be relaxed by devising an elaborate extension of the model to hide the new colored particles from existing searches at ATLAS and CMS.

The \textbf{photon portal} works in the same way as the gluon portal, except that the \Petadark couples to the Standard Model photons instead of the gluons. This portal leads to very \textbf{photon-rich} final states. Its lifetime bound is informed by collider constraints on new charged particles, which are required in the UV completion of this portal.

The \textbf{vector portal} assumes that the Standard Model photon mixes with the dark vector meson \Pomegadark, similar to the photon-$\rho$ mixing in the Standard Model. In this portal it was assumed that the \Petadark was absolutely stable, leading to a \textbf{semi-visible jet} phenomenology. The UV completion of this portal requires the introduction of an additional, elementary \PZprime vector field. The lifetime bound in Fig.~\ref{fig:lifetime} was informed by the bounds on the mass and coupling of the \PZprime from direct searches and electroweak precision observables.

In the \textbf{Higgs portal} scenario, the \Petadark decays by mixing with the Standard Model Higgs, leading to a \textbf{heavy flavor-rich} phenomenology. In this scenario, no new particles are needed in the UV completion. The $H^\dagger H$ operator does however contribute directly to the masses of the constituents of the \Petadark mesons, which leads to the lower bound in Fig.~\ref{fig:lifetime}. This bound can be evaded by fine-tuning the masses of the consituents of the \Petadark.

Finally, the \textbf{dark photon portal} is inspired by the Standard Model $\eta\to\gamma\gamma$ decay, as it assumes a light, elementary vector field (\Pphodark) which couples to the \Petadark through a chiral anomaly. The \Pphodark itself can then mix with the Standard Model photon, resulting fairly \textbf{lepton-rich} final states. With this portal, the lifetime constraint is very mild and comes exclusively from direct searches for the \Pphodark itself.

All assumptions and branching ratios are encoded in the public python script 
    \url{https://gitlab.com/simonknapen/dark_showers_tool}
which can generate \PYTHIA configuration cards for all five decay portals.
\subsection{Existing constraints}\label{constraints}
\emph{Contributors: Giuliano Gustavino, Steven Lowette,  Kevin Pedro, Pedro Schwaller,  Andrii Usachov, Carlos Vázquez Sierra}
 
The discussion so far has not only established that QCD-like dark sectors are theoretically interesting, but that they can also lead to exotic signatures for the experiments. In light of this observation, several searches at the LHC have already been carried out, with many other studies also underway. The search results are currently reported using one or more of the benchmarks discussed in ~\autoref{sec:benchmarks}. In this Section, we illustrate some of the public search results, and existing constraints on the theory landscape.  
Given that the searches rely on generic signal characteristics, it may be possible to reinterpret the results in terms of other, UV/IR coherent models.

\subsubsection{ATLAS search program}

While no direct constraints have been published so far by the ATLAS Collaboration on these models, a broad set of semi-visible jet scenarios in both $t$- and $s$-channel production processes are being studied by the collaboration as mentioned in the previous section. Besides possible dedicated searches, the recasting of previously published results in other channels might prove useful in constraining the parameter space. Indeed, existing exclusion limits obtained in analyses looking at di-jet~\cite{ATLAS:2019fgd} and \met+jet~\cite{ATLAS:2021kxv} final states should already be able to constrain a phase-space predicted by dark QCD models in different \rinv ranges. Furthermore, some studies also focus on scenarios with emerging jets, where dark hadrons have a non-negligible lifetime.

\subsubsection{CMS search for emerging jets}\label{sec:cms-emj-result}

The CMS emerging jet search~\cite{CMS:2018bvr} follows the class of models introduced in Ref.~\cite{Schwaller:2015gea}, as described in Section~\ref{sec:cms-emj-benchmark}.
The search considers the following parameter variations: $\mbifun = 400\text{--}2000\,\GeV$, $\mpidark = 1\text{--}10\,\GeV$, $\ctpidark = 1\text{--}1000\unit{mm}$.

The search requires four high-\pt jets and triggers on such events using  the scalar sum of their momenta, \HT.
Per-jet quantities indicating displaced tracks---including the 2D impact parameter, the 3D impact parameter significance, and the fraction of the \pt from prompt tracks associated with the primary vertex---are used to identify or ``tag'' jets as emerging.
The signal region definition requires either two jets tagged as emerging, or one jet tagged as emerging along with substantial missing transverse momentum.
The latter option increases sensitivity to models with larger \ctpidark values where many dark hadrons decay outside of the detector.
The misidentification rate for this tagging procedure is measured and used to estimate the QCD multijet background.
Heavy flavor jets from B hadrons are found to be misidentified more frequently, as expected because of their non-negligible decay lengths.
Multiple signal regions with different selection requirements are defined, and for each model, the signal region with the highest expected sensitivity is used.

\begin{figure*}[htb!]
\centering
\includegraphics[width=0.49\linewidth]{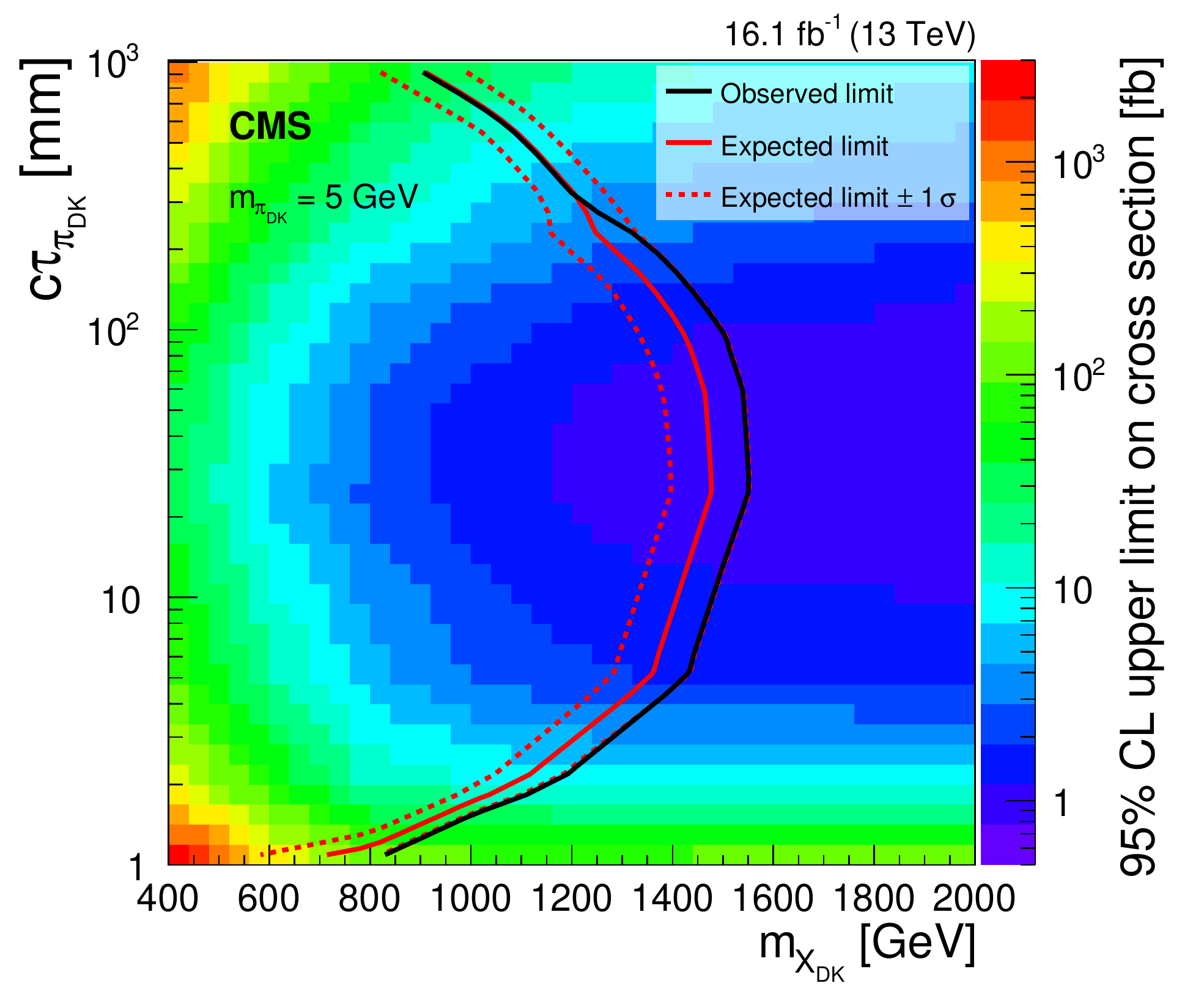}
\caption{95\% CL limits on the cross section for pair production of \Pbifun leading to emerging jets, in the plane of \ctpidark versus \mbifun (here called $\text{c}\tau_{\pi_\text{DK}}$ and $\mathrm{m_{X_{DK}}}$) where \mpidark (here called $m_{\pi_\text{DK}}$) is set to 5 GeV. Reproduced from Ref.~\cite{CMS:2018bvr}.
}
\label{fig:cms_emj_limits}
\end{figure*}

The search results are shown in Fig.~\ref{fig:cms_emj_limits}.
Using a $13\,\TeV$ dataset with $16.1\,\ifb$ of integrated luminosity, no significant excess above the SM prediction is observed.
This search excludes models with $400 < \mbifun < 1250\,\GeV$ for $5 < \ctpidark < 255\unit{mm}$ at 95\% confidence level (CL).
It is found that the limits are similar over the range of \mpidark values explored.
Work is ongoing to incorporate the remainder of the LHC Run 2 dataset, up to $138\,\ifb$, and to improve the sensitivity to other models, such as Ref.~\cite{Renner:2018fhh}, which is summarized in Section~\ref{sec:cms-emj-flavored}.

\subsubsection{CMS search for semi-visible jets}

The CMS search for semi-visible jets~\cite{CMS:2021dzg} is based on the class of models introduced in Refs.~\cite{Cohen:2015toa,Cohen:2017pzm}, as described in Section~\ref{sec:cms-svj-benchmark}.

The search requires two high-\pt wide jets and triggers on the jet \pt and the \HT.
This dijet system is combined with the missing transverse momentum to compute the transverse mass \MT,
which has a falling spectrum for the SM backgrounds, while the signal has a kinematic edge at the \PZprime mediator mass.
The QCD multijet background is rejected by requiring high values of the transverse ratio $\RT = \met/\MT$,
while the electroweak backgrounds (\ttbar, \wlnjets, \znnjets) are rejected by vetoing identified and isolated leptons (\Pge, \Pgm) and requiring a small minimum angle between the jets and the \met.
Various sources of instrumental background and misreconstruction, which introduce artificial \met, are also rejected.
Two signal regions are defined in terms of the \RT variable to provide additional sensitivity.

\begin{figure*}[htb!]
\centering
\includegraphics[width=0.49\linewidth]{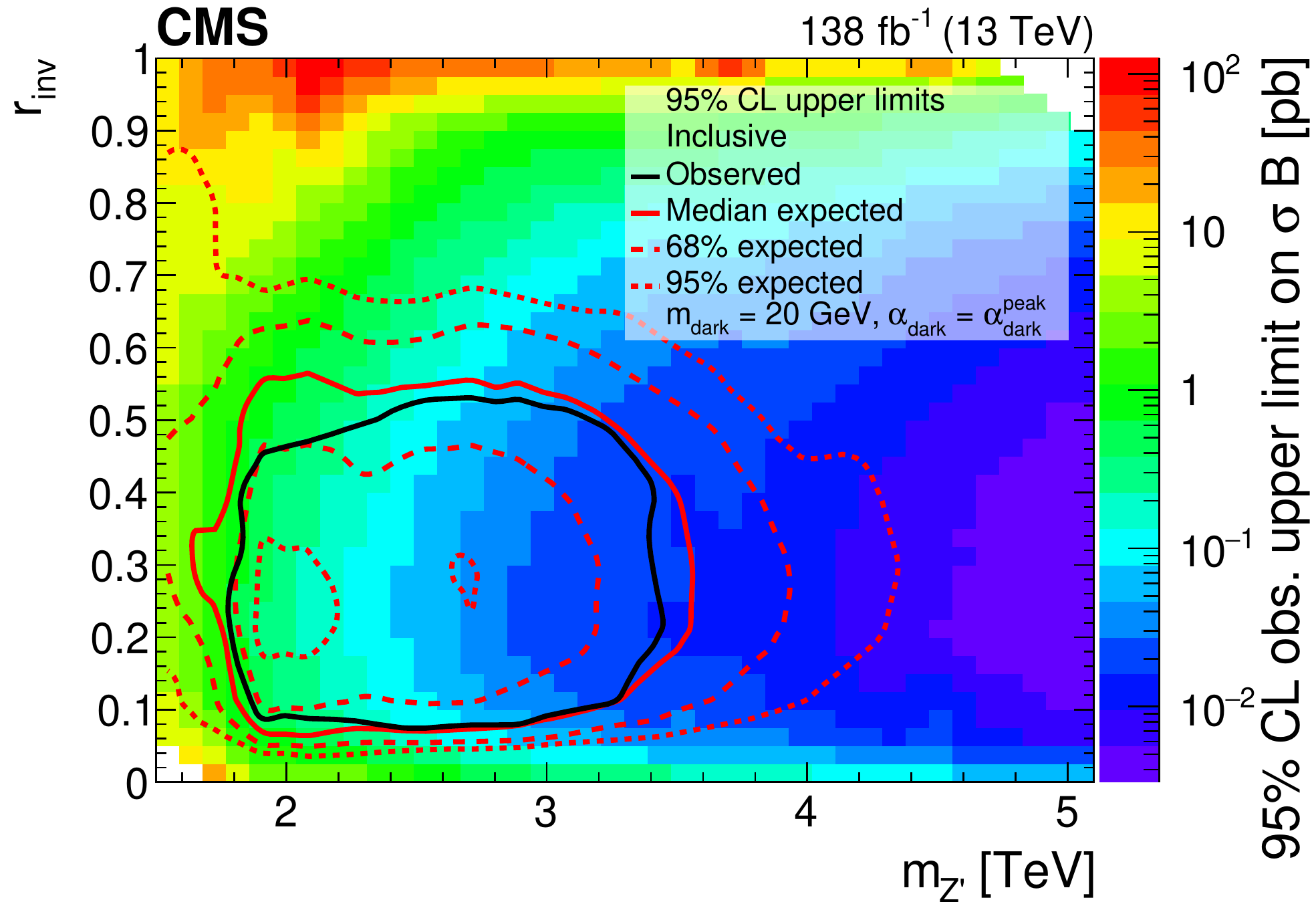}
\includegraphics[width=0.49\linewidth]{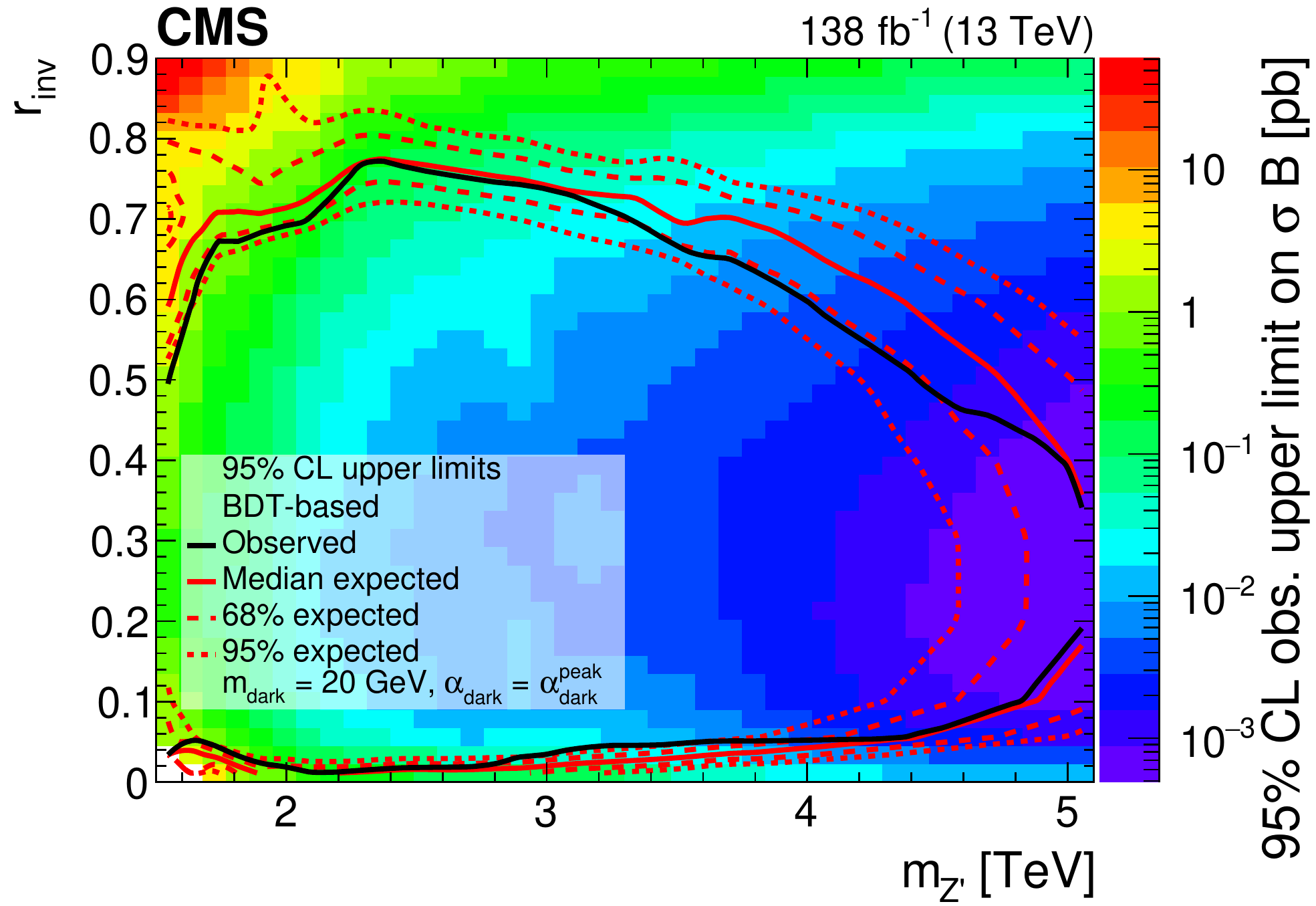}
\caption{95\% CL limits on the cross section for $\PZprime\rightarrow\text{semi-visible jets}$ in a two-dimensional plane with variations of \mZprime and \rinv from the model-independent (left) and model-dependent (right) searches, where $\mpidark=\mrhodark=20\,\GeV$ (here called $\mathrm{m_{dark}}$) and $\adark=\adarkPeak$ (here called $\alpha_\text{dark}$). Reproduced from Ref.~\cite{CMS:2021dzg}.
}
\label{fig:cms_svj_limits}
\end{figure*}

The full $13\,\TeV$ dataset of $138\,\ifb$ is analyzed and no indication of a resonance in the \MT spectrum is observed.
Both model-independent and model-dependent results are obtained, as shown in Fig.~\ref{fig:cms_svj_limits}.
The former excludes models with $1.5 < \mZprime < 4.0\,\TeV$ and $0.07 < \rinv < 0.53$ at 95\% CL, depending on the other model parameters.
The latter uses a boosted decision tree (BDT) that combines jet substructure variables to tag jets as semi-visible.
It extends the exclusions to $1.5 < \mZprime < 5.1\,\TeV$ and $0.01 < \rinv < 0.77$ for the specific models described in Section~\ref{sec:cms-svj-benchmark} that were used to train the BDT.

\subsubsection{CMS search for SIMPs as a link to signatures of trackless jets}

Dark sector models could give rise to experimental signatures where jets are formed with visible regular hadrons arising from decays of hidden-sector particles that are long-lived, and thus make these hadrons appear displaced within the jet. With sufficient displacement, jets can arise in which none of the tracks of the constituent charged hadrons can be reconstructed, thus making the jets appear neutral. Such neutral jets are extremely rare among high-momentum jets from regular standard-model quarks or gluons, and can thus be a very sensitive probe of physics beyond the standard model from dark sectors.

The CMS collaboration has performed a search for a pair of such trackless jets~\cite{CMS:2021rwb} using $16.1\,\ifb$ of integrated luminosity recorded in 2016. The signature probed consisted of a pair of back-to-back high-momentum trackless jets, where experimentally the trackless nature was sought by looking for the ratio of the jet energy carried by charged particles to the energy carried by neutral particles to be less than 5\%. To illustrate the effectiveness of this requirement in suppressing standard model QCD background jets, a background rejection of over $10^5$ in data was reported for this 5\% requirement.

This CMS search for trackless jets was inspired by and interpreted in a model proposing a new interaction through a low-mass mediator with a new dark matter fermion~\cite{Daci:2015hca}. The interaction leads to very high interaction cross sections, which are not necessarily excluded, though many model assumptions need to be made to avoid the many cosmology, particle physics and astrophysical constraints. The model considered is not a dark sector model per se, as indeed there is no decay back from a dark sector leading to missing charged hadrons, but rather the jets constitute solely of SIMP particles interacting in the calorimeters, thus generating neutral jets. The aspect of a displaced decay is thus missing, though the similar experimental signature does potentially impact dark sector searches.

The interpretation of the CMS trackless jets search in the SIMPs model has been found to be difficult at large SIMP masses, above ${\sim}100\,\GeV$. At such high masses, the modeling of the SIMP-nucleon interaction is complicated by the SIMP mass, and the approximation used in the CMS analysis, which involved treating the SIMP in the Geant simulation as a massive neutron-like object, becomes exploratory. As such, the strong exclusion limits on the SIMP pair production cross section as a function of the SIMP mass, obtained by CMS in absence of an excess of trackless jets in the analyzes data, shown in Fig.~\ref{fig:simps}, are reported with this caveat above $100\,\GeV$.

\begin{figure}
    \centering
    \includegraphics[width=0.6\linewidth]{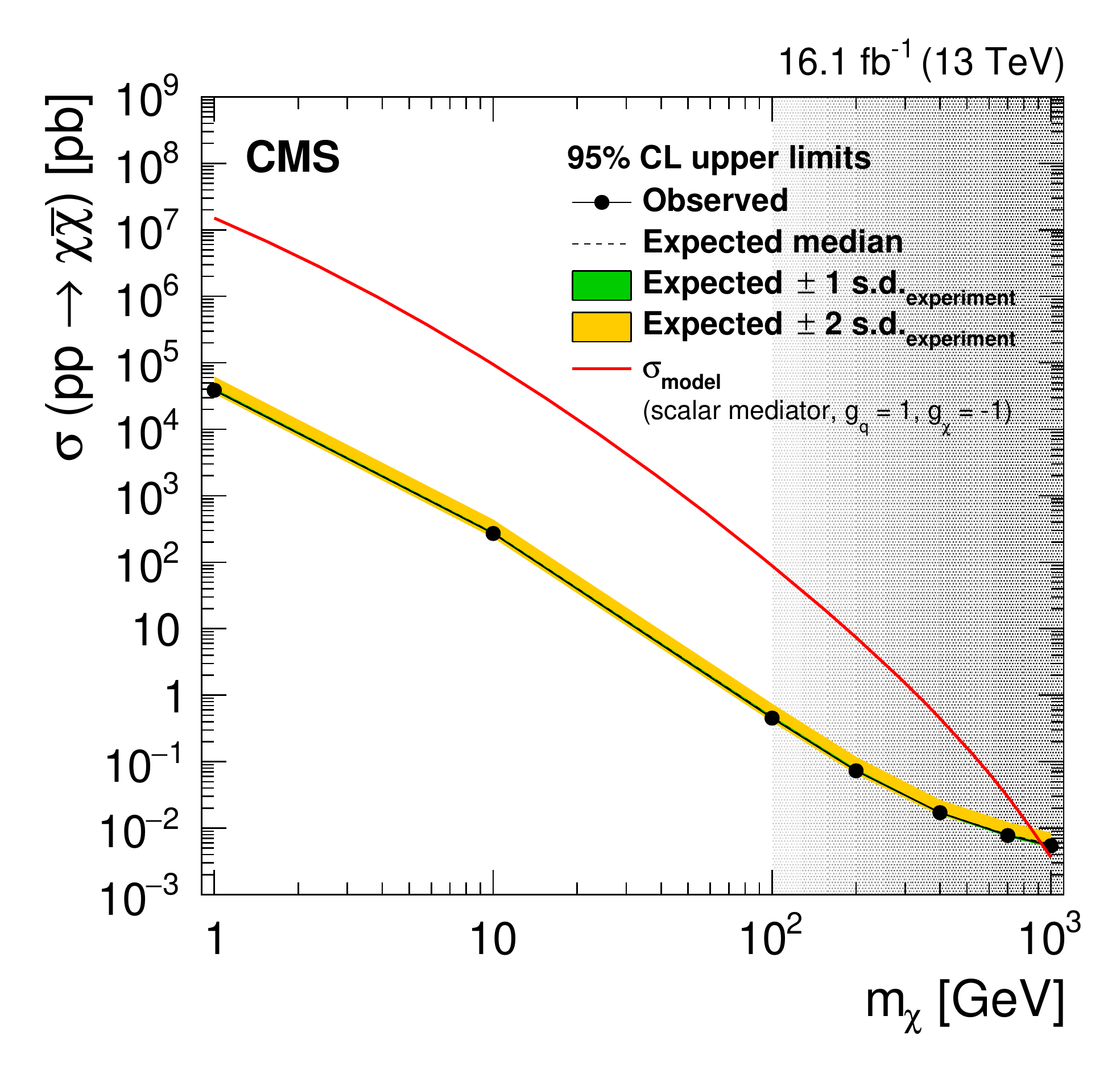}
    \caption{Cross section upper limits for the production of a pair of SIMPs (represented by $\chi$) in the context of the CMS search for trackless jets. See text for details. Figure taken from \cite{CMS:2021rwb}.}
    \label{fig:simps}
\end{figure}

\subsubsection{Collider constraints on $t$-channel models}\label{sec:existing_constraints_t_channel}
A plethora of new physics searches are ongoing at the LHC and have pushed the limits on the masses of new particles above the TeV scale in many cases. While dark showers are a spectacular and unique signature, existing new physics searches still retain some sensitivity in regions of parameter space where the signal looks SM-like. 

For $t$-channel dark sectors~\cite{Bai:2013xga,Schwaller:2015gea,Renner:2018fhh}, often the leading production mode is pair production of the heavy mediator, which gives rise to a signature with two ordinary QCD jets and two dark showers, as shown in~Fig.\ref{fig:t-channel_production}. It is clear that missing energy searches should become efficient in the limit where the dark sector particles become very long lived, while searches for prompt multijet signals can probe the regime of very short lifetimes. Due to the stochastic nature of particle decays, these searches also retain some sensitivity in the intermediate lifetime regime. 
\begin{figure}[h]
    \centering
    \includegraphics[width=0.6\linewidth]{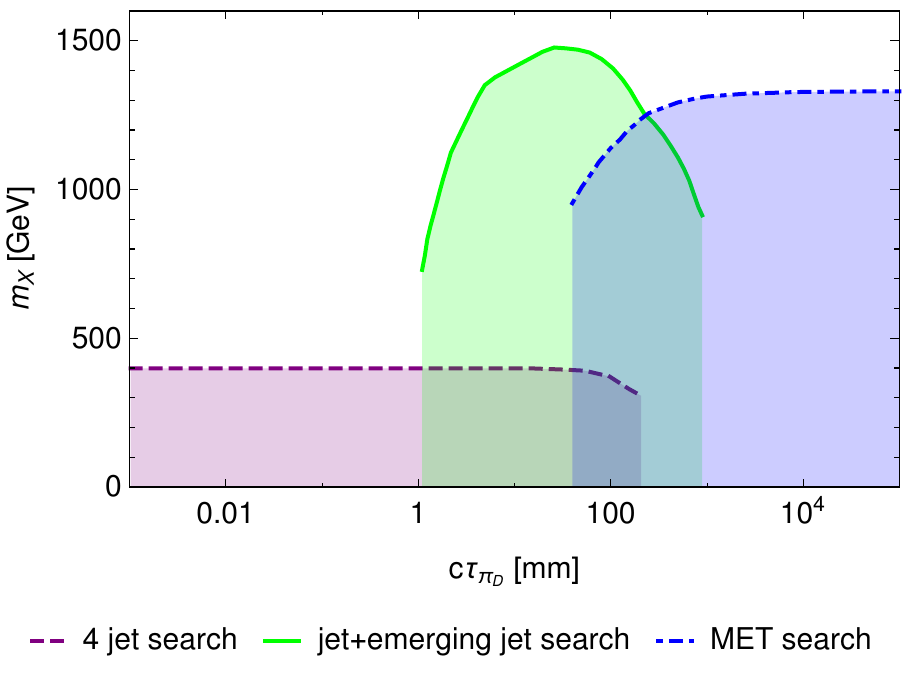}
    \caption{Constraints on the mass of a $t$-channel mediator \Pbifun (here called $X$) as a function of the dark sector particle lifetime. See text for details. Figure taken from \cite{Mies:2020mzw}.}
    \label{fig:recast_tchannel_ctau}
\end{figure}

A recast of a di-jet plus MET search~\cite{CMS:2019zmd}, a search for paired prompt di-jet resonances~\cite{ATLAS:2017jnp} and of the dedicated emerging jets search~\cite{CMS:2018bvr} was performed in Ref.~\cite{Mies:2020mzw}. As can be seen in Fig.~\ref{fig:recast_tchannel_ctau}, the dedicated emerging jets search performs best in the intermediate lifetime regime, while the recast searches can probe all the remaining range of lifetimes. As expected, mediator masses below the TeV scale are already strongly constrained. In the short lifetime regime, the constraints are weaker. This is partially because the published search uses only a limited amount of data, but also because fighting the QCD multi-jet background is hard. Another option to constrain the $t$-channel mediators in the regime of short dark sector lifetimes is through their contribution to angular correlations in dijet events~\cite{CMS:2018ucw}, when the mediator is exchanged in the $t$-channel instead. This constraint however will depend on the magnitude of the Yukawa coupling of the mediator. 

The $t$-channel mediators naturally allow couplings that break SM flavor symmetries. While most flavor violating couplings are constrained by low-energy flavor observables~\cite{Renner:2018fhh,Carmona:2021seb}, the LHC has the potential to constrain flavor changing couplings of the top quark~\cite{Carmona:2022jid}. Such scenarios could also be of interest for searches for dark showers produced in association with tops or even stemming from exotic top decays.

\subsubsection{Existing constraints and projections from LHCb}

The LHCb experiment, originally designed for heavy-flavor physics, has shown its potential as a \emph{general purpose detector} in the recent years. An excellent secondary vertex resolution, low momentum thresholds and particle identification capabilities make LHCb a natural candidate to search for dark QCD signatures in the low-mass region. These searches are now becoming part of the LHCb physics program, where world-leading constraints have already been set for hidden valley scenarios, with reinterpretations in other dark sector models as well as a large number of very encouraging sensitivity projections, described in the following paragraphs.

LHCb has published a search for low-mass resonances decaying into pairs of muons, using 5.1 fb$^{-1}$ of data collected at 13 TeV~\cite{LHCb:2020ysn}. In this article, a model-independent search of both prompt and displaced resonances $X$ is performed. For the displaced case, the secondary $X~{\to}~\mu^+\mu^-$ decay vertex is required to be transversely displaced from the primary vertex in the range $12<\rho_T<30$ mm, allowing for the resonance to become a long-lived signature in the detector. Then, limits on the cross-section $\sigma(X~{\to}~\mu^+\mu^-)$ are placed, and interpreted in various production models. One of these models in the regard of dark QCD is that of a hidden valley scenario, where constraints are set on the kinetic mixing strength $\gamma-Z_{HV}$ between a heavy hidden valley boson $Z_{HV}$ with photon-like couplings, and a photon, fixing the average multiplicity of hidden valley hadrons to a value of 10. 
These are the most stringent constraints placed up to date, for masses of a composite hidden valley vector boson, $X$, up to 3 GeV, as presented in figure~\ref{fig:LHCb_HV_lowmass}.  

\begin{figure}[h!]\begin{center}
\includegraphics[width=0.7\textwidth]{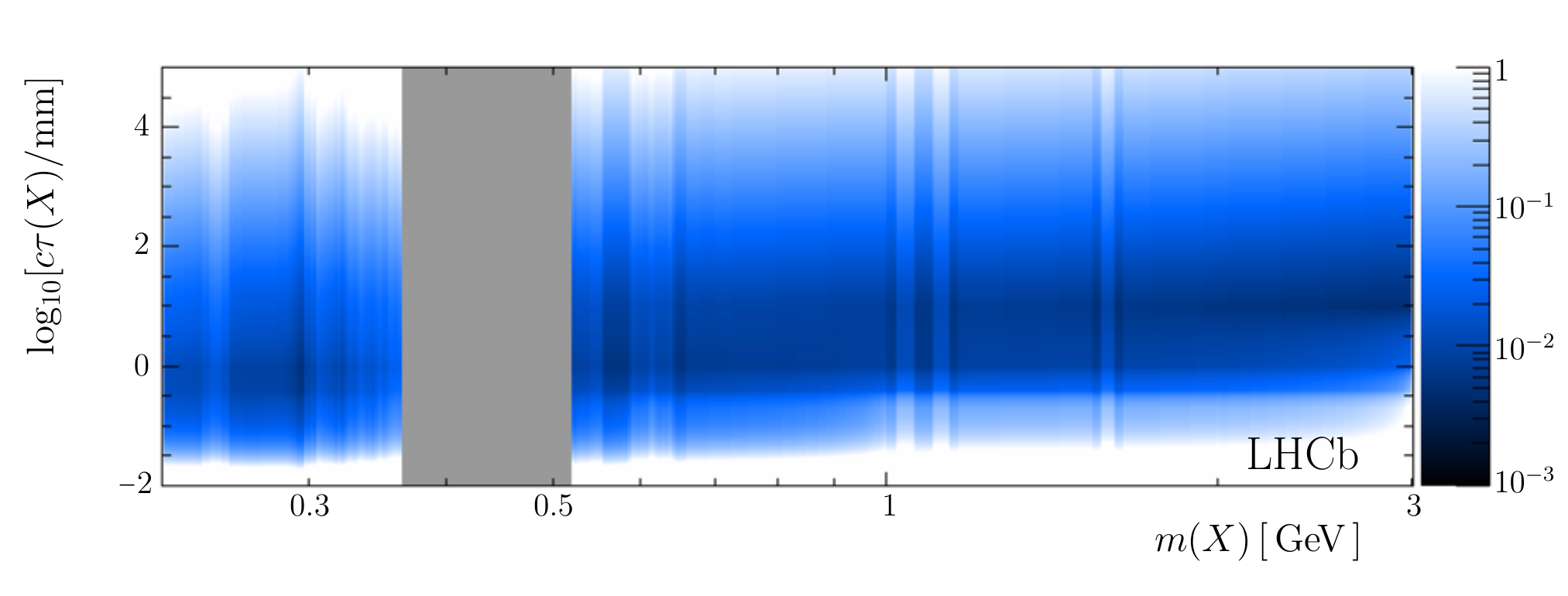}
\caption{Upper bounds at 90$\%$ C.L. on the kinetic mixing strength $\gamma-Z_{HV}$. The grey box shows a vetoed region due to the large doubly misidentified $K_S^0$ background. Figure taken from ref.~\cite{LHCb:2020ysn}.}\label{fig:LHCb_HV_lowmass}
\end{center}\end{figure}

These results have been also re-interpreted in the context of Z-initiated dark showers, assuming various benchmark scenarios~\cite{Cheng:2021kjg}. In Figure~\ref{fig:LHCb_Zinitiated}, projections for one of the scenarios are shown, showing the capabilities of LHCb to probe Z branching fractions down to 10$^{-7}$ during the high-luminosity phase. 

\begin{figure}[ht!]\begin{center}
\includegraphics[width=0.7\textwidth]{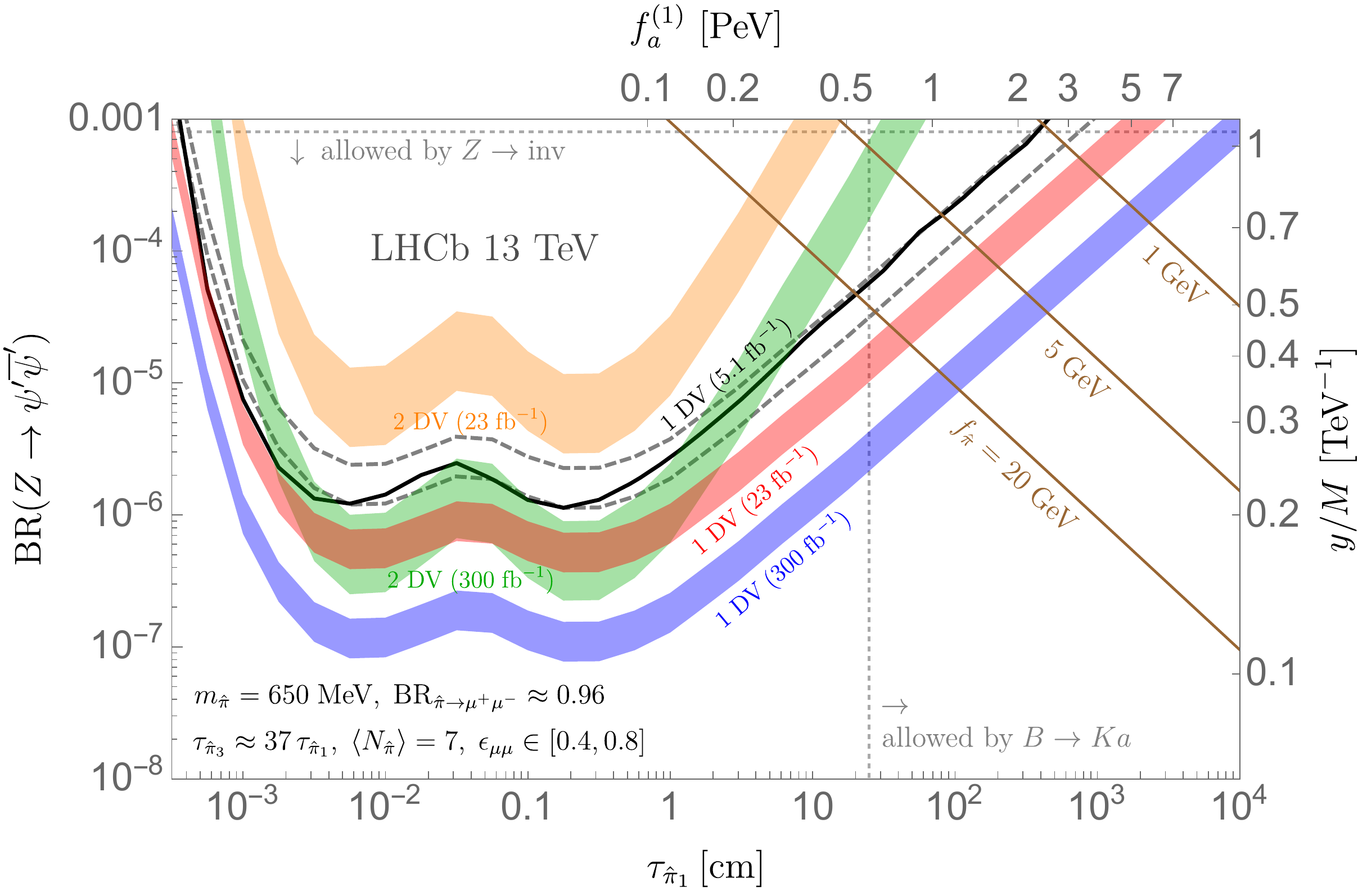}
\caption{Projections at 90\% C.L. of LHCb sensitivity to a muon rich dark showers initiated by the decay of a Z into dark quarks (here called $\psi'$). In this benchmark scenario, a dark pion mass (here called $\hat{\pi}$) of 650 MeV, an average multiplicity of 7 and a branching fraction of the dark pion into two muons of 96\%, are assumed. More details can be found in ref.~\cite{Cheng:2021kjg}.}\label{fig:LHCb_Zinitiated}
\end{center}\end{figure} 

In a more general sense, the capabilities of LHCb to probe other dark QCD models are summarized in Ref.~\cite{Borsato:2021aum}, in the context of benchmark scenarios featuring a range of dark hadron and mediator masses, for different assumptions on the average dark hadron multiplicity in the dark sector. 
The projections described in this major report show the outstanding potential of the LHCb experiment to place very stringent constraints in the low-mass range, in complementarity with ATLAS and CMS. 
\clearpage

\section{Dark sector beyond QCD-like scenarios}
\label{beyond_QCDlike_theories}
\subsection{SUEP}\label{suep}

\emph{Contributors: Cari Cesarotti, Carlos Erice, Karri Folan DiPetrillo, Chad  Freer, Luca Lavezzo, Christos Papageorgakis, Christoph Paus, Matt Strassler}

Dark showers produced in hidden valley models need not result in collimated jets like in SM QCD. Events with spherically-symmetric, large multiplicities of low momentum charged particles, are also a possible phenomenology of strongly-coupled hidden valley models. This section discusses the motivation for these so called soft-unclustered-energy patterns (SUEPs), tools available for simulation, and typical phenomenology. Experimental challenges for SUEP searches at LHC general purpose detectors are also summarized. 

\subsubsection{Theoretical Motivation}

 Although the production of quarks in a QCD-like confining theory leads inevitably to jets of hadrons, the details are not always the same.  The width of the jets, the hadron multiplicity per jet, and the jet multiplicity all depend on the value of the running coupling.  More specifically, jets arise from the partonic shower, which depends on the running 't Hooft coupling $\lambda\equiv\adark\Ncdark$ ($\alpha_s N_c$ in QCD), evaluated at and somewhat below the energy scale at which the jets are produced.  

In QCD-like theories, like in QCD, jets at lower energy, closer to the confinement scale and thus at larger $\lambda$, are broader, because gluon radiation at larger angles is more common with the large coupling.   One might then imagine that if $\lambda$ could somehow be taken large {\it without} reducing the energy of the collisions, then the jets produced might become so broad and numerous as to blur together, creating a  smooth distribution and a non-jetty final state.  In QCD-like theories this question is almost academic, since  $\lambda$ only becomes large near the confinement scale.  In $e^+e^-$ collisions near the confinement scale, only a small handful of hadrons are produced, making it hard to define jets in the first place.   Jets were only identified in $e^+e^-$ collisions at scales $\sim 10$ GeV, where $\lambda < 1$, and gluon jets only at 27 GeV. 

However, in 1998 it was shown \cite{Maldacena:1997re} that certain classes of supersymmetric conformal field theories (roughly speaking, these are theories with whose beta functions are all zero and are thus scale-invariant) are equivalent to string theories on certain curved spaces.  These theories can have arbitrary values of $\adark$ (now constant) and $\Ncdark$.   The duality can be used to compute many of  properties of these theories when  $\adark \Ncdark\gg 1\gg \adark$.  It was noted in \cite{Polchinski:2002jw} that rapid pdf evolution in this regime leads to an absence of hard partons inside hadrons.  Since similar dynamics controls jet evolution, this naturally suggests that jets will be absent in this regime as well. Several groups \cite{Strassler:2008bv,Hatta:2008tx,Hofman:2008ar} argued that there should be no jets in this limit, and it was proven convincingly in \cite{Hofman:2008ar} that the correlation function of energy operators indicates that a partonic shower in this regime, allowed to proceed over a wide range of energies, will approach a spherically symmetric distribution on average.  The distribution of momenta of these partons is not determined, however.

Note, importantly, that {\bf a spherically symmetric shower is not a consequence of conformal symmetry.}  A conformally invariant theory at small $\lambda$ will exhibit jets much like QCD itself; indeed, QCD at high energy is nearly conformally invariant, with scaling violated only by small logs (a fact which played an important role in the discovery of quarks.)  Only at when $\lambda \gg 1 \gg \adark$ are roughly spherical showers expected, and corrections to the spherical shape are of order $1/\sqrt{\lambda}$.  Thus, for events that typically differ from spherical by $< 10\%$, one probably must have $\Ncdark\gg100$.

A conformally-invariant hidden sector is generally observable only as \met (except for small rare processes discussed in the unparticle literature) as the energy will be shared down to massless partons.  Interesting hidden valley signatures arise only if the conformal invariance is broken at some scale $Q_c$ much lower than the production scale $M$.  The shower that follows production of hidden partons is converted at $Q_c$ to a large number of hidden particles of small mass $\mdark \lesssim Q_c$.  If some of these are able to decay to the SM, then this can lead to a signature of many particles which are roughly spherically distributed in some frame of reference, not necessarily the lab frame  \cite{Strassler:2008bv,Knapen:2016hky}.  This signature is defined as a Soft Uncorrelated Energy Pattern, or SUEP.  

We note that there have been disputes in the theoretical literature concerning whether the SUEP arising in this context is related directly to cascades of 5d KK states, with different points of view taken by \cite{Csaki:2008dt,Costantino:2020msc} versus \cite{Strassler:2008bv,Cesarotti:2020uod}.  From the phenomenological point of view this may not matter very much in the near term, but the issue could arise if in future KK-cascade-based generators are created and assumed (perhaps incorrectly) to describe the same phenomenon as SUEP.

If the conformal symmetry breaking involves the complete Higgsing of a gauge group, it is clear that a near-spherical shower leads to a near-spherical SUEP.  If the conformal symmetry breaking involves confinement, this is less clear and has not been proven; it may be that it is somewhat model-dependent.  We will nevertheless assume it is the case, and that SUEPs can arise from the SM decay products of dark hadrons produced from a spherical shower of dark gluons.   We will further assume that those dark hadrons can resemble those from a QCD-like spectrum, with low-lying spin-0 and spin-1 mesons and with limited roles for heavier mesons.  Baryons should have mass of order $\Ncdark\gg 1$ and will play no role.

Since techniques to calculate SUEPs, even in the few models where they are known to occur, do not yet exist,  the simulations \cite{Knapen:2016hky,Knapen2020} to study them are necessarily somewhat ad hoc.  The assumption is made that particles are produced spherically according to a thermal distribution with an unknown temperature $\Tdark$ of order $Q_c$, the conformal breaking scale; see sec.~\ref{sec:SUEP_simulations}.  (Note $\mdark$, the typical hadron mass, may be of order $Q_c$ or may be much smaller, especially for pion-like states; in the latter case, multiplicities may be surprisingly low.)   The value of $\Tdark/Q_c$ should be varied and treated as a quantifiable uncertainty.   

Less clear is how to treat the uncertainties of the thermal approximation in the first place.   A thermal model for hadronization in QCD works moderately well; perhaps this reflects the tendency for complex statistical systems to approach thermal ones.  However, there is no proof that it would work for all or even most confining theories.  One way to approach the uncertainty might be to vary the thermal spectrum in one or another way, using insights from studies of how systems equilibrate.  

 Finally, an uncertainty arises from the fact that in realistic theories the spherical approximation will suffer corrections, and at the current time those corrections have not been characterized theoretically nor incorporated into the simulation tools.  Some effort to quantify this uncertainty ought to be undertaken.

Since SUEPs and the simulation packages used to simulate them represent a certain idealized situation, real signatures may differ from this idealized model.  In this regard, the following should probably be kept in mind:
\begin{itemize}
\item Only a small number of theories of this class have been proven to exist, all of them supersymmetric and with $\Nfdark\ll\Ncdark$, where $\Nfdark$ is the number of quarks in the fundamental representation.  
\item The statement that expected distributions are spherical receives substantial corrections, of order $1/\sqrt{\lambda}$ at the confinement scale.  For an observably spherical SUEP, one may need $\Ncdark\sim 100$.    Because the particles of the hidden sector can appear in loops, coupling a mediator to a theory with $\Ncdark\sim 100$ or more can have large consequences for the mediator or even the Standard Model sector; care must be taken in defining a consistent model.
\item Event-to-event fluctuations can lead to large deviations from the SUEP idealization.   When the muliplicity of visibly-decaying hadrons is low, Poisson fluctuations are large. For instance, if sixty dark hadrons are produced near-spherically but only ten decay visibly, the observed event will be far from spherical and potentially very asymmetric.
\end{itemize}

\subsubsection{Simulation Tools}
\label{sec:SUEP_simulations}

In order to search for these novel signatures at high energy colliders, it is essential to generate events that capture the phenomenological characteristics of the energy pattern. Perturbation theory breaks down in the large coupling regime, and standard approaches to event simulation are unreliable. Novel methods are necessary to generate SUEP events. 

Several simulation methods that can generate quasi-spherical energy patterns exist, such as black hole generators \cite{Harris:2003db,Dai:2007ki,Cavaglia:2006uk} or simplified 5d models \cite{Cesarotti:2020uod}. However, these tools are not ideal for developing new analyses or triggering strategies. LHC experiments already have extensive search programs for black holes \cite{CMS:2017boz, CMS:2018ozv,CMS:2018hnz,ATLAS:2017dpx,ATLAS:2017eqx}. The latter tool does not have an obvious portal to connect to the SM. We will therefore discuss the utility of another tool in this section: the SUEP generator \cite{Knapen2020}. 

\begin{figure}
    \centering
    \includegraphics[width=0.4\textwidth]{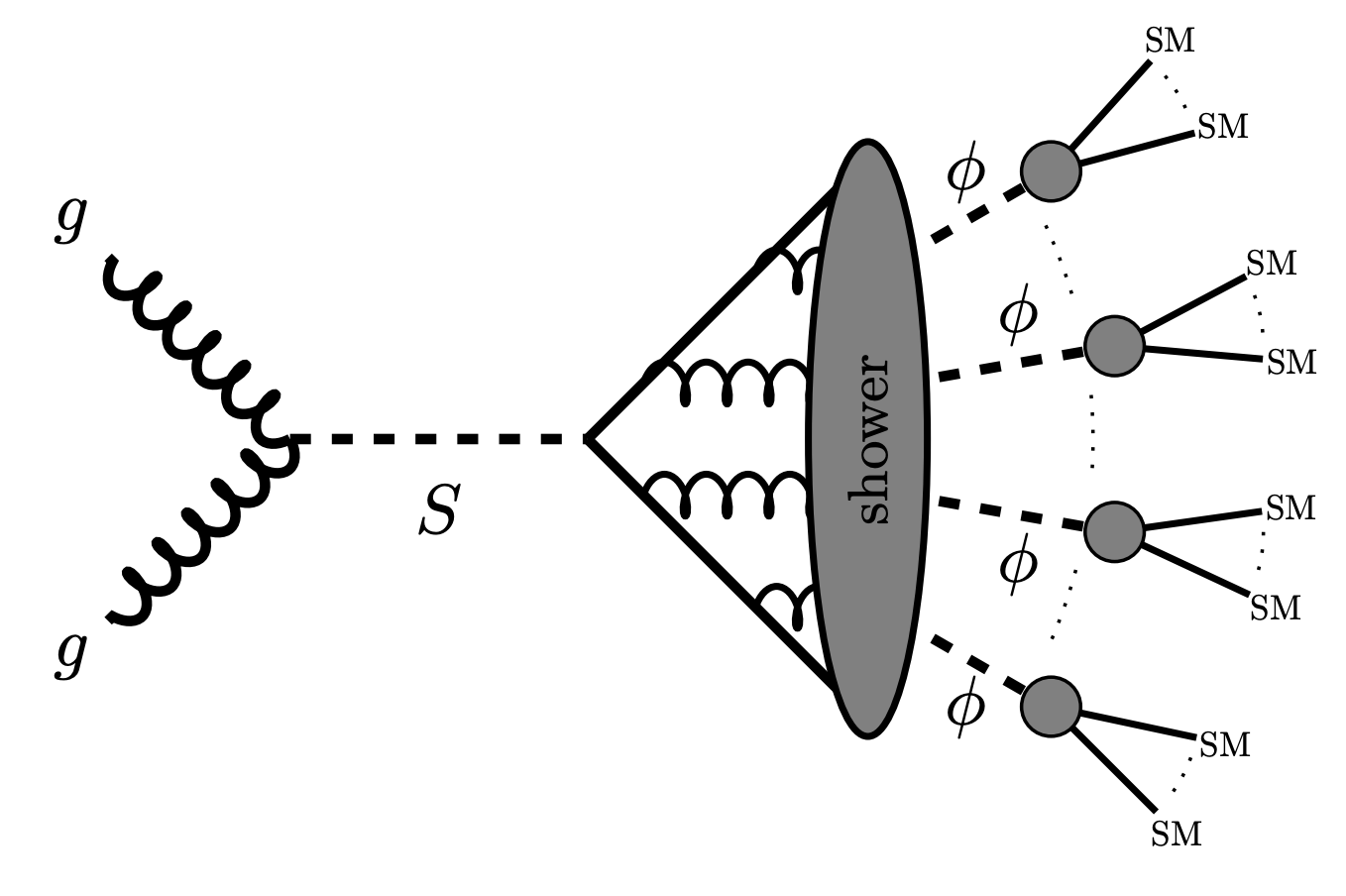}
    \caption{A visualization of a generic SUEP event. A heavy scalar $S$ accesses a confining HV and showers into light mesons (e.g. \Ppidark, here denoted $\phi$) before decaying to SM particles. Figure adapted from Ref.~\cite{Knapen:2016hky}.}
    \label{fig:suep_diag}
\end{figure}

The simplified model used by this generator is described in Ref.~\cite{Knapen:2016hky}. In this framework, a hidden valley (HV) of new physics with confining dynamics is accessed via a heavy scalar mediator. A wide class of mediators can be used to connect the SM and HV, but a scalar portal from gluon fusion was chosen to both explore a triggering `nightmare scenario' as well as study a potential rare Higgs decay mode. The scalar then decays into a high multiplicity of light HV mesons of a single flavor, of mass $\mdark$, that follow a Maxwell-Boltzmann distribution
\begin{equation}
    \frac{dN}{d^3\textbf{p}} \sim \exp \left( -\sqrt{\textbf{p}^2+\mdark^2}/\Tdark \right).
\label{eq:Boltzmannsuep}\end{equation}
An illustration of this process is shown in Figure~\ref{fig:suep_diag}. The user is free to select the temperature, \Tdark, and $\mdark$, but for reasonable results the two parameters should satisfy $\Tdark/\mdark > 1$ such that the final state is high multiplicity. Note that in a confining theory like QCD, the mass of the lightest meson can be much less than the confinement scale $\lamdark \sim \Tdark$, so $\Tdark\gg \mdark$ is a motivated and physical choice. For a fixed scalar and meson mass, a higher temperature will correspond to fewer particles with a more significant boost. The tool is only intended to work in the high multiplicity limit, which means that the user must set $m_S\gg \Tdark, \mdark$ for consistent results. A numerically small amount of momentum conservation violation may be observed in some events.

After production and showering, the dark mesons must decay back to SM for a visible signal. In this benchmark model, it is assumed that the dark mesons couple to a new U(1) gauge boson $A'_\gamma$ that kinetically mixes with SM hypercharge. The dark mesons decay into a $A'_{\gamma}A'_{\gamma}$ pair. Each $A'_\gamma$ then decays into SM particles, for example dilepton ($e^+e^-$ and $\mu^+ \mu^-$) or hadronized final states. The $A'_{\gamma}$ mass and branching ratio to Standard Model particles is configurable by the user.

This tool has the potential to inform and test diverse future analysis techniques. The phenomenology of the model depends on the choices of scalar mediator mass $m_S$, the dark meson mass $\mdark$ and temperature $\Tdark$, as well as the decay branching ratios of the new particles. Depending on the signal of interest, the user can configure these values to achieve different multiplicity or species final states. Simple extensions of the package could include different possible mediators and decay portals. 

\subsubsection{Phenomenology}

The classification of `SUEP' is on the final state signature rather than a specific type of model. A SUEP (previously called 'soft bomb'~\cite{Knapen:2016hky}) is usually an event with a high multiplicity of soft particles distributed quasi-isotropically in their rest frame. The underlying physics that produces such events can be varied. As discussed earlier, if the new particles interact strongly over a wide energy window, the shower develops by soft and isotropic emissions \cite{Strassler:2006im,Polchinski:2002jw,Hofman:2008ar}. However, SUEPs can also develop from kinematics due to phase space arguments. If a new physics model includes many unstable particles with small mass splittings \cite{Csaki:2008dt, Cesarotti:2020uod}, subsequent decays are unboosted and can approximate a spherical energy distribution for sufficiently large particle multiplicity. A common example of such event is R-parity violating SUSY \cite{Barbier:2004ez,Evans:2013jna}.

Since many different new physics scenarios can produce a SUEP signature, it is compelling to generically search for such events at colliders. As their common underlying feature is their global radiation pattern,  event shape observables can serve as useful analysis tools. By studying the global event shape, it is both possible to quantify new physics and distinguish signal from background.

Event shape observables have been used to study QCD and measure $\alpha_s$ \cite{Farhi:1977sg,Barber:1979bj,Althoff:1983ew,Bender:1984fp,Abrams:1989ez,Li:1989sn,Buskulic:1995aw,Adriani:1992gs,Braunschweig:1990yd,Abe:1994mf,Heister:2003aj,Abdallah:2003xz,Achard:2004sv,Abbiendi:2004qz,Dissertori:2008cn}. There have been several observables developed to quantify the degree to which a collider event is isotropic versus jet-like, including thrust \cite{Farhi:1977sg,Brandt:1964sa,DeRujula:1978vmq}, sphericity \cite{Bjorken:1969wi,Ellis:1976uc}, spherocity \cite{Georgi:1977sf}, and the $C$- and $D$-parameters \cite{Parisi:1978eg,Donoghue:1979vi,Ellis:1980wv}. While these observables have provided indispensable insights to QCD, they are most sensitive to deviations from dijet events rather than a robust probe of isotropy. To capture the phenomenology of a SUEP event it is necessary to define new observables which probe the opposite regime.

A new observable called event isotropy aims to study deviations from truly isotropic events \cite{Cesarotti:2020hwb}. This observable is defined using the Energy Mover's Distance (EMD) \cite{Komiske:2019fks,Komiske:2020qhg}, the particle physics application of the Earth Mover's Distance \cite{Peleg1989AUA,Rubner:1998:MDA:938978.939133,Rubner2000,Pele2008ALT,Pele2013TheTE}. Event isotropy quantifies how `far' an event is from isotropic, with smaller values indicating an event is more isotropic. Figure~\ref{fig:suep_isotropy_comp} shows event isotropy for SUEP benchmark models with different mediator masses and temperatures.

\begin{figure}[htb]
    \centering
    \includegraphics[width=0.45\textwidth]{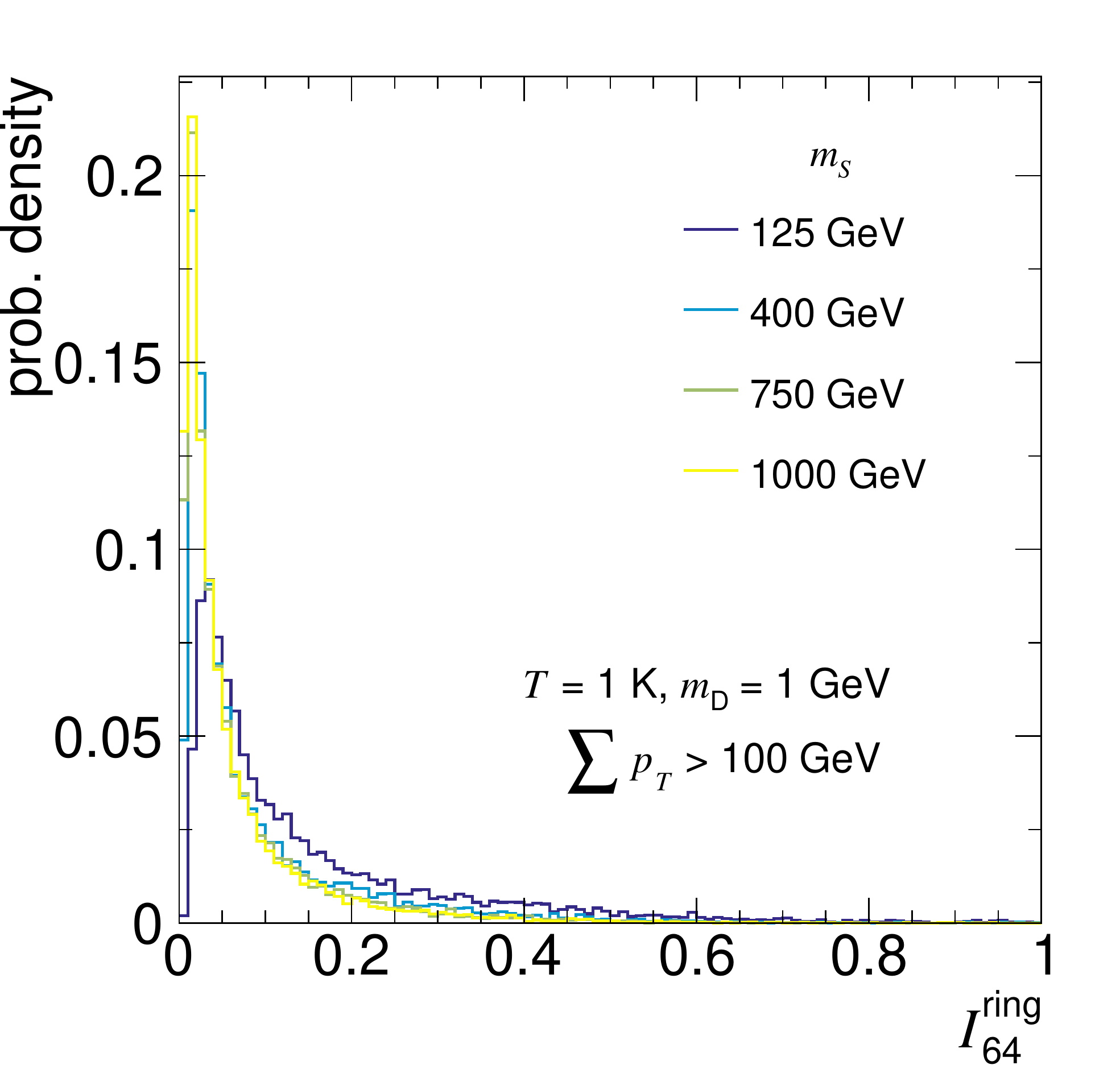}
    \includegraphics[width=0.45\textwidth]{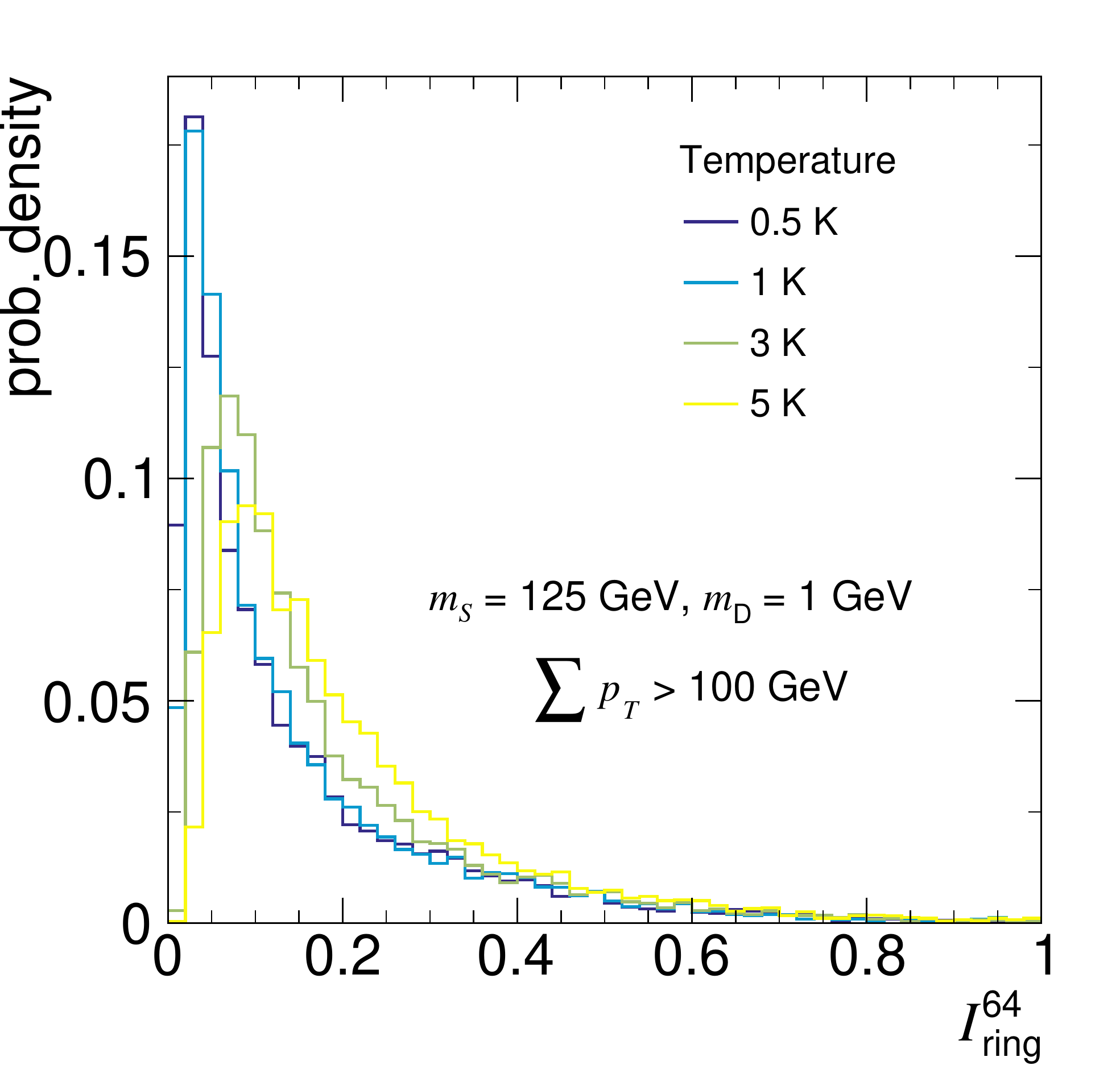}
    \caption{Isotropy distributions are shown for different SUEP mediator masses on the left and for different temperatures on the right. The isotropy is calculated for ring geometry with segmentation 64 for generator level tracks. Tracks are defined as status~$=1$ charged particles with $p_T > 0.1$~GeV and $|\eta| < 2.4$. Only events that pass the requirement $\sum p_T > 100$~GeV are kept. The production mode is Gluon Fusion (GF) with a dark photon branching fraction of $BR(A'_{\gamma} \rightarrow e\bar{e},\mu\bar{\mu},\pi\bar{\pi}) = (40,40,20)\%$.}
    \label{fig:suep_isotropy_comp}
\end{figure}

Figure~\ref{fig:suep_isotropy_corr} shows results from a related study that demonstrates how event isotropy is not strongly correlated with final state multiplicity, or reconstructed number of jets. Additionally, it correlates with canonical event shape observables much less than they correlate with each other in the quasi-isotropic regime. While there can be correlation with traditional event shapes, both types of observables can be used to better characterize the underlying physics.

\begin{figure}[htb]
    \centering
    \includegraphics[width=0.45\textwidth]{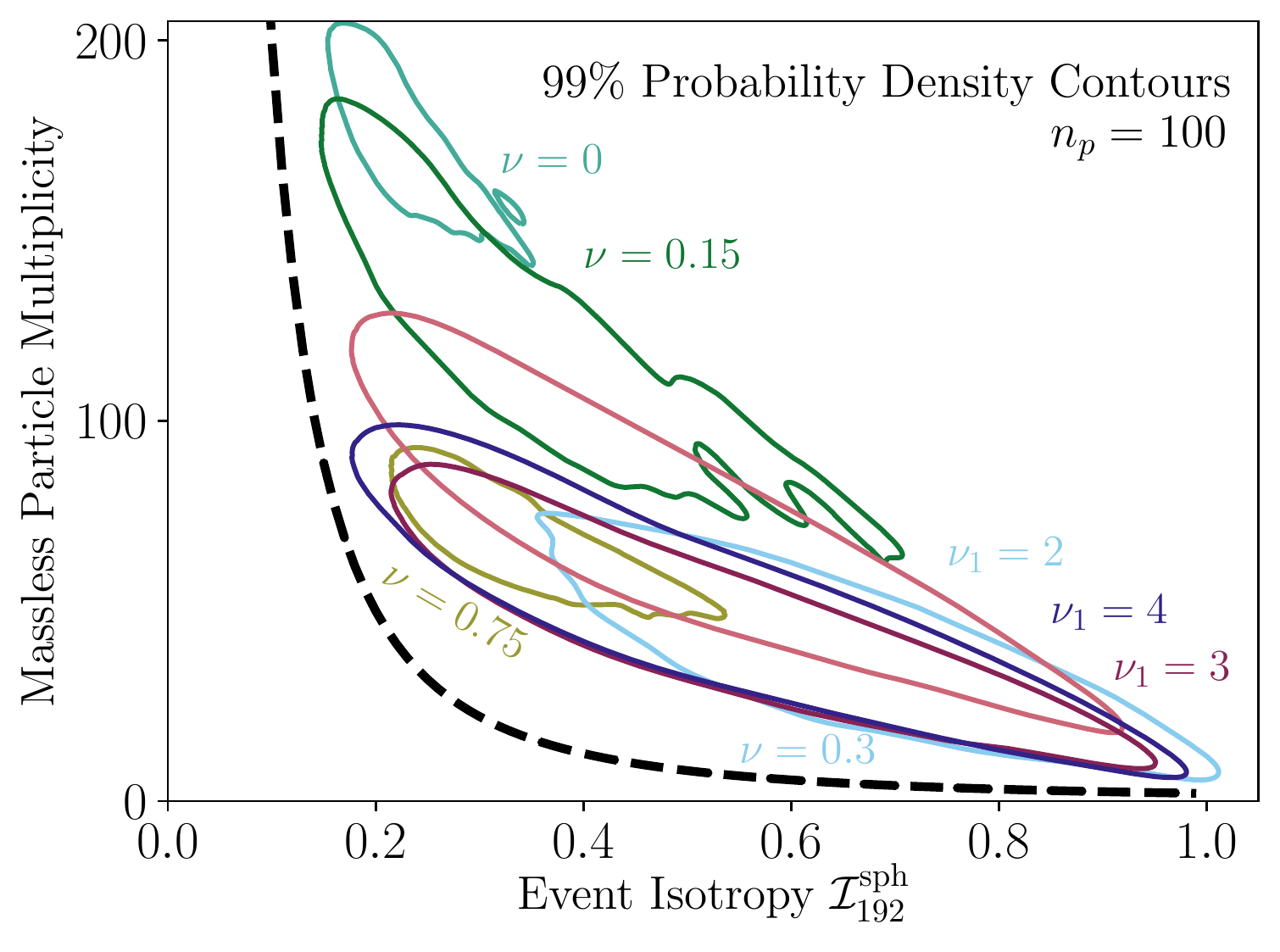}
    \includegraphics[width=0.45\textwidth]{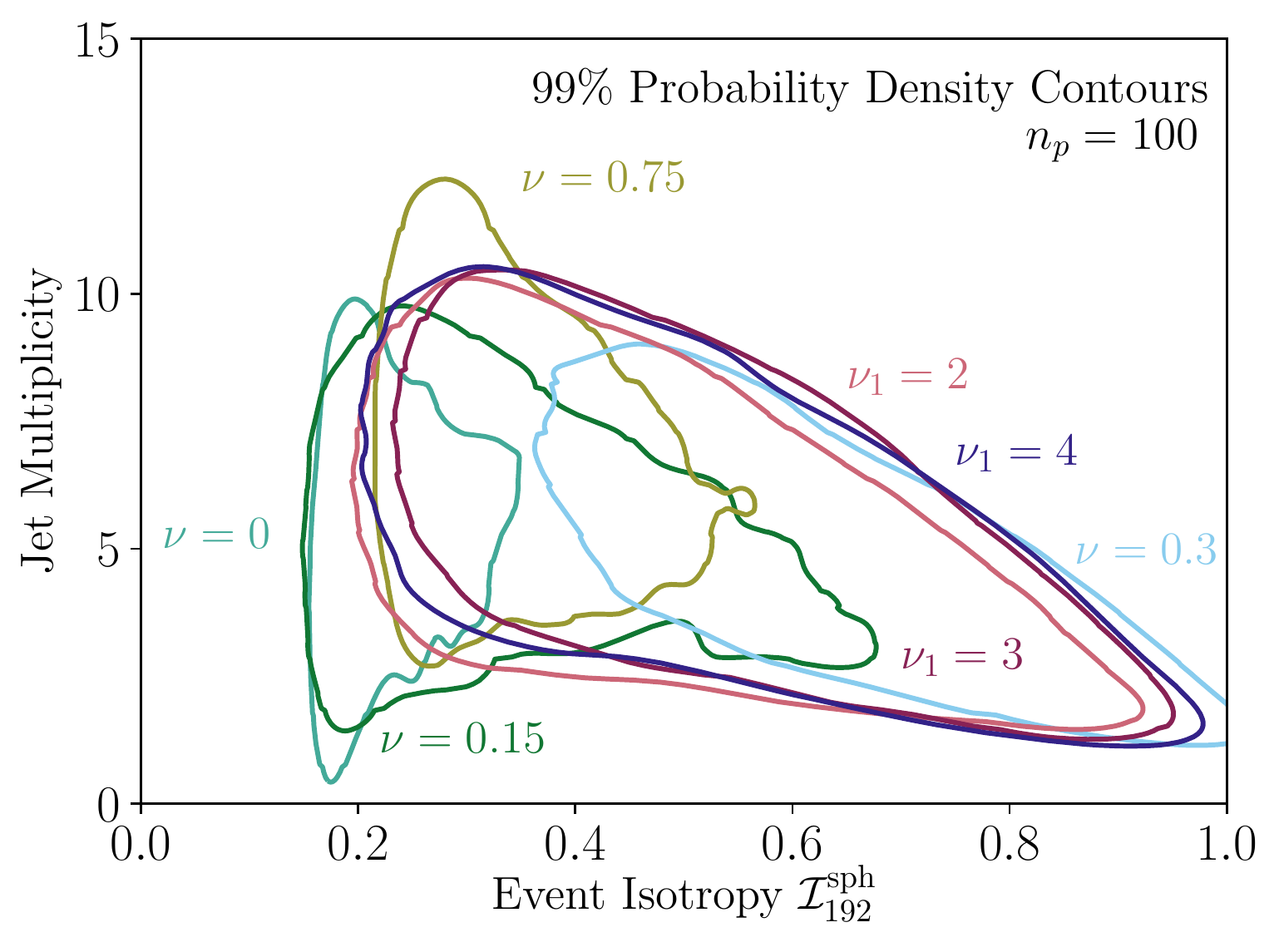}
    \includegraphics[width=0.45\textwidth]{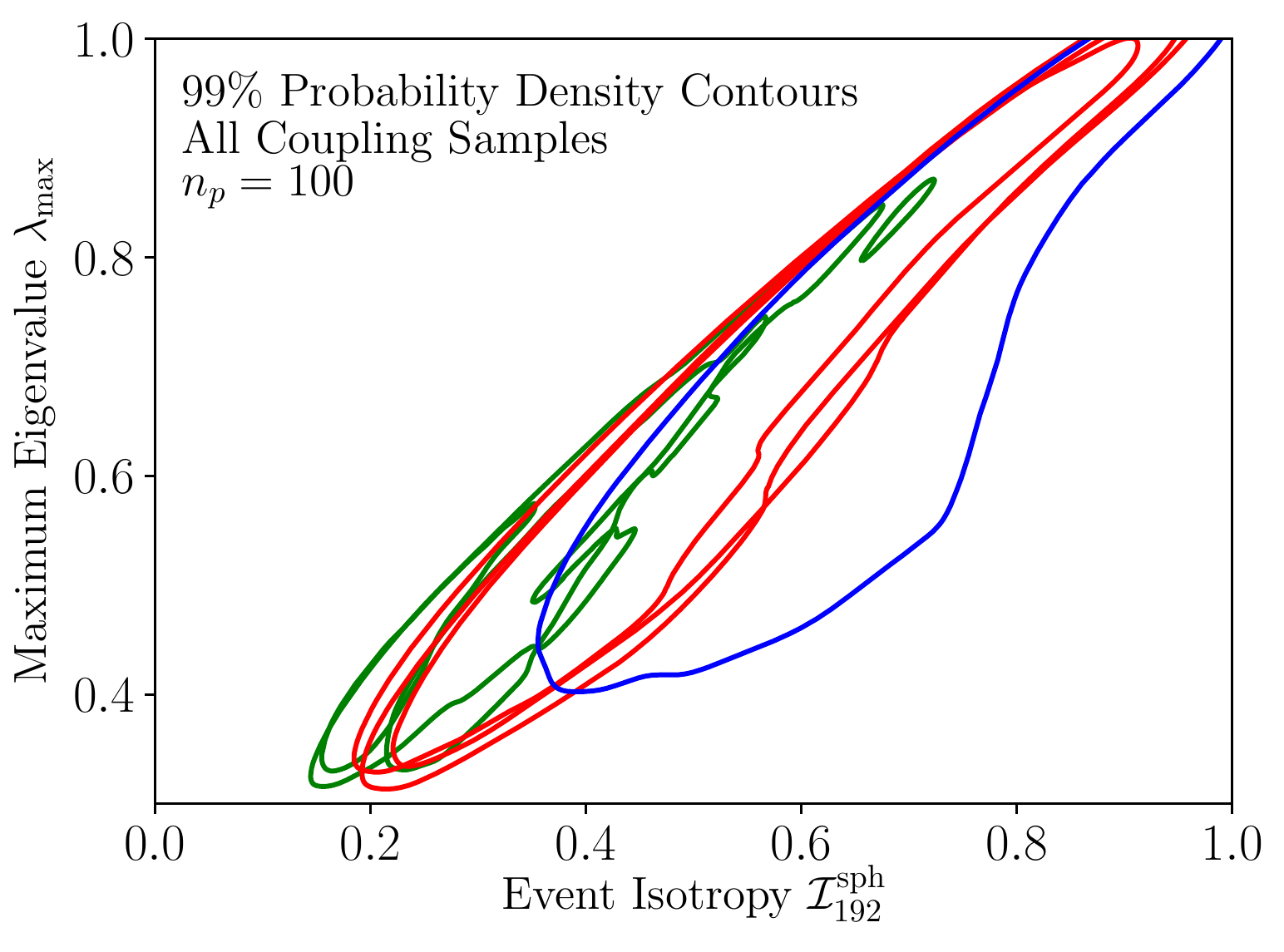}
    \includegraphics[width=0.45\textwidth]{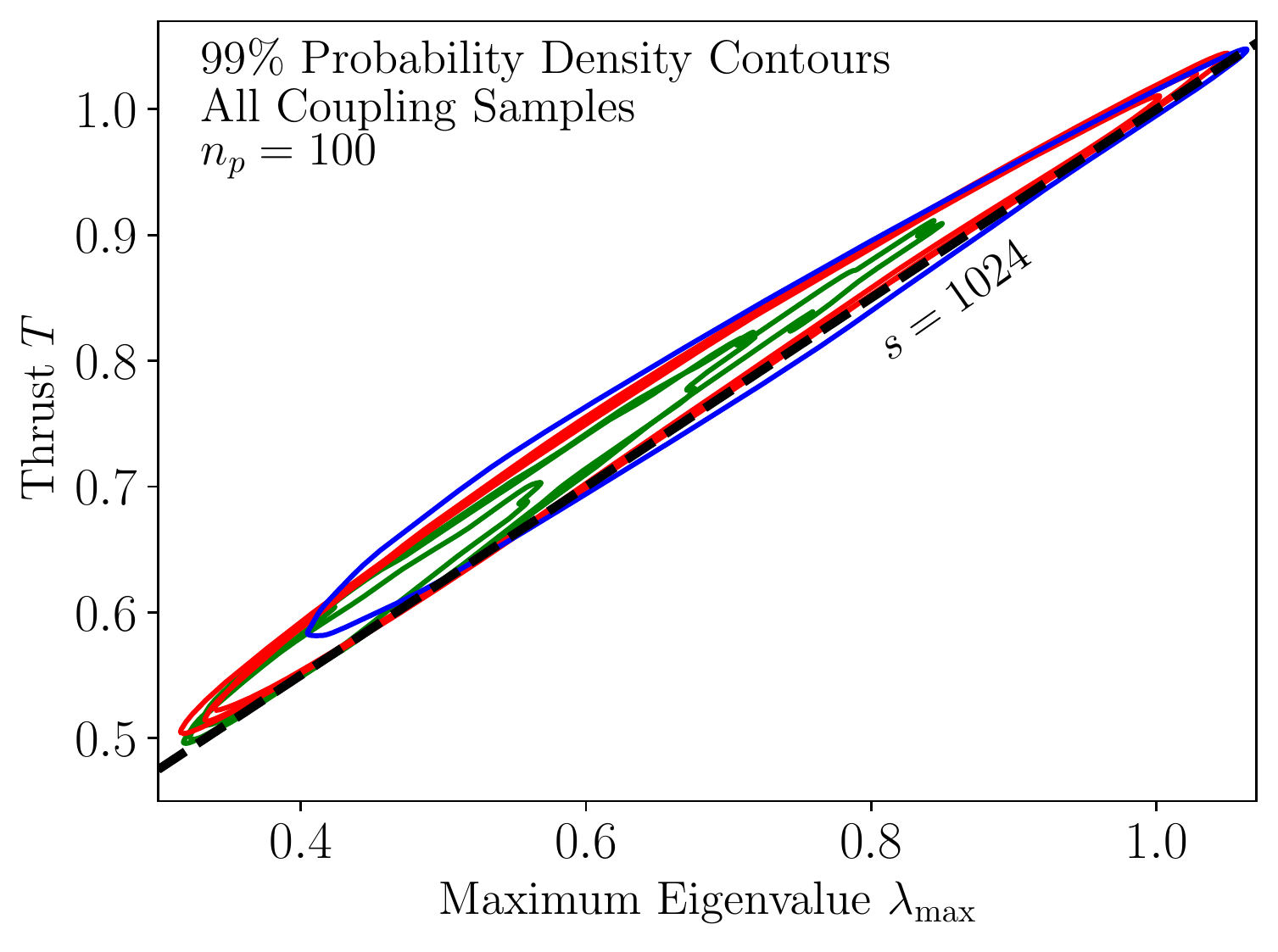}
    \caption{Correlations between event isotropy and other canonical event shape observables calculated on quasi-isotropic events generated using the framework of ref.~\cite{Cesarotti:2020uod}, such as particle multiplicity (left top), jet multiplicity (right top), and the maximum eigenvalue of the sphericity tensor (left bottom), which is closely related to the $C$ and $D$ parameter in quasi-isotropic radiation patterns. For reference, we also include the correlation plot of the same samples in thrust and maximum eigenvalue (bottom right). The contours enclose 99\% of the events for all of the samples. Figures adapted from \cite{Cesarotti:2020ngq}.}
    \label{fig:suep_isotropy_corr}
\end{figure}

\subsubsection{Experimental Aspects}

The diffuse low momentum nature of SUEP events strongly resembles soft-QCD backgrounds at the LHC. Without high momentum final state particles, SUEP events with all-hadronic final states can easily be mistaken for pile-up collisions, and pose extreme challenges for general purpose detectors such as ATLAS and CMS.

The first and most challenging step for any SUEP analysis is to identify a trigger strategy. The trigger systems of ATLAS and CMS operate in a two-step process to determine which events are saved for analysis. The first stage, Level 1, makes a fast decision incorporating coarse calorimeter and muon information. The second stage, the High Level Trigger (HLT), makes use of refined calorimeter and muon information and adds limited tracking. SUEP events typically have low efficiency for traditional triggers, which are designed to reject pile-up. However, there remain several possible analysis strategies utilizing data already collected during Run 2 of the LHC. 

In order to characterize the difficulty of observing lower mass mediators, several benchmarks points are used: mediator masses of $1000$~GeV, $750$~GeV, $400$~GeV, and $125$~GeV. Here, the choice of the $125$~GeV benchmark is motivated by the observed Higgs boson, which may itself serve as a portal to the hidden sector. The mediator is assumed to be produced via gluon fusion (GF). For the Higgs portal benchmark, associated production (ZH) also considered, where the Z boson decays leptonically. All signals are produced assuming dark mesons have a mass of 2 GeV, and the system has a temperature of 2 GeV. These dark mesons subsequently decay into a pair of dark photons. Multiple dark photon decays are explored for GF production using dark photon branching ratios; $BR(A'_{\gamma} \rightarrow u\bar{u})=100\%$, with $m(A'_{\gamma})=1$~GeV, $BR(A'_{\gamma} \rightarrow e\bar{e},\mu\bar{\mu},\pi\bar{\pi}) = (40,40,20)\%$ with $m(A'_{\gamma})= 0.5$~GeV labelled as leptonic, and $BR(A'_{\gamma} \rightarrow e\bar{e},\mu\bar{\mu},\pi\bar{\pi}) = (15,15,70)\%$ with $m(A'_{\gamma})= 0.7$~GeV labelled as hadronic.

Trigger strategies based on the scalar sum of hadronic activity ($H_{T}$) in the event can be used to target mediators produced via GF. In order to prevent extremely high background rates from QCD, ATLAS and CMS typically required $H_{T} > 500$~GeV at Level 1, and $H_{T} > 1$~TeV at HLT throughout Run 2. The vast majority of signal events which pass these requirements involve a mediator recoiling against an initial state radiation jet with high $p_{T}$. Trigger efficiency significantly decreases as the mediator mass decreases and the total energy deposition decreases. Figure~\ref{fig:ht} shows the $H_{T}$ distribution as well as the number of tracks for different SUEP mediator masses.

\begin{figure}
    \centering
    \includegraphics[width=0.4\textwidth]{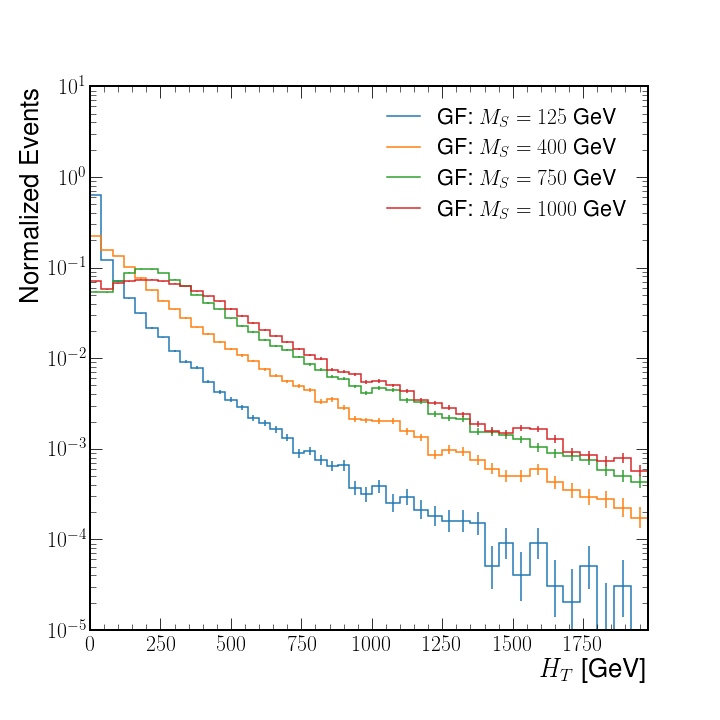}
    \includegraphics[width=0.4\textwidth]{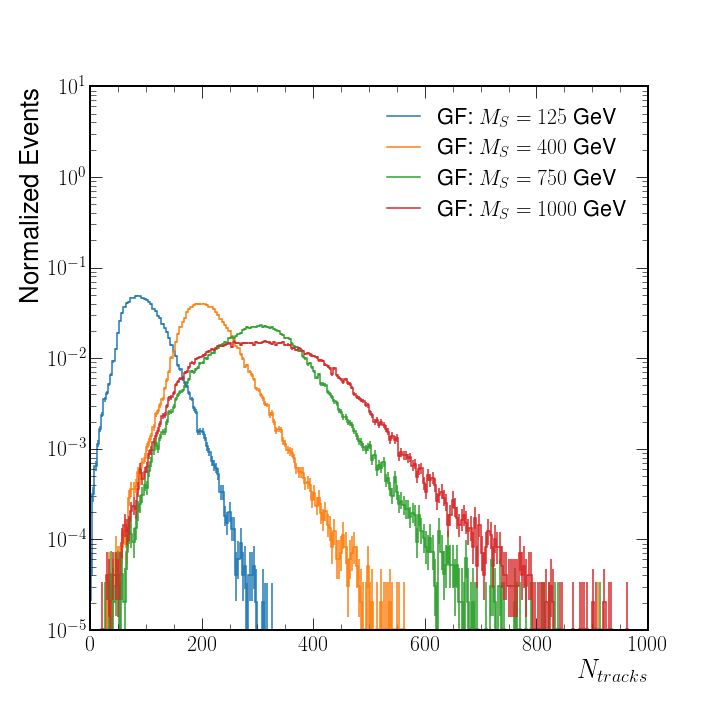}
    \caption{The $H_{T}$ distribution for different SUEP mediator masses is shown on the left. The $H_{T}$ is calculated by taking the scalar sum of $p_{T}$ for generator level jets with $p_{T}>20$ GeV and $|\eta|>4.7$. The number of tracks per event for different SUEP mediators is shown on the right. Tracks are defined as status~$=1$ charged particles with $p_T > 0.7$~GeV and $|\eta| < 2.5$. Both plots show Gluon Fusion (GF) production mechanism with a dark photon branching fraction of $BR(A'_{\gamma} \rightarrow u\bar{u})=100\%$. }
    \label{fig:ht}
\end{figure}

Depending on the production mechanism and branching fraction of the dark photon, SUEP events may contain multiple final state muons. In these cases, muon triggers designed to target vector boson or b-physics processes can provide much higher trigger efficiency than $H_{T}$ triggers, especially for lower mass mediators. Figure~\ref{fig:muon} shows the number of muons and the leading muon $p_{T}$ for several Higgs portal SUEP scenarios. Signals with larger leptonic branching ratios result in large multiplicities of moderate momentum muons, and can be targeted with tri-muon triggers. Leptons produced via associated production can be targeted with single and double-muon triggers.

\begin{figure}
    \centering
    \includegraphics[width=0.4\textwidth]{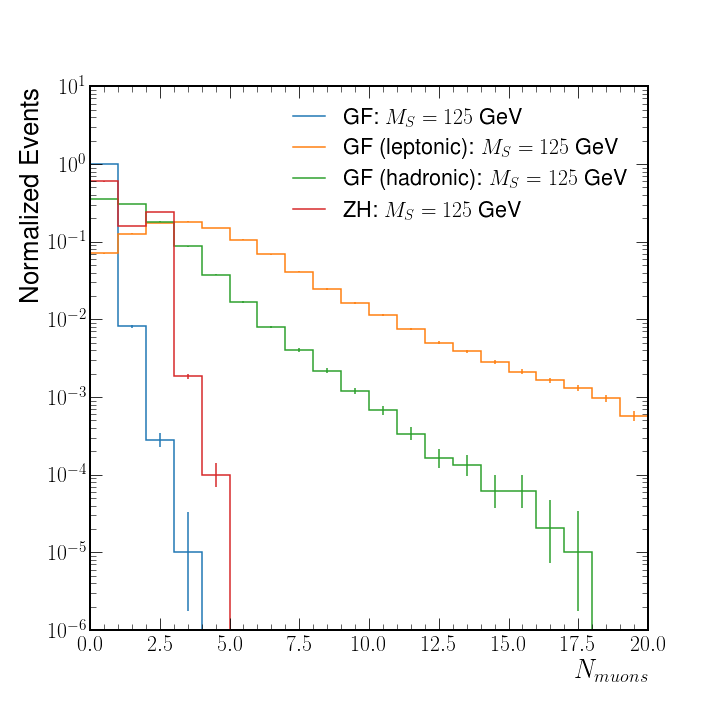}
    \includegraphics[width=0.4\textwidth]{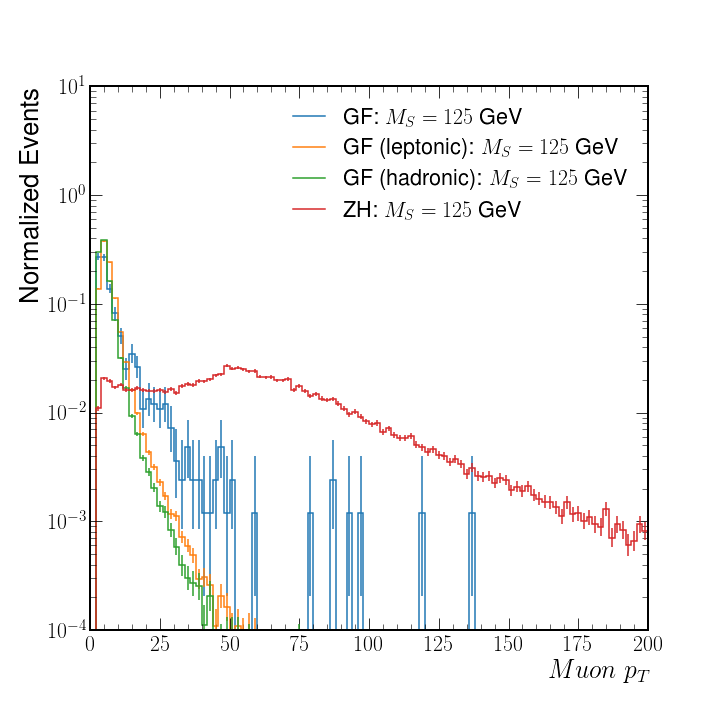}
    \caption{The number of muons at the generator level for different SUEP production mechanisms and dark photon decay branching fractions is shown on the left. Muons are selected with $p_T > 3$ GeV and $|\eta| < 2.5$. The muon with the highest $p_T$ is shown on the right for different SUEP production mechanisms and dark photon decay branching fractions.}
    \label{fig:muon}
\end{figure}

The upcoming Run 3 at the LHC offers an opportunity to design new triggers that specifically target SUEP signatures. One possibility is to use a standard $H_{T}$ or multi-jet trigger at Level 1, with a large multiplicity of tracks in the High Level Trigger. SUEP events are likely to be boosted after passing the Level 1 trigger, with SUEP decay products recoiling against Standard Model jets from initial state radiation. Figure~\ref{fig:suep_isotropy_ptcut} shows how events become less isotropic with increasing $\sum p_T$ requirements. As a result, using event shape observables would be suboptimal at HLT. In contrast, it is possible to choose an HLT track multiplicity requirement that is highly efficient for SUEP events, and also reduces backgrounds such that the $H_{T}$ requirement can be kept as low as Level 1 thresholds.

\begin{figure}
    \centering
    \includegraphics[width=0.7\textwidth]{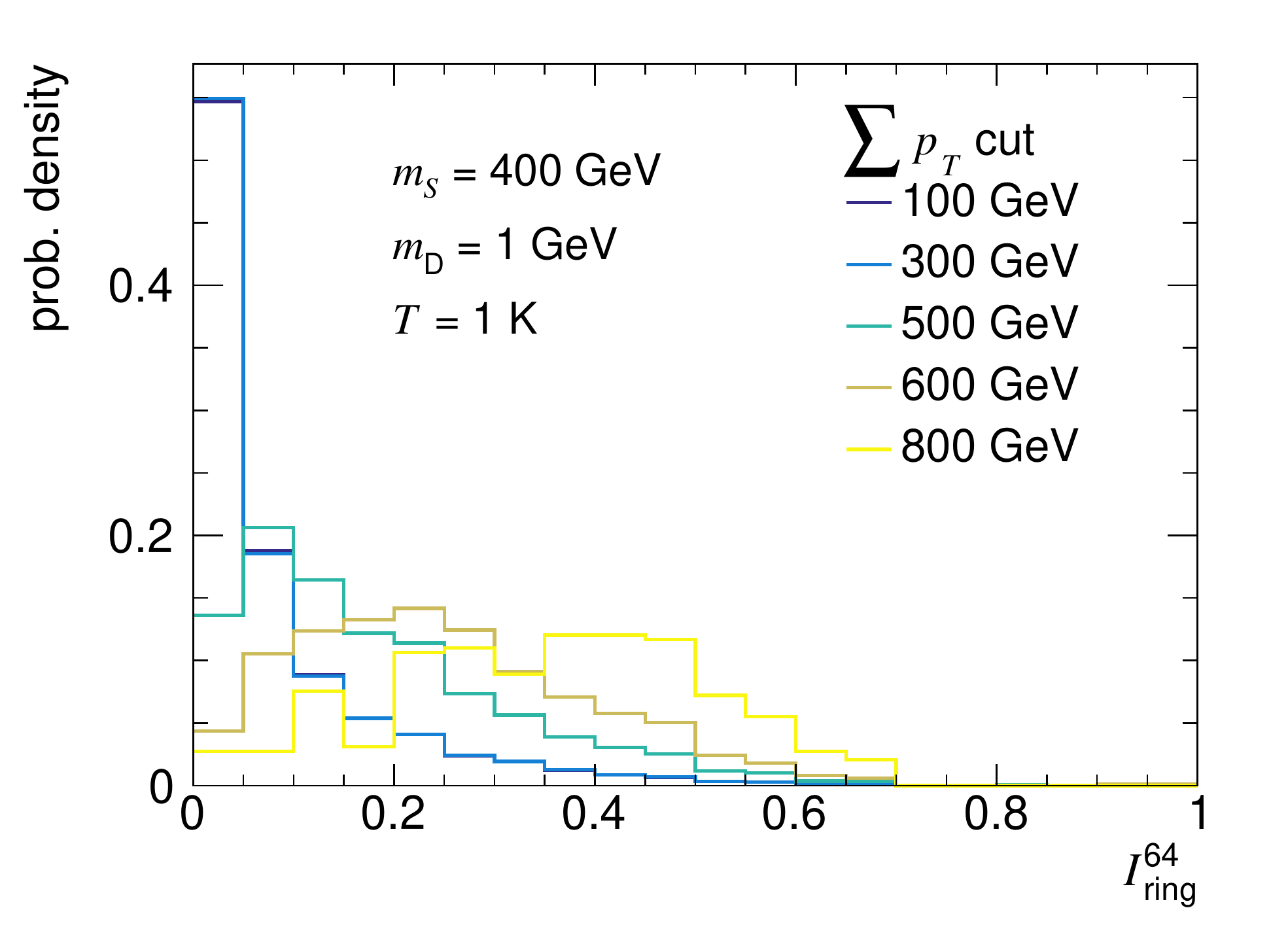}
    \caption{Isotropy distributions are shown for different values of the $\sum p_T$ event selection. The $\sum p_T$ is calculated for generator level tracks that are defined as status~$=1$ charged particles with $p_T > 0.1$~GeV and $|\eta| < 2.4$. The production mode is Gluon Fusion (GF) with a dark photon branching fraction of $BR(A'_{\gamma} \rightarrow e\bar{e},\mu\bar{\mu},\pi\bar{\pi}) = (40,40,20)\%$.}
    \label{fig:suep_isotropy_ptcut}
\end{figure}

A trigger which requires $H_T>500$~GeV and at least 150 tracks per event would yield a QCD efficiency one order of magnitude smaller than the currently employed $H_T>1050$~GeV trigger, while recovering nearly all the SUEP events that already pass the $H_T>500$~GeV selection. Note that this study assumes tracks can be reconstructed with nearly full efficiency and associated to the primary vertex.

If track reconstruction is too computationally expensive to run in the High Level trigger or has sub-optimal performance, it is possible to design a trigger which counts the number of hits in the innermost layers of the ATLAS and CMS tracking detectors. This approach was studied in Ref~\cite{Knapen:2016hky}. In addition to being less computationally expensive, hit counting is sensitive to even the softest particles in SUEP events. Charged particles with $p_{T}>\mathcal{O}(10)$~MeV can reach the innermost layer of the tracker and produce a hit. One trade-off is that hits cannot be associated to a primary vertex, and pile-up collisions increase the number of hits per event for background. Hits from SUEP events are likely to be localized in $z$ near the primary vertex, and this can be used to discriminate against pile-up. 

There are also several potential strategies to trigger on SUEP events directly at Level 1. When SUEP events are not boosted, final state particles create a band which spans all $\phi$ and are centered around a definite value of $\eta$. It may be possible to reduce thresholds by designing a trigger which looks for high $H_T$ in one $\eta$-slice compared to the rest of the event. 

Additional Level 1 strategies depend on the final state particles in the SUEP event. An all-electron or all-photon signal would create an abnormally high ratio of energy deposited in the Electromagnetic Calorimeter compared to the Hadronic Calorimeter. This method could be further refined by requiring Electromagnetic energy be centered around a single $\eta$-slice. Alternatively, SUEP events with dark matter particles in the final state, semi-visible SUEP could potentially be accessed via a missing-transverse-momentum trigger.

Both ATLAS and CMS will upgrade their detectors and trigger schemes at the High Luminosity LHC, offering further opportunities to design new SUEP triggers. The projected CMS trigger at the HL-LHC will reconstruct charged particles with $p_{T}>2$~GeV at Level 1. This new scheme would enable even more optimization of the trigger design at the L1 level. One could optimize a selection on the number of tracks at L1 to recover an even greater quantity of low $H_T$ events. This strategy would be most effective for higher temperature scenarios. 

Once a trigger strategy has been determined, it is essential to reconstruct the low-momentum charged particles associated with SUEP signatures. Standard ATLAS and CMS track reconstruction is highly efficient for charged particles which traverse roughly 8 layers of the silicon trackers \cite{ATLAS:2017kyn,CMS:2014pgm}. In CMS reconstruction is roughly $90\%$ efficient for charged particles with $p_{T} \geq 1$~GeV, and roughly $60\%$ efficient for particles with $p_{T}\sim 300$~MeV. In ATLAS, nominal track reconstruction has a minimum requirement of $p_{T} \geq 500$~MeV. Standard reconstructed tracks typically have impact parameter resolutions on the order of $100~\mu\mathrm{m}$~\cite{CMS-DP-2020-032}, which enables tracks to be associated to the primary vertex (proton-proton collision of interest). This primary vertex requirement ensures that the computed track multiplicity is not biased by nearby pile-up collisions. 

There are additional possibilities to reconstruct charged particles with even lower momenta. In ATLAS, tracks with $p_{T}>100$~MeV can be reconstructed for a small subset of events~\cite{McCormack:2718583}. These events would be processed with an additional pass of tracking, using leftover hits, in a region of interest of $z \sim 1$~mm around the primary vertex.
To access even lower momentum charged particles, $p_{T} \sim 10$~MeV, it is possible to count the number of hits in the inner most layers of the detector. While this strategy would increase the acceptance for extremely low-momentum SUEP particles, it is impossible to associate hits to a particular primary vertex. This strategy also requires accessing low level data which may not be possible due to storage constraints.

For SUEP signatures with higher temperatures, a significant number of final state Standard Model particles will have moderate momenta, $p_{T}>3$ GeV. For these scenarios, it is possible to identify final state particles as muons, electrons, or hadrons. Events with a large multiplicity of low momentum leptons would be a particularly striking signal, and this information could be used to further improve  signal to background discrimination.

After reconstruction, the unique properties of SUEP events can then be used to separate against signal from QCD background. The two most powerful observables include the characteristic high track multiplicity, and the isotropic distribution of such tracks.

For analyses using data that were collected during Run 2, trigger strategies will likely rely on the presence of additional objects produced in association with the SUEP shower, either based on a high quantity of ISR QCD or a massive gauge boson decaying to triggering objects. In both cases, a simple procedure to recover the SUEP shower can be followed. First, the triggering object can be identified with standard reconstruction techniques and subtracted from the overall event representation. Second, the remaining tracks can be boosted against the triggering object. In case the later is produced back-to-back with the mediator, these would allow to recover both the track multiplicity and spherically symmetrical distribution of the SUEP shower.

Recently, the anomaly detection techniques as a generic way to search for new physics have been incorporated to SUEP searches. A proposal based on autoencoders \cite{Barron:2021btf} shows the strength of such techniques for low mediator masses and temperatures. Additional details on this approach are presented in Section \ref{autoencodersSUEP}.

\subsection{Glueballs}\label{glueballs}

\emph{Contributors: David Curtin, Caleb Gemmell, Christopher B. Verhaaren}

In the $\Nfdark = 0$ limit for $SU(\Ncdark)$ Yang-Mills theories, the only hadronic states that form below the confinement scale are glueballs, composite gluon states. This limit is unique because there are no light degrees of freedom below the confinement scale. In the absence of such light states color flux tubes cannot break via the creation of quark-antiquark pairs. This process is essential to present QCD hadronization models~\cite{ANDERSSON198331,WEBBER1984492}, thus the usual understanding of hadronization does not directly apply to the $\Nfdark = 0$ limit. This qualitative difference has hindered efforts to study dark glueball showers in scenarios with pure Yang-Mills dynamics.

Sectors with $\Nfdark = 0$ $SU(\Ncdark)$ Yang-Mills descriptions commonly appear in neutral naturalness theories, such as Twin Higgs models \cite{Chacko:2005pe,Craig:2015pha}, Folded Supersymmetry \cite{Burdman:2006tz} and many others \cite{Barbieri:2005ri,Cai:2008au,Cohen:2018mgv,Cheng:2018gvu}. The glueballs of a hidden confining sector have also been considered as dark matter candidates \cite{Boddy:2014yra,GarciaGarcia:2015fol,Soni:2016gzf,Forestell:2016qhc}. Thus, $\Nfdark = 0$ $SU(\Ncdark)$ Yang-Mills models are motivated by possible solutions to the Little Hierarchy problem and the unknown nature of dark matter. Current ignorance of the pure-glue hadronization process has left these models in an largely unstudied corner of motivated parameter space. The hadronization process determines the final state multiplicity and energy distribution of dark glueballs, which are essential for both collider and indirect detection studies.

The properties of the glueballs themselves are relatively well-known, having been studied in lattice gauge theory \cite{Morningstar:1999rf,Yamanaka:2019aeq,Yamanaka:2019yek,Athenodorou:2021qvs,Yamanaka:2021xqh}.   In the absence of external couplings, these studies have established a spectrum of 12 stable glueball states characterized by their $J^{PC}$ quantum numbers. When considered as part of a dark sector, these glueballs can be stable or decay through a variety of portals to the SM~\cite{Juknevich:2009gg, Juknevich:2009ji}, with possibly long lifetimes on collider or cosmological time scales. This spectrum can entirely be parameterised by the confinement scale of the theory, or equivalently lightest glueball mass, $m_0 \sim 6\lamdark$. These glueball properties are also known for several $\Ncdark \neq 3$ which paves the way for studying exotic dark sectors outside the standard dark $SU(3)$ case \cite{Boddy:2014yra,Soni:2016gzf,Batell:2020qad,Kilic:2021zqu}.

Recently, efforts have been made to enable quantitative studies of pure Yang-Mills parton showers and hadronization~\cite{Curtin:2022tou}. This includes the creation of a new public python package, \texttt{GlueShower}.\footnote{\href{https://github.com/davidrcurtin/GlueShower}{github.com/davidrcurtin/GlueShower}} 
This package allows users to simulate dark glueball showers produced from an initial pair of dark gluons. 
\texttt{GlueShower} combines a perturbative pure-glue parton shower with a self-consistent and physically motivated parameterization of our ignorance regarding the unknown glueball hadronization behaviour. 
Two qualitatively different hadronization possibilities are included: a more physically motivated jet-like assumption, and a more exotic plasma-like option that accounts for the possibility that color-singlet gluon-plasma-states are created by hypothetical non-perturbative effects far above the confinement scale. Each such plasma-ball then decays isotropically to glueballs in its restframe, somewhat akin to dark hadron production in SUEP scenarios~\cite{Strassler:2008bv,Knapen:2016hky,Barron:2021btf}. 
Within each hadronization option, two nuisance parameters control the hadronization scale and the hadronization temperature, mostly controlling the glueball multiplicity and relative abundance of different glueball species respectively.

The study~\cite{Curtin:2022tou} defines a set of 4 benchmark points for these nuisance parameters in each option to represent the range of physically reasonable glueball hadronization possibilities. For phenomenological studies, the range of predictions spanned by these hadronization benchmarks can be interpreted as a theoretical uncertainty on predictions for glueball production. Despite the wide range of possibilities for hadronization that are considered, most glueball observables are predicted within an $\mathcal{O}(1)$ factor. This is in large part due to the modest hierarchy between the glueball mass and the confinement scale, which makes most inclusive glueball observables dominantly dependant on the physics of the perturbative gluon shower. However, exclusive production of certain glueball states can vary by up to a factor of 10 depending on hadronization assumptions, a range that accurately represents our current theoretical uncertainty. 

In summary, to support the increasing interest in dark showers, efforts are being made to ensure the possibility space is covered. Until recently the $\Nfdark = 0$ limit has been largely ignored due to difficulties in studying pure-glue hadronization. However, such hidden sectors are motivated by both naturalness concerns and as a possible dark matter particle. The \texttt{GlueShower} tool aims to effectively facilitate studying the $\Nfdark = 0$ limit, even without a full understanding of the underlying non-perturbative hadronization physics. Despite the current theoretical limitations, this demonstrates that quantitative studies of and searches for glueball signatures can be reliably conducted if the underlying  uncertainties are accurately accounted for.
\clearpage

\section{Simulation tool limitations and how to build consistent benchmarks from the underlying physical parameters for semi-visible jets}
\label{sec:model_building}
\subsection{Consistent parameter setting and roadmap for improving on the simulation of dark showers}\label{parameters}

\emph{Contributors: Suchita Kulkarni, Se\'an Mee, Matt Strassler}

The non-perturbative nature of QCD-like strongly interacting scenarios makes it impossible to set consistent UV and IR parameters based purely on perturbative analysis. In this section, we address this problem, going beyond existing efforts in the literature.  First, we sketch the importance of lattice calculations to set low energy bound state masses given UV parameters.  Second, we illustrate the importance of portal phenomenology and associated symmetry breaking patterns, and use chiral Lagrangian techniques to set the interactions of the low energy bound states among themselves and to the SM final states. Finally, we comment on the hadronization parameters necessary for LHC phenomenology and make some observations for a subset of them. After this, we turn to the simulation of benchmark models consistent with the above observations.  We describe the recent improvements to the \PYTHIA8 Hidden Valley module, and present a few benchmarks which are used in later sections for studies.

\subsubsection{UV scenarios: SM extension with non-Abelian gauge groups}
\label{sec:mass_spec}
We suppose the Standard Model is extended with a new sector, consisting of an additional non-Abelian gauge group $SU(\Ncdark)$ with $\Nfdark$  degenerate Dirac fermions $\PqdarkA$ in the fundamental representation, with current mass $\mqdark$.  We will refer to the new sector as the ``dark sector,'' though we note this is fully equivalent to a ``hidden valley'' as defined in~\cite{Strassler:2006im}.  This sector has $SU_L(\Nfdark) \times SU_R(\Nfdark)\times U(1)_B$ global symmetry broken by the mass term to a diagonal $SU(\Nfdark)\times U(1)$\footnote{There is also an axial $U(1)$ which is broken by $\mqdark$ and an anomaly, though the anomaly disappears in the large $\Ncdark$ limit.}.  We will assume $\Ncdark,\Nfdark$ are such that the theory confines and has a chiral $\PqdarkA\PaqdarkA$ condensate, which in the absence of $\mqdark$ would spontaneously break the chiral symmetries and lead to Nambu-Goldstone bosons.  Instead, as in QCD itself, we have pseudo-Nambu-Goldstone bosons with masses that are proportional to $\sqrt{\mqdark}$. 

If the only connection between this sector and the SM is through a mediator (or ``portal'') which is either massive or ultra-weakly coupled, then typical confining Hidden-Valley-type phenomenology inevitably results. Specifically, production of the ``dark quarks'' leads to production of dark hadrons.  These will be collimated in jets if the theory is QCD-like and the invariant mass of the produced $\Pqdark\Paqdark$ pair is far above the dark confinement scale.  We will specifically consider a mediator in the form of a heavy $U(1)^\prime$  leptophobic $\PZprime$ mediator between SM quarks and the dark quarks. These are the same models introduced in sec.~\ref{schannel}, used widely in the semi-visible jet searches and are also considered in the dark matter working group~\cite{Boveia:2016mrp}.   The process $\Pq\Paq \to \PZprime \to \Pqdark\Paqdark$ allows dark hadrons to be produced, while the $Z^\prime$ also allows some dark hadrons to decay to SM quark-antiquark pairs, leading to an all-hadronic signal. 
We will assume throughout that  $\mZprime$ is much larger than the confinement scale, by at least $\sim 30$, so that the physics in the dark sector actually leads to jets of dark hadrons.\footnote{If the two scales are too close, then the physics is analogous to $e^+e^-\to$ QCD hadrons at a few GeV: multiplicities are of order 4 to 6 and no jetty structure is seen.} Because a fraction of the dark hadrons are typically stable, at least on LHC-detector time scales, the observable signal usually consists of at least two relatively fat jets with considerable substructure, and often a high multiplicity of SM hadrons, along with roughly collinear \met.  These ``semi-visible jets'' (SVJ), introduced in sec.~\ref{schannel}, are the target of the searches in question here.

The mediator's couplings to the dark sector will break the $SU(\Nfdark)\times U(1)$ flavor symmetry to a smaller subgroup $G_f$. Without this breaking, the majority of the dark hadrons would be charged in the adjoint of $SU(\Nfdark)$ and would be unable to decay to a SM final state, which would be a singlet under this symmetry.  (Note that decays with both dark and SM particles in the final state would still be permitted.) If the couplings are vector-like, we assign the dark quarks charges $Q_\alpha$ under the $U(1)^\prime$, and define the charge matrix ${\bf Q}$ as a diagonal matrix with eigenvalues $Q_\alpha$; the group $G_f$ is the subgroup of $SU(\Nfdark)\times U(1)$ which commutes with ${\bf Q}$. (In chiral models the left- and right-handed quarks have different charges $Q_\alpha$ and $\tilde Q_\alpha$, and get their masses from a Higgs field with charge $Q_\alpha-\tilde Q_\alpha$. We will not consider such models in detail here.)  The precise choice of ${\bf Q}$ has a significant impact on the phenomenology.

The ultraviolet Lagrangian for the hidden sector is  
\begin{equation}
    \mathcal{L_D}\subset -G_{D,\mu\nu}G_D^{\mu\nu} + \Paqdark i \slashed{D} \Pqdark - \mqdark \Paqdark \Pqdark
\end{equation}
where $G_{D,\mu\nu}$ is the gauge field strength tensor and $\mqdark$ is the current mass of the dark quarks. This part of the theory has two discrete parameters $\Ncdark$ and $\Nfdark$, and two continuous parameters, the running gauge coupling $\adark(\mu)$, with $\mu$ a renormalization scale, and the ``current'' quark mass $\mqdark$.  Since neither of these parameters has direct contact with the observable phenomena, we replace them with the confinement scale $\lamdark$, or some proxy for it, such as the one-loop dimensional transmutation scale, and the mass $\mpidark$ of the light pseudoscalar mesons $\Ppidark$.

The interaction of this sector with the SM via  a $\PZprime$ mediator takes the form
\begin{equation}
    \label{eqn:lag}
    \mathcal{L_{{\rm int}}}\subset -e_D \PZprime_\mu \sum_{\alpha}\PaqdarkA Q_\alpha \gamma^{\mu} \PqdarkA - \gq\,\PZprime_\mu \sum_{i} \Paq_i\gamma^{\mu}\Pq_i 
\end{equation}
if the $\PZprime$ couplings are vectorlike and the $\PqdarkA$ have charge $Q_\alpha$.  (If the couplings to the hidden sector are chiral, separate charge $Q_\alpha$ and $\tilde Q_\alpha$ must be assigned for left- and right-handed hidden quarks.)
We will assume all SM quarks have the same charge under $U(1)^\prime$ for simplicity, though in realistic models one must account for the differences, which can affect observables, such as the SM heavy flavor fractions in the SVJs. 

One other issue of importance is whether the $\PZprime$ and the $\Pqdark$ obtain their masses from the same source, in such a way that the longitudinal polarization of the $\PZprime$ mixes with the $\Ppidark$ from chiral symmetry breaking.  Similar mixing of the SM charged pion with the $\PW$ bosons allows the classic decay $\pi\to\mu\nu$.  By analogy, the $\PZprime$ mixing with the $\Ppidark$ affects the decays of the latter to the SM.

\subsubsection{From ultra-violet theories to infrared parameters}

The $SU(\Ncdark)$ confines at around the scale $\lamdark$ and various dark hadronic bound states are produced.  In the exact $SU(\Nfdark)$ limit, the lightest hadrons consist of the spin-0 flavor-adjoint $\Ppidark$, which are pseudo-Nambu-Goldstone bosons (PNGBs), the spin-1 flavor-adjoint $\Prhodark$, the spin-1 flavor-singlet $\Pomegadark$, and the spin-0 flavor-singlet $\Petaprimedark$.  In general $\mpidark<\mrhodark\lessapprox \momegadark$, while the $\Petaprimedark$ mass depends on the anomaly, which scales like $\sqrt{\Nfdark/\Ncdark}$ when it is dominant. Thus  for $\Nfdark$ flavors, the theory contains $\Nfdark^2 -1$ mass-degenerate pions and an equal number of degenerate rho mesons, along with an omega which will be slightly heavier than the rhos, and an eta-prime which may be near-degenerate with the pions for $\Nfdark\ll \Ncdark$ but is much heavier for $\Nfdark\sim \Ncdark$.  In addition there are baryons and antibaryons with mass of order $\Ncdark \lamdark$ (bosons for $\Ncdark$ even, fermions otherwise), except for $\Ncdark=2$ in which case they are exactly degenerate with the pions and are themselves PNGBs. We note from these remarks that that $\Nfdark=1$ and $\Ncdark=2$ are special cases, which must be treated with care.

There are ambiguities in defining the scale $\lamdark$.  One way to define it is via the running gauge coupling $\adark(\mu)$ at one loop.  As pointed out by 't Hooft, physics in the large $\Ncdark$ limit depends mainly on $\adark(\mu)\Ncdark$, up to $1/\Ncdark$ corrections.  Lattice results show that $\Ncdark=3$ is already close to $\Ncdark=\infty$, and moreover the physics of a QCD-like shower has very small $1/\Ncdark$ corrections, a fact that the \PYTHIA showering routines take advantage of.  With this in mind, the one-loop running coupling can be written in a form familiar from QCD: 
\begin{equation}
    \label{eq:one_loop_lambda}
    \adark(\mu^2) \Ncdark = \displaystyle\left[{\displaystyle\frac{1}{2\pi}\left(\frac{11}{3}-\frac23 \,\frac{\Nfdark}{\Ncdark}\right) \log\left(\frac{\mu}{\lamdark}\right)}\right]^{-1}. 
\end{equation}
This form emphasizes that the physics of this theory is really a function of $\Nfdark/\Ncdark$, with $1/\Ncdark$ corrections, for large $\Ncdark$. To this end in  fig.~\ref{eq:one_loop_lambda}(left), we show the running of $\alpha_D$ at one lop for several values of $\Ncdark/\Nfdark$ for a fixed $\Lambda_D = 1$ GeV. 

Although the one-loop formula for $\lamdark$ is currently used in the \PYTHIA8 HV module and by us throughout this document, this situation should be viewed as temporary.  Inevitably, this method indexes non-pertubative hadronic masses to a scale which is perturbative, and this may pose challenges for interpreting results from lattice gauge theory in which $\lamdark$ is often defined non-perturbatively, for instance through the string tension. Moreover, the connection between this one-loop estimate of $\lamdark$ and the physical confinement scale becomes less and less accurate as $\Nfdark/\Ncdark$ increases; two-loop effects become important, with the effect that non-perturbative definitions of $\lamdark$ will be much smaller than the one-loop definition.  It seems likely a two-loop perturbative definition would be significantly closer to non-perturbative ones. To illustrate for this effect in fig.~\ref{eq:one_loop_lambda}(right) we demonstrate the effect of two loop correction for running of $\alpha_D$ (solid lines) in comparison with the corresponding one loop correction (dashed lines), for two different values of $\Nfdark = 3,6$ with fixed value of $\Ncdark = 3$ keeping $\Lambda_D = 1$ GeV. In order to derive this result, we have used the procedure as described in ~\cite{vanRitbergen:1997va} with beta function as defined in ~\cite{Prosperi:2006hx}. 

In fact for sufficiently large $\Nfdark/\Ncdark$ the theory will no longer confine because its running coupling reaches an infrared fixed point. The values for $\Ncdark,\Nfdark$ where this occurs are not precisely known. For various $\Ncdark$, estimates of the value of $\Nfdark$ above which the theory is believed not to confine are computed in e.g.~\cite{Lee:2020ihn} and tabulated in table~\ref{tab:infrared_fixed}.  These results suggest that theories with $\Nfdark < 3\Ncdark$ likely confine, and we will refer to them as ``QCD-like'' theories.  

Again, we have at this time only used the one-loop formula above to define $\lamdark$, and we use the same definition for the parameter {\tt Lambda} in \PYTHIA8, which is perhaps reasonable for $\Nfdark/\Ncdark\sim 1$ or below.  But it is important to note that for $\Nfdark/\Ncdark\to 3$,  the one-loop running is inaccurate and at a minimum a two-loop formula (within which the fixed points at large $\Ncdark, \Nfdark$ can be observed) ought to be used.  

\begin{table}[h!]
    \centering
    \begin{tabular}{ |c |c |c |c |c|}
        \hline
        $\Ncdark$ & 3 & 4 & 5 & 6\\ 
        \hline
        $\Nfdark$ & 9 & 13 & 16 & 18\\  
        \hline
    \end{tabular}
    \caption{Values of $\Ncdark, \Nfdark$ which lead to asymptotically free $SU(\Ncdark)$ theories.  It is important to stay well below these values in order to remain within the QCD-like scenarios of current interest to us. 
}
    \label{tab:infrared_fixed}
\end{table}

\begin{figure}[h!]
    \centering
        \includegraphics[width=0.45\textwidth]{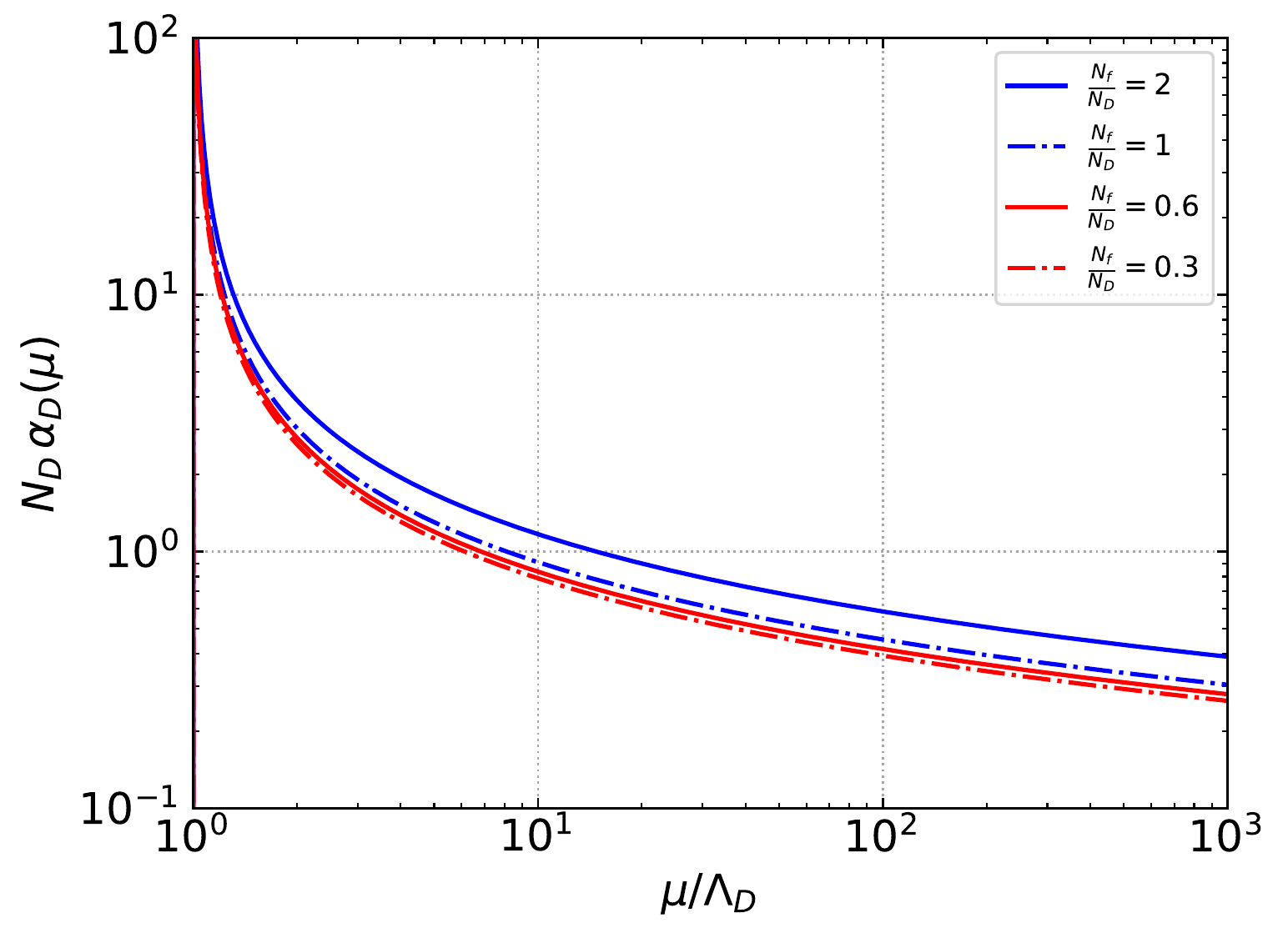}
        \includegraphics[width=0.45\textwidth]{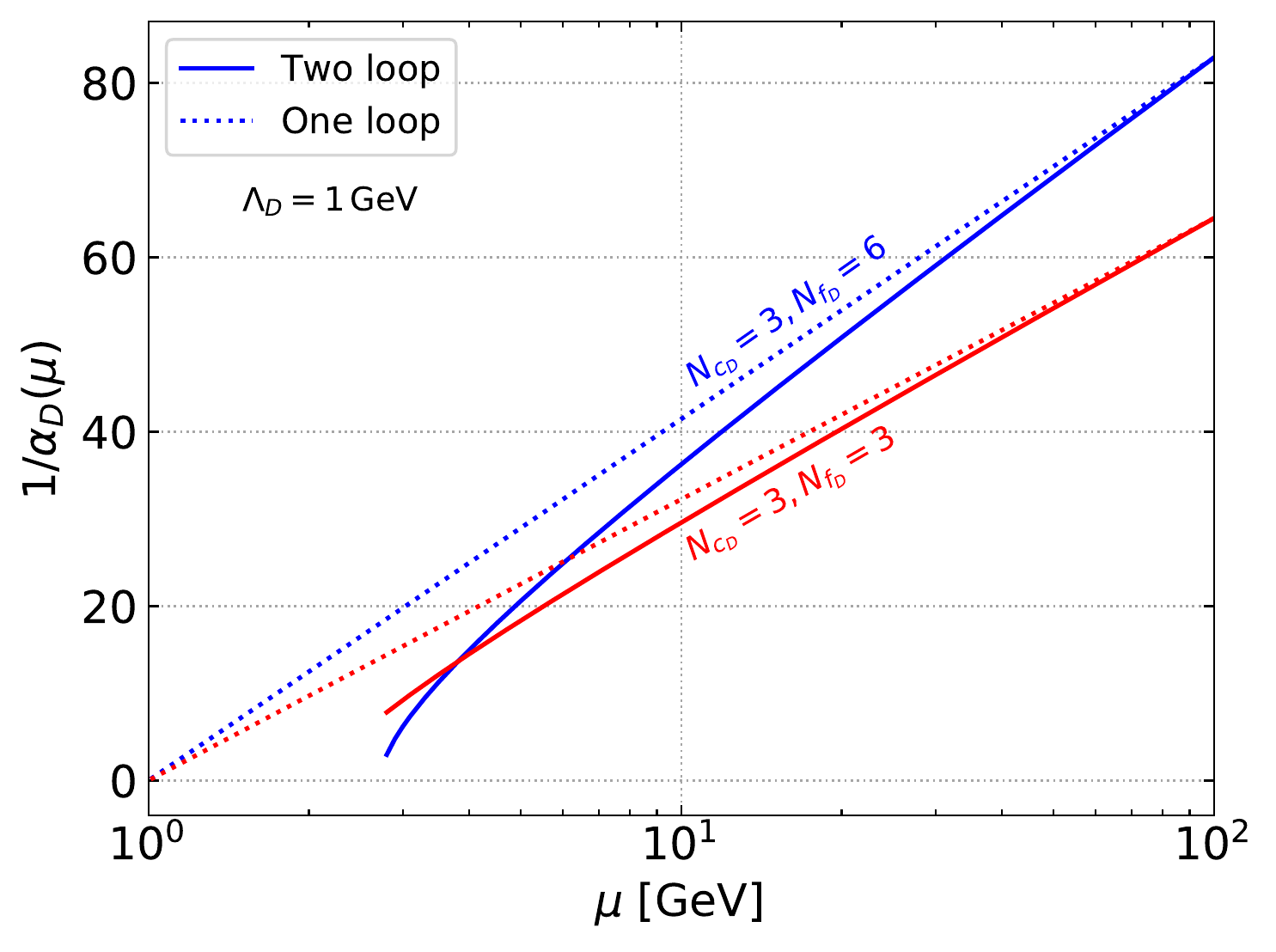}
    \caption{Left: The 't Hooft coupling $\Ncdark\,\adark(\mu)$, computed at one loop, as a function of $\mu/\lamdark$ for several values of $\Nfdark/\Ncdark$. Right: Two loop running of $\adark(\mu)$ for $\Ncdark = 3, \Nfdark = 3, \,6\, \rm{and}\, \lamdark = 1\,\rm{GeV}$. }
    \label{fig:alphaD_running}
\end{figure}

The masses and couplings of the low-lying bound states are a direct consequence of UV parameters, but are not calculable analytically. Rough estimates of these quantities can be obtained by combining lattice gauge theory calculations with the chiral Lagrangian for spin-0 mesons, extended by including spin-one mesons as though they were flavor-symmetry gauge bosons. It is convenient to set the overall scale of the dark hadrons using a non-perturbative definition of the confinement scale, which we will call $\tlamdark$ and specify later, and to express all other dimensionful dark-hadronic quantities in terms of this parameter. Once we have fixed $\tlamdark$ as the overall hadronic scale, and specified the ratio
$$\frac{\mpidark}{\tlamdark}\propto\sqrt{\frac{\mqdark}{\tlamdark}}$$
(where the relation $\mpidark\sim \sqrt{\mqdark}$ follows from the chiral Lagrangian), everything else should be computable in principle. 

Such lattice calculations for mass degenerate fermions in the fundamental representation of $SU(N)$ gauge theories are available in abundance, albeit in the quenched approximation, see e.g.~\cite{DeGrand:2012hd, Bali:2007kt, Bali:2008an, DelDebbio:2007wk}. A particularly useful resource is~\cite{Bali:2013kia}, which summarises the spectrum of mesons in the large-$N$ limit of QCD-like theories. These calculations can be used to determine the ratios of the dark hadron masses as a function of the hidden sector parameters.

Using lattice calculations and fits plotted in Figure 19 of~\cite{Fischer:2006ub}, we can relate the dark quark mass to the dark pion mass and the dark rho mass.  We express these in terms of $\tlamdark$ defined as the chiral limit ($m_\pi\to0$) of the $\rho$ mass divided by 2.37.  (In terms of the physical units used in Figure 19 of~\cite{Fischer:2006ub}, this  puts the analogue of $\tlamdark$ for physical QCD at 300 MeV.) These relations are concretely shown in fig.~\ref{fig:lattice_fits_plots}.  Our analytic fits to the curves shown are
\begin{equation}
    \label{eq:lattice_fits}
    \displaystyle\frac{\mpidark}{\tlamdark} = 5.5\,\sqrt{\displaystyle \frac{\mqdark}{\tlamdark}} \qquad \qquad \qquad 
    \displaystyle\frac{\mrhodark}{\tlamdark} = \sqrt{5.76 + 1.5
    \displaystyle\frac{\mpidark^2}{\tlamdark^2}} \, .
\end{equation}
The fit functions and coefficients shown in Eq.~\ref{eq:lattice_fits} are appropriate for small $\mpidark/\tlamdark$, though they work far beyond this expectation, and begin to differ from lattice computations by $>10\%$ only for $\mpidark/\tlamdark > 2.3$.

\begin{figure}[h!]
    \centering
        \includegraphics[width=0.45\textwidth]{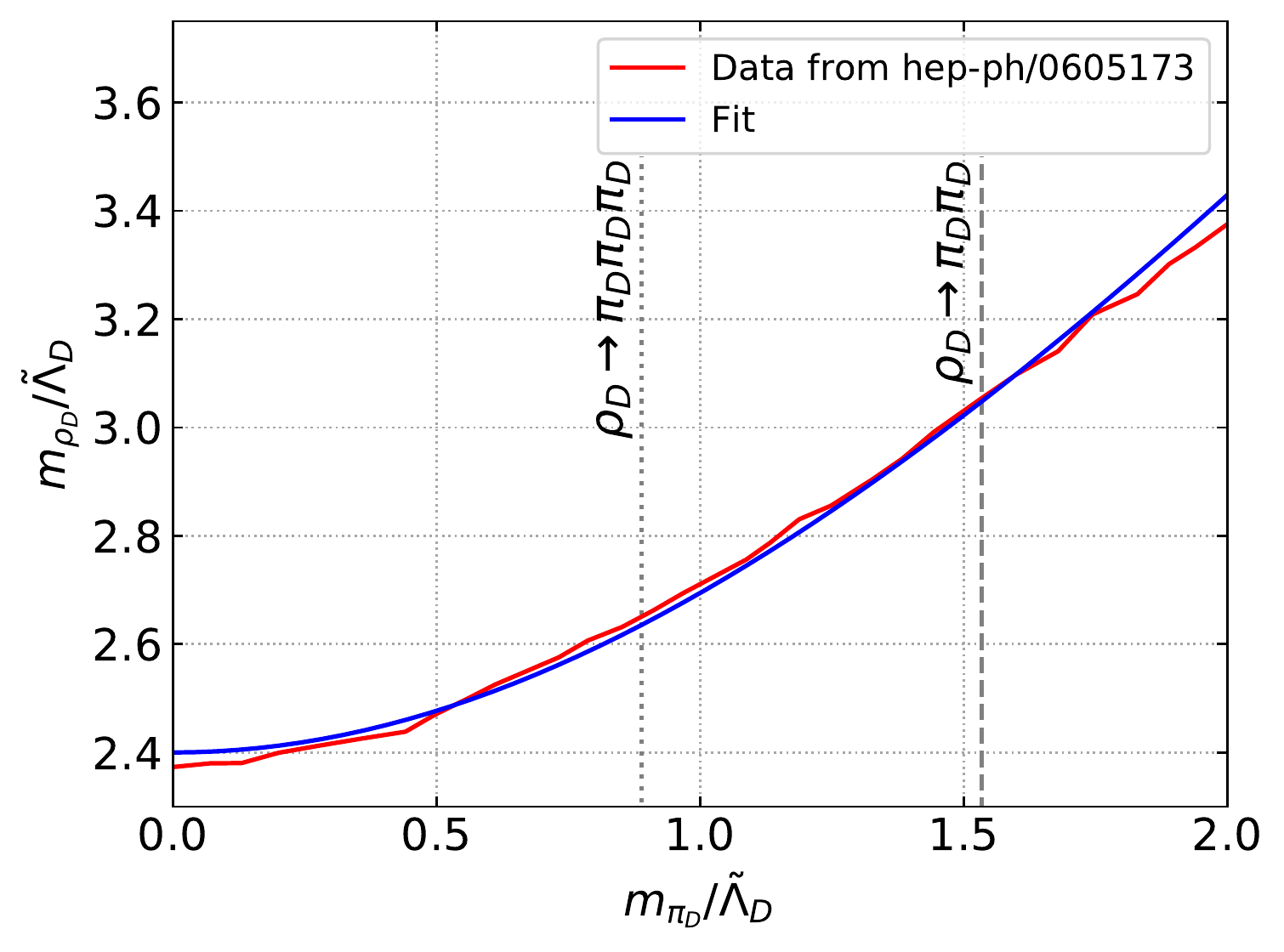}
        \includegraphics[width=0.42\textwidth]{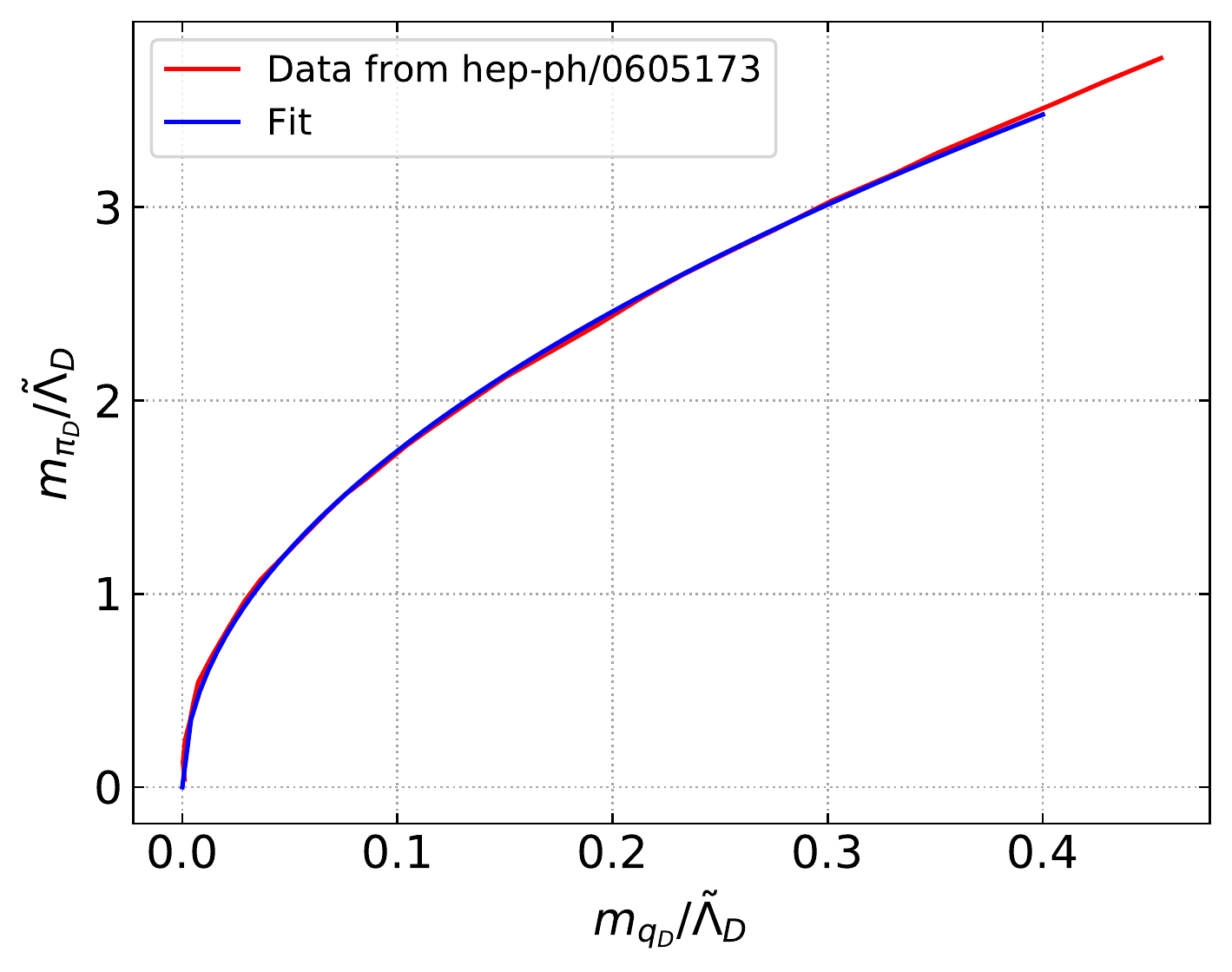}
    \caption{Fits given in Eq.~\ref{eq:lattice_fits} for the $\Prhodark$ mass (left) and $\Ppidark$ mass (right) to results from lattice simulations~\cite{Fischer:2006ub}. The left panel also indicates the kinematic thresholds for $\Prhodark$ to decay to $\Ppidark\Ppidark$ and $\Ppidark\Ppidark\Ppidark$.}
    \label{fig:lattice_fits_plots}
\end{figure}

The relation between the perturbative $\lamdark$ of Eq.~\ref{eq:one_loop_lambda} (or its higher-loop version) and the non-perturbative $\tlamdark$, defined in terms of the chiral limit of $\mrhodark$ (or some other similar definition), is not established.  Although they are proportional, the proportionality depends on $\Ncdark$ and $\Nfdark$ (mostly on $\Nfdark/\Ncdark$.)  {\bf In what follows, we will assume they are the same.}  The uncertainties and inaccuracies that result from this choice can only be reduced in future through more careful matching between perturbative and non-peturbative quantities.

The spin-1 singlet is expected to be nearly degenerate with the spin-1 adjoint hadrons, so there is little importance in giving it a different mass\footnote{Its decays are potentially another matter, but we have not yet attempted to treat them carefully.}. By contrast, the spin-0 singlet is expected to be heavier than the spin-0 adjoint, possibly by a large amount if $\Nfdark\sim \Ncdark$, due to the axial anomaly.  Some analysis of QCD hadrons using the chiral Lagrangian, which we will present elsewhere, suggests 
\begin{equation}
\label{eq:etap_mass}
\metaprimedark^2 \approx \mpidark^2 + \frac{\Nfdark}{\Ncdark}(3\tlamdark)^2 \ .
\end{equation}
Thus for $\Nfdark\sim \Ncdark$, as in SM QCD, the splitting of the spin-0 singlet from the adjoint is large.
Note that the factor of $3$ in front of $\tlamdark$ depends on the precise definition of the corresponding $\tilde\Lambda_{QCD}$, and will retain some uncertainties until this definition is handled more carefully. In any case, our current benchmark models presented below do not account for the anomaly term, and instead treat the $\Petaprimedark$ as degenerate with the $\Ppidark$.  However, the new version of \PYTHIA8 includes a parameter {\tt HiddenValley:}{\tt separateFlav}, described in sec.~\ref{sec:hadronization_param} below; when it is set ``on,'' the $\Petaprimedark$ mass can be set separately from that of the $\Ppidark$ states.

For the scenarios we are interested in, a hard process leads to production of dark quarks which shower over a wide energy range; the shower then subsequently hadronizes.  Hadronization in the dark sector is a far more challenging problem, because neither lattice calculations nor effective field theory methods are applicable to this process. All we know of it arises from studies of QCD data, and in particular through the use of phenomenological models (such as the Lund string model used in \PYTHIA, or the clustering model used in {\tt Herwig}) whose parameters are tuned to fit experimental results.  Currently there is no theoretical insight into how these models or their parameters should be adjusted for different values of $\Ncdark$, $\Nfdark$, or $\mpidark/\lamdark$.  Consequently we take existing parameterizations from data as a starting point.  One must vary these parameters within reason to obtain a sense of uncertainties.

In the present context, we are using \PYTHIA8's HV module, whose four main parameters  {\tt HiddenValley:}{\tt aLund}, {\tt HiddenValley:}{\tt bmqv2}, {\tt HiddenValley:}{\tt rFactqv}, {\tt HiddenValley:}{\tt sigmamqv} parallel those of the corresponding QCD hadronization routine.  The dimensionless HV parameters are set to exactly the same values as the dimensionless QCD parameters, while those with dimensions are scaled by the ratio of constituent quark mass parameter {\tt mqv} = {\tt 4900101:m0} (which is {\it not} the current quark mass $\mqdark$ but rather a phenomenological parameter, of order $\lamdark$ for small $\mqdark$) to the constituent quark mass in QCD, which for u, d quarks is $330$ MeV.  There are other parameter tunes proposed by \PYTHIA8 experts, see for example the Monash tune~\cite{Skands:2014pea}.  It is probably wise to try two or more tunes that are known to work in QCD as a means of estimating a minimum systematic error from this source.  However, we have not studied this, and so further investigation is needed before an informed recommendation can be made.

There are three other hadronization parameters that are currently in use in the HV module.  About the parameter {\tt HiddenValley:probVector}, which gives the probability that a new meson formed in the hadronization should be assigned to spin-1 rather than spin-0, we have two pieces of information.  Were spin-0 and spin-1 mesons mass-degenerate (appropriate for $\mpidark/\lamdark\gg 1$ and bordering on unphysical for the Lund model), we would expect probVector=0.75 based on spin counting (three spin-1 states versus one spin-0 state.)  Data from QCD, with $m_\pi/\Lambda_{QCD}\sim 0.5$, suggests use of probVector=0.5, downweighting spin-1 presumably because of phase space.\footnote{see settings for {\tt StringFlav:mesonUDvector} in \PYTHIA8. It should be noted this value in the SM is subject to the tune used, for example the default \PYTHIA8 value is 0.62, which is reduced to 0.5 in Monash tune.}.  From this we learn that the appropriate probVector is a slowly increasing function of $m_\pi/\Lambda_{QCD}$.
It would be reasonable to choose a phase-space-motivated functional form for this function, with a smooth $\mpidark\to 0$ limit, but we have not made an effort to do this.  Little is known about the limit $\mpidark\to 0$; it is not even clear that the Lund model is accurate there.

When the parameter {\tt HiddenValley:}{\tt separateFlav} (included in the new version of \PYTHIA8  and described in sec.~\ref{sec:hadronization_param})\, is set ``on'', the parameter {\tt HiddenValley:}{\tt probKeepEta1} downweights the probability of producing a singlet $\Petaprimedark$ meson relative to other diagonal mesons.  This should be set to 1 when $\Ncdark\gg \Nfdark$ since in the large-$\Ncdark$ limit (with $\Nfdark$ fixed) the axial anomaly is negligible and the $\Petaprimedark$ is like the adjoint-flavor bosons.  Conversely it should be set to a small value when $\Nfdark$ is of order or greater than $\Ncdark$ and the $\Petaprimedark$ is heavy, as it is in QCD; in \PYTHIA8, the corresponding QCD parameter {\tt StringFlav:etaPrimeSup} is set by default to 0.12.

Finally, the option of allowing baryons in hadronization for $\Ncdark=3$ can be controlled with the parameter {\tt HiddenValley:probDiquark}; this determines the likelihood of pair-producing diquarks, which, for $\Ncdark=3$ {\it only}, combine with a quark to form a baryon.   We have not validated this parameter and recommend that for now baryons (at most a 10\% effect, which is probably smaller than hadronization uncertainties) should not yet be used.

\subsubsection{Decays of dark hadronic bound states}
The dark hadronic bound states are either stable, undergo decays within the dark sector, or decay to final states that include SM particles. The decay patterns depend on the charge matrix ${\bf Q}$ and the dark hadron mass hierarchy. In particular, when the mass $\mrhodark$ is larger than twice  $\mpidark$, the $\Prhodark$ decay to $\Ppidark\Ppidark$. In the regime where such decays are not allowed some of the $\Prhodark$ may decay back to the SM via mixing with the $\PZprime$. The details of these decay modes however are determined by the group algebra and need careful treatment. We outline below the salient considerations in setting such decay modes.

\noindent\textbf{Region 1: Dark sector decays $2\mpidark < \mrhodark$ }\

When the decay channel $\Prhodark\to\Ppidark\Ppidark$ is open, it dominates all other decays since $g_{\Prhodark\Ppidark\Ppidark}$ is large compared to any other coupling.  The width 
\begin{equation}\label{eq:rho_to_pi}
\Gamma(\Prhodark^a\to\Ppidark^b\Ppidark^c ) \propto |f^{abc}|^2 \ 
\frac{g_{\Prhodark\Ppidark\Ppidark}^2}{16\pi} \mrhodark \ ,
\end{equation}
where $f^{abc}$ are the structure constants of $SU(\Nfdark)$,
is non-zero for all $\Prhodark$ mesons, and large unless $\Ncdark$ is enormous\footnote{The fate of the spin-1 singlet is a separate issue that we do not discuss here.}.

Without any mixing between the $\PZprime$ and the $\Ppidark$, the latter is stable and invisible, so we assume that we are considering a model where such mixing occurs, allowing the decay
\begin{equation}\label{eq:pi_to_SM}
\Ppidark \to (\PZprime)^* \to \Pq\Paq
\end{equation}
This mixing typically arises because a Higgs field (whose scalar is assumed too heavy to be of interest here) gives mass to both the $\PZprime$ and the quarks $\Pqdark$, typically along with some additional flavor violation.  The exact details of the mixing and corresponding lifetimes depends on precise model-building.

Because the decay in Eq.~\ref{eq:pi_to_SM} is helicity-suppressed, the width for this process is of order $|y_q|^2 \mpidark^5/\mZprime^4$ or smaller, where $y_q$ is the Yukawa coupling of the Standard Model quark; the heaviest kinematically-accessible quarks dominate.  Note this width is parametrically small and low-mass $\Ppidark$ will have displaced decays.  To determine if a particular $\Ppidark$ decays promptly, its lifetime needs to be calculated in a consistent leptophobic model, but to our knowledge the relevant model building has not been done. 

Furthermore, we do not treat the flavor singlet $\Petaprimedark$ in detail here as our analysis is not yet complete.  For $\Nfdark\gtrsim \Ncdark$ it is heavy, as in QCD, and rarely produced.
Similarly, since baryons are only available for $\Ncdark=3$, where they are a small effect, we do not discuss them here.

We conclude by noting that one should keep in mind that the $\PZprime$ charge assignments also determines the decay branching fractions of the $\PZprime$.  Especially when dark hadron multiplicities are small in $\PZprime$ decays, this introduces a small but significant correlation in the flavors of the dark hadrons,  This will become more important in future studies with non-degenerate quark masses.

\begin{figure}[h!]
    \centering
        \includegraphics[width=\textwidth, ]{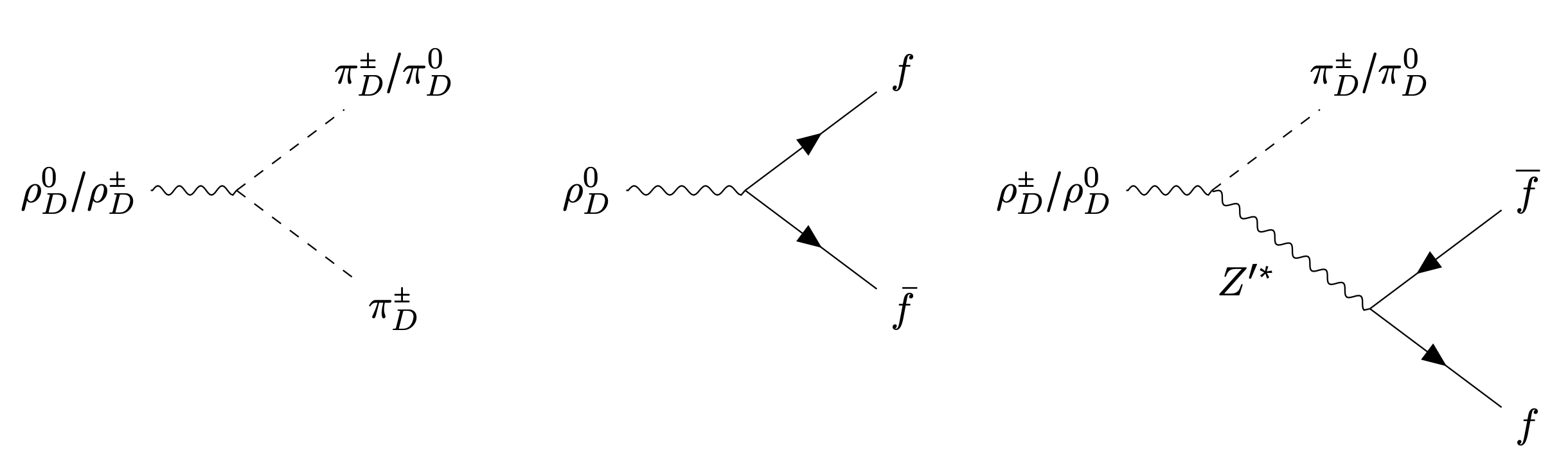} 
    \label{fig:meson_decays}
    \caption{Decay modes of diagonal and off-diagonal dark rho mesons for regime 1 (left hand side) and regime 2 (middle and right hand side). }
\end{figure}

\noindent\textbf{Region 2: Dark sector decays $2\mpidark > \mrhodark$ }\

In this case we will assume that there is no mixing between the $\PZprime$ and the $\Ppidark$ --- that the quarks and the $\PZprime$ get their masses from separate sources.  The hidden pions charged under $G_f$ are then stable and invisible.  The precise fate of the other singlets needs further investigation.  Some may be stable due to a discrete symmetry.  Others are obviously unstable due to standard flavor anomalies, though the details depend on the matrix ${\bf Q}$.  Decays to one or two SM $\Pq\Paq$ pairs have small widths because of powers of $1/\mZprime^2$ factors along with loop factors or phase space factors.  In general these particles will be very long-lived on LHC detector time-scales, and thus a source only of \met.  This statement may however be model-dependent and so one must be careful to compute the lifetimes for these states in a particular model.  We assume here that all $\Ppidark$ are LHC-detector stable.

The decays of the $\Prhodark$, however, can be observed.  First, there can be mixing between the $\PZprime$ and the $\Prhodark$ mesons which are singlets under the group $G_f$.  These decays
$$
\Prhodark\to(\PZprime)^* \to \Pq\Paq
$$
are not helicity suppressed and are thus faster than the corresponding $\Ppidark$ decays that we discussed in the previous section. 

For those $\Prhodark$ that are non-singlet under $G_f$, flavor symmetry would not prevent the decay
$$
\Prhodark\to \Ppidark + (\PZprime)^* \to \Ppidark \  \Pq\ \Paq
$$
The $\Prhodark$ and the $\Ppidark$ in this decay have the same $G_f$ quantum numbers, while the $\Pq\Paq$ are a flavor singlet. This decay would be prohibited by the naive symmetry $\Ppidark\to -\Ppidark$ in the chiral Lagrangian, but this symmetry is violated by the usual chiral anomaly that mediates $\pi_0\to\gamma\gamma$ decay in QCD, and allows a $\Prhodark\Prhodark\Ppidark$ coupling in this context. Mixing between a $G_f$-singlet $\Prhodark$ and a $\PZprime$ then induces a $\Prhodark \PZprime \Ppidark$ coupling, which permits this decay to proceed.  Specifically 
$$
\Gamma(\Prhodark^a\to  \Ppidark^b \Pq\Paq) \propto
|d^{abc}{\rm Tr}(T^c{\bf Q})|^2 \propto |{\rm Tr}(\{T^a , T^b\} \,{\bf Q})|^2
$$
where $d^{abc}$ appears in the anti-commutator~\cite{Haber:2019sgz}
$$
\{T^a, T^b\} = \frac{1}{\Ncdark}\delta^{ab} + d^{abc} T^c
$$

Importantly, however, Tr$(\{T^a , T^b\} \,{\bf Q})$ can vanish.  If this occurs, then this decay channel is not available.  For instance, if $T^a$ is the matrix whose $(\alpha,\beta)$ entry is 1 and whose other entries are all zero, then the above trace is proportional to $Q_\alpha+Q_\beta$.  Equal and opposite eigenvalues in ${\bf Q}$ then assure that the corresponding $\Prhodark$ does not decay via the anomaly.  Although this does not guarantee that this particle is stable against decay via higher-order processes, it does mean that it has a very long lifetime and is likely LHC-stable. 

Because this decay has a 3-body phase space and because by assumption $\mrhodark - \mpidark < \frac{1}{2} \mrhodark$, this decay is heavily suppressed, and will lead to displaced vertices if $\mrhodark/\mZprime$ is too small.  In the limit  $\lamdark\ll \mZprime$ as we have assumed in this section,

\begin{equation}
\label{eq:rho_three_body}
\begin{aligned}
    \Gamma(\Prhodark^a \to \Ppidark^b q \bar{q})  &= \displaystyle\frac{\left|{\rm Tr}\left(\{T^a,T^b\}{\bf Q}\right)\right|^2 e^2_D g^2_D \Ncdark^2 \mrhodark^{11}}{5898240\,\mZprime^4 \,\pi^7 f^6_{\Ppidark}}F\left(\frac{\mpidark^2}{\mrhodark^2} \right)\\
    \\
   {\rm where} \ F(x)\equiv & \ 1-15x-80 x^2+80 x^3+15 x^4-x^5-60 (x+1) x^2 \log (x) \ .
\end{aligned}
\end{equation}

In addition, we have used 

\begin{equation}
    |g_{\Prhodark\Prhodark\Ppidark}| = \frac{\Ncdark g^2_{\Prhodark\Ppidark\Ppidark}}{8\pi^2 f_{\Ppidark}}
\end{equation}
and 
$g_{\Prhodark\Ppidark\Ppidark} = m_{\Prhodark}/(\sqrt{2}f_{\Ppidark})$. The former relationship and in particular the factor of $\Ncdark$ arises from $SU(\Ncdark)$ symmetry~\cite{Gudino:2011ri} while the latter is KSFR relationship~\cite{Kawarabayashi:1966kd, Riazuddin:1966sw}, and $e_D, \gq$ are defined in Eq.~\ref{eqn:lag}.\footnote{see also~\cite{Berlin:2018tvf, Lee:2015gsa} for a discussion of this decay. We note here that we disagree with formula as given in~\cite{Berlin:2018tvf}.} For these decays to be prompt, the ratio $\mrhodark^{11}/\mZprime^4$ must not be too small. It should further be noted that $f_{\Ppidark}$ also includes a mild $\Ncdark$ dependence~\cite{Bali:2007kt}.

As above, we do not discuss the $\Petaprimedark$,  the $\Pomegadark$ or baryons here.

\subsubsection{Updates and inputs for {\tt PYTHIA 8} hidden valley module}
\label{sec:hadronization_param}
The \PYTHIA8 hidden valley module has received an update in version 8.307, after having been stable for some years.  One update is substantive; the previous versions were overproducing very soft hidden hadrons (mainly pions) at low \pt.  This bug fix slightly affects many plots, as we will see in Sec.~\ref{sec:phenovalid}; for example it affects the total multiplicity of hadrons.  Fortunately, the methods used in previous SVJ analyses are not very sensitive to this effect, which leaves total visible energy in a jet, the \met aimed in its direction, and most substructure variables roughly unchanged.  The possible exception is for  variables which are not infrared safe, most notably \ptd; see~\ref{sec:phenovalid}. 

The other main change has been to increase the flexibility of the module.   Depending on a newly introduced flag {\tt separateFlav}, the simulation in each regime may proceed in two ways.  An imperfect but often sufficient simulation, which was already available in \PYTHIA 8.150, is available with {\tt separateFlav=off}; in this case the full adjoint multiplets of spin-0 and spin-1 mesons are each simplified into two states, one flavor-diagonal and one flavor-off-diagonal.  This division is not consistent with most choices of ${\bf Q}$, as it requres that $G_f=U(1)^{\Nfdark}$.  However, as long as all dark hadrons are stable (on LHC timescales) or decay promptly, it is possible to mock up other choices of ${\bf Q}$, where for instance only a fraction of the flavor-diagonal states would decay visibly, by assigning the flavor-diagonal meson a probability to decay to a visible SM state and a corresponding probability to decay invisibly.

Alternatively, the setting {\tt separateFlav=on} allows full control over all the spin-0 and spin-1 states; separate lifetimes and decay modes can be assigned to each.  The particle ID number for the spin-0 (spin-1) meson with quark $i$ and anti-quark $j$ is 4900ij1 (4900ij3), at least for $i\neq j$.  For $i=j$ the situation is more complicated since the diagonal mesons are flavor mixtures; for example, with $\Nfdark=3$, the pion is a $u\bar u-d\bar d$ state and the $\eta$ is a $u\bar u+d\bar d-2s\bar s$ state (ignoring normalizations).  Typically one may order the diagonal flavor-adjoint mesons in a canonical way, through the increasing number of quark flavors appearing in their wavefunctions (or equivalently by the increasing number of non-zero entries in the corresponding diagonal $SU(\Nfdark)$ generator). The flavor-singlet state is always $(1/\sqrt{\Nfdark})\sum_\alpha \PqdarkA \bar \PaqdarkA$ and is always assigned particle 4900FF1 (4900FF3) where $F\equiv \Nfdark$.  This is important because this state has special status, see below.

This setting then requires the user to create a full decay table for of order $\Nfdark^2$ dark hadrons. Although we will comment on the settings for this below, we have not yet automated this task, so at this time we have no benchmarks for   {\tt separateFlav=on}.  It also permits the hidden quarks to have different masses, but we have not yet validated this capability and more studies are needed.

Other changes, not utilized below, are in the treatment of the flavor singlets, especially for spin-0, and baryons for $\Ncdark=3$.  Since the flavor singlets can be given different masses with {\tt separateFlav=on}, this allows for a more accurate spectrum.  The masses of the singlets should be assigned to the spin-0 and spin-1 particle ID codes 4900FF1 and 4900FF3. This is especially important for spin-0 because the singlet can have a much larger mass than the adjoint due to the axial anomaly, as for the $\eta^{\prime}$ in QCD.  As we mentioned above, an additional parameter  {\tt probKeepEta1}, which can be chosen between 0 and 1, has been added; this reduces the probability of producing of the $\Petaprimedark$ relative to other spin-0 mesons in the hadronization process.  Meanwhile the routines for producing baryons in the SM sector have been activated for the HV sector as well, but only work for $\Ncdark=3$.  (For $\Ncdark>3$ this is not a concern since baryon production would be highly suppressed. For $\Ncdark=2$ a special routine must be written, because baryons, antibaryons and mesons are all degenerate; this is why the current HV module should not be used for $\Ncdark=2$, at least not without careful consideration of how to reinterpret its results). For now, only one type of diquark is produced, that of $\PqdarkO\PqdarkO$.  All the baryons produced are assumed to have spin 3/2 and to have one of the $\Nfdark$ quark flavors $i$ combined with a single flavor of diquark, with particle ID code 490i114. For {\tt separateFlav=off}, all of these states are conflated into the state with $i=1$.

We have mentioned that \PYTHIA8's hadronization routine cannot simulate a theory with $\Nfdark=1$ or $\Ncdark=2$, but it may fail for other reasons.  For any choices of $\Nfdark$ and $\Ncdark$, one should avoid overly small or large values of $\mpidark/\lamdark$. At small values approaching the chiral limit, theoretical understanding of hadronization is lacking, and the Lund string model used in \PYTHIA8 may not function in any case; meanwhile at large values other hadrons (glueballs, in particular) will become as important as pions or rhos, but are not included in the Lund string model.  To be conservative, we suggest limiting studies to  $0.25 < \mpidark/\lamdark< 2$ until there has been further theoretical work on this issue.

We now turn to the \PYTHIA8 parameters that must be set to simulate the models discussed above.  We begin with those that are independent of whether {\tt separateFlav=off} or  {\tt separateFlav=on}. 
\begin{itemize}
    \item {{\tt HiddenValley:Ngauge, HiddenValley:nFlav}} -  These are $\Ncdark$ and $\Nfdark$; the former should always be set {\bf greater than 2} and the latter should always be set {\bf greater than 1}.  (For $\Ncdark=2$ or $\Nfdark=1$, \PYTHIA8 is currently missing essential dark hadrons and gives an inaccurate simulation.)
    \item Constituent dark quark mass {\tt 4900101:m0} - The quark mass defined in \PYTHIA8 is the constituent quark mass, not the current quark mass.   This quantity has never been given a theoretical definition, but may be roughly defined by $m_{q_{const}} \approx \mqdark + \mathcal{O}(1)\times\lamdark$.  For definiteness we will use this relation with the coefficient fixed to 1, namely $m_{q_{const}} \equiv \mqdark + \lamdark$.
    \item Confinement scale {\tt HiddenValley:Lambda} - This can be defined in multiple ways, but we take it for now to be the scale at which the running gauge coupling constant diverges at 1-loop order, since currently the {\tt PYTHI8} HV module has implemented the running coupling at one loop.  As we have mentioned above, further consideration of this definition is warranted. The associated behaviour is illustrated in Fig.\ref{fig:alphaD_running}.
    \item Lund model hadronization parameters must be set: {\tt HiddenValley:aLund}, {\tt HiddenValley:bmqv2}, {\tt HiddenValley:rFactqv}, {\tt HiddenValley:sigmamqv}; see sec.~\ref{sec:mass_spec}. The effects of these parameters on the underlying phenomenology have not yet been investigated, so we make no specific recommendations for them beyond the existing default settings. 
    \item Certain hadronization parameters must be set, such as {\tt HiddenValley:probVector}; see sec.~\ref{sec:mass_spec}.
    \end{itemize}
    
    Next,  if  {\tt separateFlav=off}, only two additional parameters must be defined. 
    \begin{itemize}
    \item Dark pion mass {\tt 4900111:m0, 4900211:m0}. These are the masses of the bound state spin-0 multiplets; they should always be taken equal\footnote{The spin-0 states also include the flavor-singlet $\Petaprimedark$, which, as discussed in Eq.~\ref{eq:etap_mass}, can be relatively heavy. However there is no way to take this into account for {\tt separateFlav=off}.}. Within the chiral regime, these may be related to the confinement scale $\lamdark$ and current quark mass $\mqdark$ via Eq.~\ref{eq:lattice_fits}. However, we advise taking this observable as an input parameter and viewing $\mqdark$, which is scheme-dependent, as an output. 

    \item Dark rho mass {\tt 4900113:m0, 4900213:m0}. These are the masses of the bound state spin-1 multiplets, and should always be taken equal\footnote{Although the spin-1 states include the flavor-singlet  $\Pomegadark$, analogous to the spin-0 $\Petaprimedark$, it is expected to be close in mass to the other  spin-1 states, as noted in sec.~\ref{sec:mass_spec}.}. Within the chiral regime, these are related to the confinement scale $\lamdark$ and the dark pion mass {\tt 4900111:m0}  using Eq.~\ref{eq:lattice_fits}.
\end{itemize}
     In addition, decay channels and lifetimes for these four states must be defined by the user.
     
     If instead {\tt separateFlav=on}, then even for the mass-degenerate case, all spin-0 and all spin-1 mesons must have separately defined masses, {\tt 4900ij1:m0, 4900ij3:m0} for $\Nfdark\geq i\geq j\geq1$.  Note the flavor singlets have particle ID codes 4900iis with $i=\Nfdark$ and $s=1,3$; the user may wish to change {\tt probKeepEta1} which can be used to suppress the the spin-0 singlet production.  Again the user must define all lifetimes and decay channels, now for a much larger set of particles.  Depending on the model, it may be very important to ensure that the flavor structure of the decays is precisely specified, as is emphasized in the earlier Eqs.~\ref{eq:rho_to_pi}, ~\ref{eq:rho_three_body}.

\subsubsection{Proposed Benchmarks}

We have created benchmarks for the purpose of the studies in sec.~\ref{sec:phenovalid}. We are implicitly assuming dark hadron lifetimes are short enough to be considered prompt, as appropriate for the SVJ signatures.  For low-mass dark hadrons, this is far from obvious. Lifetimes need to be calculated in the context of complete models, but constructing such models is no simple matter in the context of a leptophobic $\PZprime$ because of potential $U(1)'$ gauge anomalies that would make the theory inconsistent.   We are not aware of any complete calculations of dark hadron lifetimes in this context, so we must warn the user that some of these benchmarks, especially those with light $\Ppidark$, may not be realizable theoretically.

Let us first note what all the benchmarks have in common.  In each case
\begin{itemize}
    \item $\mZprime=1$ TeV;
    \item We take {\tt separateFlav=off}.
\end{itemize}
Versions of the benchmarks with {\tt separateFlav=on} would be more accurate in their treatment of flavor-singlets, but will have to be created at a later time.

We have several benchmarks with $\mpidark<\frac12 \mrhodark$.
\begin{itemize}
    \item All have $\Ncdark=\Nfdark=3$.
    \item All have $\mpidark=0.6 \lamdark$ (and thus $\mrhodark=2.6\ \lamdark$ by Eq.~\ref{eq:lattice_fits}.)
    \item Because of this choice, the parameter {\tt probVec} is taken to be $0.5$, since $\mpidark/\lamdark$ is similar to its value used in real-world QCD.  
    \item Three choices of $\lamdark$ are considered: 5 GeV, 10 GeV and 50 GeV.
    \item For each of these, the number of stable diagonal spin-0 mesons is $k=0$, 1 or 2, with $3-k$ decaying to the SM; since the six off-diagonal pions are stable in this model, this gives $\rinv=(6+k)/9$
    \item The dark pions are assumed to decay promptly and only to $c\bar c$ (charm being the heaviest kinematically-allowed SM quark for the smaller values of $\lamdark$.)
\end{itemize}
The choice of $k$ depends on mixing among the singlet and diagonal adjoint pions and the $\PZprime$.  The details, especially the interplay between mixings and lifetimes, require careful model-building.  We are not aware of any papers in which this has been done. 

Note that the use of {\tt separateFlav=off} means that we do not treat the $SU(\Nfdark)$ flavor singlets separately from the other mesons.  For $\Ncdark=3$ the $\Petaprimedark$ is much heavier than the other states, and this is not correctly modeled.  In particular, it leads to a small correction to \rinv.  For $\Ncdark\gg \Nfdark$, the splitting between the flavor adjoint and singlet states becomes small, so the use of {\tt separateFlav=off} is less problematic there.

For $\mpidark<\frac12 \mrhodark$, we have so far defined only one benchmark
\begin{itemize}
    \item $\Ncdark=3$, $\Nfdark=4$
    \item $\lamdark=10$ GeV, $\mpidark=17$ GeV, $\mrhodark= 31.8$ GeV
    \item The parameter {\tt probVec} is taken to be $0.58$, in between the values of 0.5 (as used for QCD) and 0.75 (as appropriate for $\mpidark\approx \mrhodark$.)
    \item ${\bf Q}= \{-1,2,3,-4\}$, a choice that ensures that no $\Prhodark$ are stable.
    \item All spin-0 mesons are assumed to be stable on LHC-detector time-scales
    \item All diagonal spin-1 mesons (including the singlet, which we do not treat carefully) decay to all available SM $\Pq\Paq$ pairs 
    \item All off-diagonal spin-1 mesons decay to SM $\Pq\Paq$ plus an invisible spin-0 meson.
\end{itemize}

Table~\ref{tab:sig_category} summarises our current benchmarks.

\begin{table}[th!]
\centering
 \begin{tabular}{| c |c | c | c |c | c | c |c |} 
 \hline
 Regime & $\Ncdark, \Nfdark$ & $\lamdark$ & $\bf{Q}$ & $\mpidark$ & $\mrhodark$ & Stable & Dark hadron\\
 & & [GeV] & & [GeV] & [GeV] & dark hadrons & decays \\
 \hline
 \multirow{3}{*} {$\mpidark < \mrhodark/2$ }&  \multirow{2}{*} {3,3} & 5 & Various & 3 & 12.55 & $0/1/2 \Ppidark^0$ & $\Prhodark^{0/\pm}\to \Ppidark^{0/\pm}\Ppidark^{\mp}$ \\
 & & & & & & & $\Ppidark^0\to c\bar{c}$ \\
 \cline{2-8}
 & 3,3 & 10 & Various & 6 & 25 & 0/1/2 $\Ppidark^0$ & $\Prhodark^{0/\pm}\to \Ppidark^{0/\pm}\Ppidark^{\mp}$ \\
  & & & & & & & $\Ppidark^0\to c\bar{c}$ \\
 \cline{2-8}
 & 3,3 & 50 & Various & 30 & 125.5 & 0/1/2 $\Ppidark^0$ & $\Prhodark^{0/\pm}\to
 \Ppidark^{0/\pm}\Ppidark^{\mp}$ \\
  & & & & & & & $\Ppidark^0\to b\bar{b}$ \\
\hline
$\mpidark > \mrhodark/2$ & 3,4 & 10 & (-1,2,3,-4) & 17 & 31.77 & All $\Ppidark$ & $\Prhodark^0\to \Pq\Paq$ \\
 & & & & & & & $\Prhodark^{\pm}\to\Ppidark^{\pm} \Pq \Paq$ \\
 \hline 
 \end{tabular}
\caption{Current benchmarks for $\mpidark > \mrhodark/2$ and $\mpidark < \mrhodark/2$ regimes. In the former case all $\Ppidark$ are stable and source of \met, while for the later, the $\Prhodark$ mesons decay to $\Ppidark$ which further decay to $c\bar{c}$ final states at the LHC. The benchmarks assume that the decays of $\Prhodark, \Ppidark$ are prompt.} \label{tab:sig_category}
\end{table}

To compose benchmarks with {\tt separateFlav=on}, there are a number of additional steps needed. For $\mpidark<\frac12 \mrhodark$, the decays $\Prhodark^a \to \Ppidark^b\Ppidark^c$ need to be correctly programmed. For example, for $\Nfdark=3$, $\Prhodark^3$, the diagonal member of the rho isotriplet (particle ID 4900113), decays to spin-0 bosons $\Ppidark^{ij}$ (particle ID 4900ij1) in the following pattern:
$$
\Prhodark^3 (4900113) \to \Ppidark^{12}\Ppidark^{21} \ (66\%)\ ,
\Ppidark^{13}\Ppidark^{31} \ (16\%)\ , 
\Ppidark^{23}\Ppidark^{21} \ (16\%)\ , 
$$
the 4:1:1 branching ratios reflecting the relative isospins-squared of these spin-0 states.  All of these details need to be correctly laid out in the \PYTHIA decay table in order that spin-0 mesons be produced in the right abundances.  It is also important to decide how to treat the singlet states, especially the spin-0 singlet whose mass and production rate in hadronization may be quite different from the others.  Finally, all the spin-0 meson decays and lifetimes must be separately entered into the \PYTHIA decay table.

For $\mpidark>\frac12 \mrhodark$, similar efforts are required to ensure that the flavor structure of the diagonal and off-diagonal $\Prhodark$ decays are correctly implemented in the decay table.  

\subsubsection{Final remarks}

Before we proceed with phenomenological studies using the benchmarks proposed in this section, we would like to emphasize that we have only laid out an initial road for defining  consistent phenomenology in the context of semi-visible jets. We have considered the leptophobic $\PZprime$ SM--DS portal widely used in the semi-visible jets literature, and pointed out the crucial role of charge assignments in determining the phenomenology, though we have not worked out the details. 
Dark hadron masses may potentially be extracted from a combination of lattice simulation of the hidden sector and and general theoretical considerations.  But their decay channels and lifetimes are highly model-dependent, and calculating them involves careful consideration of the detailed charge assignments of the $\PZprime$, its mixing with various dark hadrons, and the spectrum and interactions of the dark hadrons  (including anomalies) as obtained from symmetry considerations and the chiral Lagrangian.  These sometimes intricate calculations must be performed in each model, unless an over-arching theoretical treatment, covering all models in this class, can be given. 

In the context of semi-visible jets, the lifetimes of the various states are particularly important.  This signature is defined to be one in which all objects either are stable, producing \met, or decay promptly to SM-hadronic final states. Long-lived particles with lifetimes greater than a few centimeters and less than 10 meters (in the lab frame) would move the signature into a different regime, outside the semi-visible jet framework.  It is therefore imperative to identify all unstable dark hadrons and calculate their lifetimes correctly.  We have estimated lifetimes and have moderate confidence that all particles in our benchmarks decay promptly or are stable on LHC detector scales, but we have not by any means done a thorough analysis. 

We would also like to note that there are still significant issues with hadronization that we have not begun to address.  We have made a few observations about the hadronization parameters used in the \PYTHIA8 HV module, but have neither attempted to explore the impact of their uncertainty on the underlying phenomenology, nor made concrete statements about what ranges of values they might take.  These questions, and even deeper ones about how hadronization models perform in other regimes, such as the chiral limit ($\mpidark\ll \lamdark$), must be left for future studies.
\subsection{Improvements on the {\tt PYTHIA8} Hidden Valley Module and their validation}\label{newpythia}

\emph{Contributors: Guillaume Albouy, Cesare Cazzaniga, Annapaola de Cosa, Florian Eble, Marie-Hélène Genest, Nicoline Hemme, Suchita Kulkarni, Stephen Mrenna, Ana Peixoto, Akanksha Singh, Torbjörn Sjöstrand, Matt Strassler}

\subsubsection{Sample generation}
The signal process considered for the validation of new Hidden Valley (HV) Module of \PYTHIA 8 \cite{2010,2011} consists of semi-visible jets \cite{Cohen:2015toa} produced in the $s$-channel via a heavy \PZprime mediator. A set of signal samples has been produced with different versions of \PYTHIA 8 for proton-proton collisions (and also electron-positron collisions for completeness) at the benchmark centre-of-mass energy of 13 TeV (1 TeV). Namely, in order to test the new implementation of the HV Module we have produced three main groups of samples as illustrated in Table \ref{table:signal_MC_samples} for different dark sector color charge $\Ncdark$ and dark quark flavor $\Nfdark$ choices.  
\vskip0.1cm
\begin{table}[!htbp]
\centering
\begin{tabular}{*4c}
\toprule
\multicolumn{4}{c}{\bfseries{SVJ Monte Carlo samples categories}} \\
\midrule
\multicolumn{4}{c}{$p p \to \PZprime \to \Pq \Paq \quad [\sqrt{s}=13 \ \text{TeV}] $} \\ \hline
Sample name  & \PYTHIA 8 version & Flavor (FV) splitting & Simulated events \\
1 & 8.245  & OFF & $50 \times 10^3$  \\
2 &  8.307   & OFF & $50 \times 10^3$    \\
3 & 8.307   & ON & $50 \times 10^3$   \\ 
\bottomrule
\end{tabular}
\caption{SVJ MC samples generated with \PYTHIA 8 for a \PZprime mass $\mZprime = 1 $ TeV. For all these categories, only the decay of the \PZprime to dark quarks is simulated.}
\label{table:signal_MC_samples}
\end{table}
\vskip0.1cm
In particular, the first type of samples have been produced with an older \PYTHIA 8.245 release \cite{2015}\footnote{We do not expect major changes to the outcome of our study even if we would have used \PYTHIA 8.306, the immediate predecessor of \PYTHIA 8.307 which we validate in this section.}, while the second one have been generated with the new \PYTHIA 8 version 8.307 \cite{py8-update}. \\
In the new \PYTHIA 8 release, it is now possible to set the masses of all 8 dark quarks and associated 64 mesons for each pseudo-scalar and vector multiplet individually. Even if this allows to consider mass split scenarios, we consider only mass degenerate dark quarks since a consistent treatment for UV to IR settings in mass split scenario is not yet available. As an outcome of this choice, flavor symmetry leads to mass degenerate pseudo-scalar and vector multiplets. However, it is still crucial to have the possibility to set all dark mesons properties individually since the lifetimes of these different states can differ according to the model and mediator. Following these necessities, compared to \PYTHIA 8.245 (8.306) release, in the newest version a more detailed handling of dark hadrons is possible with the setting {\tt HiddenValley:separateFlav = on}. As shown in Table \ref{table:signal_MC_samples}, a third sample has been added in order to test this new option. In particular, using the flavor splitting option, each of the quark and meson flavors are shown explicitly. The quark names now are $\PqdarkI$, with $i \in \{ 0, \cdots, \Nfdark \}$. Similarly, meson names are $\PpidarkIJ$ and $\PrhodarkIJ$, where $i = j$ are the flavor-diagonal mesons, and else $i > j$, with $j$ representing the anti-quark. The identity codes then are $4900ij1$ for pseudo-scalars and $4900ij3$ for vectors. An anti-meson comes with an overall negative sign, and here $i$ gives the anti-quark. The data tables by default contain identical properties for all diagonal mesons in a multiplet. All nondiagonal mesons of a multiplet are also assumed to be identical and stable by default. 

An advantage of the SeparateFlav on option, is the possibility of setting masses (as well as decay modes) of spin-0,1 flavour singlets differently than the corresponding multiplets. As discussed in section~\ref{parameters}, the exact computation of the flavor singlet mass with respect to the flavor multiplets, especially for spin-0 states, is an open question. There are indications that spin-0 singlets tend to be heavier than their multiplet counterparts, and therefore for these states a suppression of the production rate is also expected. For this reason, the option {\tt HiddenValley:probKeepEta1} can be set in \PYTHIA 8.307 in order to specify the suppression factor for the spin-0 flavor singlets production rates. This feature has been tested in the validation procedure, but we do not report plots related to this.   

Fixing the color charge to $\Ncdark=3$ using the option {\tt HiddenValley:Ngauge  = 3}, two configurations for the number of flavors $\Nfdark=3,8$ have been considered in this study, using the setting {\tt HiddenValley:Nflav}. $\Nfdark = 3$ corresponds to the smallest possible configuration with more particles than the triplet representation used in \PYTHIA 8.245 , while $\Nfdark = 8$ is the maximal number of flavors implemented in \PYTHIA HV module. Choosing these two values, we thus test the extremes of the flavor configurations. The hidden valley partners $F_D$ of the SM particles (charged both under both SM and hidden valley group) are assumed to be decoupled in our case, such that they will not produce interleaved showers between the hidden sector and the SM \cite{2010,2011}. While in the \PYTHIA 8.245 release only dark mesons originating from string fragmentation are implemented, in \PYTHIA 8.307 tested in this study an option to produce dark baryons has been added with the line {\tt HiddenValley:probDiquark = on}. With this option, it is possible to set the probability that a string breaks by "diquark-antidiquark" production rather than quark-antiquark one. This then leads to an adjacent baryon-antibaryon pair in the flavor chain. Currently only one kind of diquark is implemented, implying at most eight different Delta baryons $\PdeltadarkI$ if  {\tt separateFlav = on}. In the validation procedure of \PYTHIA 8.307 of the HV module we have considered decoupled Delta baryons. 

A minimal number of input parameters have to be specified in \PYTHIA 8 when the Hidden Valley module is called with the option {\tt HiddenValley:fragment = on}. In particular, the masses of the dark hadrons have to be fixed as well as the dark sector hadronization scale $\lamdark$ (set to 10 GeV in this study). Furthermore, the masses of pseudo-scalar states are set to 6 GeV, and the masses of the vector mesons are chosen to be $25$ GeV.  These settings correspond to $\Ppidark/\lamdark = 0.6$ same as that considered in benchmarks in section ~\ref{parameters}. The final states configuration that we chose for our study is simply a fully invisible signature where all the dark hadrons are considered to be stable. A further relevant setting which must be specified  is  the running of the dark sector coupling $\adark$ which can be switched on with the option {\tt HiddenValley:alphaOrder = 1}. 

For the purposes of this study, for efficient MC generation, we consider a simplified scenario where the \PZprime mediator decays only to dark quarks, even if in a real physics case the non-vanishing coupling to SM-quarks contributes to the branching ratios. By default the $\PZprime$ mediator nominal width $\Gamma_{\PZprime}$ of the $\PZprime$ boson is set to $20$ GeV and the mass $\mZprime = 1 $ TeV. Figure \ref{fig:Zp lineshape} shows the invariant mass distribution for the $\PZprime$ boson using the dark quarks before parton shower and hadronization in the hidden sector. The distribution deviates from the Breit-Wigner showing an excess of events in the low mass tail. This effect can be explained from the factorisation theorem considering that parton distribution functions blows up for low transferred momentum fractions for the SM incoming partons. Since we are only interested in typical events where a $\PZprime$ boson is created to have a consistent comparison between the different samples, we choose to cut away the low mass tail requiring the generated invariant mass of the $\PZprime$ to be within the range $[800,1200]$ GeV.

\begin{figure}
    \centering
    \includegraphics[width=0.70\textwidth]{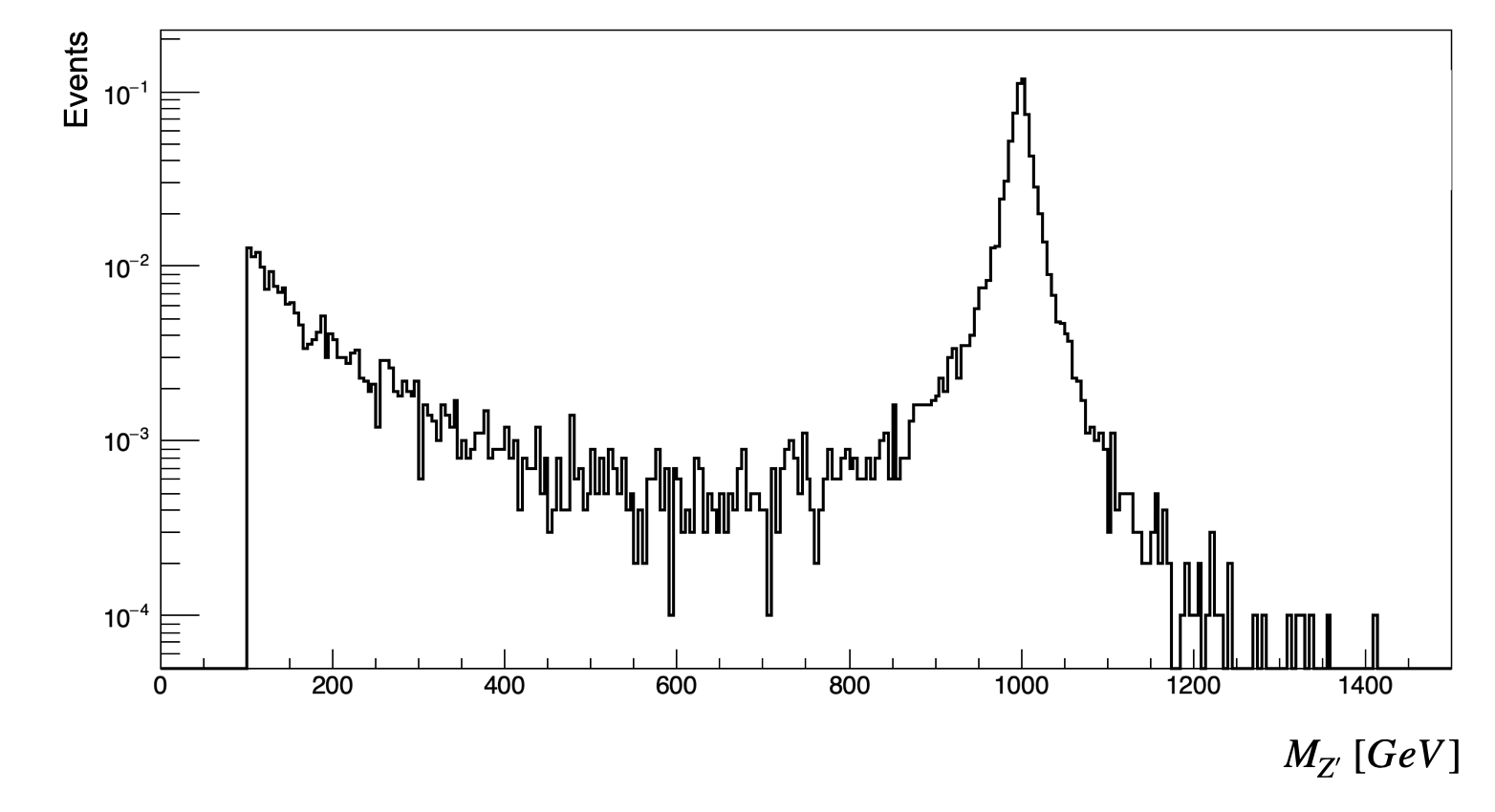}
    \caption{Distribution of the \PZprime mediator mass from dark quarks (nominal value set in the simulation $\mZprime = 1$ TeV and $\Gamma_{\PZprime}=20$ GeV).  }
    \label{fig:Zp lineshape}
\end{figure}
    
\subsubsection{Validation plots}
\textbf{{\tt PYTHIA} triplet implementation}\\
In a two flavor theory there are 4 spin-0 states (and 4 spin-1 states); 1 diagonal and 2 off-diagonals, which make up the triplet, and an additional singlet. In the current \PYTHIA 8 release there are only 3 PIDs for dark pions (3 PIDs for $\Prhodark$ mesons), which signify the positive and the negative off-diagonal and the diagonal dark pion. However, the singlet is still produced in \PYTHIA 8 and shares the same PID as the diagonal dark pion. With SeparateFlav off option, it is thus impossible to separate out the singlet: it is produced with the same probability as that of the diagonal dark pion. As the singlet is considered to be another diagonal dark pion in \PYTHIA 8, the ratio of diagonal to off-diagonal dark pions is 1:1 for $\Nfdark=2$. In other words, \PYTHIA 8 will create an even amount of diagonal and off-diagonal dark mesons, and hence the PID for the diagonal dark pions ($\Prhodark$ mesons), 111 (113), is equally as likely as the PIDs for off-diagonal dark pions ($\Prhodark$ mesons) when considered together, 211 (213) and -211 (-213). This is clearly illustrated in Figure \ref{fig:validation_PID_2f_FVsplitOff}. Similarly, in a theory with $\Nfdark=3$ there is an octet and a singlet, or 3 diagonal and 6 off-diagonal dark mesons. In the current \PYTHIA 8 release there are also only 3 PIDs for these 9 dark pions (and 3 PIDs for 9 $\Prhodark$ mesons), following the same logic as for $\Nfdark=2$. The ratio of diagonal to off-diagonal dark mesons is now 1:2 so \PYTHIA creates twice as many off-diagonal dark mesons as diagonal ones. In this situation, 111 (113) is only half as likely as 211 (213) and -211 (-213) together, see Figure \ref{fig:validation_PID_3f_FVsplitOff}. 
\begin{figure}[h!]
    \centering
    \begin{subfigure}[b]{0.49\textwidth}
        \centering
        \includegraphics[width=\textwidth]{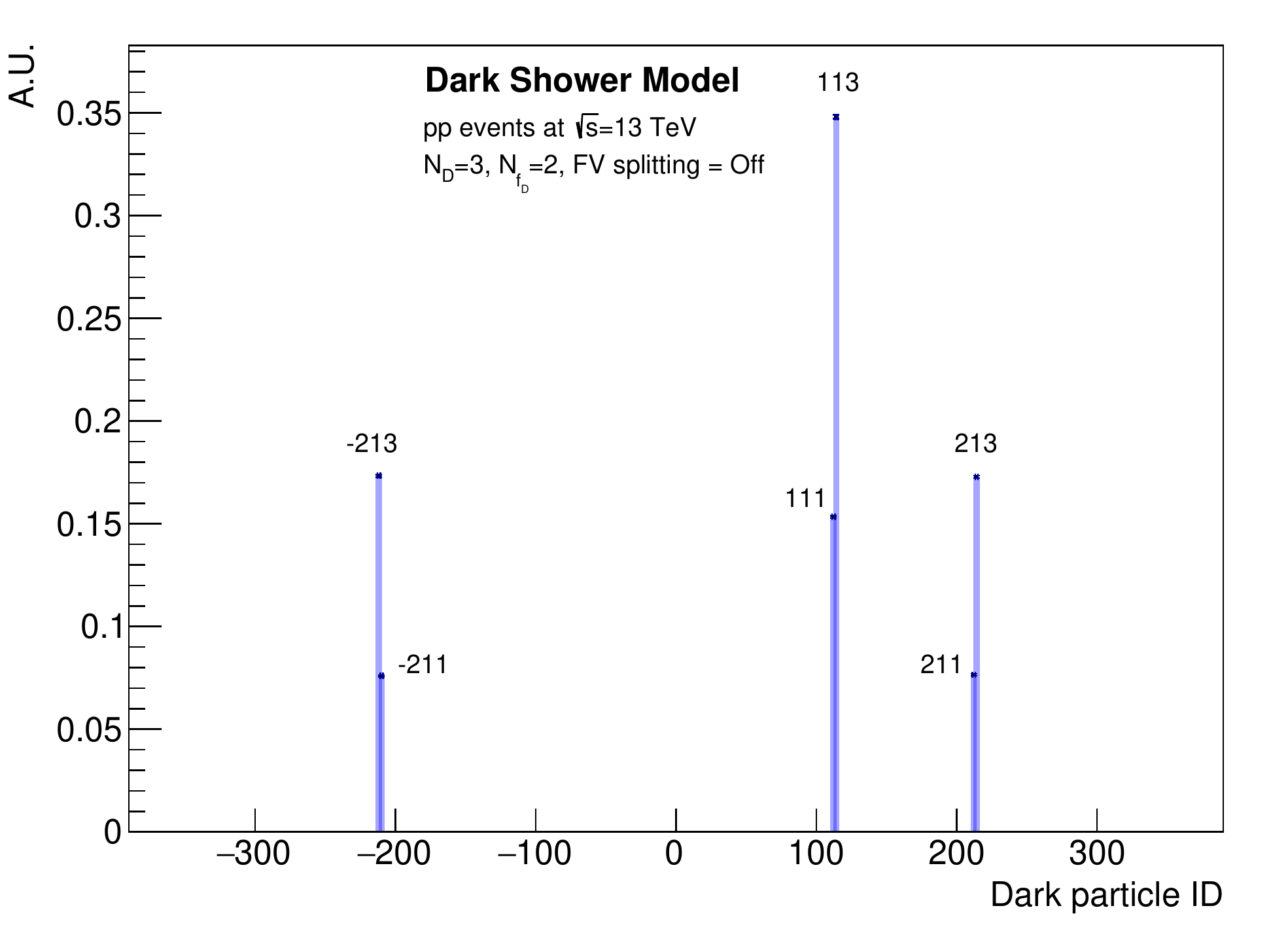}
        \caption{}
        \label{fig:validation_PID_2f_FVsplitOff}
    \end{subfigure}
    \begin{subfigure}[b]{0.49\textwidth}
        \centering
        \includegraphics[width=\textwidth]{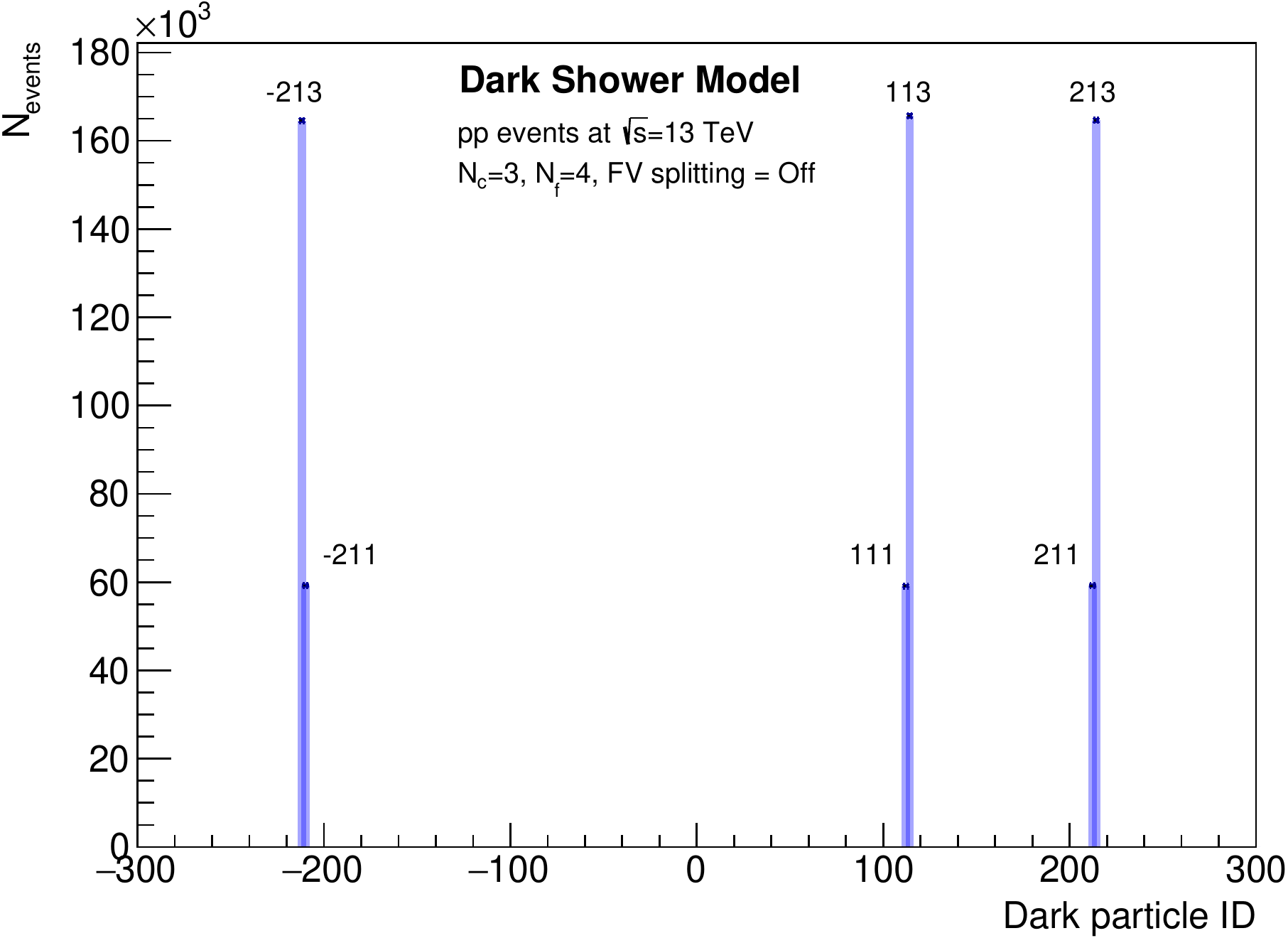}
        \caption{}
        \label{fig:validation_PID_3f_FVsplitOff}
    \end{subfigure}    
    \caption{PdgID distributions for final-state dark particles without the 4900 prefix for (a) $\Nfdark=2$ model and (b) $\Nfdark=3$ model. FV splitting is turned off to simulate as the standard \PYTHIA 8 version. The parameter $\text{probVec}=0.75$ for both models.}
\end{figure}

The \PYTHIA 8.307 includes individual PIDs for all the multiplets and singlets, as well as a new parameter called {\tt HiddenValley:probKeepEta1}, which determines the probability to create the singlet state. This probability is set relative to the probability of producing spin-0 multiplets. The default setting is 1, but it can be set to 0 such that the singlet is not produced at all. 

The handling of dark PIDs affects the expected value of \rinv. In \PYTHIA 8.245 release it is not possible to turn off the production of the singlet state and so this must be taken into account in the calculation of \rinv. Take as an example an $\Nfdark=2$ model with diagonal $\Prhodark$ mesons (113) promptly decaying to the SM through a vector portal. Firstly, the $\text{probVec}=0.75$ parameter dictates that $3/4$ of the dark mesons will be $\Prhodark$ mesons, of which half are diagonal. This means that 3/8 of the dark mesons will be unstable, while the remaining 3/8 off-diagonal $\Prhodark$ mesons and 1/4 dark pions are stable, resulting in a ratio of stable dark mesons to all dark mesons of 5/8 or 0.625. The value of \rinv can be calculated at the generator level by counting separately final-state, stable dark mesons and all dark mesons (including decayed $\Prhodark$ mesons) in the event and taking the ratio of these two sums. The distribution of \rinv for such a model with FV splitting turned off can be seen in Figure \ref{fig:validation_rinv_2f_FVsplitOff}. 
\begin{figure}[h!]
    \centering
    \includegraphics[width=0.55\textwidth]{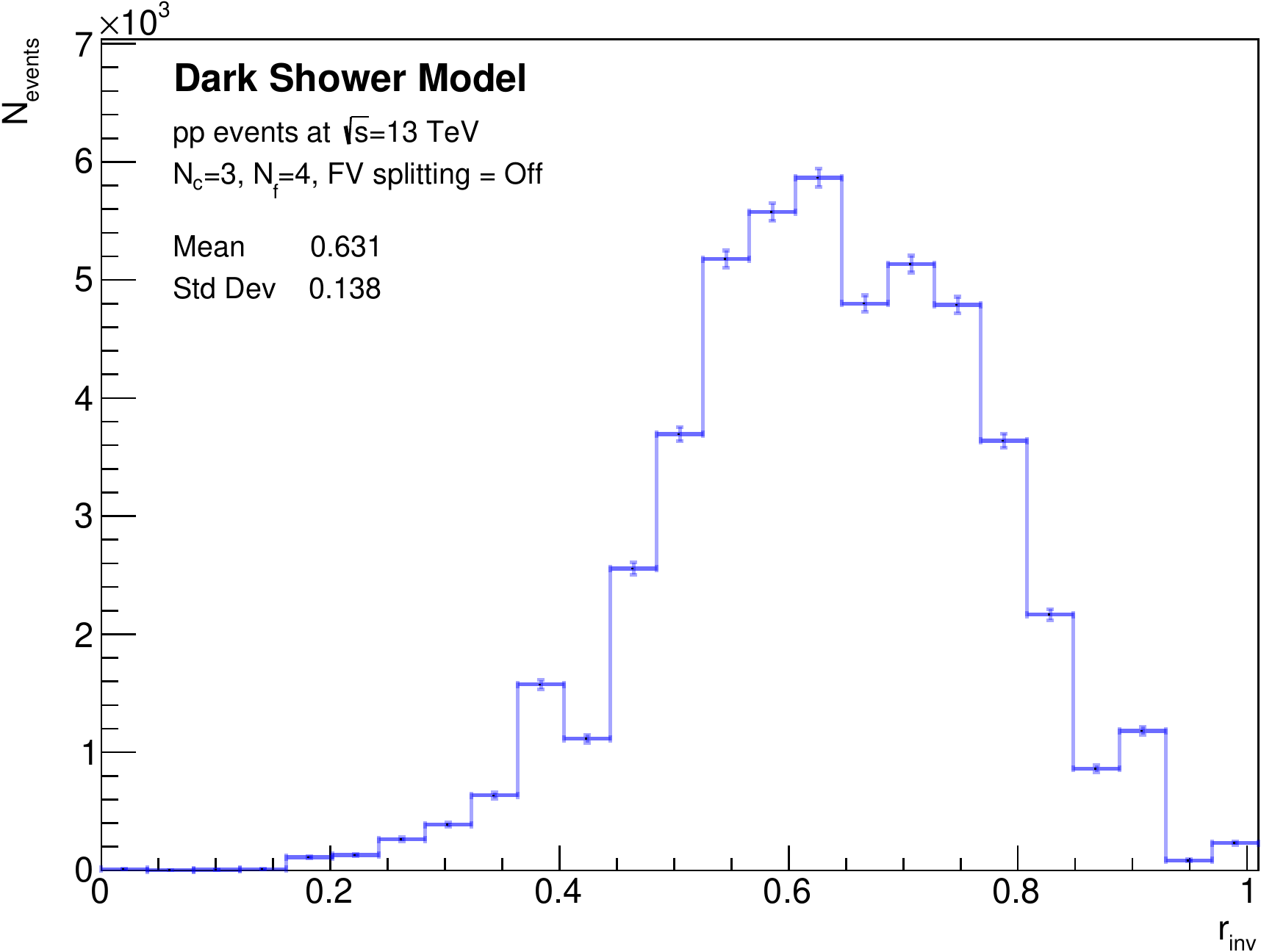}
    \caption{Distribution of \rinv for a model with $\text{probVec}=0.75$ and unstable diagonal $\Prhodark$ mesons.}
    \label{fig:validation_rinv_2f_FVsplitOff}
\end{figure}


\textbf{{\tt PYTHIA} full n-plate implementation}\\

The validation of the new \PYTHIA HV module was performed through a phenomenological analysis of the distinct variables obtained for three different cases: HV module in \PYTHIA 8.245 and in \PYTHIA 8.307 with either {\tt HiddenValley:separateFlav = off} or {\tt HiddenValley:separateFlav = on}. All dark mesons were set to be stable. The distributions of angular and kinematic variables of the different final state particles were compared for those three cases. Although the most important variables are related to the dark pion and $\Prhodark$ mesons, the missing transverse energy and the produced jets were also considered for this validation study. The reconstruction of the jets is done by clustering the generator level objects obtained after parton-shower and hadronization, using the radius parameter $\Delta R$=1.4. As the dark mesons are set to be stable, the jets we study in this section are therefore not a result of hadronization in the dark sector. They originate e.g. from initial state radiation and subsequent hadronization in the SM sector. As mentioned before, the validation of the \PYTHIA 8.307 HV module is executed for two specific models: $\Nfdark =3$ and $\Nfdark =8$ (with $\Ncdark =3$ for both).
For simplicity, only the results from the $pp$ analysis are shown, as the same conclusions were obtained for the $e^+e^-$ study. 

\textbf{$\Ncdark= 3$, $\Nfdark =3$ model}\\

Changing the $\Nfdark$ value from 2 to 3 results in additional PIDs being produced, as can be seen by comparing the dark pions and $\Prhodark$ mesons particle ID shown in Figures \ref{fig:validation_PionsPID_Nc3Nf3} and \ref{fig:validation_RhosPID_Nc3Nf3}, respectively,  to the ones shown in Figures \ref{fig:validation_PID_2f_FVsplitOff} and \ref{fig:validation_PID_3f_FVsplitOff}. One can see that dark pions and $\Prhodark$ with ID 311 (313) and -311 (-313) are produced when setting $\Nfdark$ to 3 and SeparateFlav on. For the case where {\tt HiddenValley:separateFlav = on} in particular, a total of 9 pseudo-scalar and 9 vector particles are identified, with the same production rates for all states in a given multiplet. The distributions of the multiplicity and the transverse momentum of the dark hadrons are represented in Figures \ref{fig:validation_DarkHadronsMult_Nc3Nf3} and \ref{fig:validation_DarkHadronsPt_Nc3Nf3}. A lower multiplicity of these dark hadrons and a softer transverse momentum can be seen for the {\tt HiddenValley:separateFlav = on} scenario compared with \PYTHIA 8.245. These changes are expected with the new HV module due to the bug fix related to the newly implemented $p_\text{T}$ suppression for mini-string fragmentation, as discussed in Section~\ref{sec:hadronization_param}. An overall agreement can be found for the {\tt HiddenValley:separateFlav = on} and {\tt HiddenValley:separateFlav = off} with the new HV module. Concerning the specific case of the dark pions, the corresponding multiplicity and transverse momentum can be found in Figures \ref{fig:validation_PionsMult_Nc3Nf3} and \ref{fig:validation_PionsPt_Nc3Nf3}. With similar conclusions as for the dark pions, the distributions of the same variables corresponding to the  $\Prhodark$ mesons are shown in Figures \ref{fig:validation_RhosMult_Nc3Nf3} and \ref{fig:validation_RhosPt_Nc3Nf3}. From Figures \ref{fig:validation_PionsMult_Nc3Nf3} and \ref{fig:validation_RhosMult_Nc3Nf3}, it can be concluded that the pseudo-scalars have lower multiplicity with respect to vector mesons. The difference between the distributions for the diagonal and off-diagonal dark pions and $\Prhodark$ mesons was studied. The multiplicity and the transverse momentum of the diagonal and off-diagonal dark hadrons were consistent with the previous conclusions, with an agreement between the new and old HV modules with the different {\tt HiddenValley:separateFlav} options. For completeness, the distributions of the missing transverse energy and the minimum azimuthal angle between jets and missing transverse energy can also be found in Figures \ref{fig:validation_MET_Nc3Nf3} and \ref{fig:validation_DeltaPhiMETJet_Nc3Nf3}.  The latter shows that the missing transverse energy is recoiling against jets, as expected in the fully invisible scenario investigated here. The use of the new HV module does not have any impact on the event kinematics, as expected.

\begin{figure}
    \centering
    \begin{subfigure}[b]{0.49\textwidth}
        \centering
        \includegraphics[width=\textwidth]{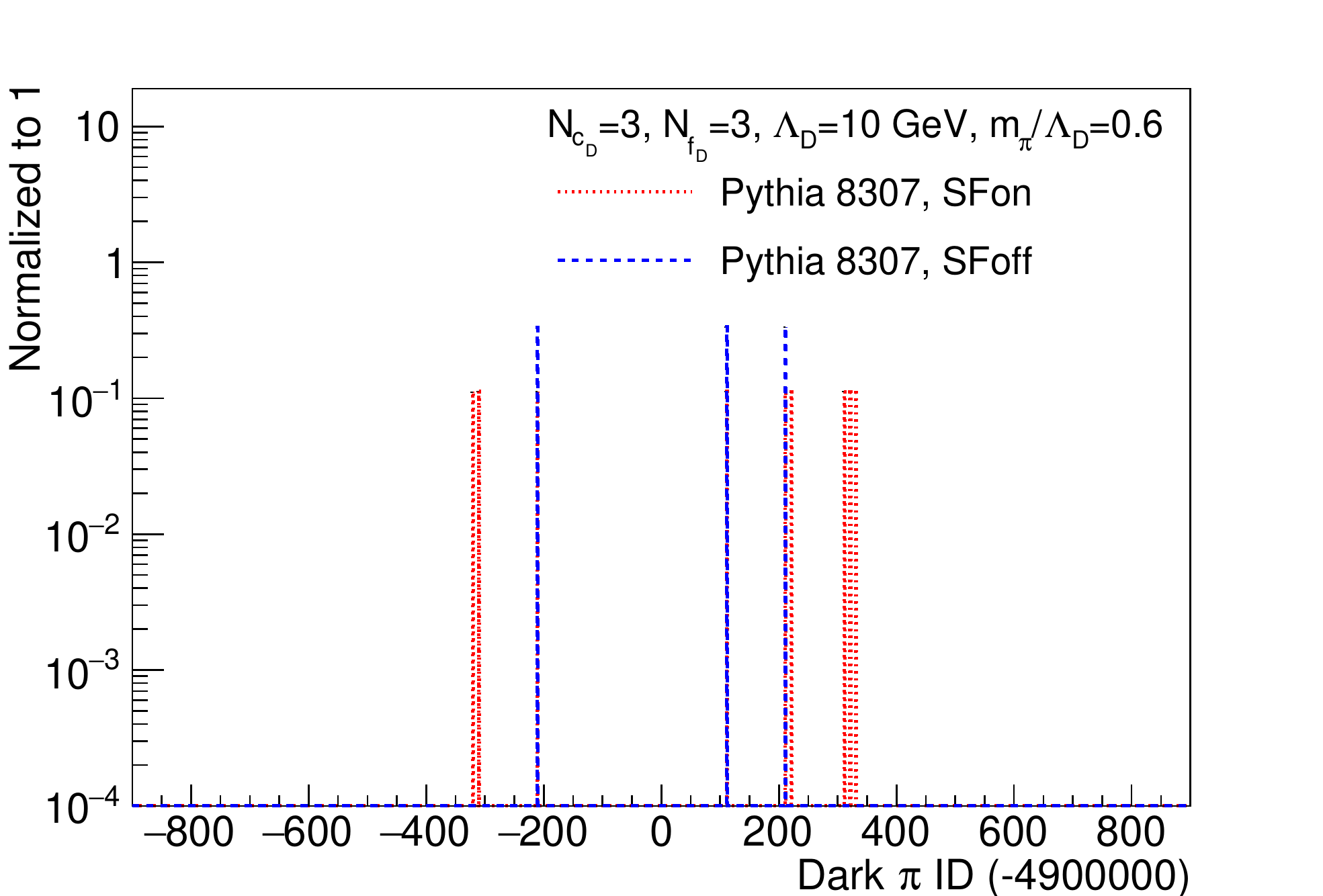}
        \caption{}
        \label{fig:validation_PionsPID_Nc3Nf3}
    \end{subfigure}
    \begin{subfigure}[b]{0.49\textwidth}
        \centering
        \includegraphics[width=\textwidth]{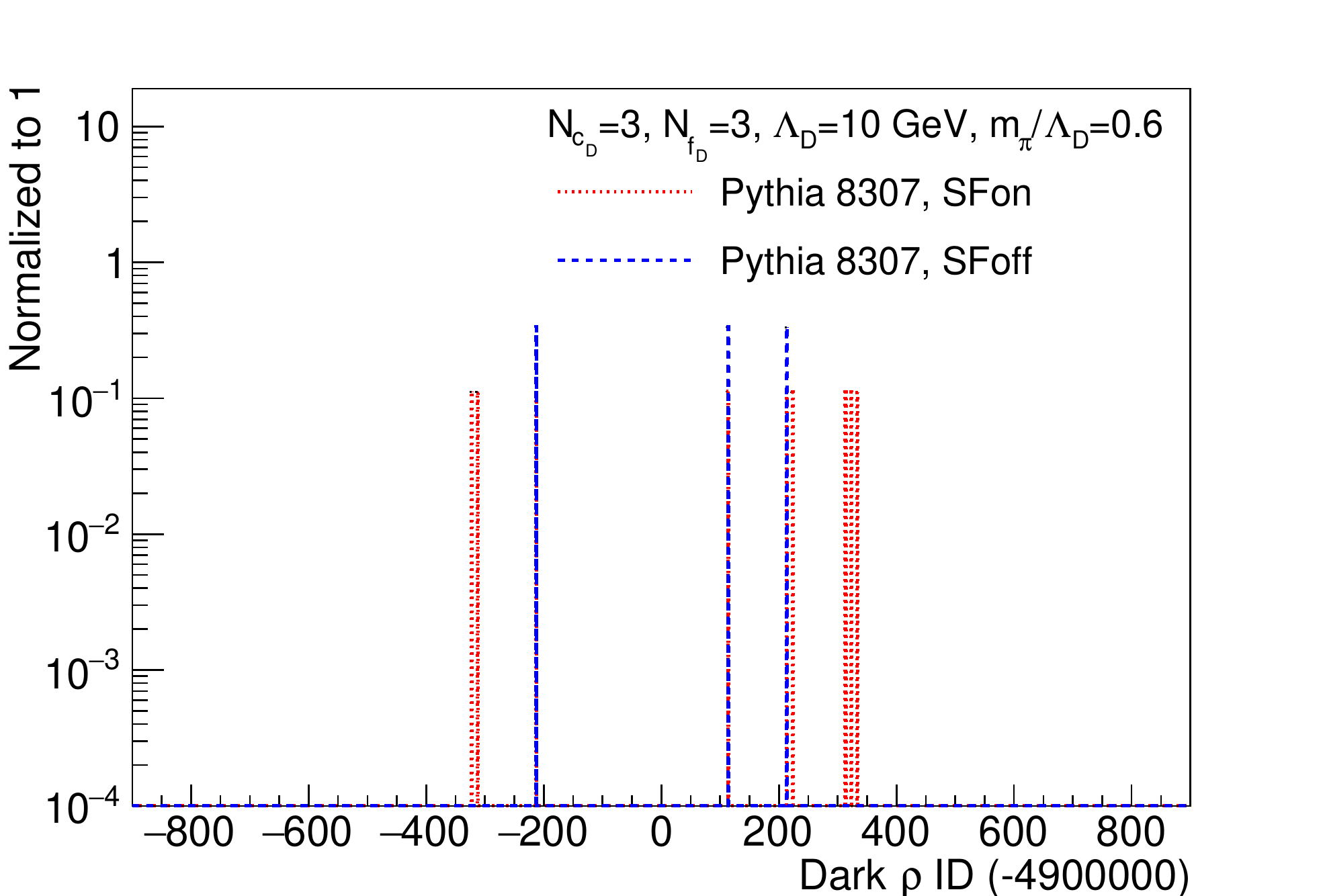}
        \caption{}
        \label{fig:validation_RhosPID_Nc3Nf3}
    \end{subfigure}    
    \caption{$\Ncdark = 3$, $\Nfdark =3$ model: (a) PdgId distribution for dark pions, (b) PdgId distribution for $\Prhodark$ mesons.}
\end{figure}

\begin{figure}
    \centering
    \begin{subfigure}[b]{0.49\textwidth}
        \centering
        \includegraphics[width=\textwidth]{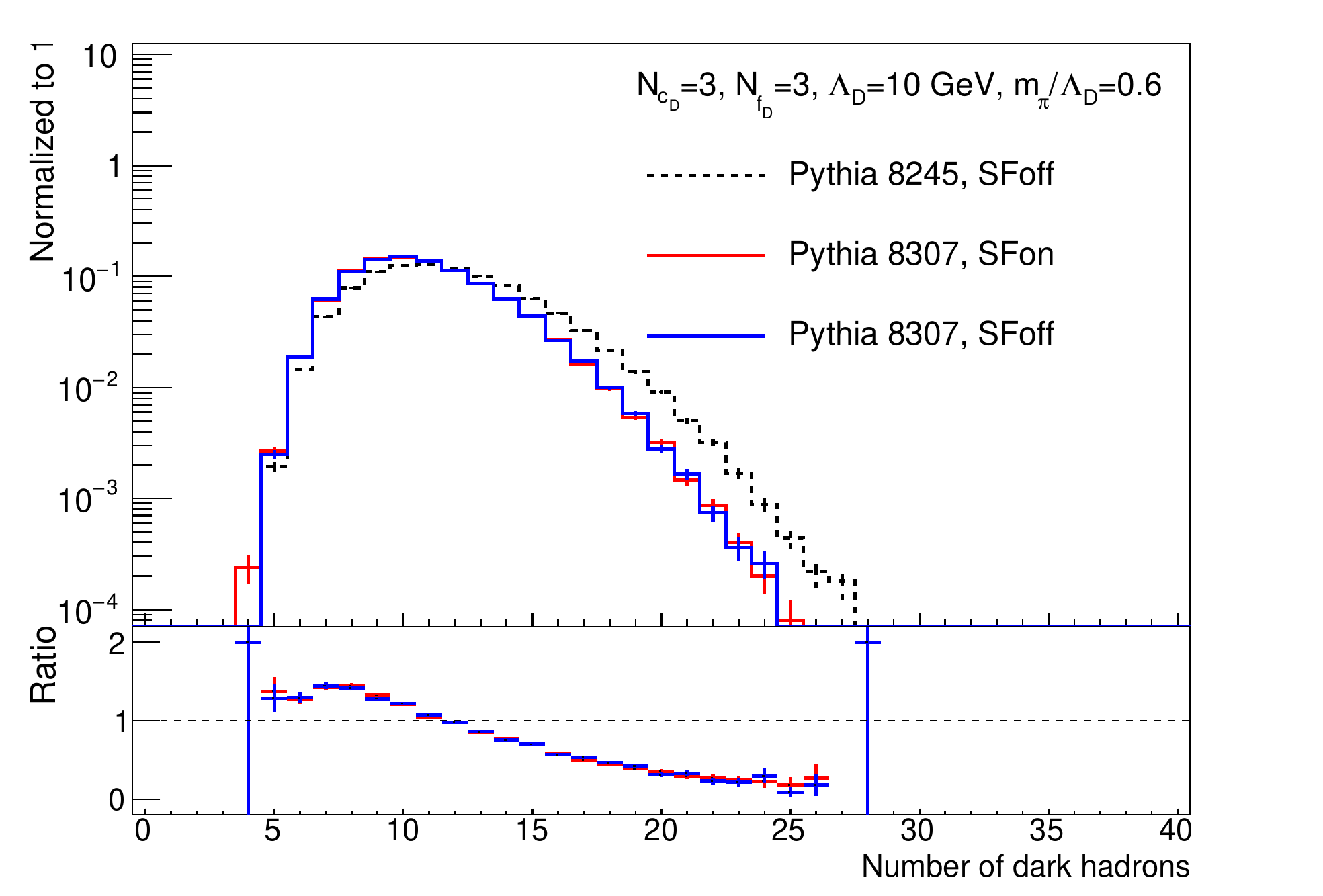}
        \caption{}
        \label{fig:validation_DarkHadronsMult_Nc3Nf3}
    \end{subfigure}
    \begin{subfigure}[b]{0.49\textwidth}
        \centering
        \includegraphics[width=\textwidth]{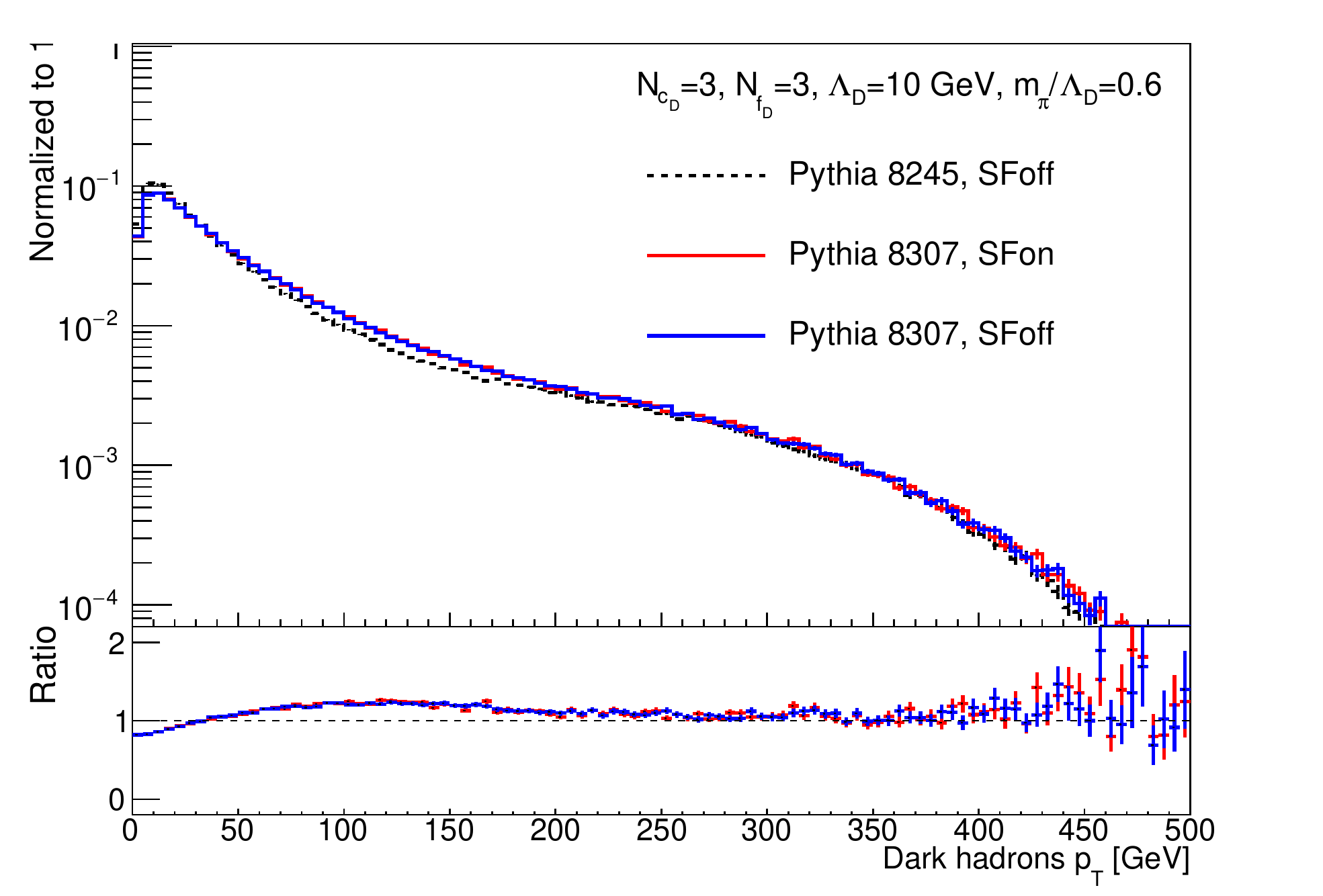}
        \caption{}
        \label{fig:validation_DarkHadronsPt_Nc3Nf3}
    \end{subfigure}    
    \caption{$\Ncdark = 3$, $\Nfdark =3$ model: (a) Distribution of the number of dark hadrons, (b) dark hadrons $p_T$ distribution. The bottom panels show the ratio of the distributions of the new HV module scenarios to the nominal one.}
\end{figure}

\begin{figure}
    \centering
    \begin{subfigure}[b]{0.49\textwidth}
        \centering
        \includegraphics[width=\textwidth]{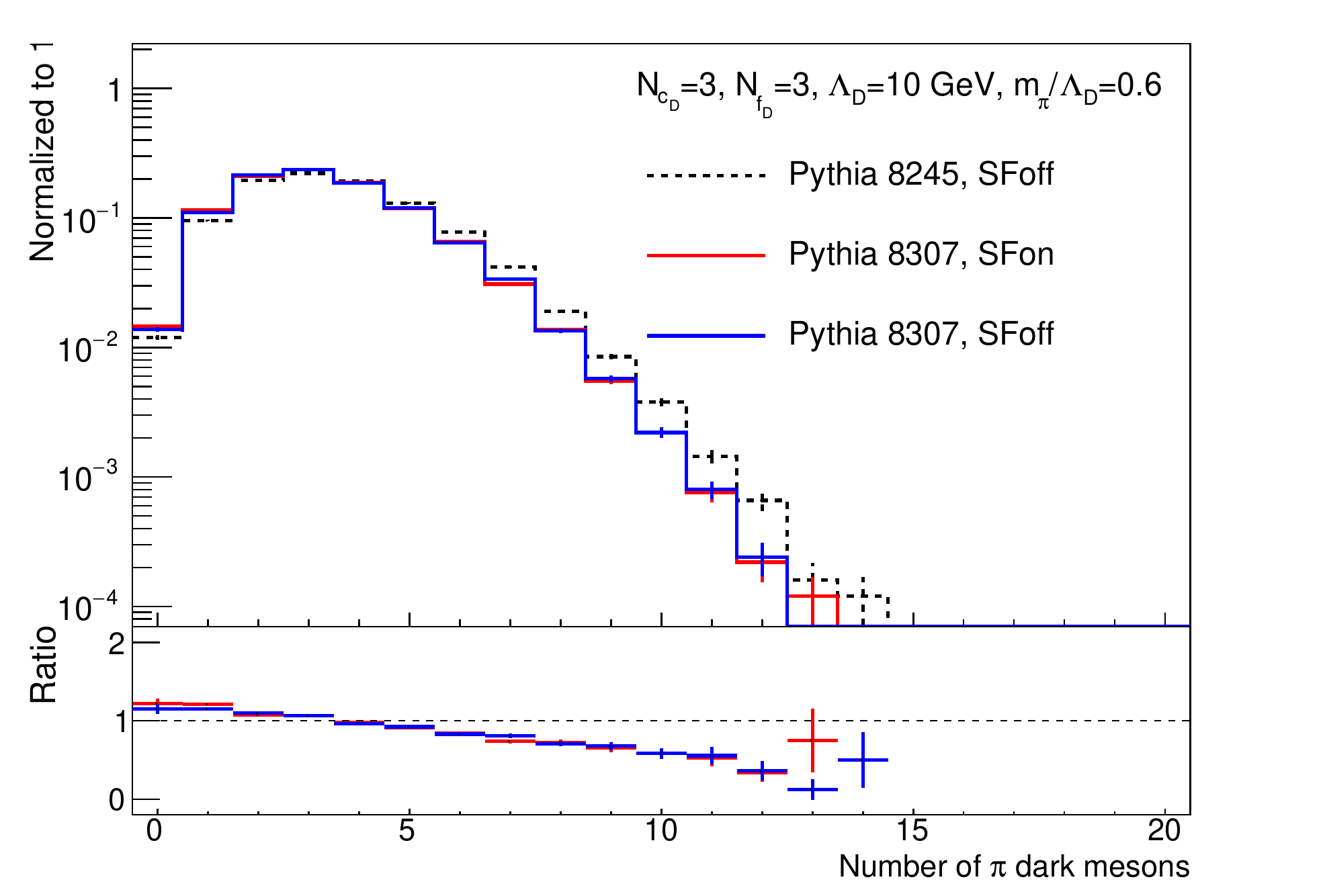}
        \caption{}
        \label{fig:validation_PionsMult_Nc3Nf3}
    \end{subfigure}
    \begin{subfigure}[b]{0.49\textwidth}
        \centering
        \includegraphics[width=\textwidth]{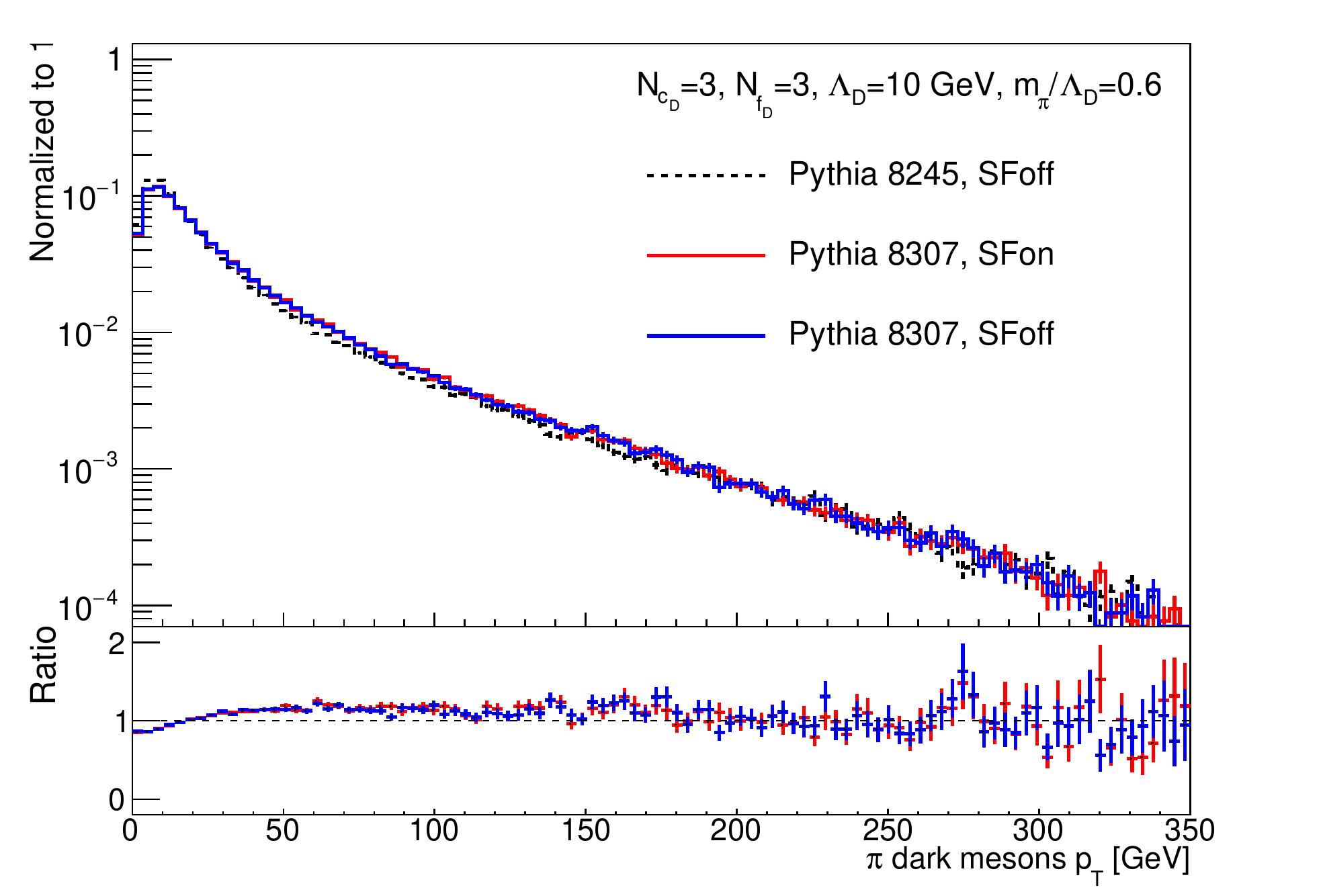}
        \caption{}
        \label{fig:validation_PionsPt_Nc3Nf3}
    \end{subfigure}    
    \caption{$\Ncdark = 3$, $\Nfdark =3$ model: (a) Distribution of the number of dark pions, (b) dark pion $p_T$ distribution. The bottom panels show the ratio of the distributions of the new HV module scenarios to the nominal one.}
\end{figure}

\begin{figure}
    \centering
    \begin{subfigure}[b]{0.49\textwidth}
        \centering
        \includegraphics[width=\textwidth]{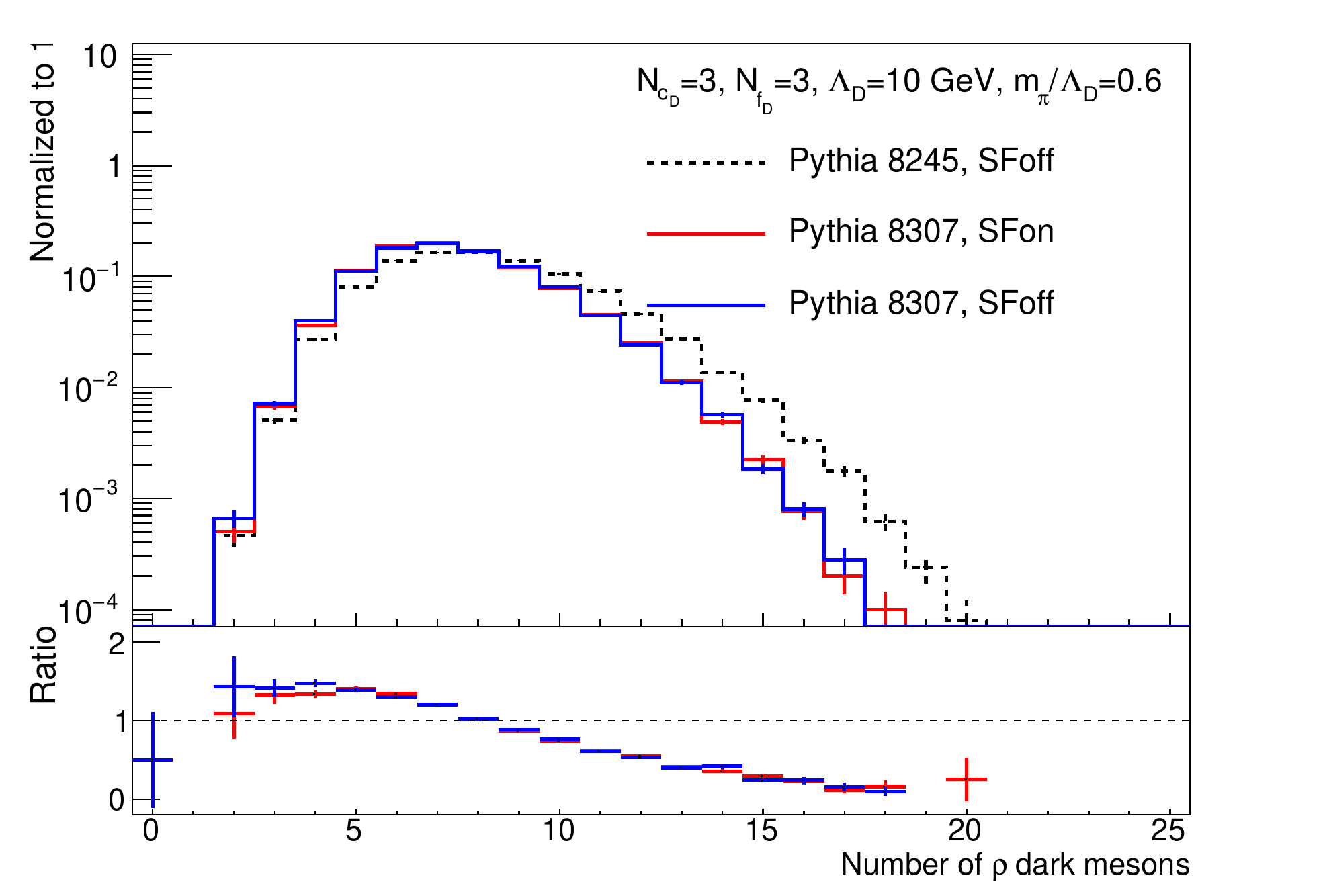}
        \caption{}
        \label{fig:validation_RhosMult_Nc3Nf3}
    \end{subfigure}
    \begin{subfigure}[b]{0.49\textwidth}
        \centering
        \includegraphics[width=\textwidth]{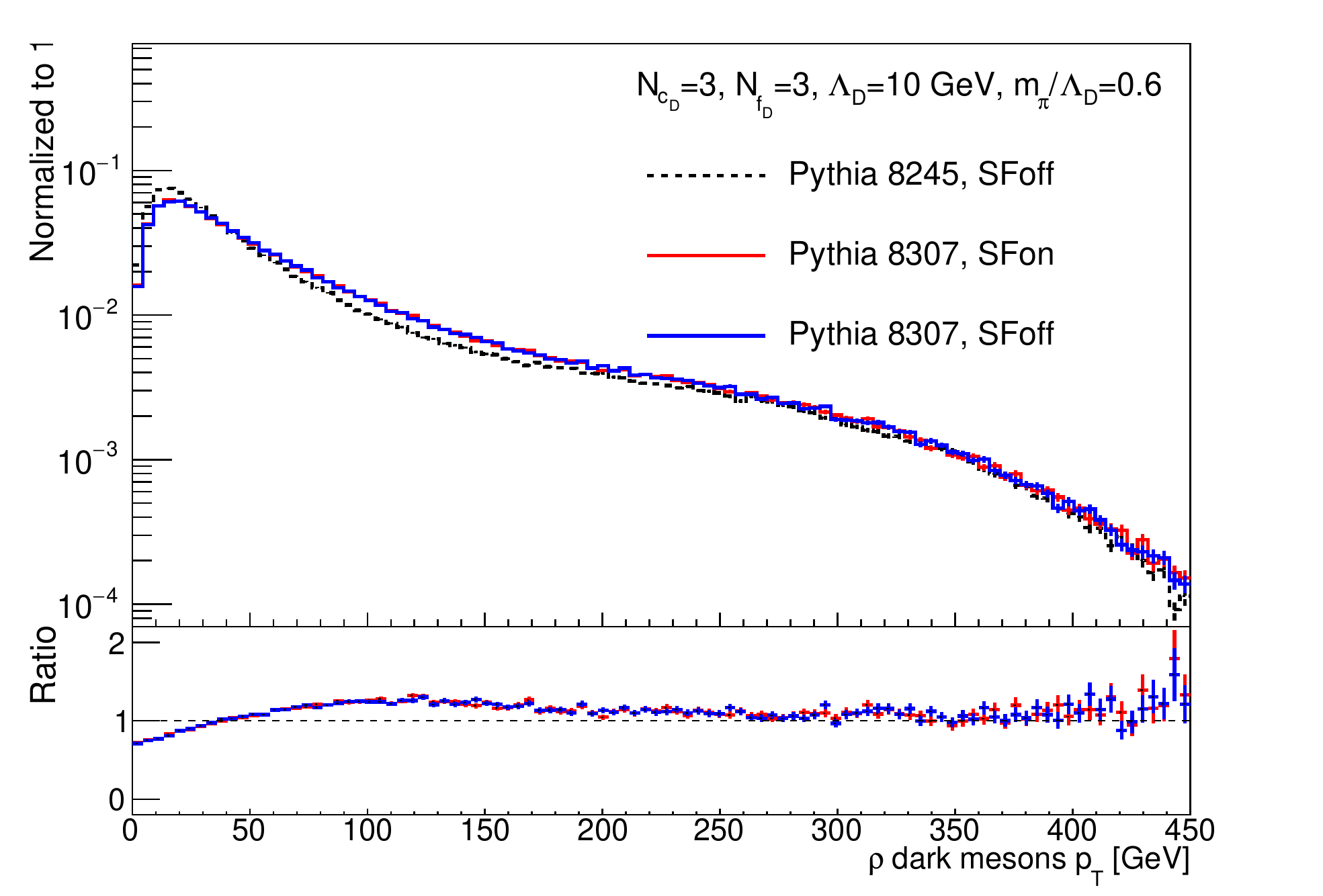}
        \caption{}
        \label{fig:validation_RhosPt_Nc3Nf3}
    \end{subfigure}    
    \caption{$\Ncdark = 3$, $\Nfdark =3$ model: (a) Distribution of the number of $\Prhodark$ mesons, (b) $\Prhodark$ meson $p_T$ distribution. The bottom panels show the ratio of the distributions of the new HV module scenarios to the nominal one.}
\end{figure}

\begin{figure}
    \centering
    \begin{subfigure}[b]{0.49\textwidth}
        \centering
        \includegraphics[width=\textwidth]{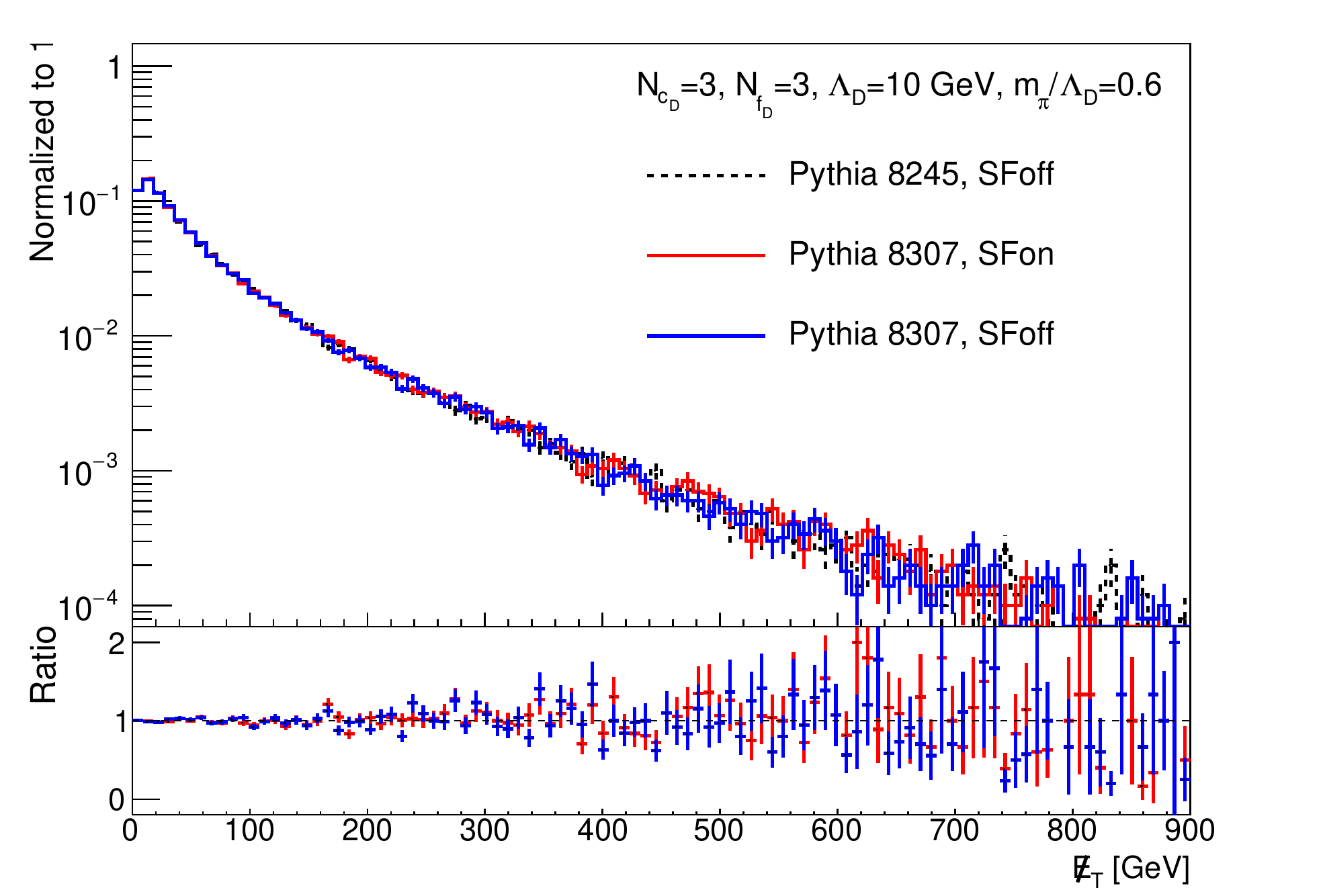}
        \caption{}
        \label{fig:validation_MET_Nc3Nf3}
    \end{subfigure}
    \begin{subfigure}[b]{0.49\textwidth}
        \centering
        \includegraphics[width=\textwidth]{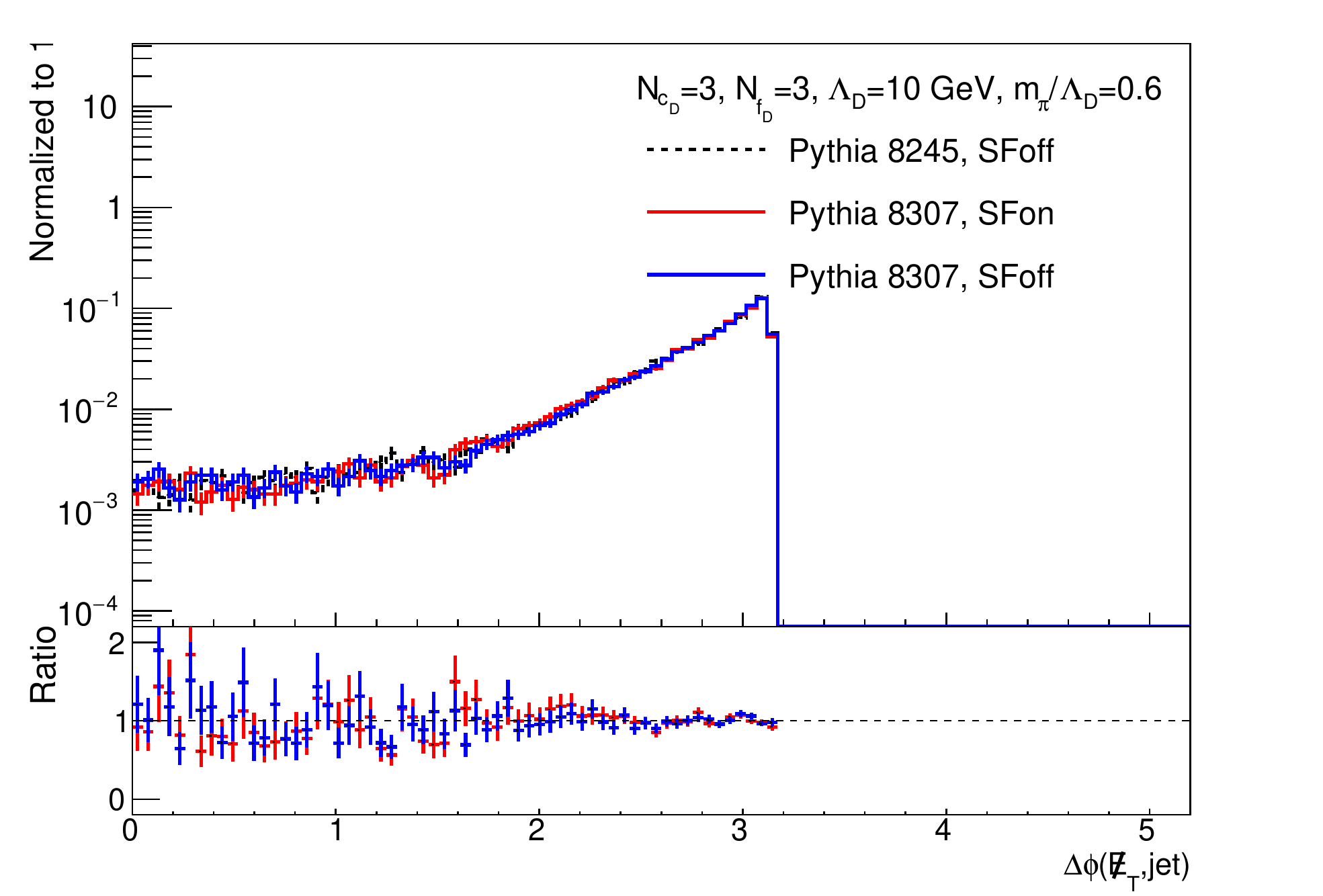}
        \caption{}
        \label{fig:validation_DeltaPhiMETJet_Nc3Nf3}
    \end{subfigure}    
    \caption{$\Ncdark = 3$, $\Nfdark =3$ model: (a) Distribution of the generator level missing transverse energy \met, (b) Distribution of the minimum azimuthal angle between jets and \met. The bottom panels show the ratio of the distributions of the new HV module scenarios to the nominal one.}
\end{figure}

\textbf{$\Ncdark = 3$, $\Nfdark =8$ model}\\

Setting $\Nfdark =8$ brings a whole new set of PIDs both for dark pions and $\Prhodark$ mesons, as confirmed in Figures \ref{fig:validation_PionsPID_Nc3Nf8} and \ref{fig:validation_RhosPID_Nc3Nf8}. For the case with {\tt HiddenValley:separateFlav = on}, additional PIDs from 311 (313) and -311 (-313) to 811 (813) and -811 (-813) are produced with a total of 64 dark pions or $\Prhodark$ mesons, with the same production rate for all states in each multiplet. The multiplicity and transverse momentum of the dark pions are shown in Figures \ref{fig:validation_PionsMult_Nc3Nf8} and \ref{fig:validation_PionsPt_Nc3Nf8} and similarly, for the $\Prhodark$ mesons in Figures \ref{fig:validation_RhosMult_Nc3Nf8} and \ref{fig:validation_RhosPt_Nc3Nf8}. In agreement with the previous model analyzed, a lower multiplicity and a softer transverse momentum of the dark hadrons are observed with the new \PYTHIA 8 HV module. The same conclusions stand when looking at diagonal and off-diagonal dark pions and $\Prhodark$ mesons separately. The missing transverse energy and the minimum azimuthal angle between jets and missing transverse energy can be found in Figures \ref{fig:validation_MET_Nc3Nf8} and \ref{fig:validation_DeltaPhiMETJet_Nc3Nf8}. Once again, these distributions agree for the three cases considered. 

\begin{figure}
    \centering
    \begin{subfigure}[b]{0.49\textwidth}
        \centering
        \includegraphics[width=\textwidth]{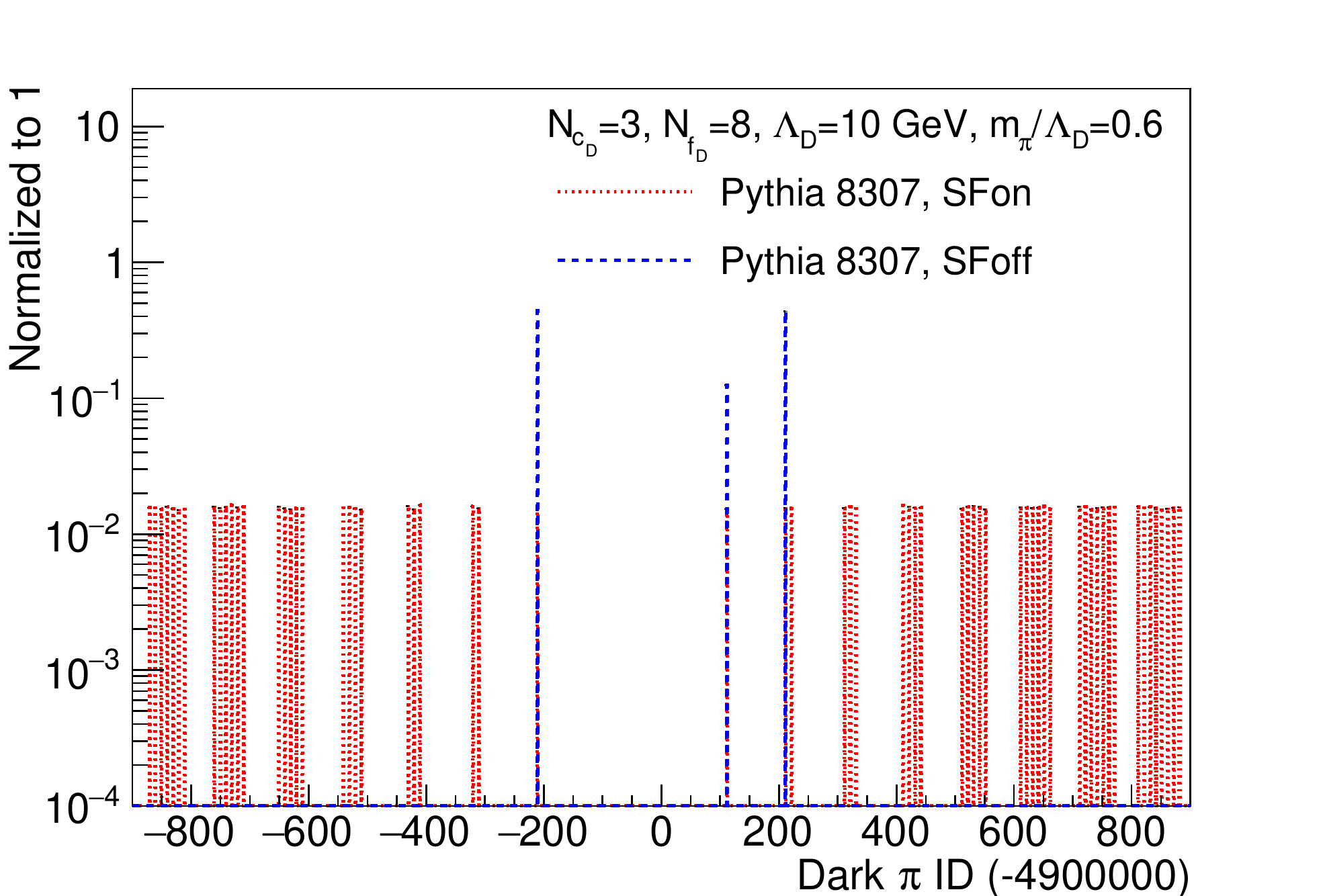}
        \caption{}
        \label{fig:validation_PionsPID_Nc3Nf8}
    \end{subfigure}
    \begin{subfigure}[b]{0.49\textwidth}
        \centering
        \includegraphics[width=\textwidth]{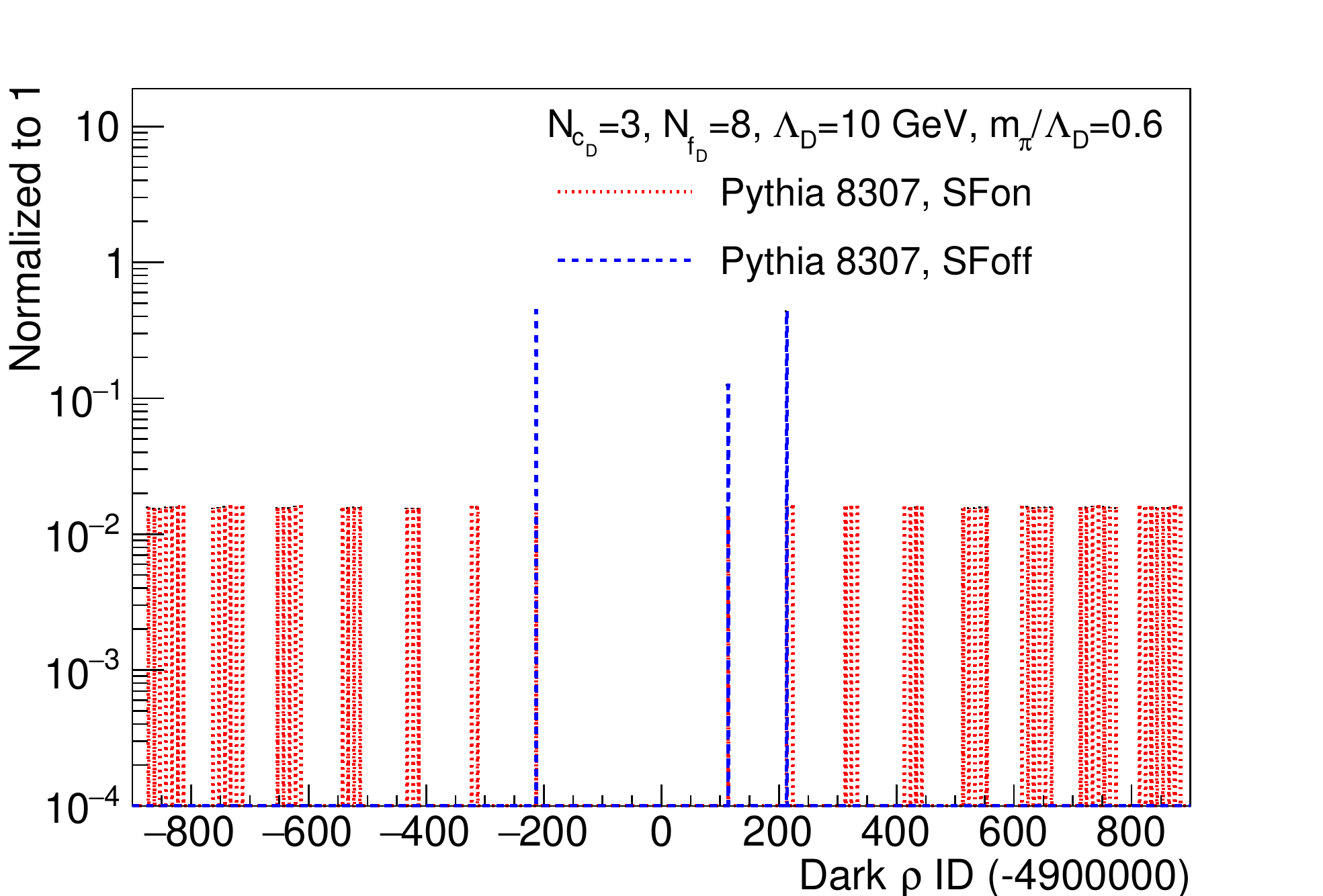}
        \caption{}
        \label{fig:validation_RhosPID_Nc3Nf8}
    \end{subfigure}    
    \caption{$\Ncdark = 3$, $\Nfdark =8$ model: (a) PdgId distribution for dark pions, (b) PdgId distribution for $\Prhodark$ mesons.}
\end{figure}

\begin{figure}
    \centering
    \begin{subfigure}[b]{0.49\textwidth}
        \centering
        \includegraphics[width=\textwidth]{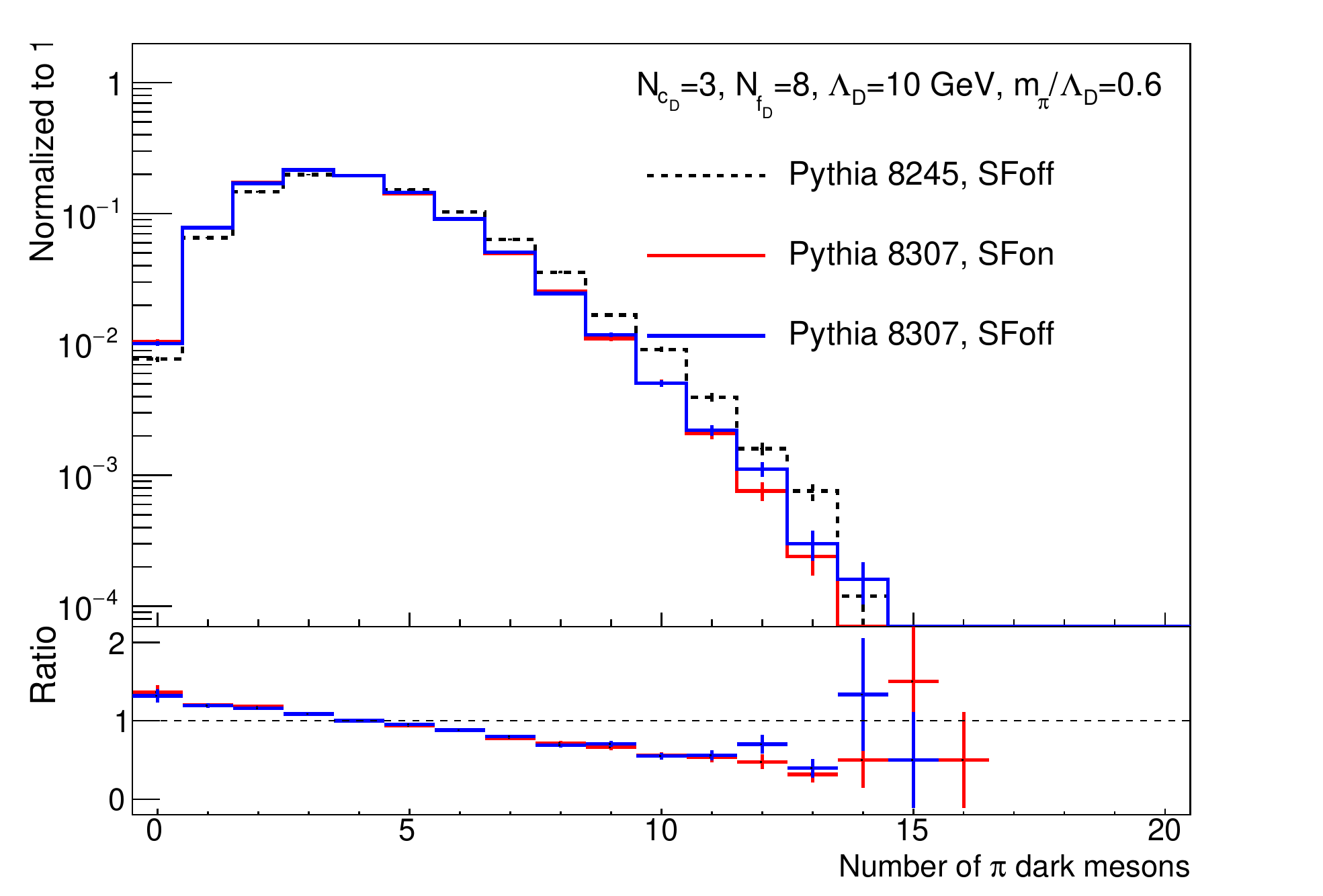}
        \caption{}
        \label{fig:validation_PionsMult_Nc3Nf8}
    \end{subfigure}
    \begin{subfigure}[b]{0.49\textwidth}
        \centering
        \includegraphics[width=\textwidth]{plots/Nc_3_Nf_3_GenPionsPt_Normalizedto1_Log.pdf}
        \caption{}
        \label{fig:validation_PionsPt_Nc3Nf8}
    \end{subfigure}    
    \caption{$\Ncdark = 3$, $\Nfdark =8$ model: (a) Distribution of the number of dark pions, (b) dark pion $p_T$ distribution. The bottom panels show the ratio of the distributions of the new HV module scenarios to the nominal one.}
\end{figure}

\begin{figure}
    \centering
    \begin{subfigure}[b]{0.49\textwidth}
        \centering
        \includegraphics[width=\textwidth]{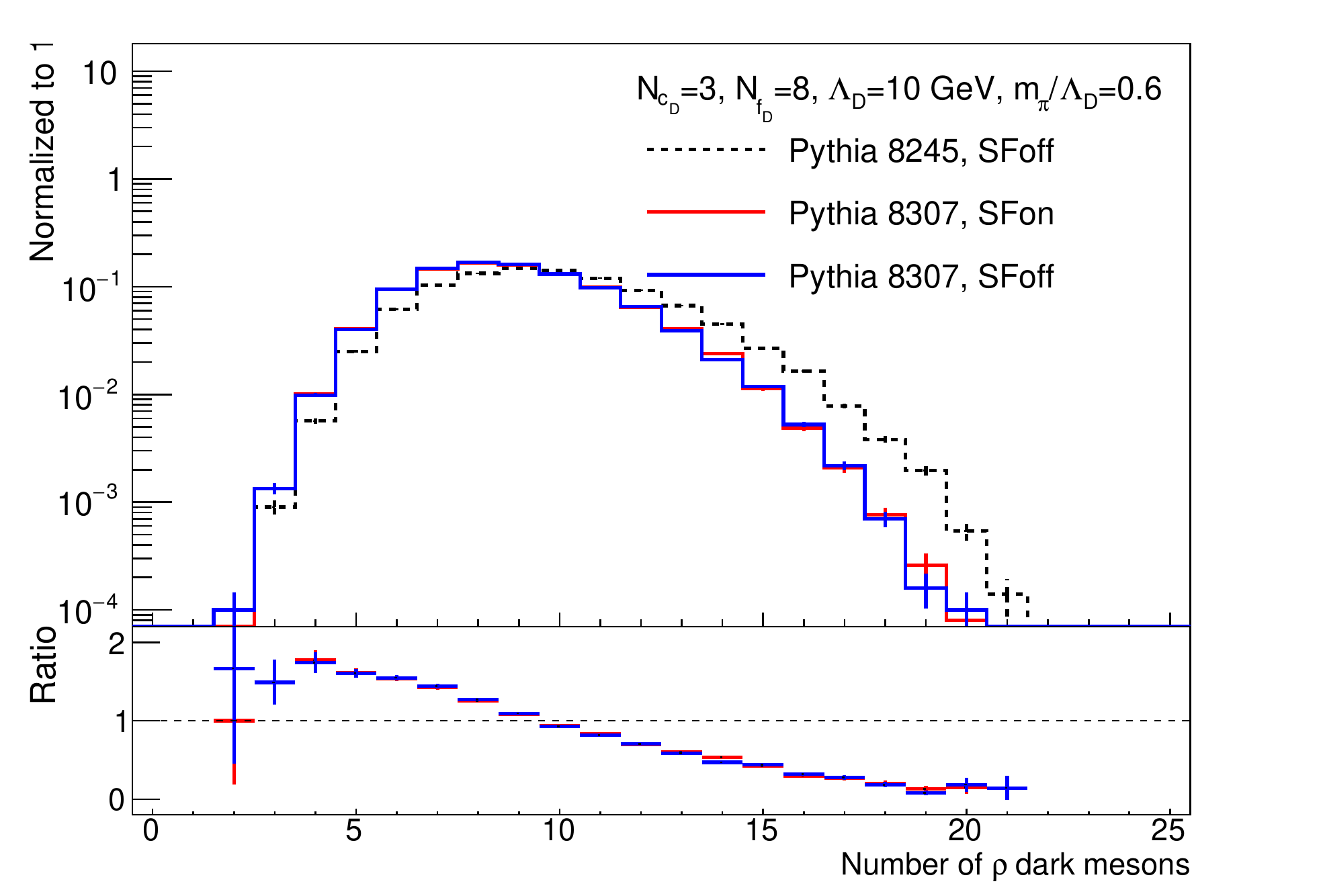}
        \caption{}
        \label{fig:validation_RhosMult_Nc3Nf8}
    \end{subfigure}
    \begin{subfigure}[b]{0.49\textwidth}
        \centering
        \includegraphics[width=\textwidth]{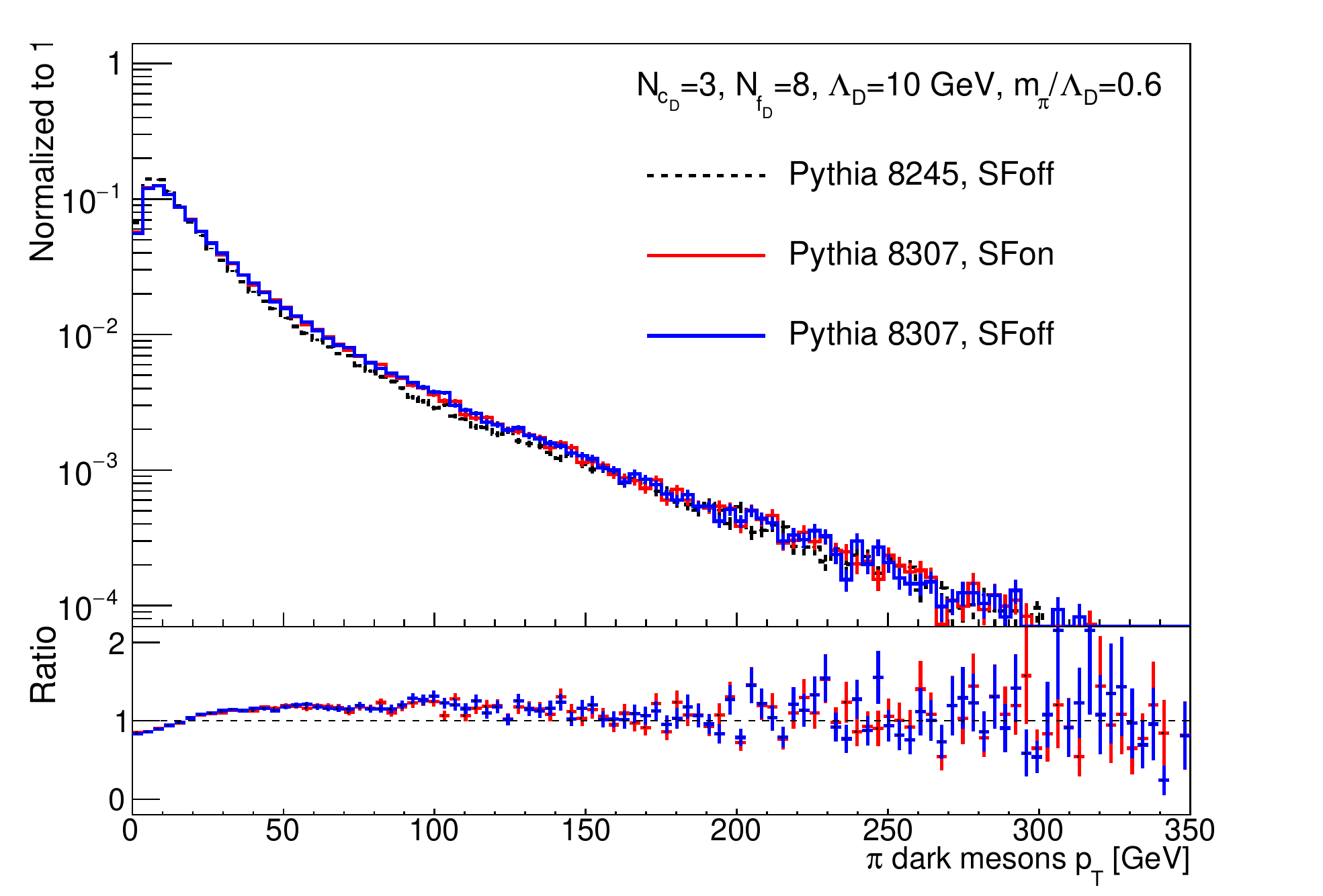}
        \caption{}
        \label{fig:validation_RhosPt_Nc3Nf8}
    \end{subfigure}    
    \caption{$\Ncdark = 3$, $\Nfdark =8$ model: (a) Distribution of the number of $\Prhodark$ mesons, (b) $\Prhodark$ meson $p_T$ distribution. The bottom panels show the ratio of the distributions of the new HV module scenarios to the nominal one.}
\end{figure}

\begin{figure}
    \centering
    \begin{subfigure}[b]{0.49\textwidth}
        \centering
        \includegraphics[width=\textwidth]{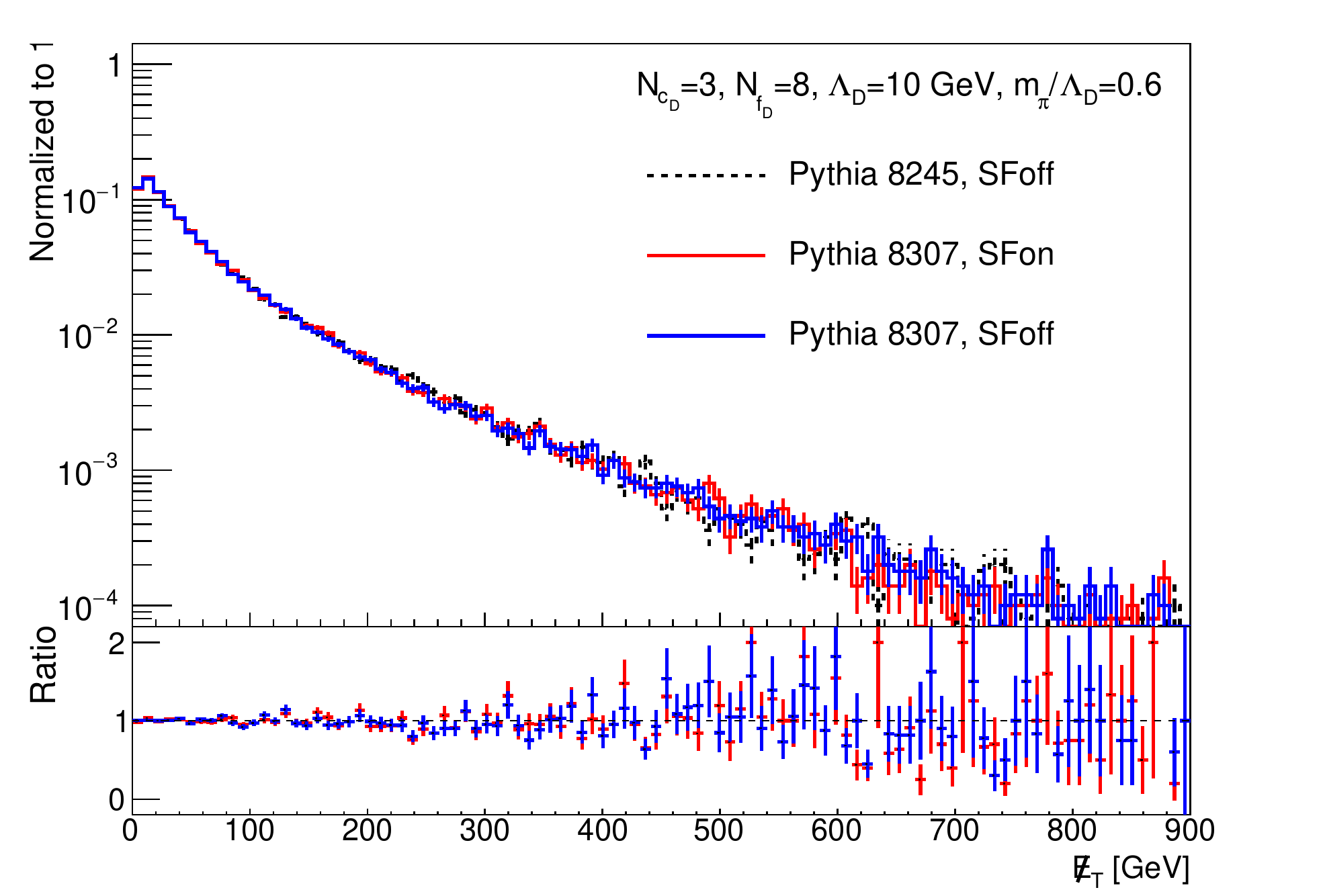}
        \caption{}
        \label{fig:validation_MET_Nc3Nf8}
    \end{subfigure}
    \begin{subfigure}[b]{0.49\textwidth}
        \centering
        \includegraphics[width=\textwidth]{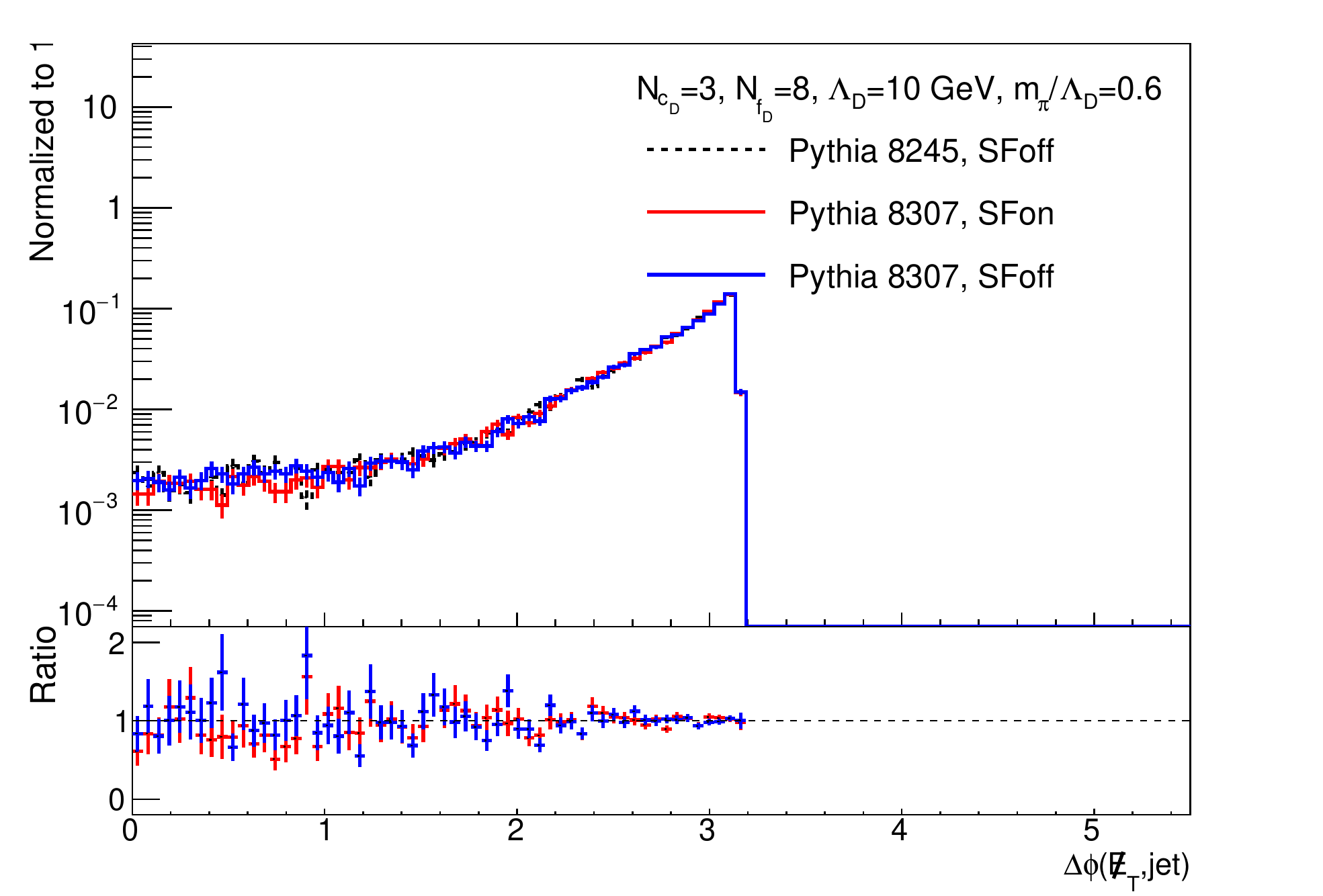}
        \caption{}
        \label{fig:validation_DeltaPhiMETJet_Nc3Nf8}
    \end{subfigure}    
    \caption{$\Ncdark = 3$, $\Nfdark =8$ model: (a) Distribution of the generator level missing transverse energy \met, (b) Distribution of the minimum azimuthal angle between jets and\met. The bottom panels show the ratio of the distributions of the new HV module scenarios to the nominal one.}
\end{figure}

Through this validation, we thus highlight some of the differences between \PYTHIA 8.245 and \PYTHIA 8.307 hidden valley module. We also demonstrate that switching  SeparateFlav on  or off does not lead to physics differences in the production rates or kinematics of the events, but  allows to access additional meson PIDs whose masses and branching ratios can be manipulated according to theory predictions.

\subsection{Phenomenological studies of jet substructure observables}\label{sec:phenovalid}

\emph{Contributors: Cesare Cazzaniga, Florian Eble, Aran Garcia-Bellido, Nicoline Hemme, Nukulsinh Parmar}

In this section we exemplify the kinematic distributions resulting from benchmarks proposed in section~\ref{parameters}, focusing on the benchmark with $\Lambda_D = 10$ GeV and $\Nfdark = 3$, and belonging to the regime $\mpidark < \mrhodark/2$ for which the $\Prhodark \to \Ppidark\Ppidark$ decay mode is open. The mass of $\PZprime$ boson is set to 1 TeV. We then consider either 1, 2 or 3 diagonal pions decaying to SM particles. The $\Ppidark$ mesons decay to $c\bar{c}$ as this is the heaviest allowed fermion pair. We simulate this signal using $\PYTHIA8.307$ with {\tt HiddenValley:separateFlav = off} and pass it through {\tt DELHPES3} using the HL-LHC card. Jets are clustered using FastJet~\cite{Cacciari:2011ma, Cacciari:2005hq} using \mbox{anti-kt} algorithm~\cite{Salam:2008}. We produce 50k events for the distributions shown in section~\ref{sec:kine}, and 500k events for those shown in~\ref{sec:JSS}.
\subsubsection{Basic kinematic distributions}\label{sec:kine}
As the $\Prhodark$ mesons all decay within the dark shower in this benchmark, they are not included in the calculation of \rinv. Figure \ref{fig:JSS_rinv} shows the \rinv parameter distribution. As expected, the 1-$\Ppidark$ decay model has a an average \rinv of $\simeq \frac{8}{9}$ as all $\Prhodark$ mesons decay to $\Ppidark$ and only 1 of the 9 $\Ppidark$ is unstable. The 2-$\Ppidark$ decays has a mean of $\rinv\simeq \frac{7}{9}$ and for the 3-$\Ppidark$, a mean of $\rinv\simeq \frac{2}{3}$ is obtained.

\begin{figure}[h!]
    \centering
    \includegraphics[width=0.35\textwidth]{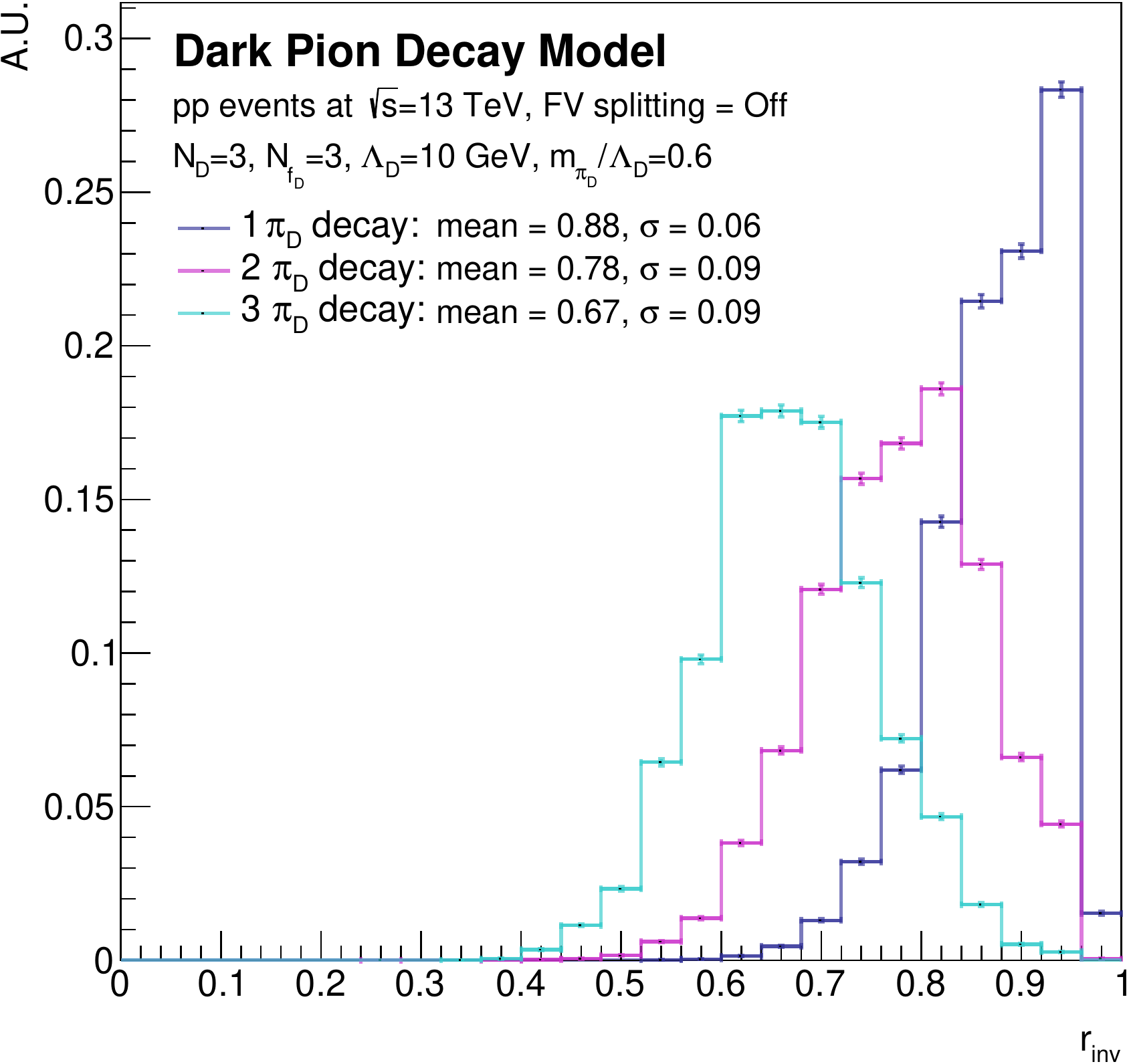}
    \caption{Comparison of \rinv for 1-, 2- and 3-\Ppidark decay models.}
    \label{fig:JSS_rinv}
\end{figure}

In Figure \ref{fig:JSS_basickinematicdistributions}, some basic kinematic variable distributions are compared for the 3 dark pion decay models. These generator-level distributions are computed with jets of radius $R=0.4$ and  $\pt > 25$ GeV. The $\pt$ distribution in Figure \ref{fig:JSS_leadjetpT} shows that more dark pion decays result in a higher average lead jet $\pt$, as expected when more dark particles decay to SM particles that can be detected. The \met distribution shown in Figure \ref{fig:JSS_2jetsmetMET} reveals very similar values between the 3 different models, which may seem contrary to what one would expect, i.e. more SM-decaying pions might be expected to result in lower \met; however, while more stable dark pions truly gives higher missing or invisible energy in the system, the additional invisible particles may be evenly distributed between the two back-to-back jets and therefore not appear in the detector as  additional \met.

Figure \ref{fig:JSS_jetmetmT} shows the distributions of the transverse mass, $\MT$, of the leading and sub-leading jet and the \met. As can be seen, having more SM-decaying dark pions generally yields a higher transverse mass. As the \met remains relatively stable but the jet $\pt$ increases with the number of unstable dark pions, this results in higher $\MT$ values.

\begin{figure}[h!]
    \centering    
    \begin{subfigure}[b]{0.313\textwidth}
        \centering
        \includegraphics[width=\textwidth]{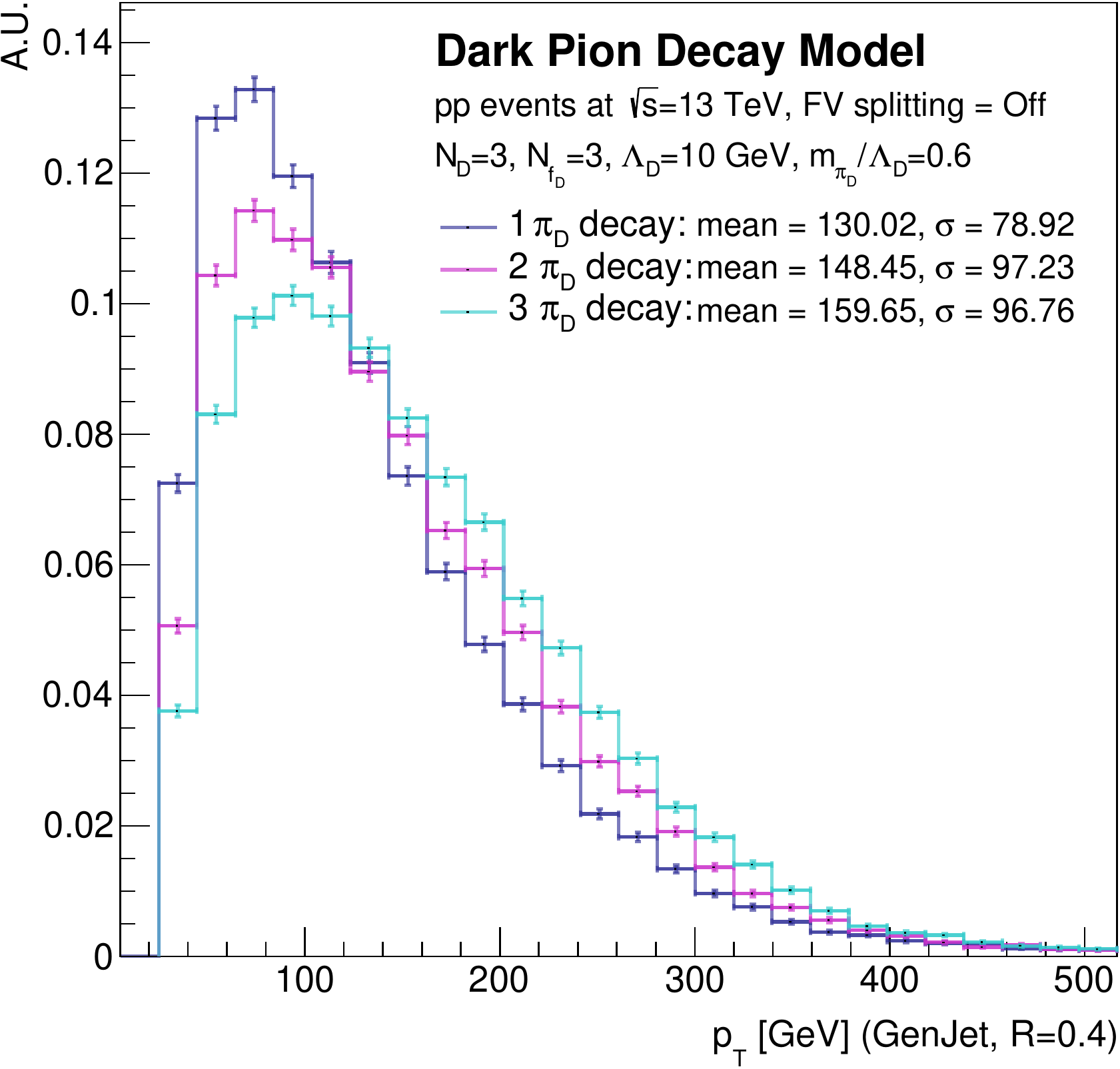}
        \caption{}
        \label{fig:JSS_leadjetpT}
    \end{subfigure} 
    \hfill
    \begin{subfigure}[b]{0.32\textwidth}
        \centering
        \includegraphics[width=\textwidth]{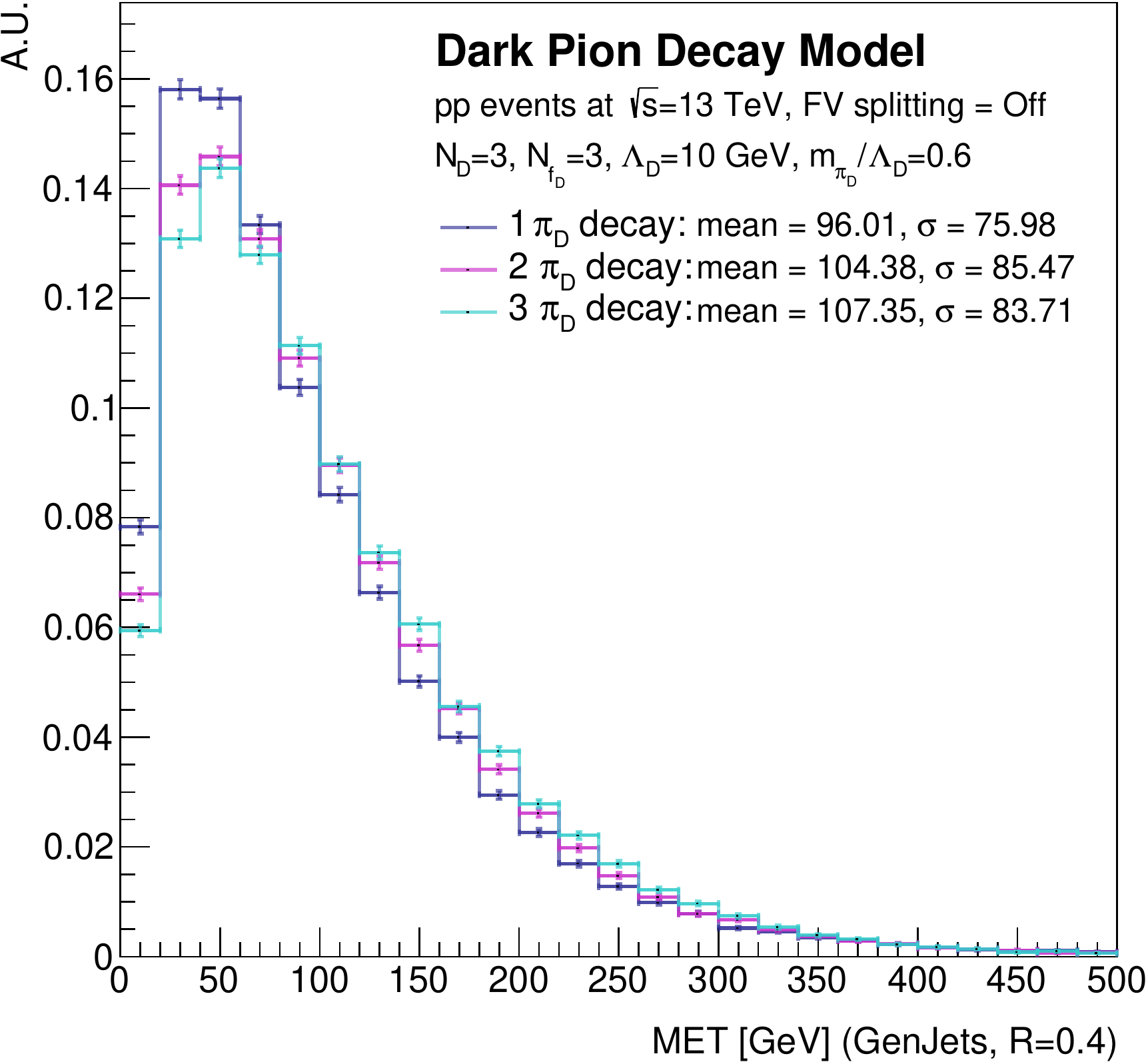}
        \caption{}
        \label{fig:JSS_2jetsmetMET}
    \end{subfigure} 
    \hfill
    \begin{subfigure}[b]{0.32\textwidth}
        \centering
        \includegraphics[width=\textwidth]{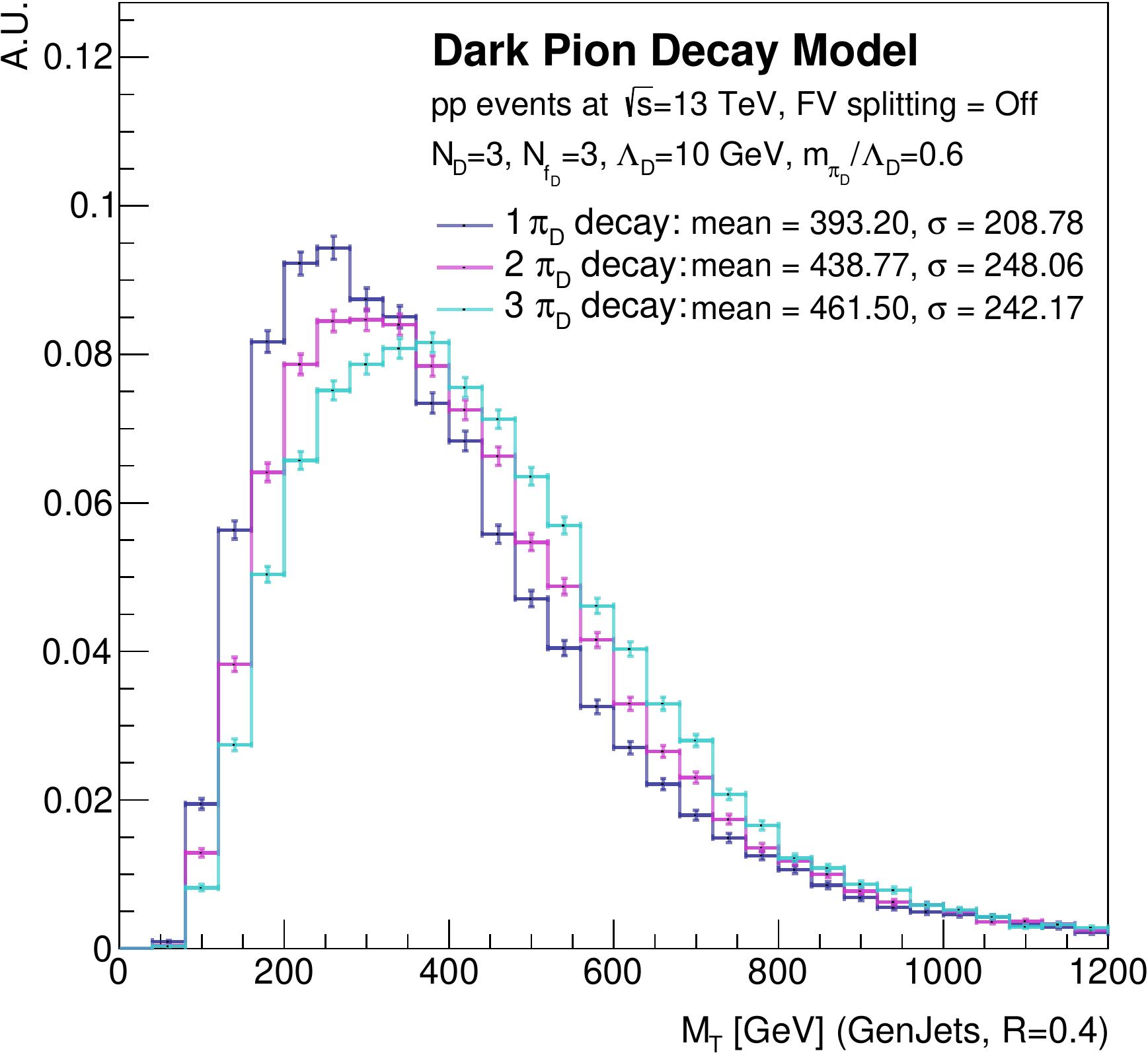}
        \caption{}
        \label{fig:JSS_jetmetmT}
    \end{subfigure}
    
    \caption{Comparison of kinematic variable distribution for 1, 2 and 3 dark pions decay: (a) $\pt$ distribution of the leading jet in the event (b) \met distribution (c) Distribution of $\MT$ of the leading and sub-leading jets plus \met.}
    \label{fig:JSS_basickinematicdistributions}
\end{figure}

\subsubsection{Jet Substructure consistency}
\label{sec:JSS}
Experimental searches and phenomenological studies for dark showers exploit jet substructure (JSS) observables to tag jets as dark jets~\cite{CMS:2021dzg, Canelli:2021aps,Kar:2020bws, Cohen:2020afv, Park:2017rfb}. Comparisons of jet suCohen:2020afvbstructure variables of interest, between the former and the new Hidden Valley \PYTHIA modules, between different dark vector meson production probabilities, and between different number of unstable dark pions \Ppidark, are presented in this section. In the \PYTHIA 8 Hidden Valley module~\cite{HV:2010,HV:2011} the probability to produce a dark vector meson can be changed by setting the parameter {\tt HiddenValley:probVector}. There is no precise theoretical prediction for the fraction of dark vector mesons produced after string fragmentation in the hidden sector. Assuming a mass degeneracy between vector and pseudo-scalar states, it is reasonable to fix {\tt HiddenValley:probVector = 0.75} as pseudo-scalars have 1 degree of freedom while vector mesons have 3 degrees of freedom. However, generically the $\Prhodark$ mesons and dark pions are not mass degenerate, hence the production rate of pseudo-scalars is enhanced compared to mass-degenerate scenarios due to the larger phase space available for lighter states. In this specific case, a reasonable value  is {\tt HiddenValley:probVector = 0.5}, very much like in QCD.  

For this study, generator-level jets have been clustered with the inclusive \mbox{anti-kt} algorithm~\cite{Salam:2008}, choosing a cone size $R=0.8$ and a minimum \pt~of 200~GeV. Jets were clustered from all visible SM particles and jet constituents were used for computing the jet substructure. The JSS observables studied here are the generalized angularities $\lambda^{\kappa}_{\beta}$, the $N$-subjettiness~\cite{Thaler_2011} $\tau_N$ and jet major and minor axes.

Generalized angularities are presented in Fig.~\ref{fig:generalized angularities} and are defined from the constituents \mbox{$i \in \{ 1, \cdots , N \}$} carrying momentum fraction $z_i$ inside a jet of cone size $R$ as:

\begin{equation}
    \lambda^{\kappa}_{\beta} = \sum_{i \in jet} z_i^{\kappa} \biggr(\frac{\Delta R_{i,jet}}{R} \biggr)^{\beta} 
\end{equation}

$N$-subjettiness $\tau^{\beta}_{N}$ are designed to count the number of subjets inside a jet. In specific, $N$-subjettiness is defined as:
\begin{equation}
\tau^{\beta}_{N} = \sum_{i}p_{\textrm{T},i} \textrm{min}({R^{\beta}_{1,i}, R^{\beta}_{2,i}, R^{\beta}_{3,i}, \cdots, R^{\beta}_{N,i}})
\end{equation}
where the sum is over the jet constituents, and $R^{\beta}_{N,i}$ is the distance between the $N$th subjet and the $i$th constituent of the jet. $\tau^{\beta}_{N}$ measures departure from N-parton energy flow: if a jet has N subjets, $\tau^{\beta}_{N-1}$ should be much larger than $\tau^{\beta}_{N}$. Originally, $\tau^{\beta}_{N}$ have been introduced in order to identify hadronically-decaying boosted objects and reject QCD background. In those studies, the angular parameter $\beta$ has been fixed to 1 as done in previous studies for boosted objects  discrimination \cite{Thaler:2010tr}.

The shape of the jet can be approximated by an ellipse in the $\eta - \phi$ plane. The major and minor axes are the two principal components of this ellipse and are defined from the following symmetric matrix $M$:

\begin{equation}
M = \begin{pmatrix}
\sum_{i} p_{\textrm{T},i}^{2}\Delta\eta_{i}^{2} & - \sum_{i} p_{\textrm{T},i}^{2}\Delta\eta_{i}\Delta\phi_{i} \\
- \sum_{i} p_{\textrm{T},i}^{2}\Delta\eta_{i}\Delta\phi_{i} & \sum_{i} p_{\textrm{T},i}^{2}\Delta\phi_{i}^{2}
\end{pmatrix}
\end{equation}
where the sum runs over all constituents of the jet and $\Delta \eta$, $\Delta \phi$ are the differences in $\eta$ and $\phi$ with respect to the jet axis. The major and minor axes are defined from the eigenvalues $\lambda_1$ and $\lambda_2$ of $M$ as:
\begin{equation}
    \sigma_{\textrm{minor,major}} = \sqrt{\frac{\lambda_{1,2}}{\sum_i p_{\textrm{T},i}^{2}}}
\end{equation}

Generalized angularities belong to the category of jet shape variables and they have been originally built to measure the quantity of radiation inside a jet in order to discriminate between jets initiated by quarks and those initiated by gluons~\cite{Larkoski:2014,Larkoski:2014pca}. Indeed, for the gluon jets the values of the generalized angularities are usually expected to be larger since gluons are expected to radiate more due to the larger color factor. In the same way, these observables have been used in analyses to discriminate between SM jets and dark jets~\cite{CMS:2021rwb}. In particular, the dark jets are expected to be wider than SM jets due to the double hadronization process and the mass splitting between the dark bound states and the SM quarks. 

\begin{figure}[htb!]
\centering
\includegraphics[width=0.4\linewidth]{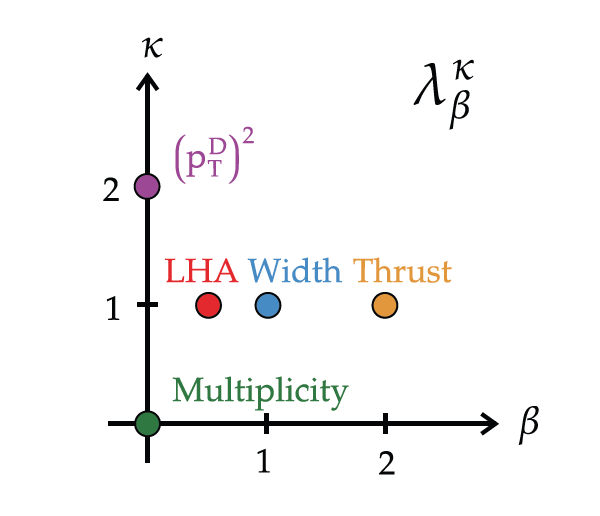}
\caption{Visualization of the space of the generalized angularities $\lambda^{k}_{\beta}$. Adapted from \cite{Larkoski:2014}.}
\label{fig:generalized angularities}
\end{figure}

We first start by comparing JSS observables between the old and new  \PYTHIA Hidden Valley modules. Comparison of quark-gluon discriminant variables and $N$-subjettiness variables are shown in Figs.~\ref{fig:qg_old_vs_new_hv_module} and~\ref{fig:nsubjettiness_old_vs_new_hv_module}. Some systematic differences are observed for the jet transverse momentum dispersion \ptd, due to the different number and different \pt spectrum of the dark mesons \Ppidark and \Prhodark in the new \PYTHIA Hidden Valley module. \mbox{$N$-subjettiness} are smaller with the new module when decreasing \rinv and looking for high number of subjets.
No large systematic difference is observed for the other substructure variables.

\begin{figure}[htb!]
\centering
\includegraphics[width=0.9\linewidth]{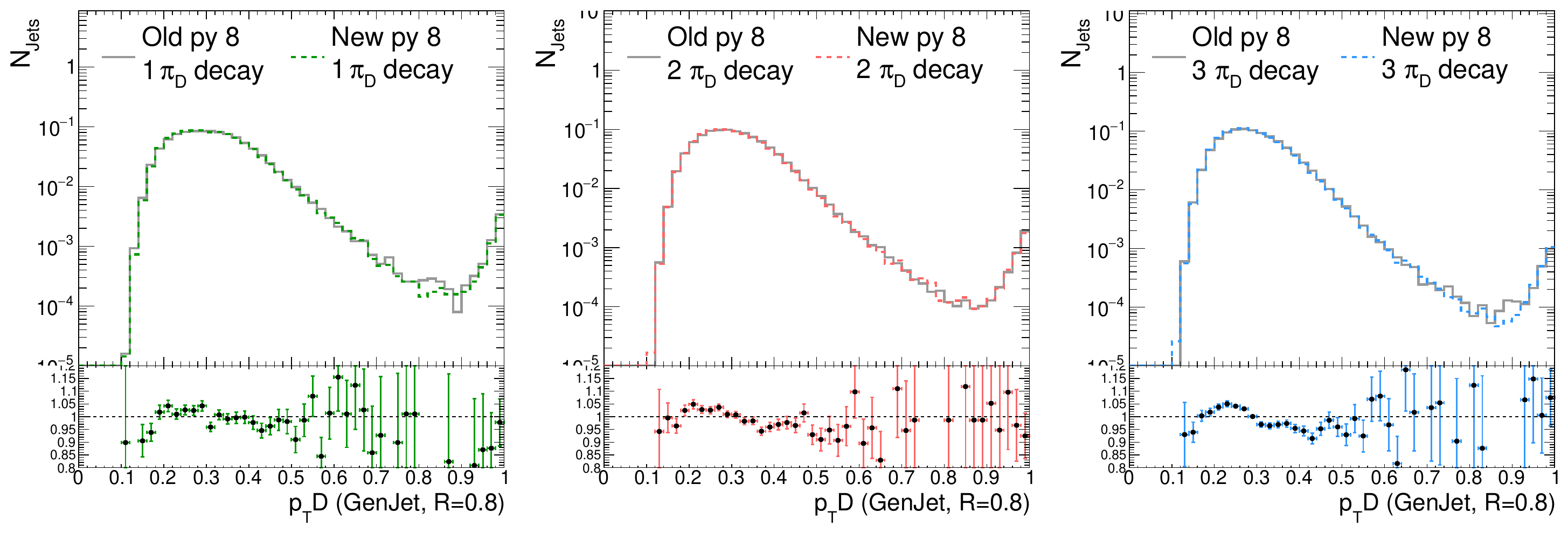}
\includegraphics[width=0.9\linewidth]{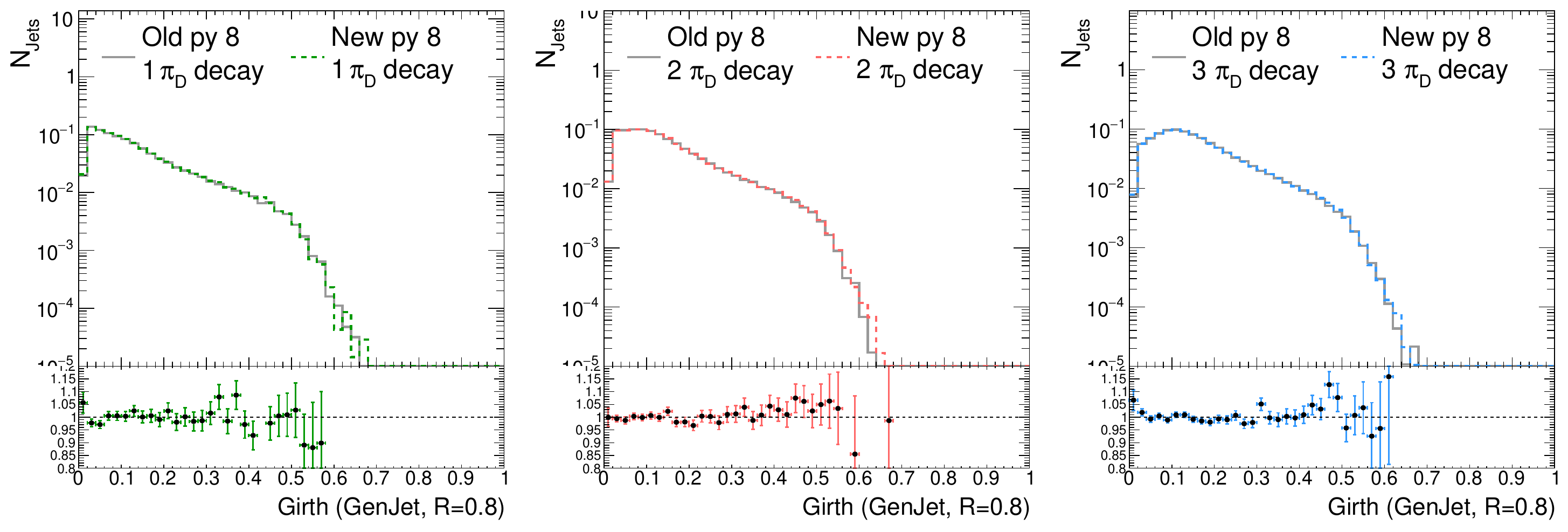}
\includegraphics[width=0.9\linewidth]{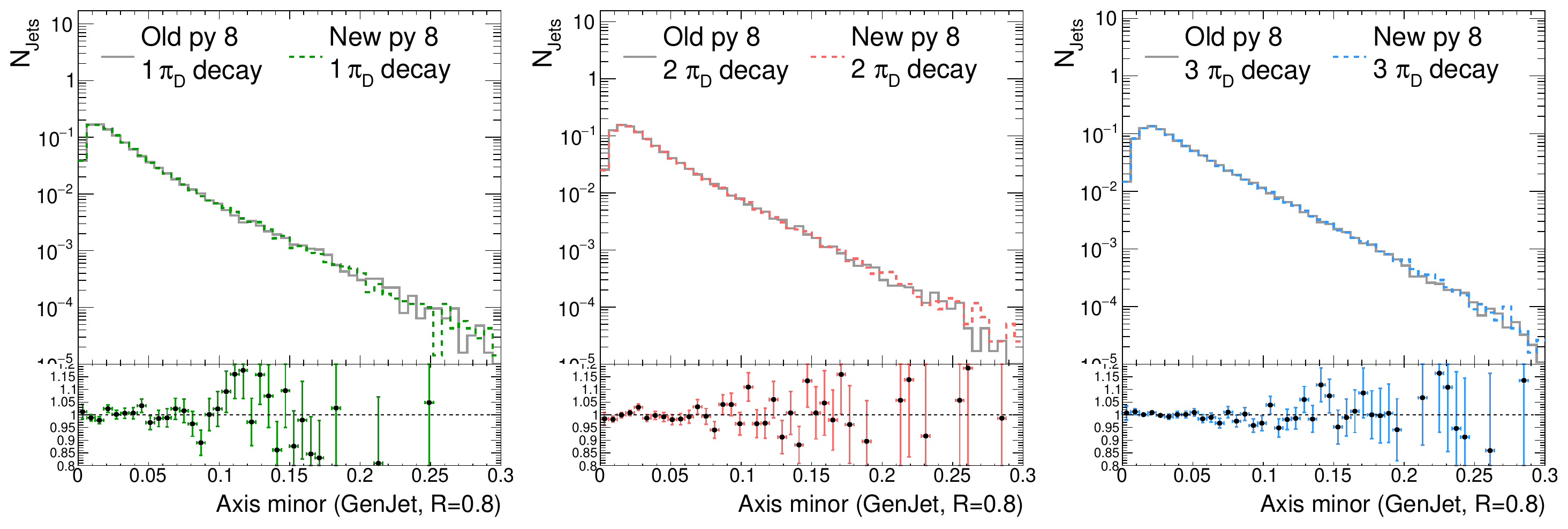}
\includegraphics[width=0.9\linewidth]{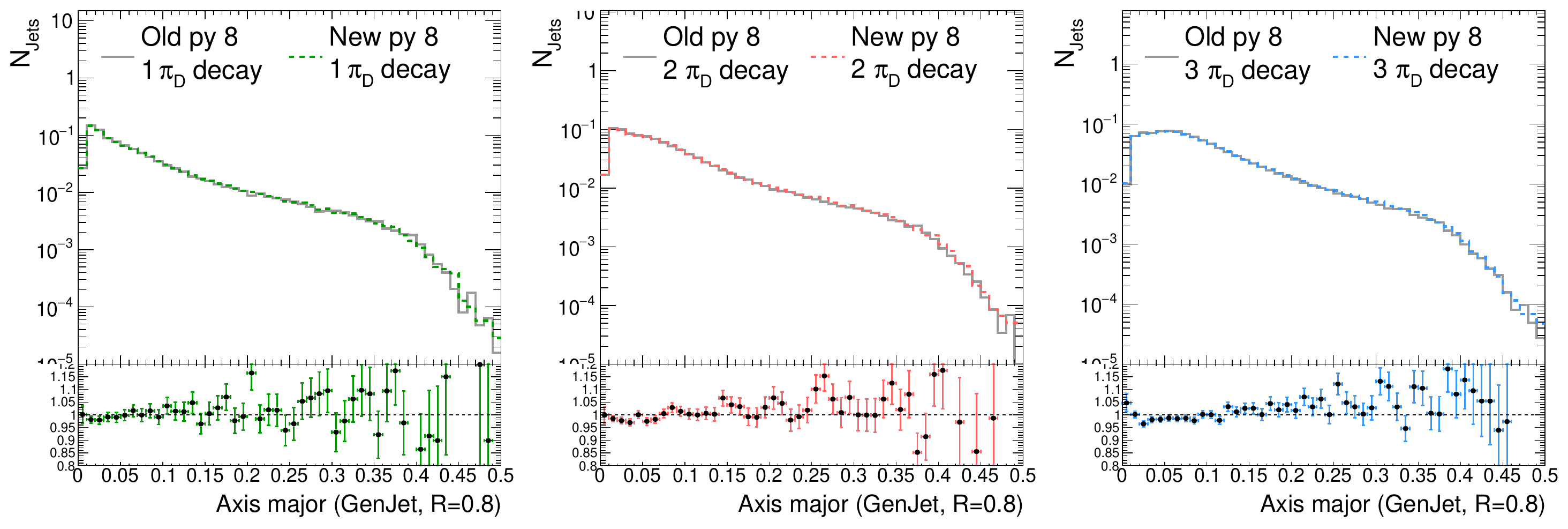}
\caption{Comparison of quark-gluon discriminant variables between old (grey solid line) and new (colored dashed line) Hidden Valley modules. The comparison is made for 1 (left), 2 (middle) and 3 (right) dark pions decay. The plotted ratio is the ratio of new to old \PYTHIA module.}
\label{fig:qg_old_vs_new_hv_module}
\end{figure}

\begin{figure}[htb!]
\centering
\includegraphics[width=0.9\linewidth]{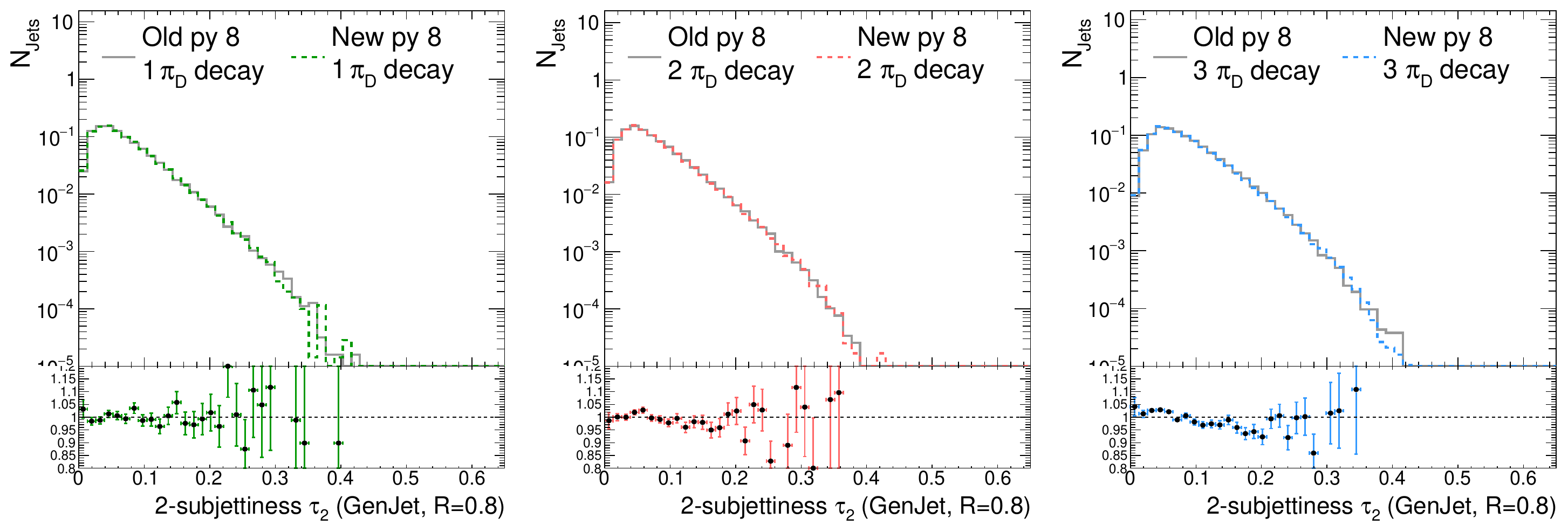}
\includegraphics[width=0.9\linewidth]{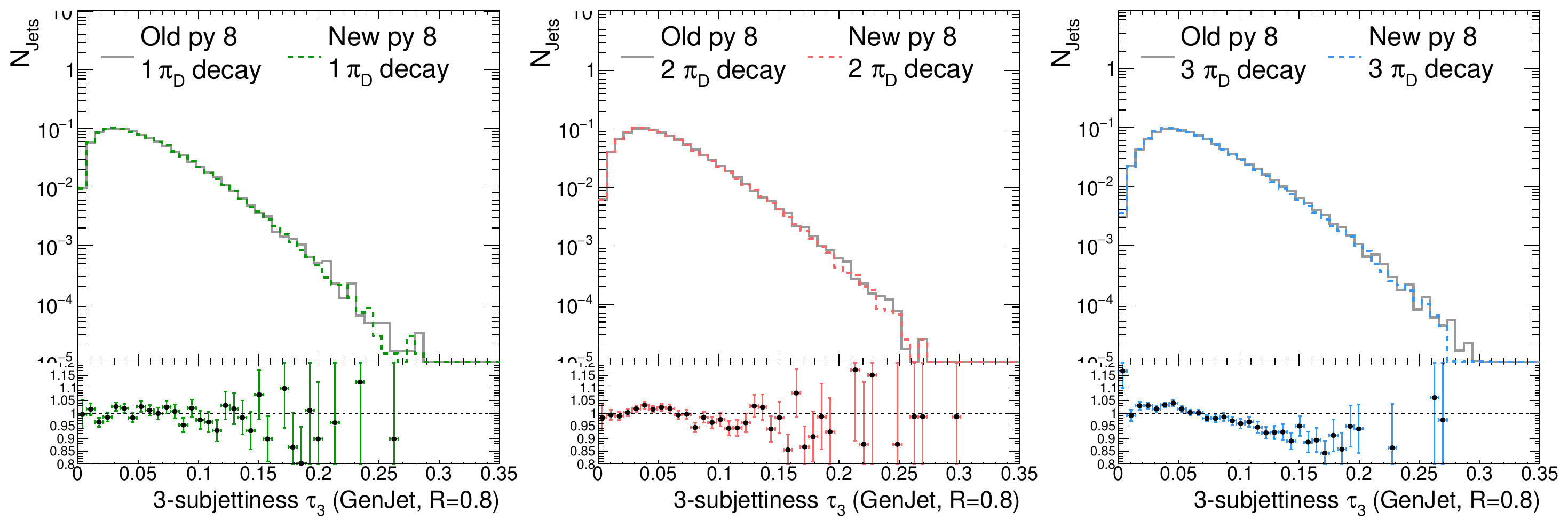}
\includegraphics[width=0.9\linewidth]{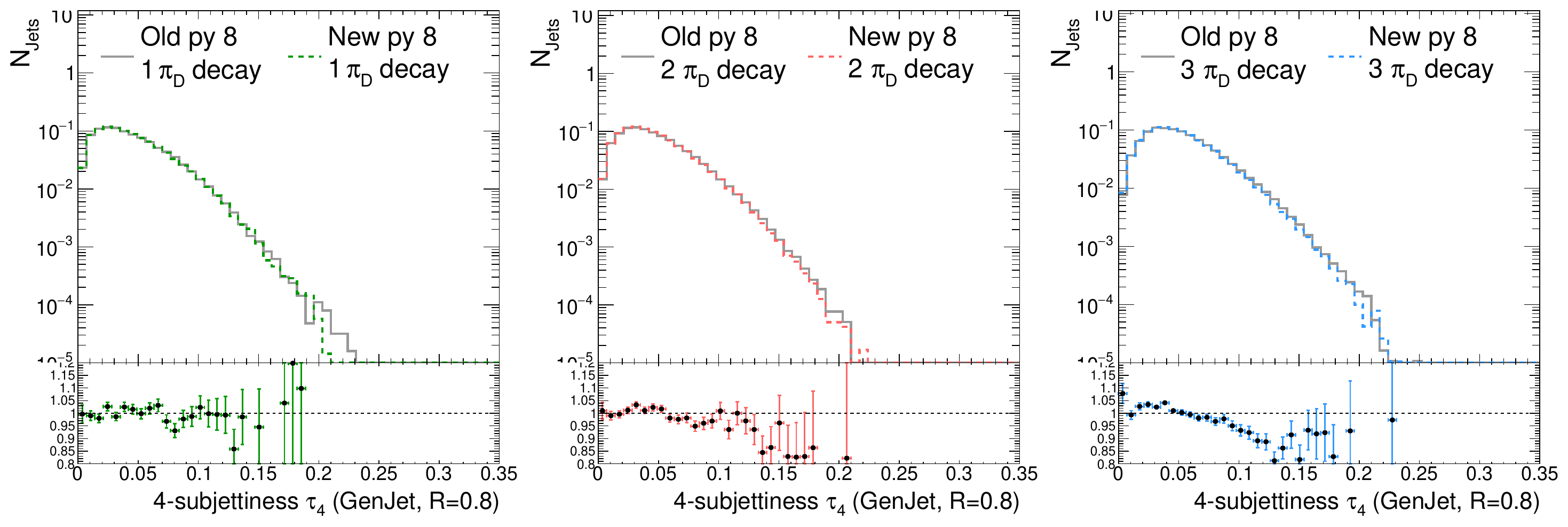}
\caption{Comparison of $N$-subjettiness variables between old (grey solid line) and new (colored dashed line) Hidden Valley modules. The comparison is made for 1 (left), 2 (middle) and 3 (right) dark pions decay. The plotted ratio is the ratio of new to old \PYTHIA module.}
\label{fig:nsubjettiness_old_vs_new_hv_module}
\end{figure}

Next, we studied the differences in the JSS observables between two dark vector meson production fractions: 50\% and 75\%. 
Comparison of quark-gluon discriminant variables, $N$-subjettiness and number of constituents are shown in Figs.~\ref{fig:qg_probVector_0p5_vs_0p75}, ~\ref{fig:nsubjettiness_probVector_0p5_vs_0p75} and~\ref{fig:multiplicity_probVector_0p5_vs_0p75}. Some systematic differences are observed for all variables. It is clear that the number of constituents in jets is lower for higher vector meson fraction. Jets with large number of soft constituents are characterized by low \ptd~while \ptd~is higher for jets where just a few constituents carry most of the momentum. The fact that \ptd~is higher for higher vector dark meson fractions is certainly an effect of the lower number of constituents. Jet girth, axes and $N$-subjettiness are all smaller in the case of \texttt{probVector=0.75} compared to \texttt{probVector=0.5}. This indicates that jets are narrower since with larger values of vector mesons fraction we observe a harder $p_{\mathrm{T}}$ spectrum for the dark hadrons decaying visibly.

\begin{figure}[htb!]
\centering
\includegraphics[width=0.9\linewidth]{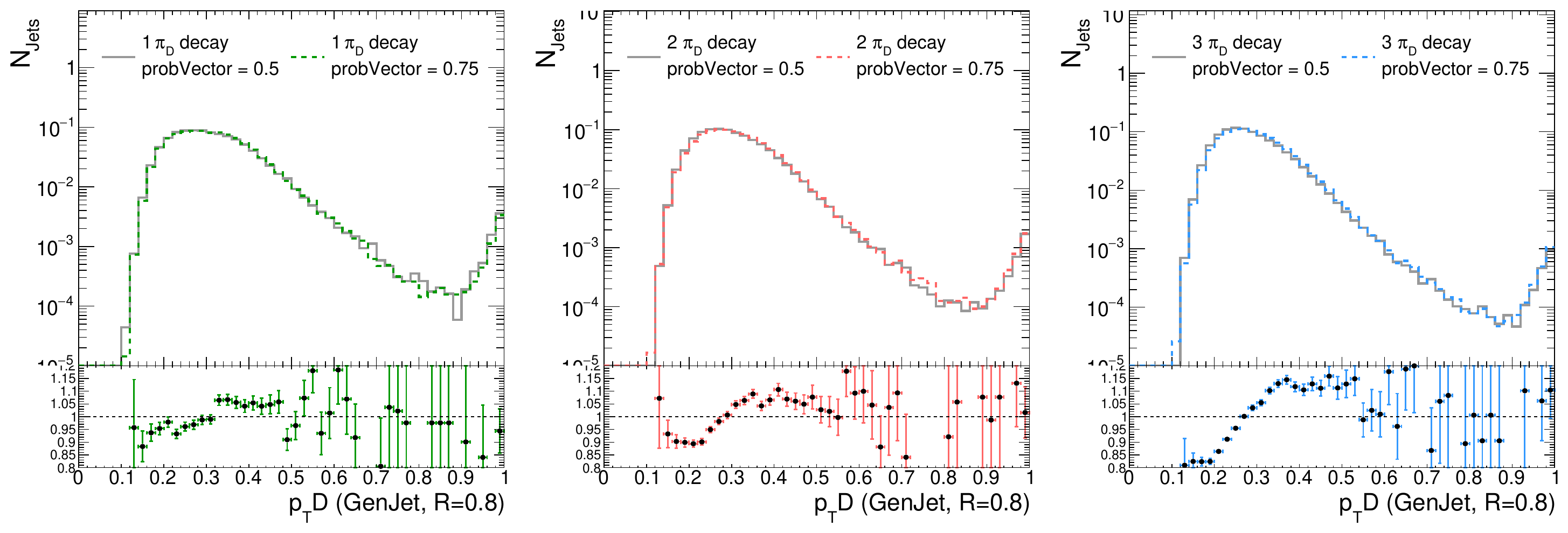}
\includegraphics[width=0.9\linewidth]{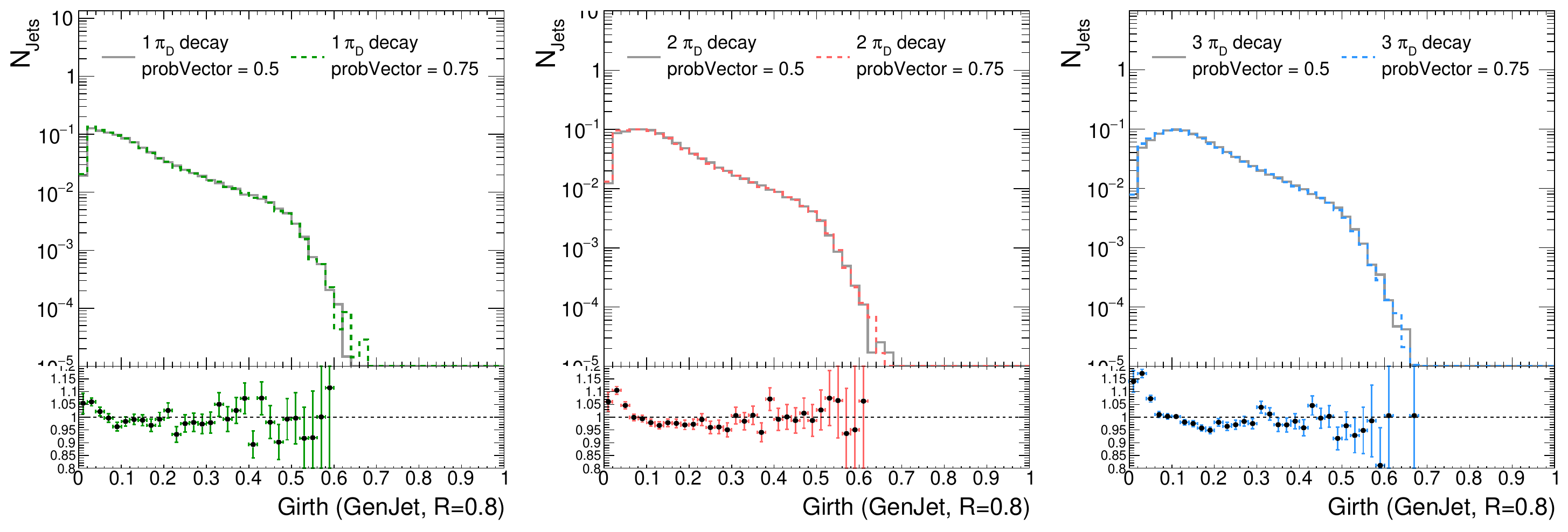}
\includegraphics[width=0.9\linewidth]{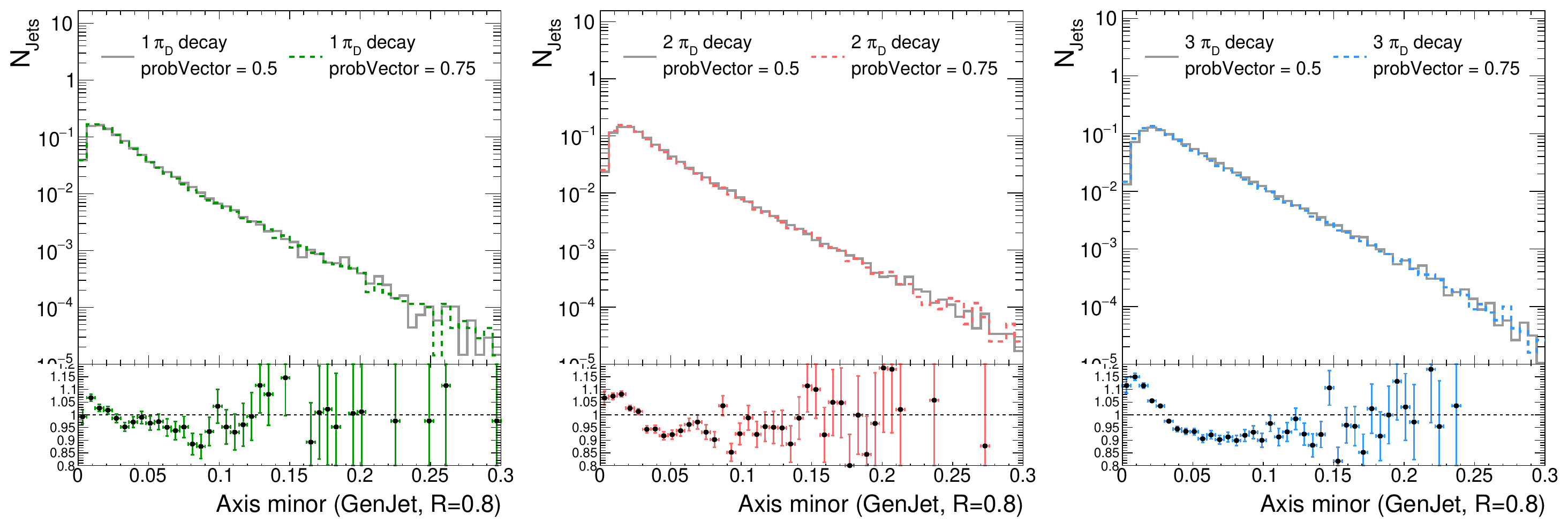}
\includegraphics[width=0.9\linewidth]{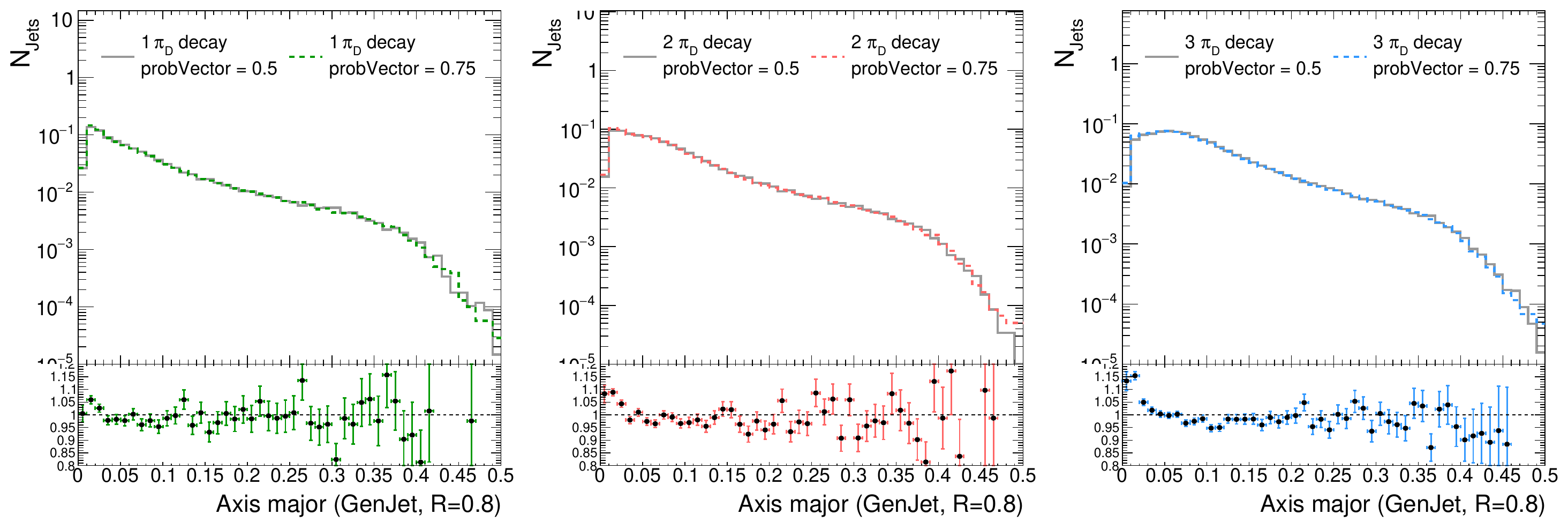}
\caption{Comparison of quark-gluon discriminant variables between \texttt{probVector=0.5} (grey solid line) and \texttt{probVector=0.75} (colored dashed line). The comparison is made for 1 (left), 2 (middle) and 3 (right) dark pions decay. The plotted ratio is the ratio of \texttt{probVector=0.75} to \texttt{probVector=0.5}.}
\label{fig:qg_probVector_0p5_vs_0p75}
\end{figure}

\begin{figure}[htb!]
\centering
\includegraphics[width=0.9\linewidth]{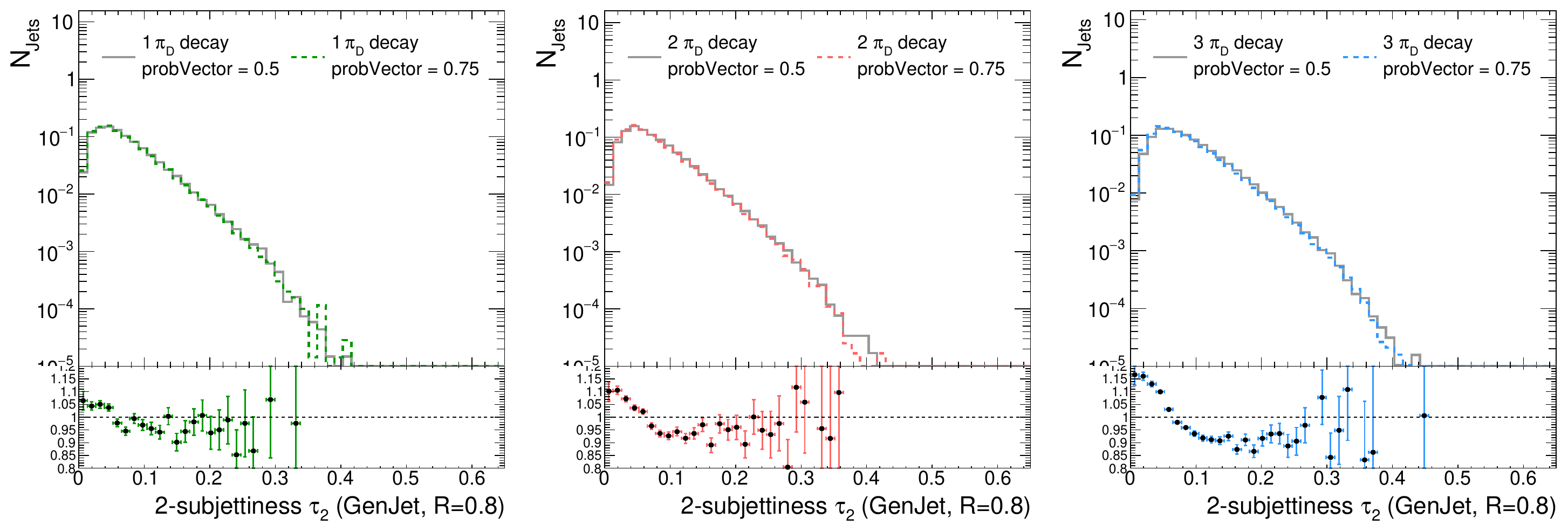}
\includegraphics[width=0.9\linewidth]{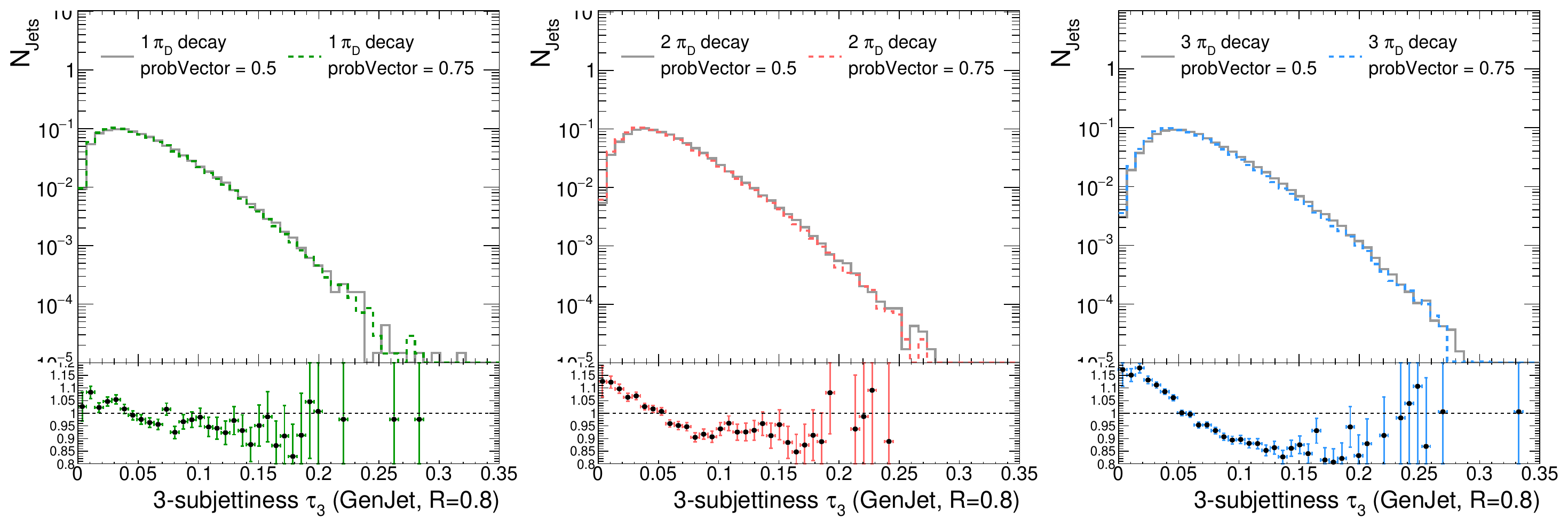}
\caption{Comparison of $N$-subjettiness between \texttt{probVector=0.5} (grey solid line) and \texttt{probVector=0.75} (colored dashed line). The comparison is made for 1 (left), 2 (middle) and 3 (right) dark pions decay. The plotted ratio is the ratio of \texttt{probVector=0.75} to \texttt{probVector=0.5}.}
\label{fig:nsubjettiness_probVector_0p5_vs_0p75}
\end{figure}

\begin{figure}[htb!]
\centering
\includegraphics[width=0.9\linewidth]{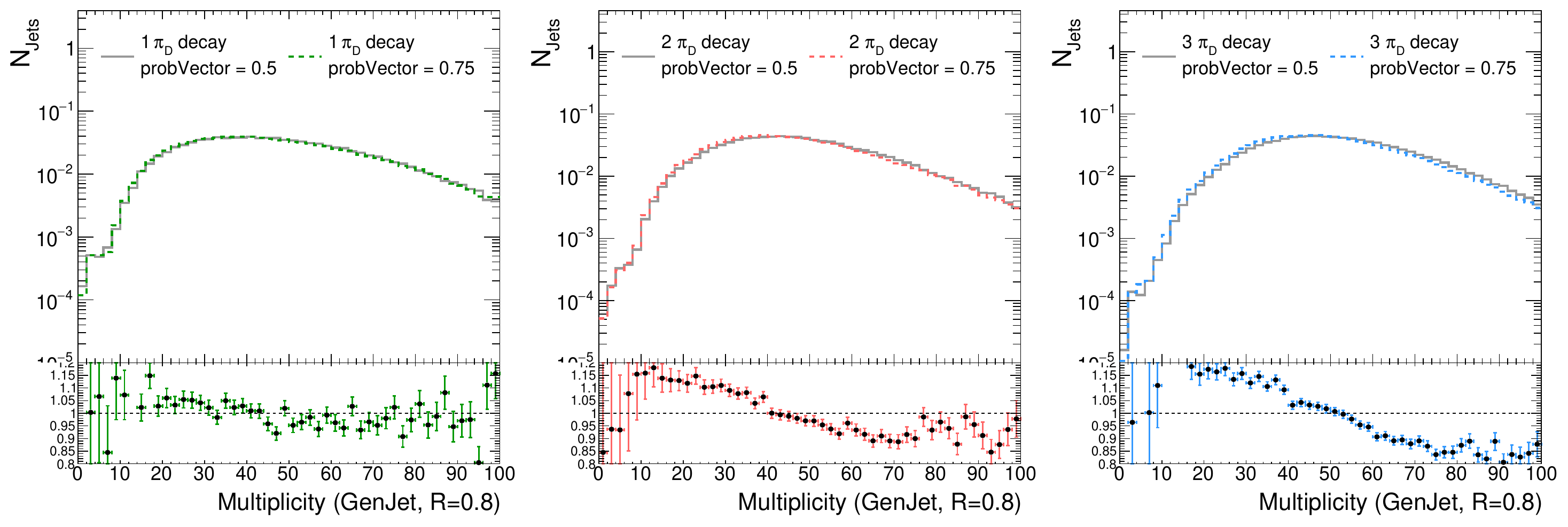}
\caption{Comparison of number of constituents in jets between \texttt{probVector=0.5} (grey solid line) and \texttt{probVector=0.75} (colored dashed line). The comparison is made for 1 (left), 2 (middle) and 3 (right) dark pions decay. The plotted ratio is the ratio of \texttt{probVector=0.75} to \texttt{probVector=0.5}.}
\label{fig:multiplicity_probVector_0p5_vs_0p75}
\end{figure}

We then studied how the number of unstable diagonal $\Ppidark$ mesons affects the jet substructure. Plots of quark-gluon discriminant, number of constituents and photon energy fraction for different number of unstable diagonal dark pions are provided in Fig.~\ref{fig:number_of_unstable_diag_pions}. Multiplicity is higher for lower \rinv, which is expected as the multiplicity is directly related to the number of unstable dark pions. Major and minor axes as well as girth are higher for lower \rinv, suggesting that the jet is wider.

In conclusion, we have noticed that the variation of the hidden sector parameters such as \texttt{probVector} can impact JSS distributions at generator level leading to a harder spectrum for the dark hadrons and consequently narrower jets. Notably, only two benchmark points for the vector meson fraction have been investigated, and further studies are encouraged to understand better the impact of the parameters of the hidden sector on the observable JSS distributions. 

\begin{figure}[htb!]
\centering
\includegraphics[width=0.3\linewidth]{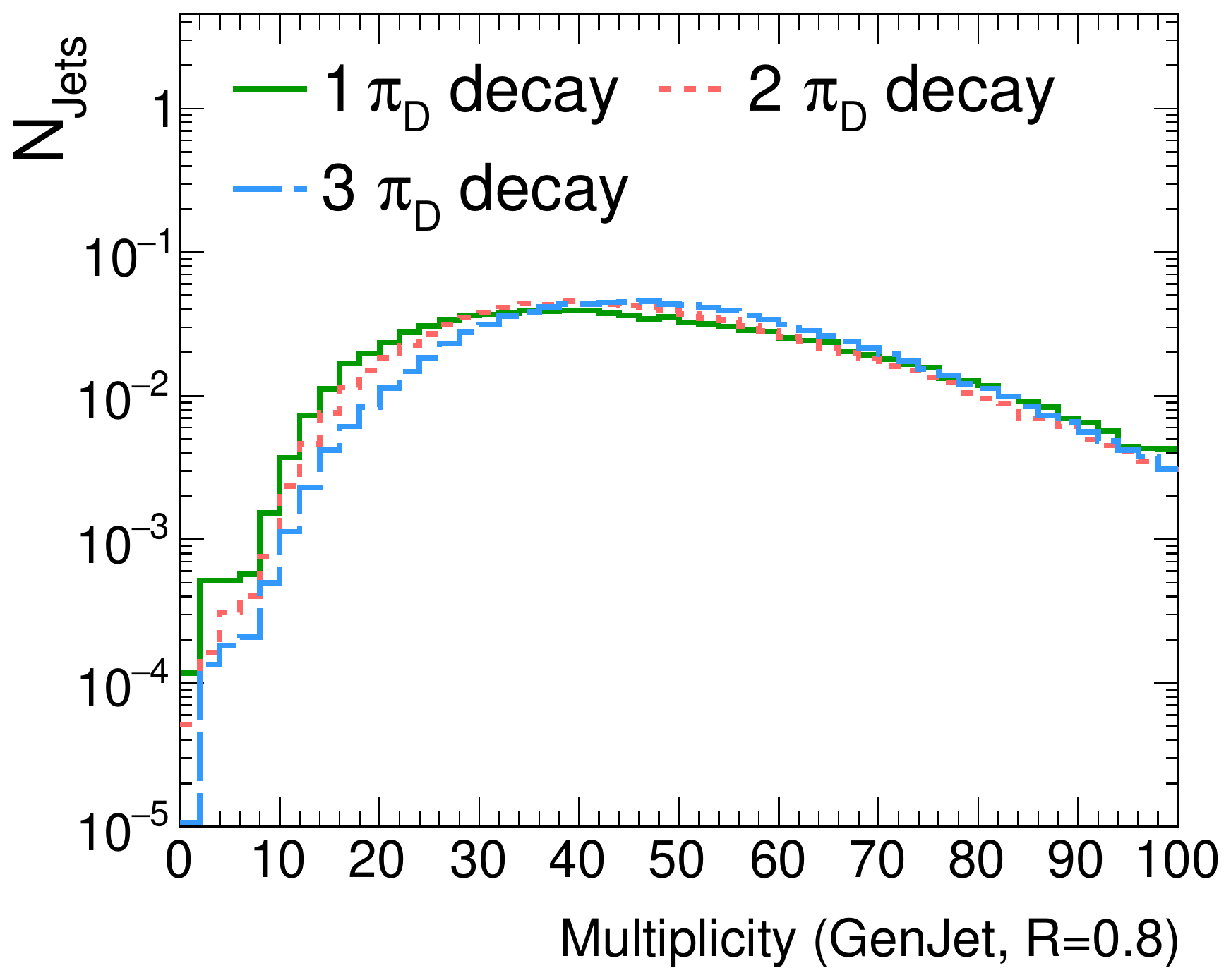}
\includegraphics[width=0.3\linewidth]{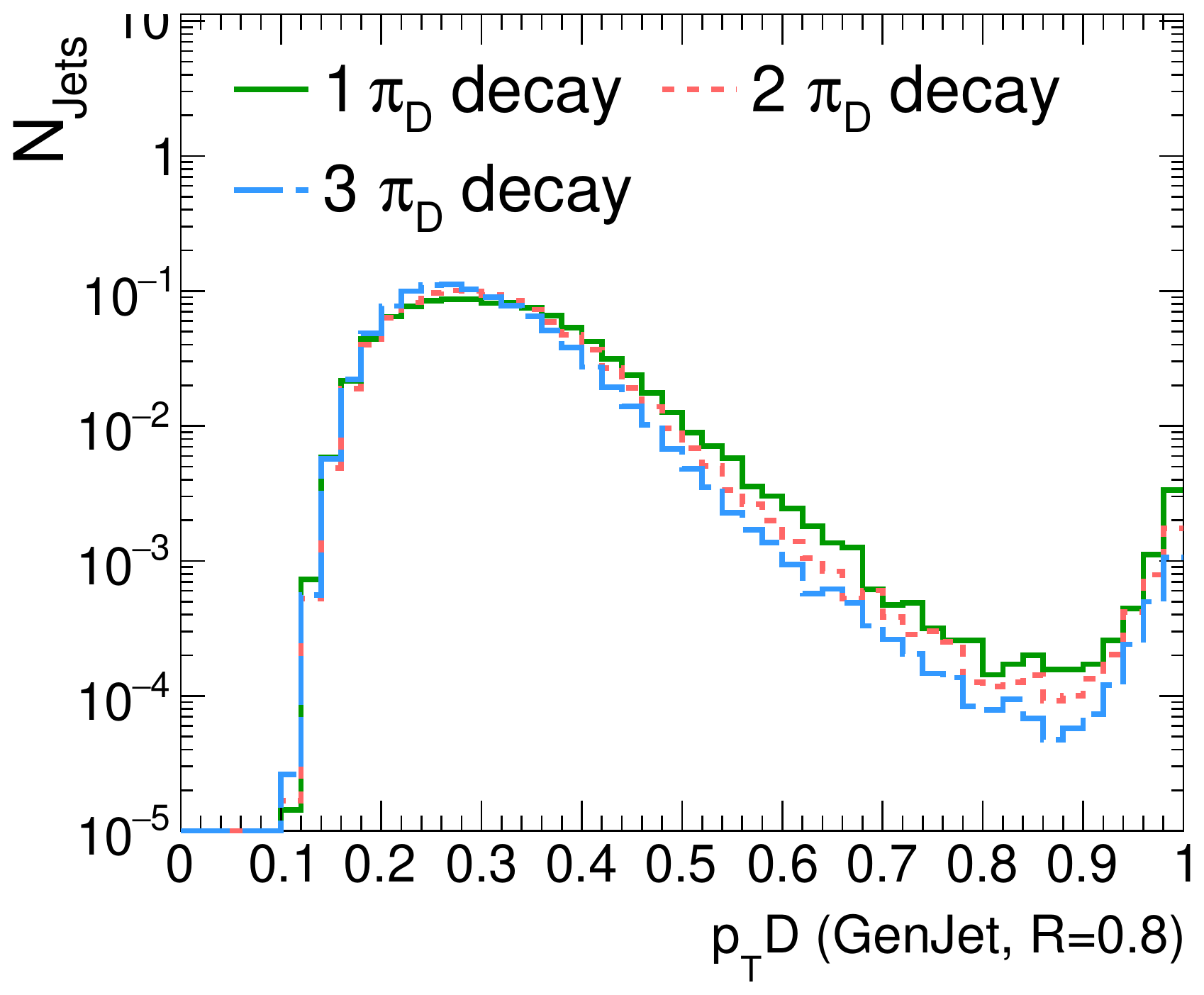}
\includegraphics[width=0.3\linewidth]{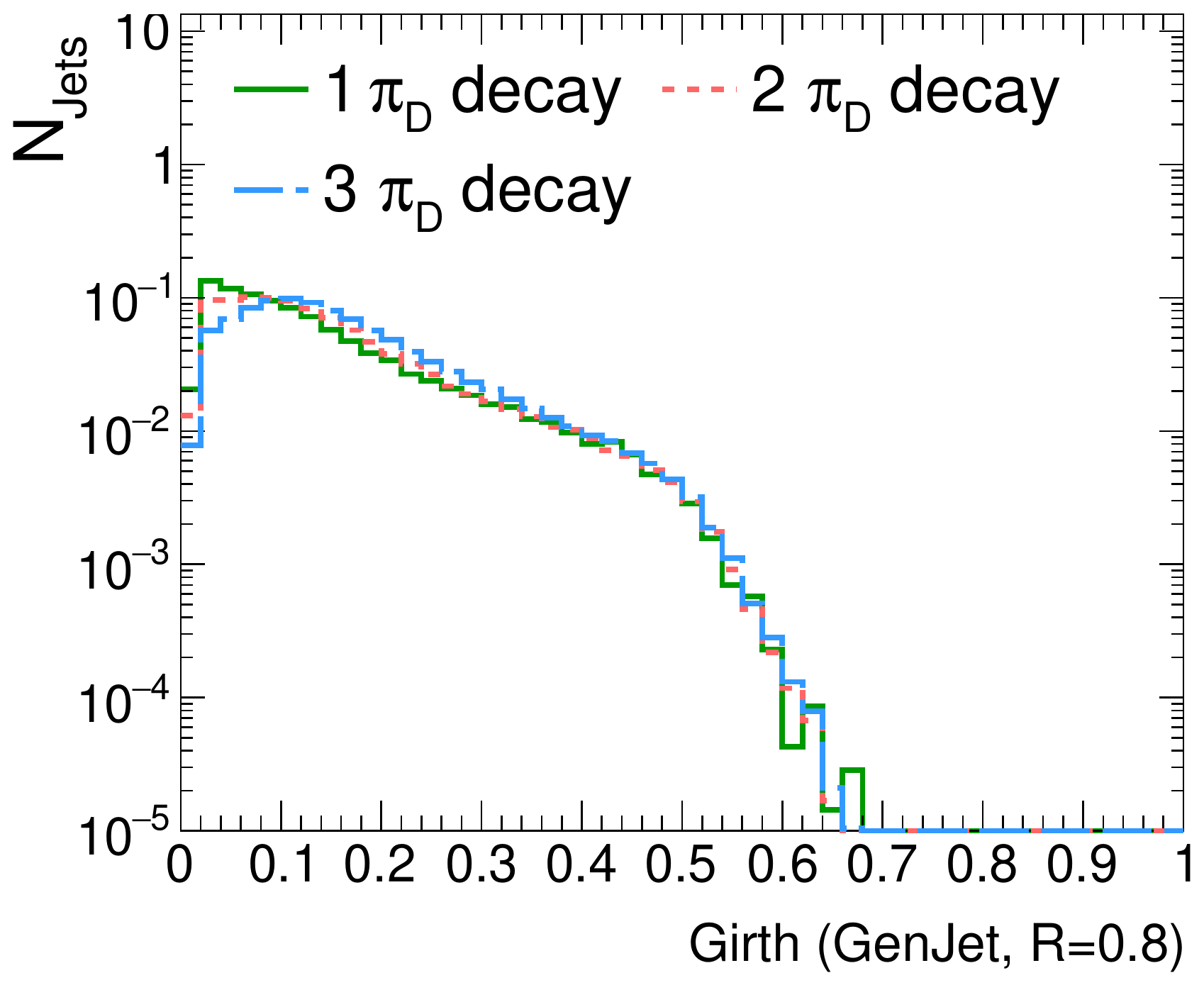}
\includegraphics[width=0.3\linewidth]{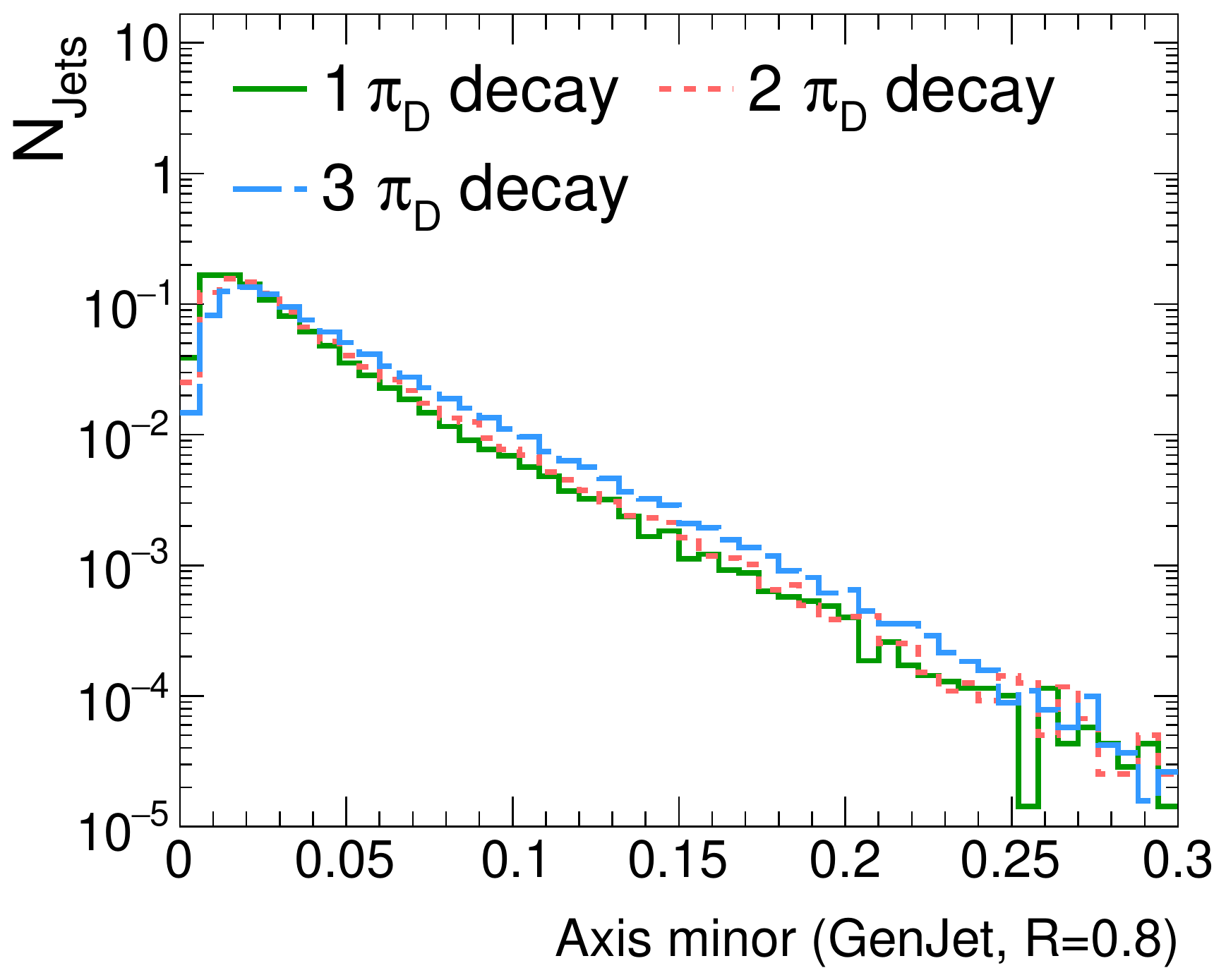}
\includegraphics[width=0.3\linewidth]{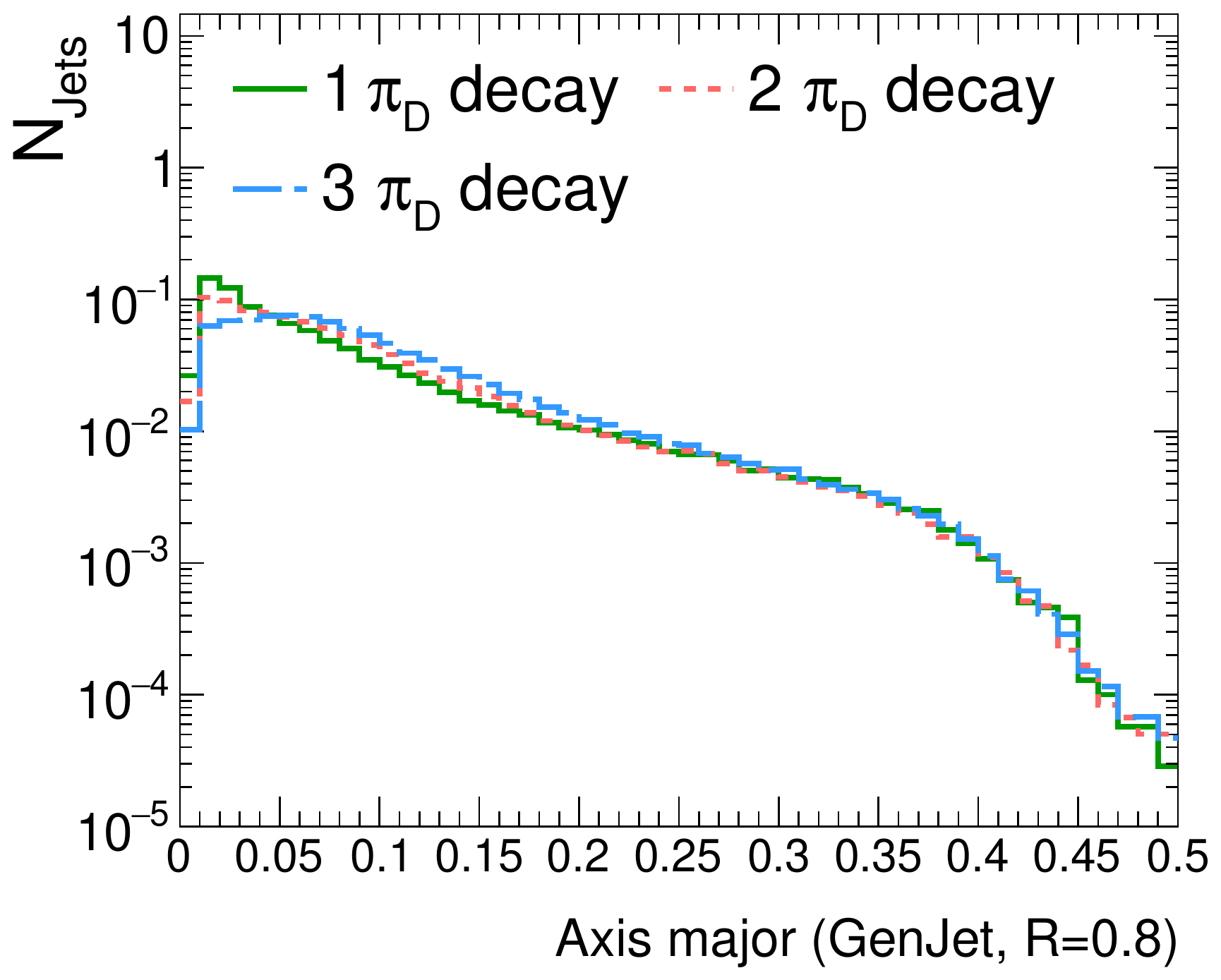}
\includegraphics[width=0.3\linewidth]{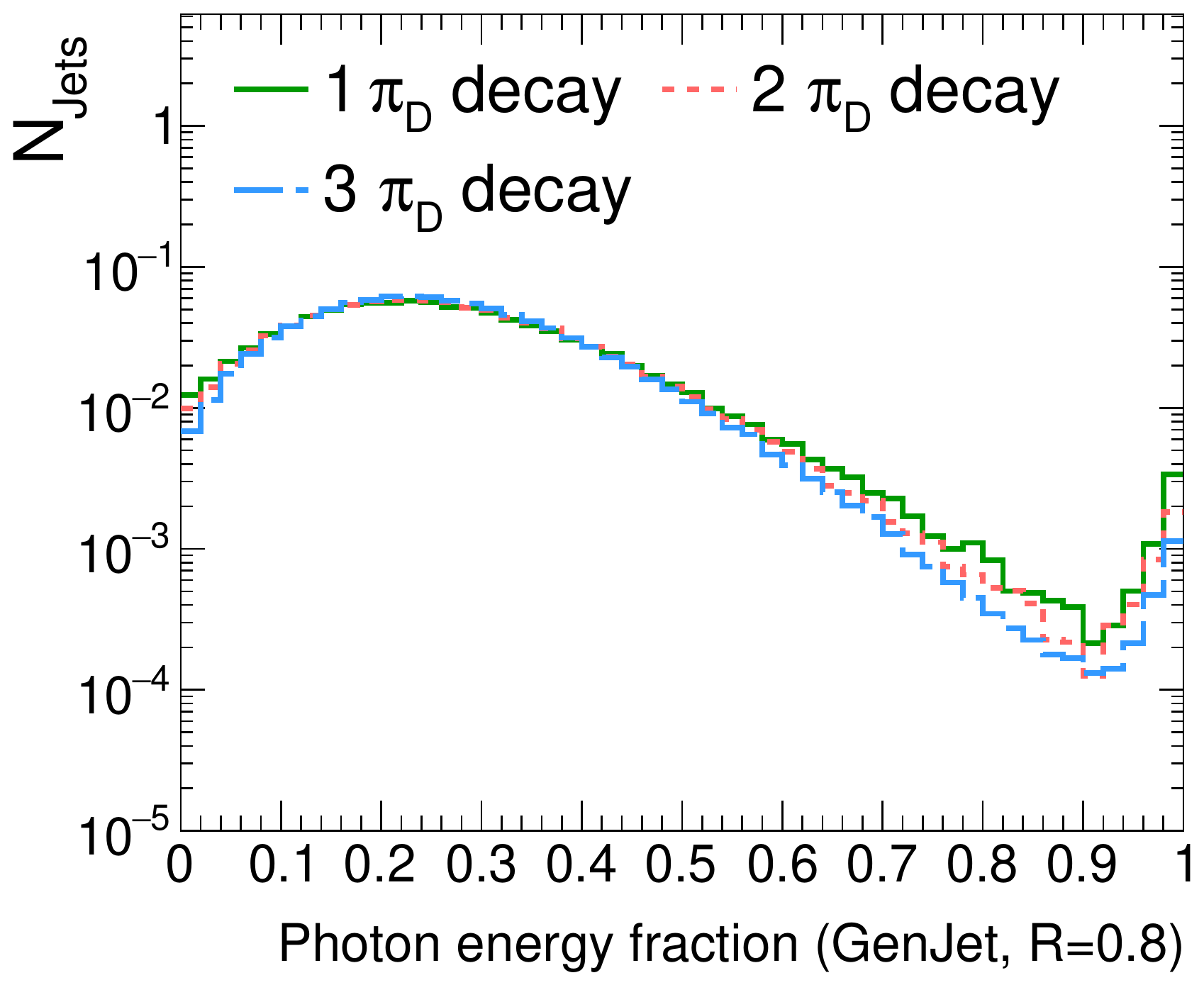}
\caption{Comparison between different number of unstable diagonal dark pions.}
\label{fig:number_of_unstable_diag_pions}
\end{figure}

\subsubsection{Infrared-collinear safety of JSS observables}
Traditional calculations in perturbative quantum chromodynamics are based on an order- by-order expansion in the strong coupling $\alpha_s$. Observables that are calculable in this way are known as “safe”~\cite{PhysRevLett.39.1436}. As it is well-known, divergences of different nature can appear in the perturbative series. For the ultraviolet divergences appearing in loop diagrams, since QCD is a renormalisable theory, such infinities can be consistently cured. Moreover, real-emission diagrams exhibit singularities in particular corners of the phase-space. More specifically, the singular contributions have to do with collinear splittings of massless partons and emissions of soft gluons off both massless and massive particles. Virtual diagrams also exhibit analogous infra-red and collinear (IRC) singularities and theorems~\cite{PhysRev.52.54,Kinoshita:1962ur,PhysRev.133.B1549} assure that such infinities cancel at each order of the perturbative series, when real and virtual corrections are added together, thus leading to physical transition probabilities that are free of IRC singularities. An observable $\mathcal{O}(\{p_i \})$ calculated from a system of particles with momenta $p_i$ is defined to be infrared safe if adding a soft particle with momentum $\epsilon$ the following relation holds:

\begin{equation}
    \mathcal{O}(\{p_i \}) = \lim_{\epsilon \to 0}  \mathcal{O}(\epsilon,\{p_i \})  
\end{equation}

Instead, if we consider a particle $p_1$ splitting into 2 particles $p_1 \to p^{(a)}_1 + p^{(b)}_1$ with angle between them $ \theta_{1,ab}  \to 0$, the observable $\mathcal{O}(\{p_i \})$ is said to be collinear safe if:

\begin{equation}
    \mathcal{O}(\{p_i \}) = \lim_{\theta_{1,ab}  \to 0}  \mathcal{O}(p^{(a)}_1, p^{(b)}_1 ,\{p_i \})  
\end{equation}

To check IRC safety of JSS observables, we computed them at different stages of the shower/hadronization going from the dark sector to the SM sector. For IRC unsafe observables, large fluctuations in the showering process are expected, while IRC safe observables should be more stable during the evolution. Therefore for collinear splittings or soft emissions happening during the parton shower, the IRC unsafe observables will tend to diverge from the original value calculated in previous stages of the showering. Due to this feature, the IRC unsafe observables if not validated on data can introduce important model dependence in analyses exploiting them in supervised classifiers . This is particularly relevant in the case of dark shower studies where the MC-data agreement for signal cannot be assessed, and therefore there is no real control on IRC unsafe observables due to the unknown details of the hidden sector (for example the dark hadronization scale $\lamdark$). Specifically, given an observable $\mathcal{O}(\{p_i \})$, changes in the Hidden sector parameters such as $\lamdark$ are expected to produce a power law scaling for IRC safe observables given the jet pt $p_{\mathrm{T}j}$:
$\langle \delta O_{safe} \rangle \sim (\lamdark/p_{\mathrm{T}j})^{\alpha}$. 
On the other hand, the scaling is logarithmic in the case of IRC unsafe observables:
$\langle \delta O_{unsafe} \rangle \sim \log(\lamdark/p_{\mathrm{T}j})$.
This means that depending on the hadronization scale of the dark sector, the IRC unsafe observables can undergo large fluctuations for $\lamdark \ll \pt$, which means that without knowing $\lamdark$ these observables are correctly described by the parton shower but are also dependent of the details of the hidden sector. \\
In this study we test IRC safety of JSS observables by calculating them at 3 levels in the evolution of the shower: dark sector hadrons, SM quarks and SM hadrons. As previously mentioned, we expect the IRC safe observables to fluctuate less in the evolution. For the test we consider two generalized angularities, namely \ptd~which is an IRC unsafe observable, and the jet girth, which is IRC safe. The collinear unsafety of \ptd~is due to its dependence on the squared of the transverse momenta of the jet constituents. Therefore, taking a particle with transverse momentum $p_{1,\mathrm{T}}$, if the particle splits into 2 particles with transverse momenta $p^{(a)}_{1,\mathrm{T}}$ and $p^{(b)}_{1,\mathrm{T}}$, \ptd~becomes:
\begin{equation}
  (p_{\textrm{T}}D)^2 \propto (p_{1,\mathrm{T}})^2 = (p^{(a)}_{1,\mathrm{T}})^2 + (p^{(b)}_{1,\mathrm{T}})^2 + p^{(a)}_{1,\mathrm{T}} p^{(b)}_{1,\mathrm{T}} cos(\theta_{12})
\end{equation}
Therefore, we expect \ptd~to fluctuate more during the showering compared to the jet girth.  
Our results for the test of IRC safety for JSS observables is presented in Fig.~\ref{fig:irc_safety_test}. The plots show the following ratios for the tested JSS observable: unstable dark hadrons vs SM quarks, SM quarks vs SM hadrons and  unstable dark hadrons vs SM hadrons. We expect the distributions of the ratios for the collinear unsafe observable calculated at different steps of the shower to differ from unity. For a fair comparison between the same observable calculated at different stages of the showering we consider only jets with a multiplicity of SM quarks which is twice the dark hadrons one. Moreover, because the girth of jets with one constituent is a special case as girth is close to 0, we consider only jets with a number of unstable dark hadrons strictly larger than one. The main result of this study is that even if IRC unsafe observables are expected to be described quite well by the parton shower, the application of IRC unsafe observables in the context of dark shower searches should be carefully validated in control regions by comparing Monte-Carlo and data. Secondly, as the dark hadronization scale is unknown, the effect of changing \lamdark on JSS observables must be evaluated. The usage of such variables especially in Hidden Valley searches can lead to important limitations in terms of interpretability of the results due to their strong dependence on the unknowns of the Hidden sector.  

\begin{figure}[htb!]
\centering
\includegraphics[width=0.3\linewidth]{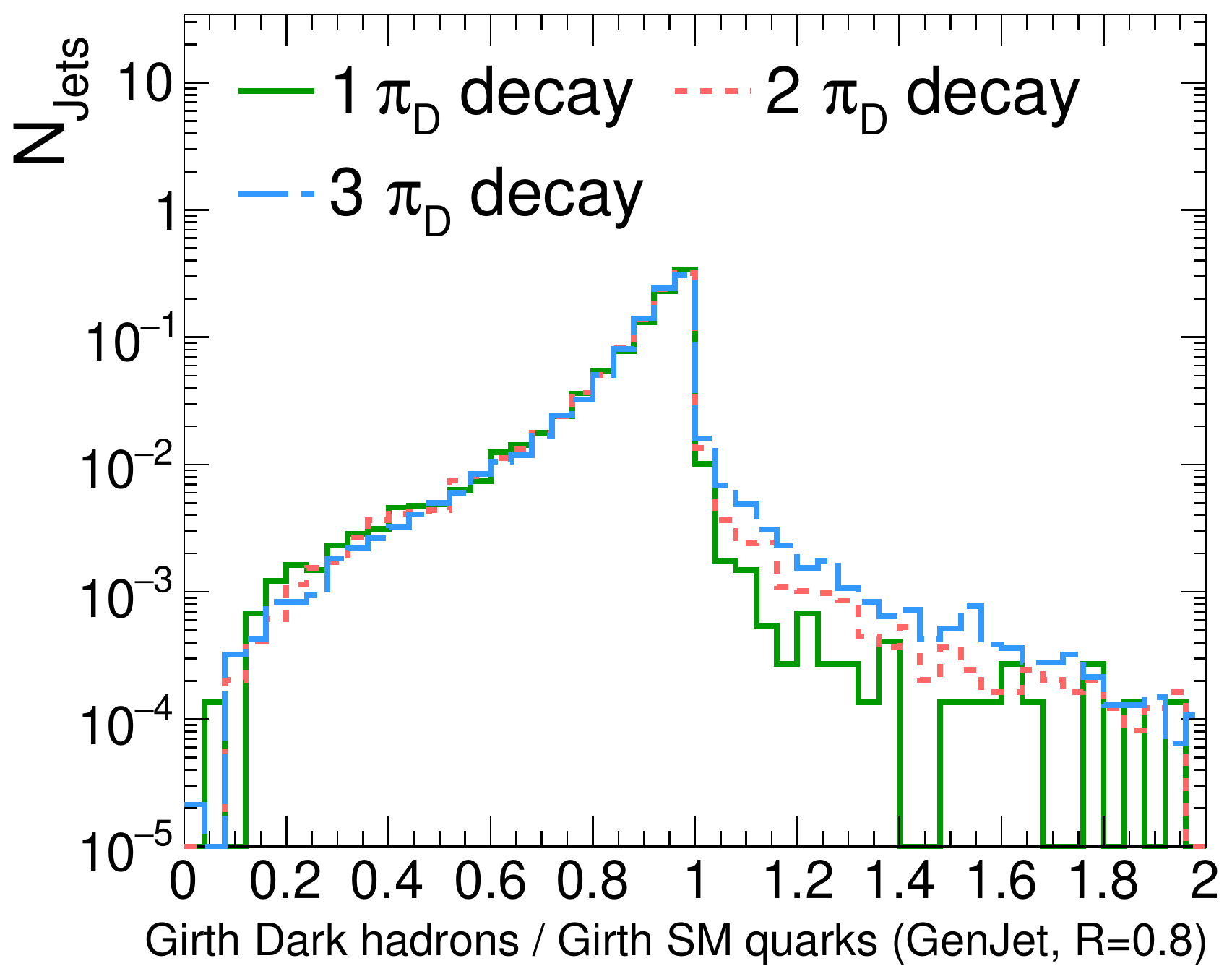}
\includegraphics[width=0.3\linewidth]{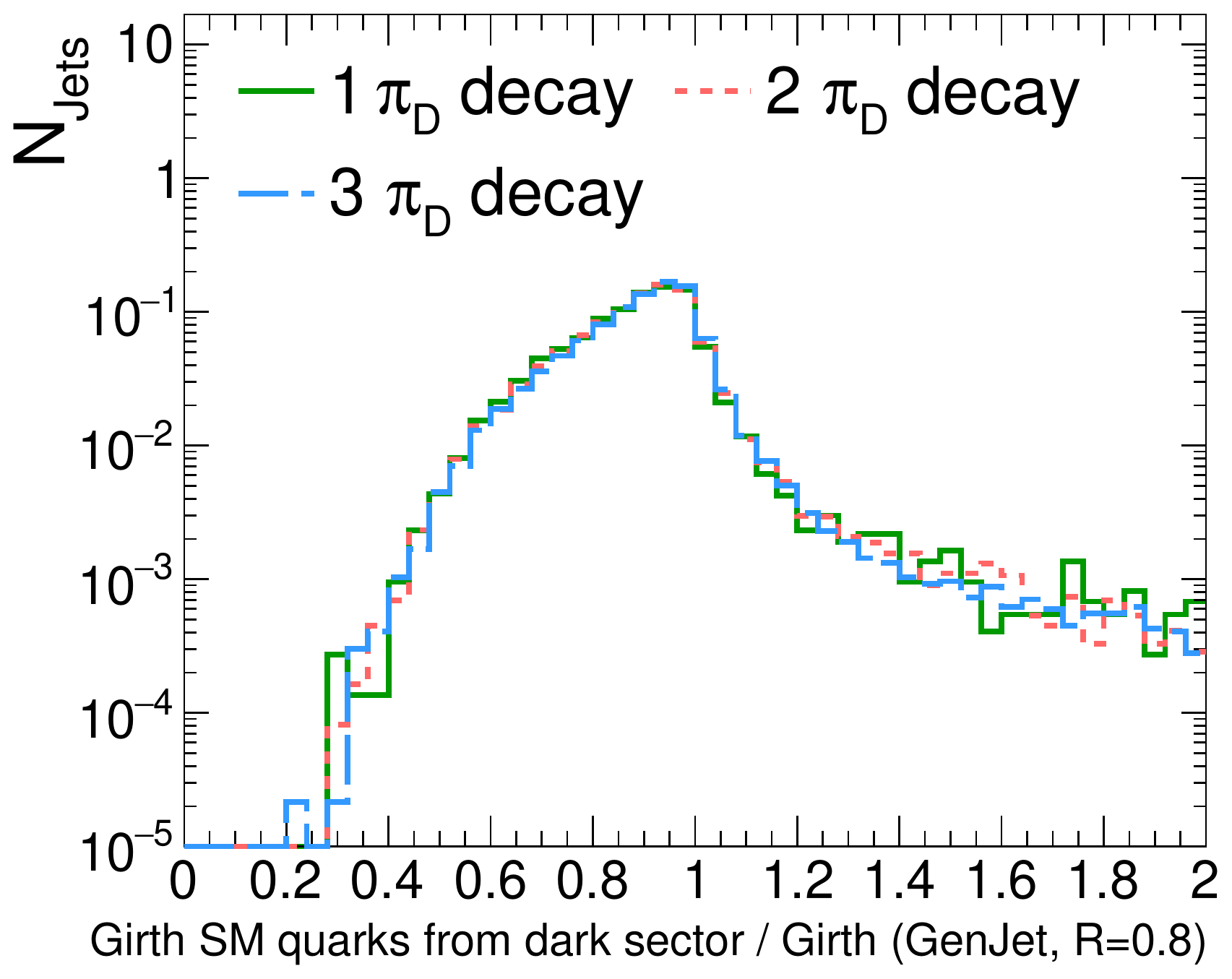}
\includegraphics[width=0.3\linewidth]{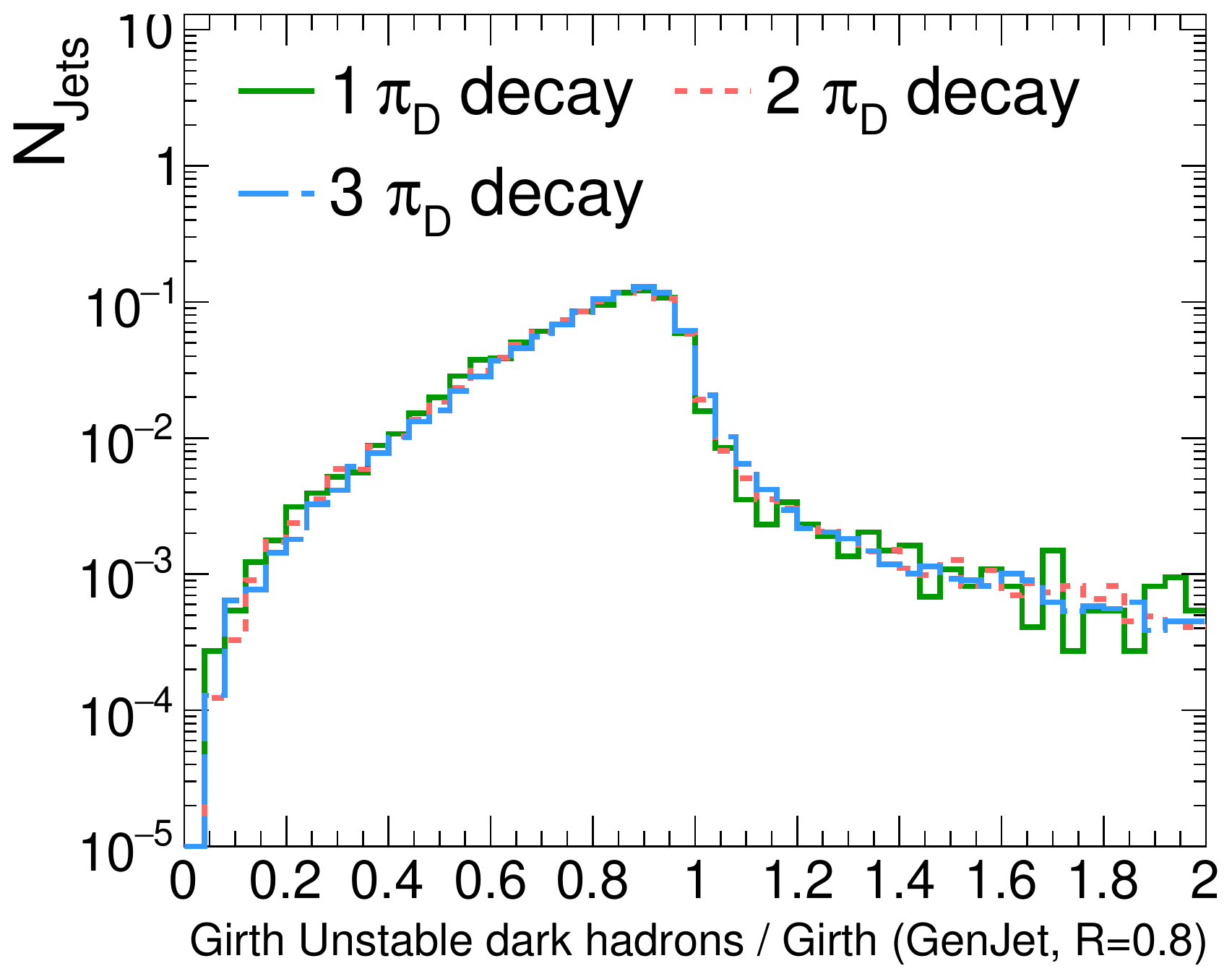}
\includegraphics[width=0.3\linewidth]{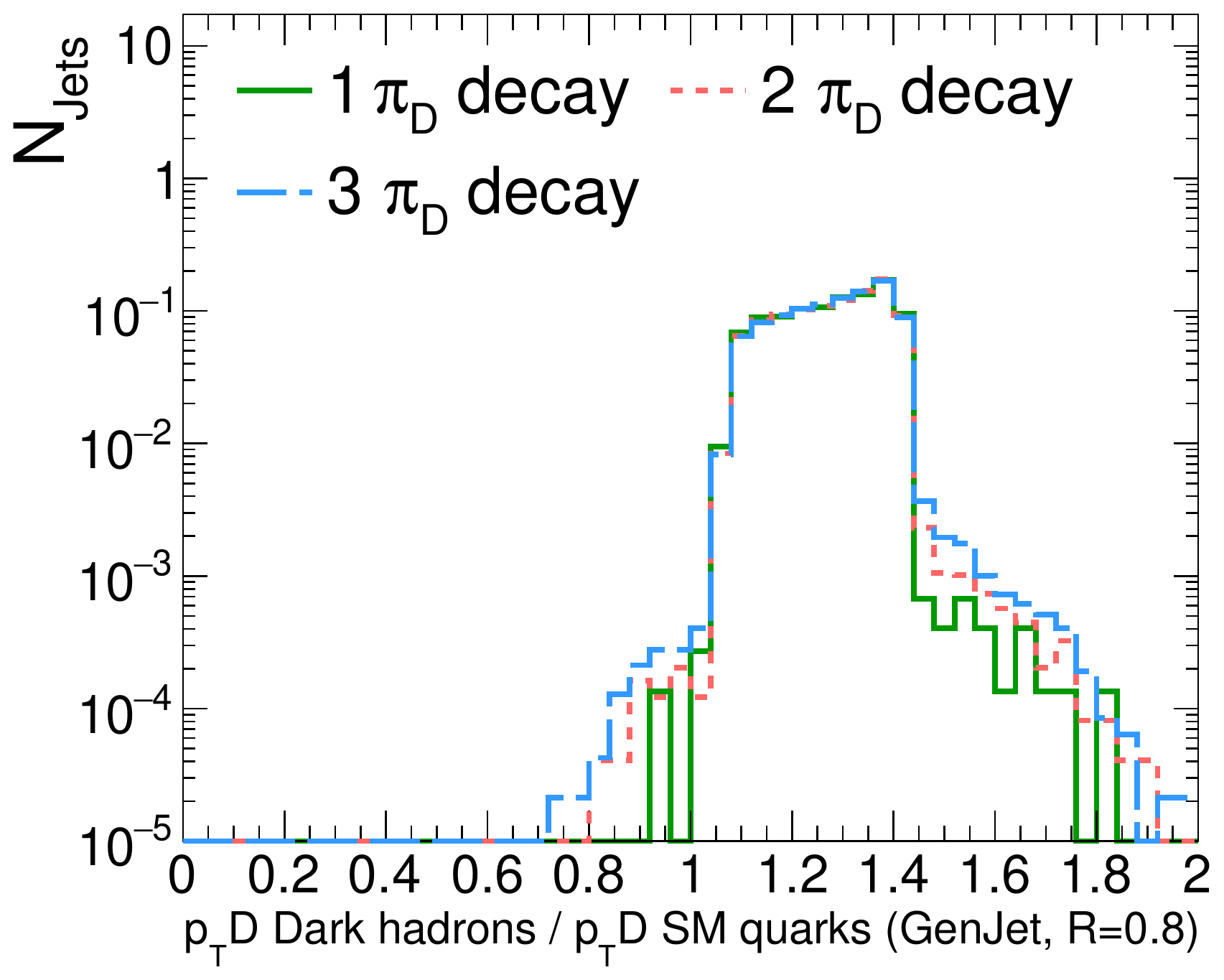}
\includegraphics[width=0.3\linewidth]{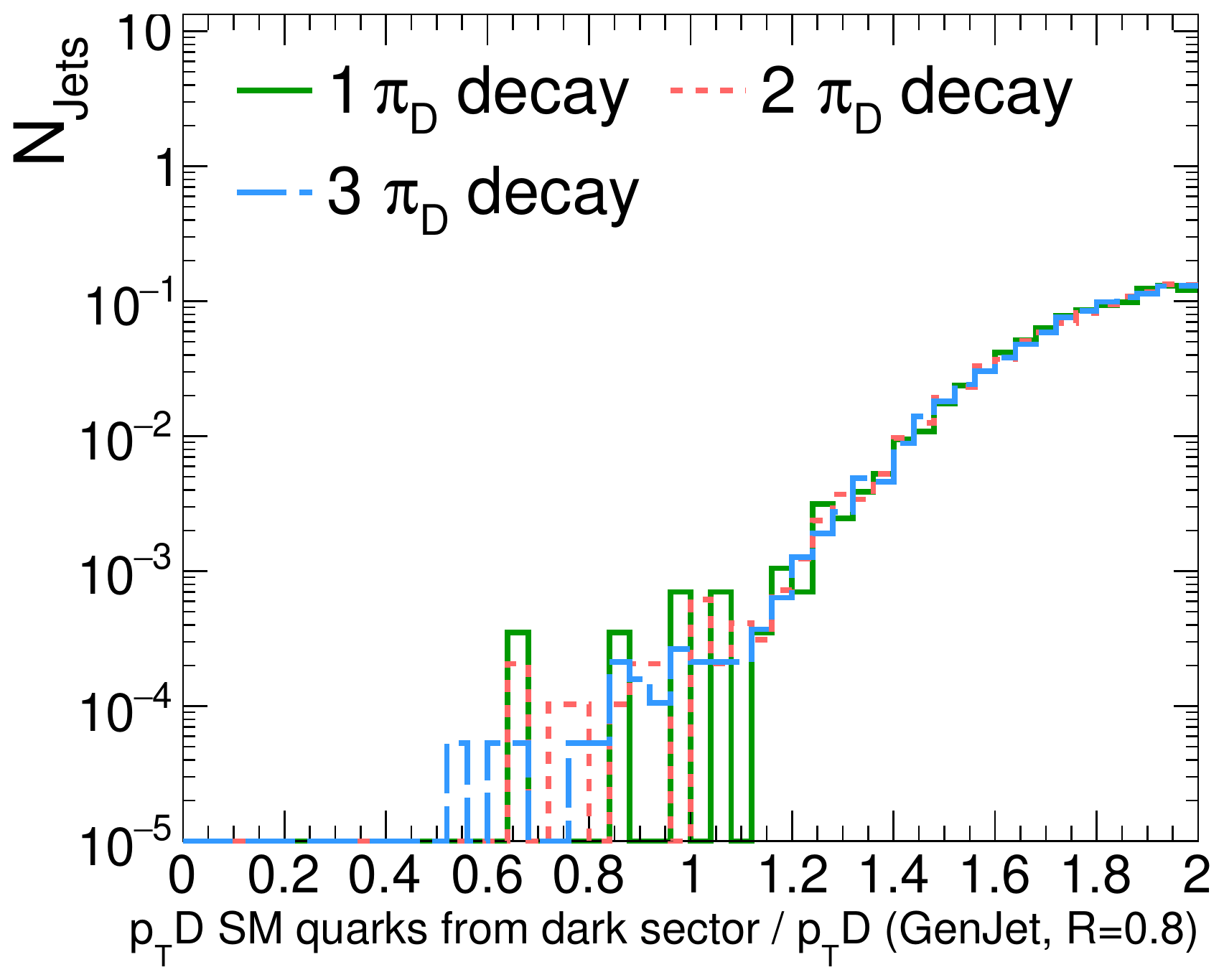}
\includegraphics[width=0.3\linewidth]{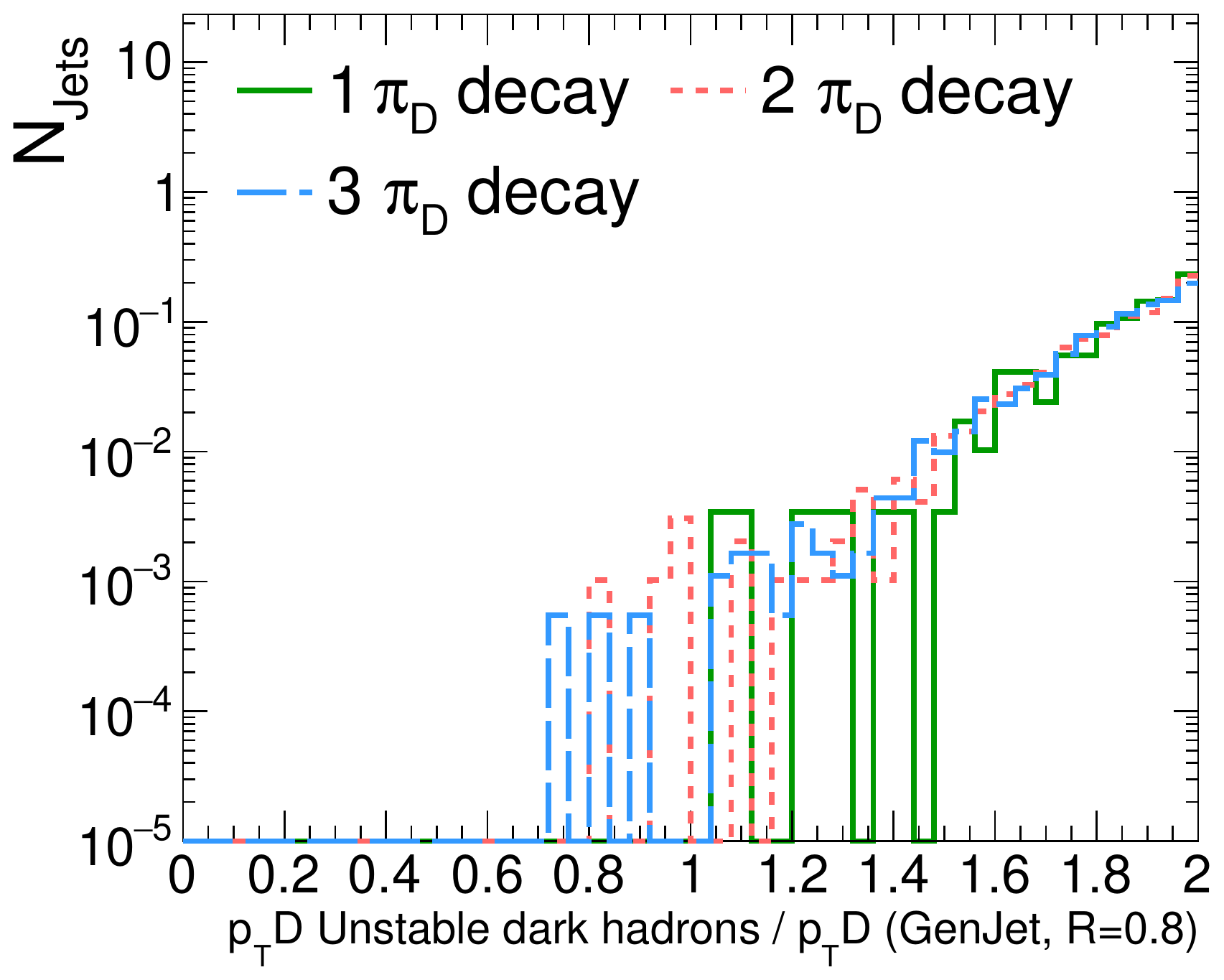}

\caption{Test of IRC safety. Ratio of jet substructure variables (top: girth, bottom: \ptd) computed at three different levels: unstable dark hadrons, SM quarks from dark hadrons and SM hadrons. Large variations are observed for \ptd, which is IRC-unsafe, while ratios of girth, which is IRC-safe, peak at 1.}
\label{fig:irc_safety_test}
\end{figure}

\subsubsection{Study of JSS observables after jet reconstruction in \Delphes}
After checking how the different parameters of the model affect the generator-level jets, we perform a similar study at reconstructed level using \Delphes output. This is important to understand the impact of detector effects on the JSS observables that can be used by the experiments to tag dark jets efficiently. 
\Delphes was configured for a CMS-like detector at the HL-LHC, and in particular Particle Flow candidates have been clustered with four different distance parameters, $R$ = 0.4, 0.8, 1.0 and 1.2, using FastJet~\cite{Cacciari:2011ma,Cacciari:2005hq}. Jets with larger radius help in containing more of the radiation of the dark jet. 
Jets are required to have at least two tracks, and $\vert \eta \rvert < 2.5$ and a minimum $p_{T}$ for clustering of 25 GeV. 
Figures~\ref{fig:reco_probvec_a} and~\ref{fig:reco_probvec_b} show the difference between the samples with \texttt{probVector=0.5} and \texttt{0.75} when the jets are clustered with $R=0.8$.

\begin{figure}[htb!]
\centering
\includegraphics[width=0.3\linewidth]{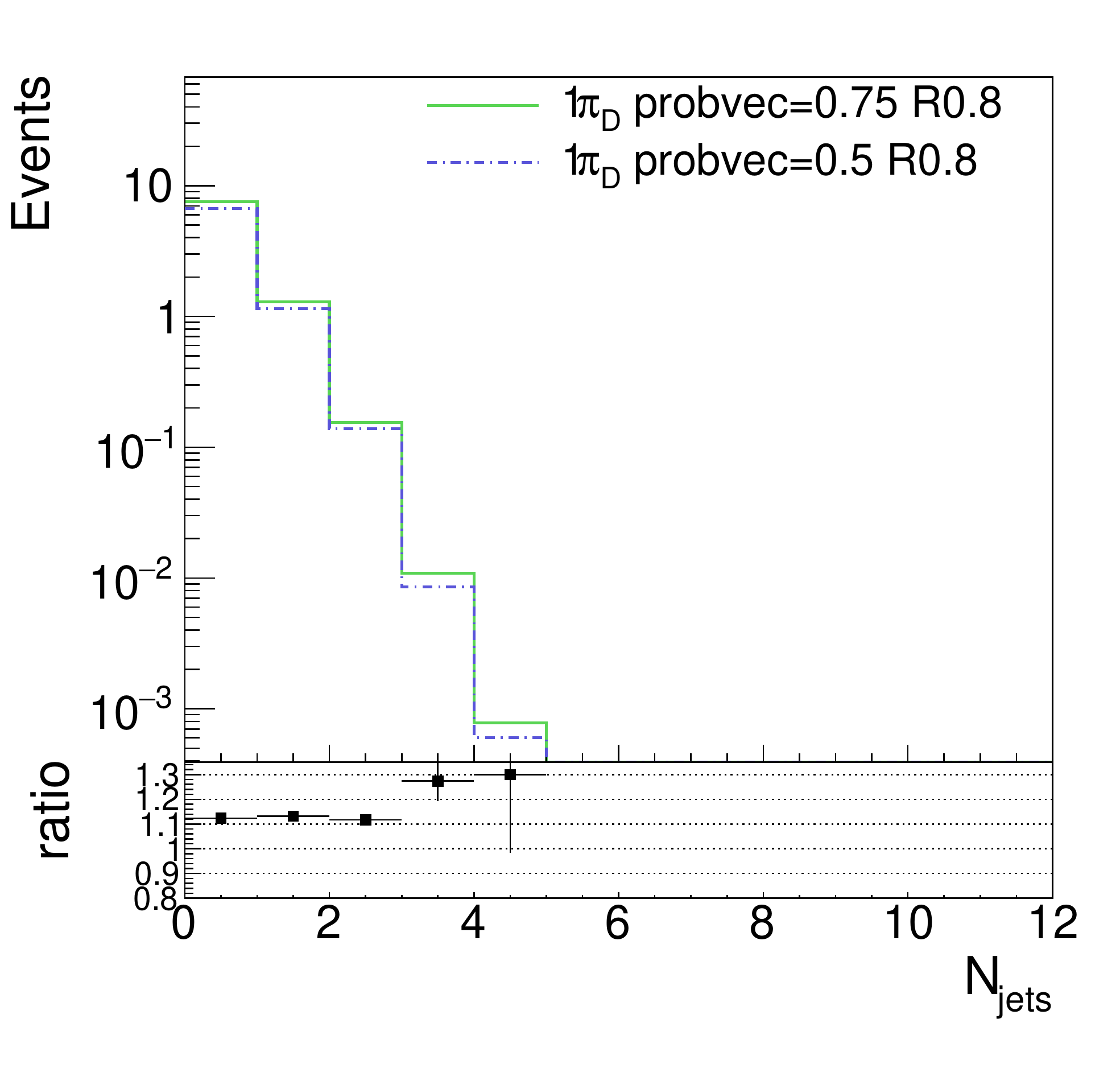}
\includegraphics[width=0.3\linewidth]{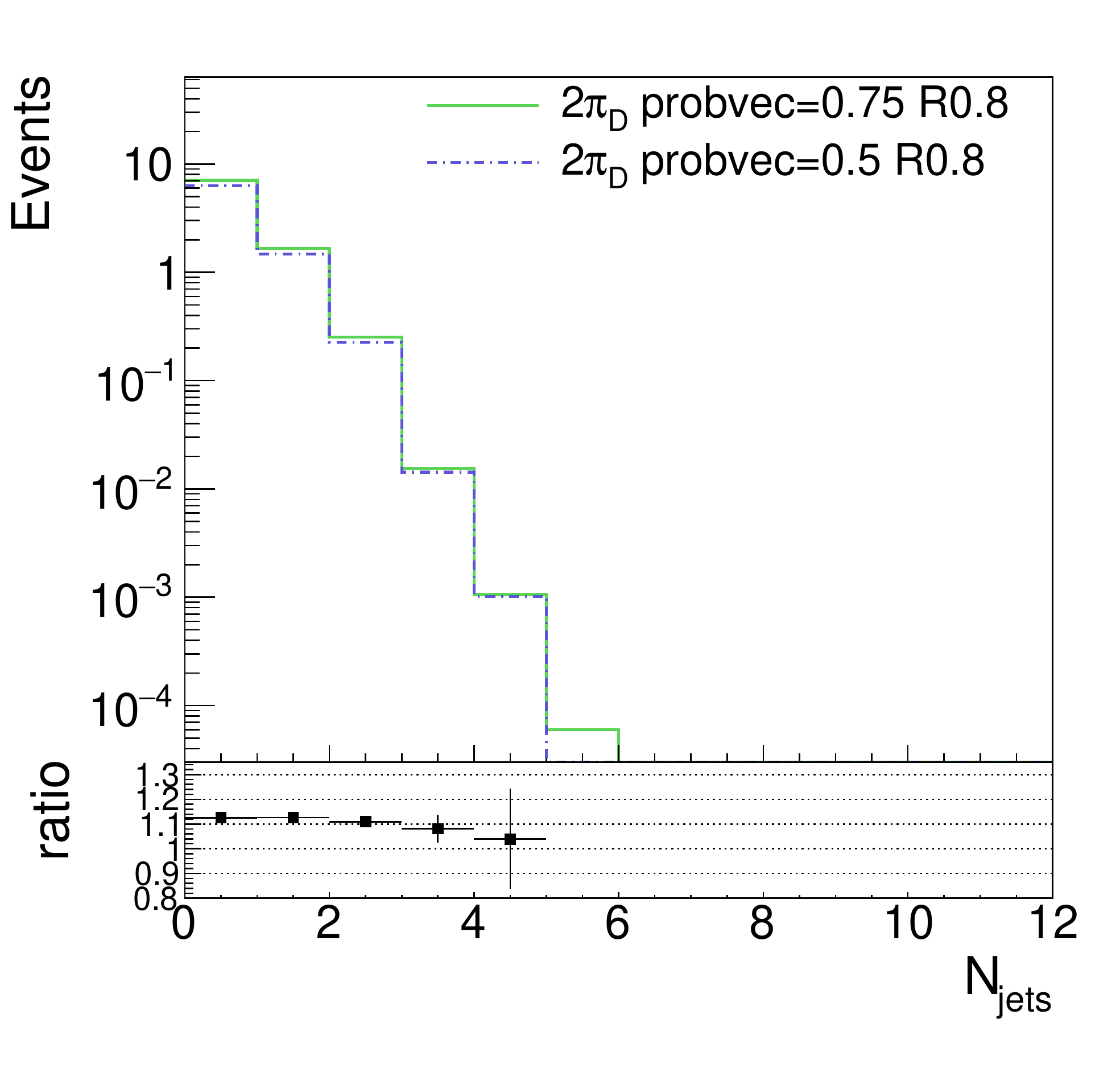}
\includegraphics[width=0.3\linewidth]{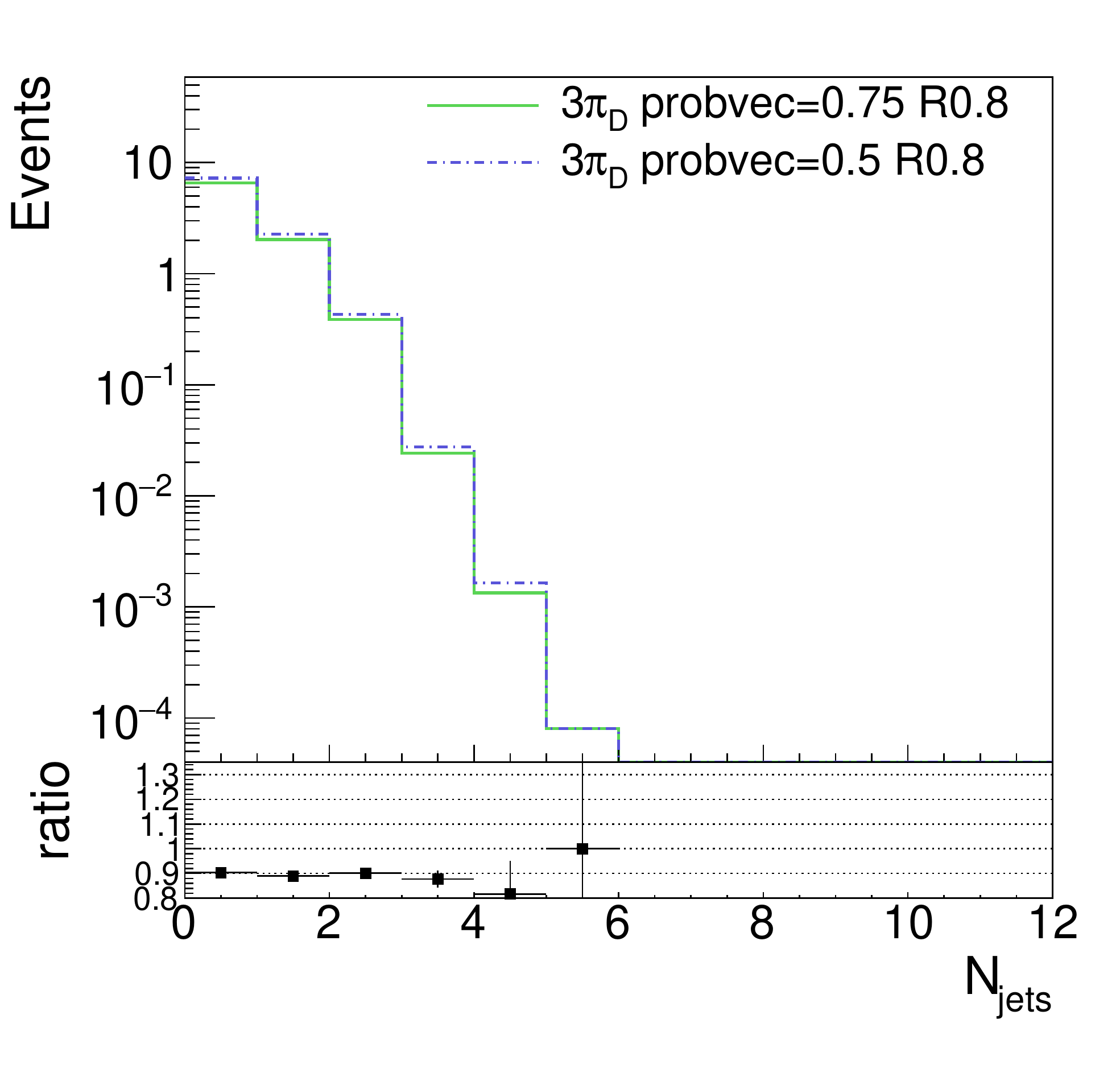}
\includegraphics[width=0.3\linewidth]{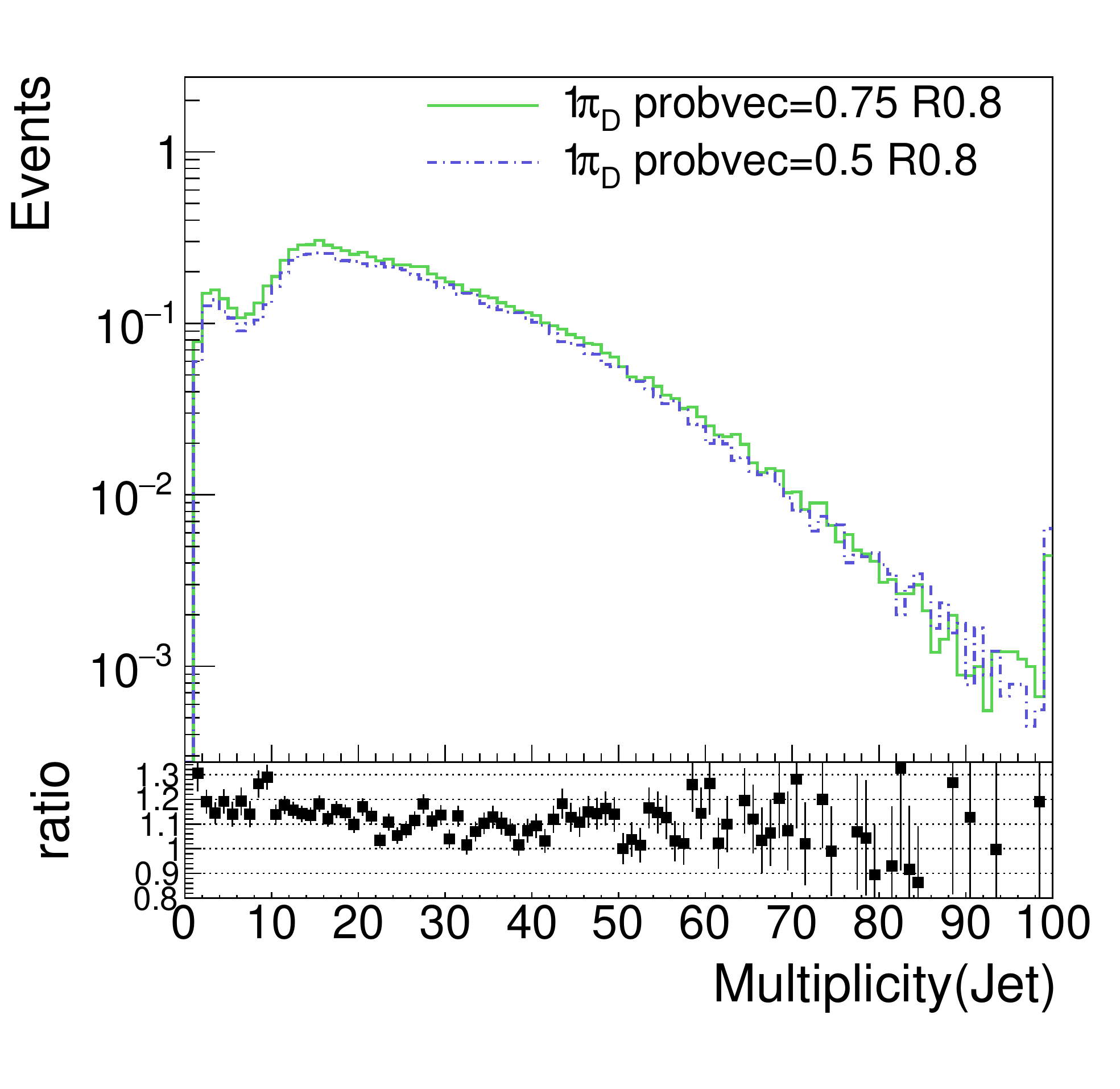}
\includegraphics[width=0.3\linewidth]{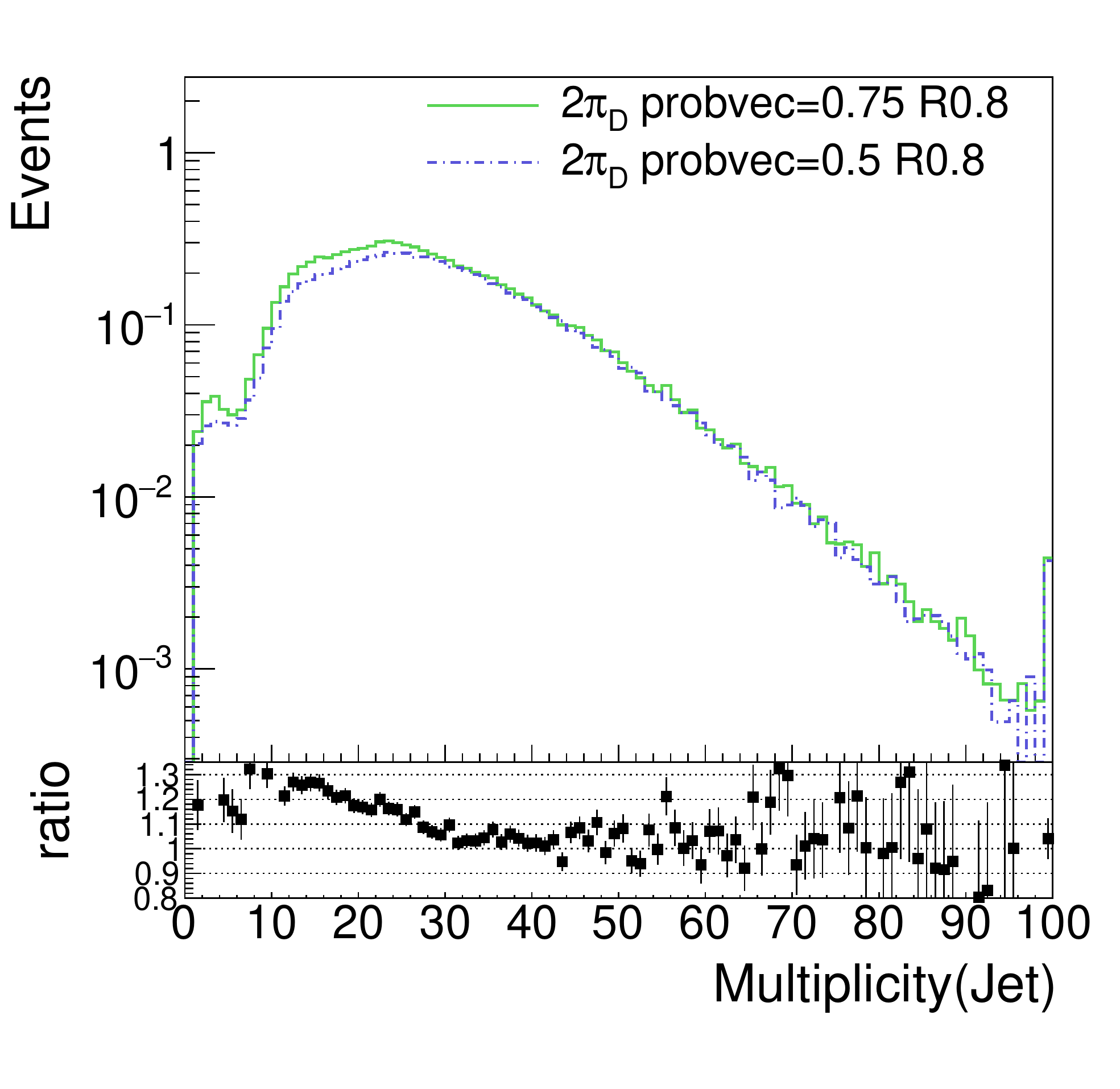}
\includegraphics[width=0.3\linewidth]{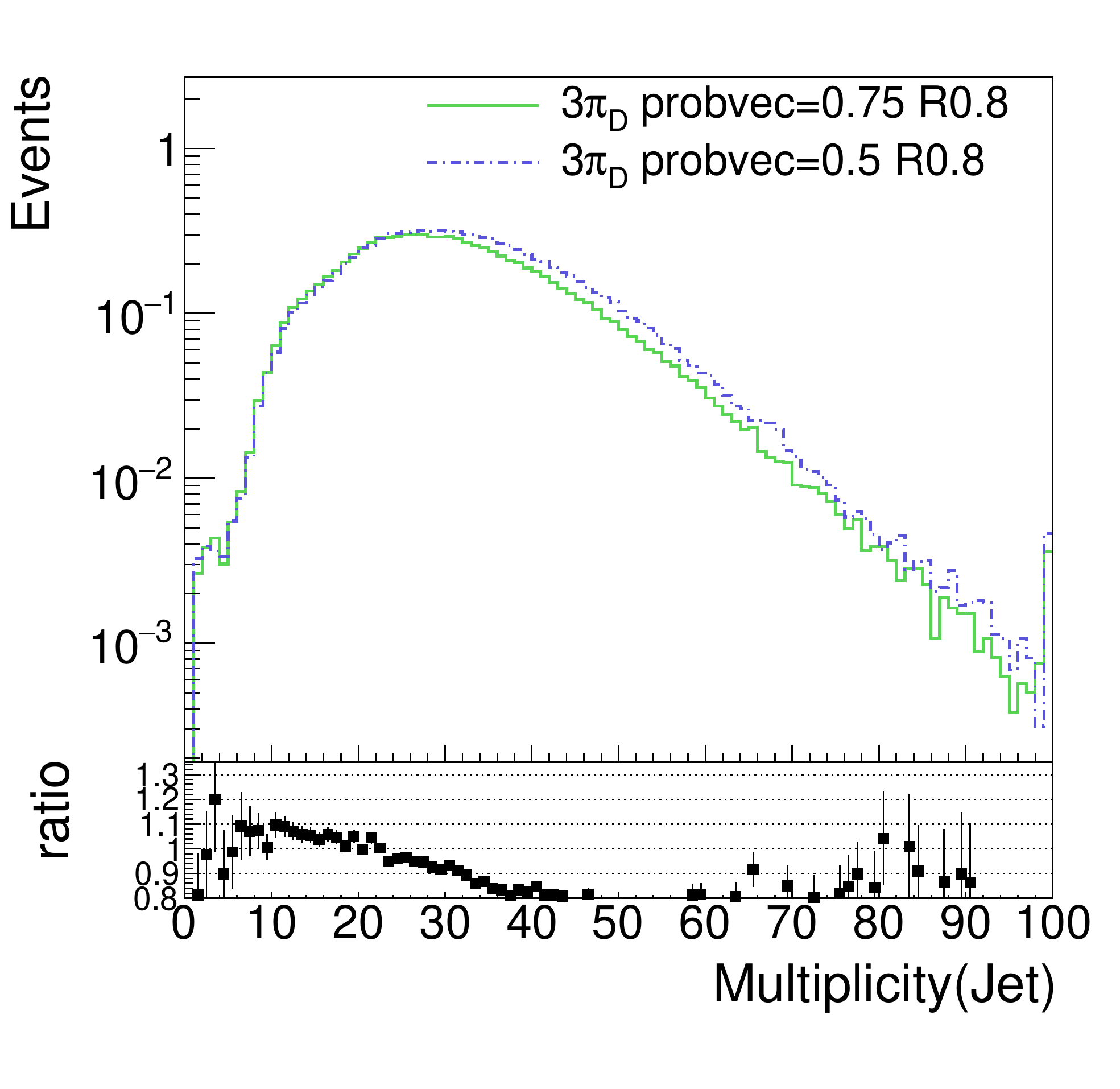}
\includegraphics[width=0.3\linewidth]{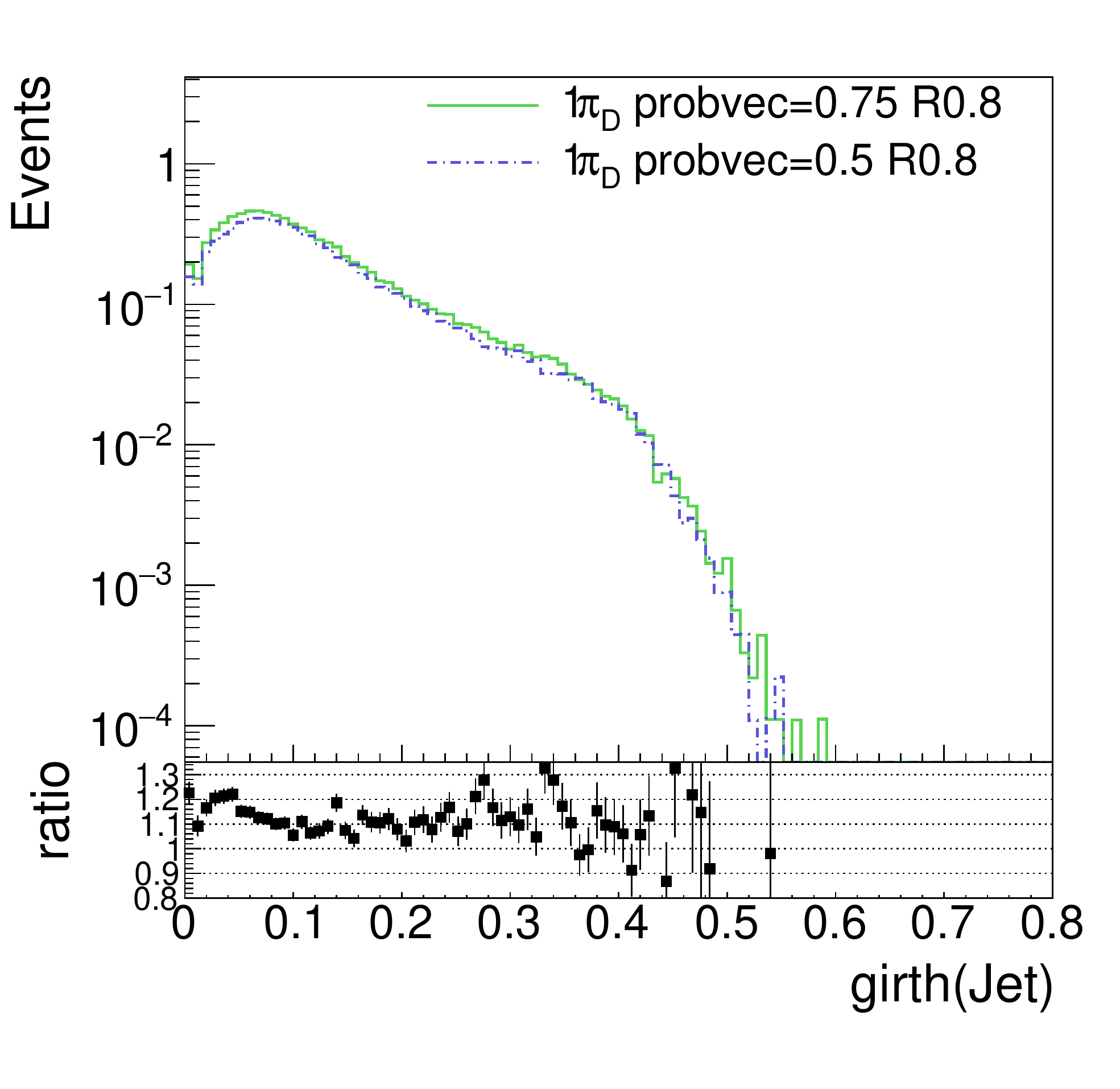}
\includegraphics[width=0.3\linewidth]{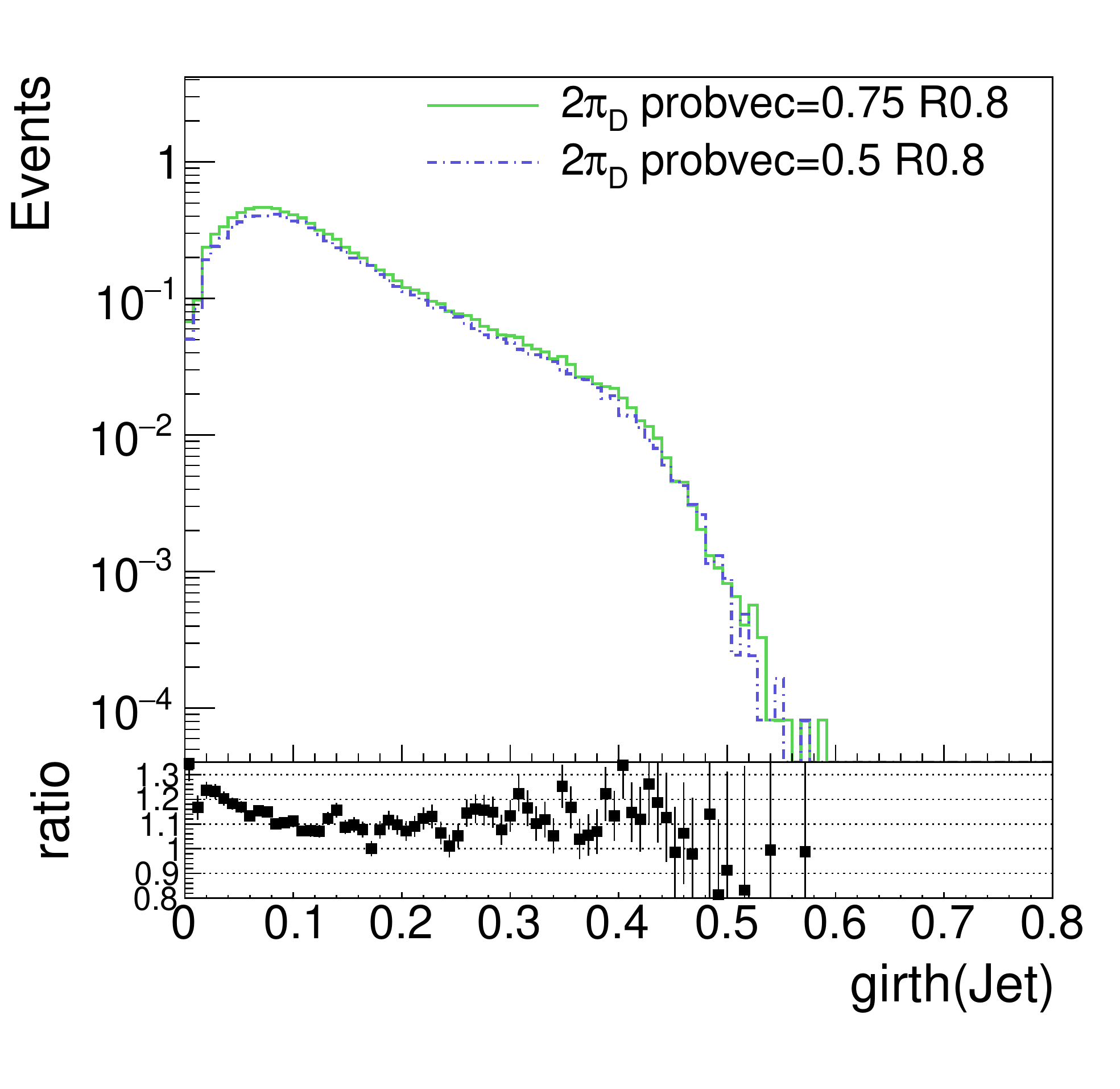}
\includegraphics[width=0.3\linewidth]{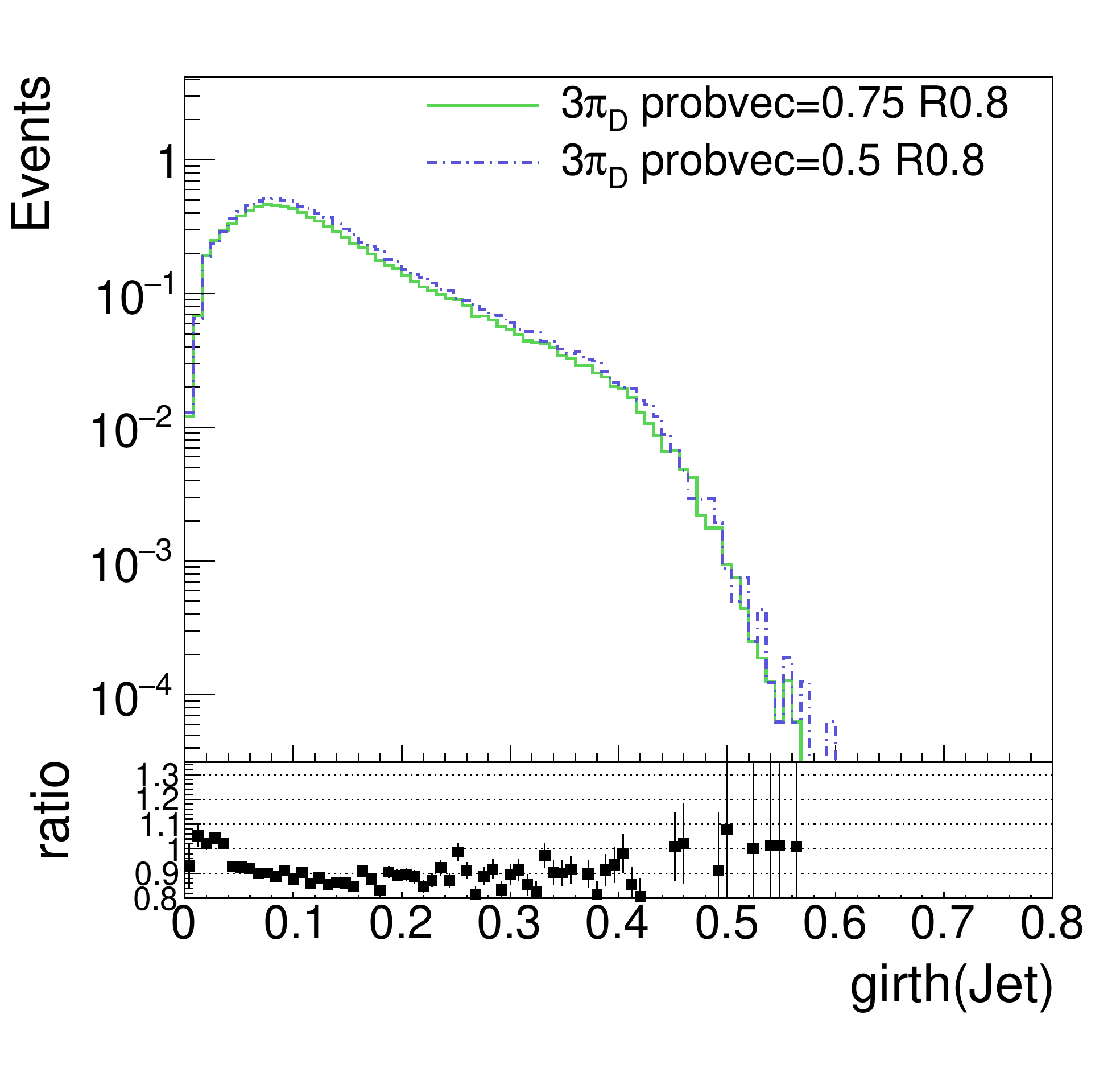}
\caption{Comparison of reco-level variables between \texttt{probVector=0.5} and \texttt{probVector=0.75} for different number of unstable diagonal dark pions. 
The first row shows the number of reconstructed jets, the second row shows the constituent multiplicity of all jets, and the third row shows the girth of all jets in the event. The plotted ratio is the ratio of \texttt{probVector=0.75} to \texttt{probVector=0.5}.}
\label{fig:reco_probvec_a}
\end{figure}

\begin{figure}[htb!]
\centering
\includegraphics[width=0.3\linewidth]{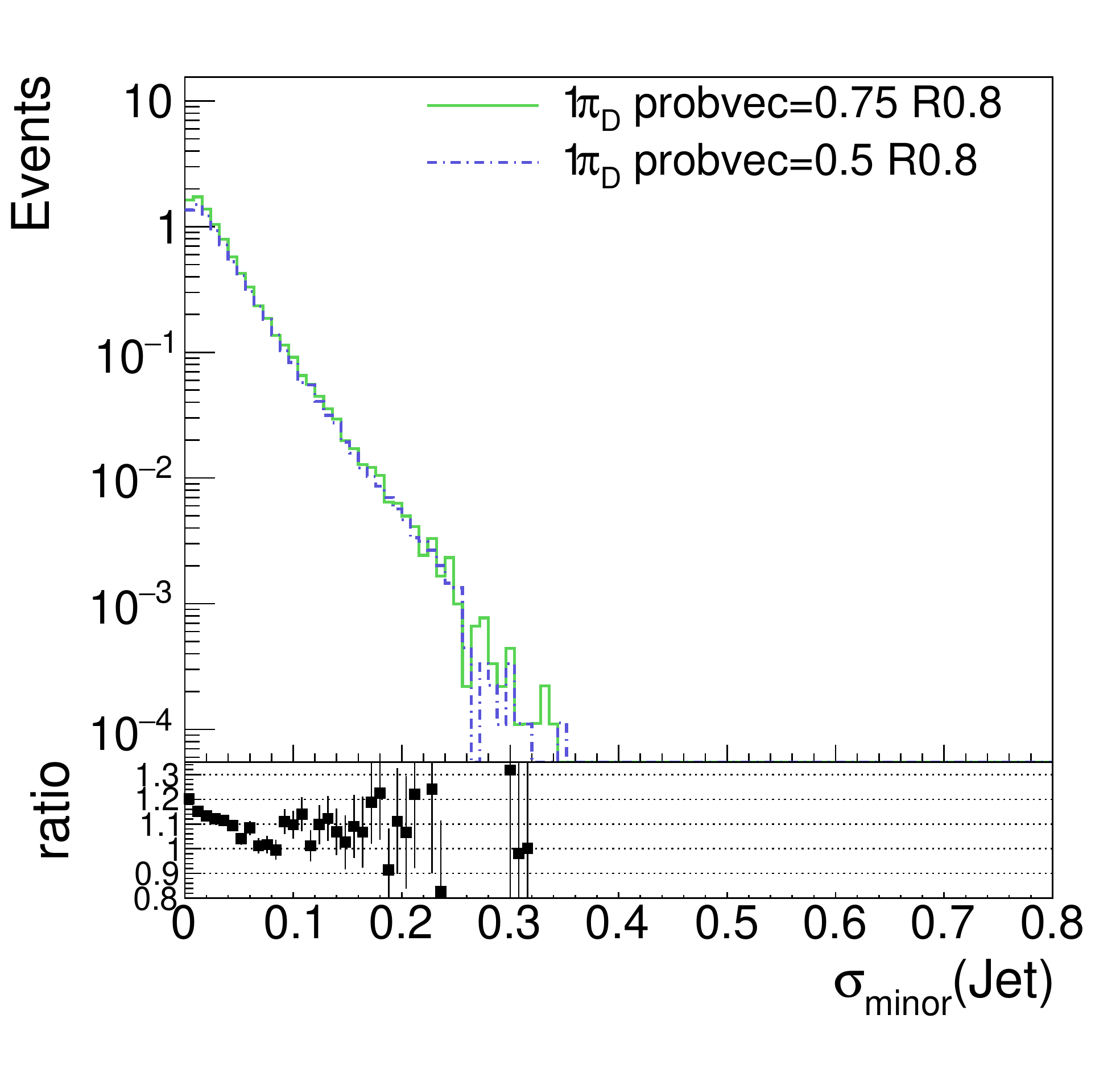}
\includegraphics[width=0.3\linewidth]{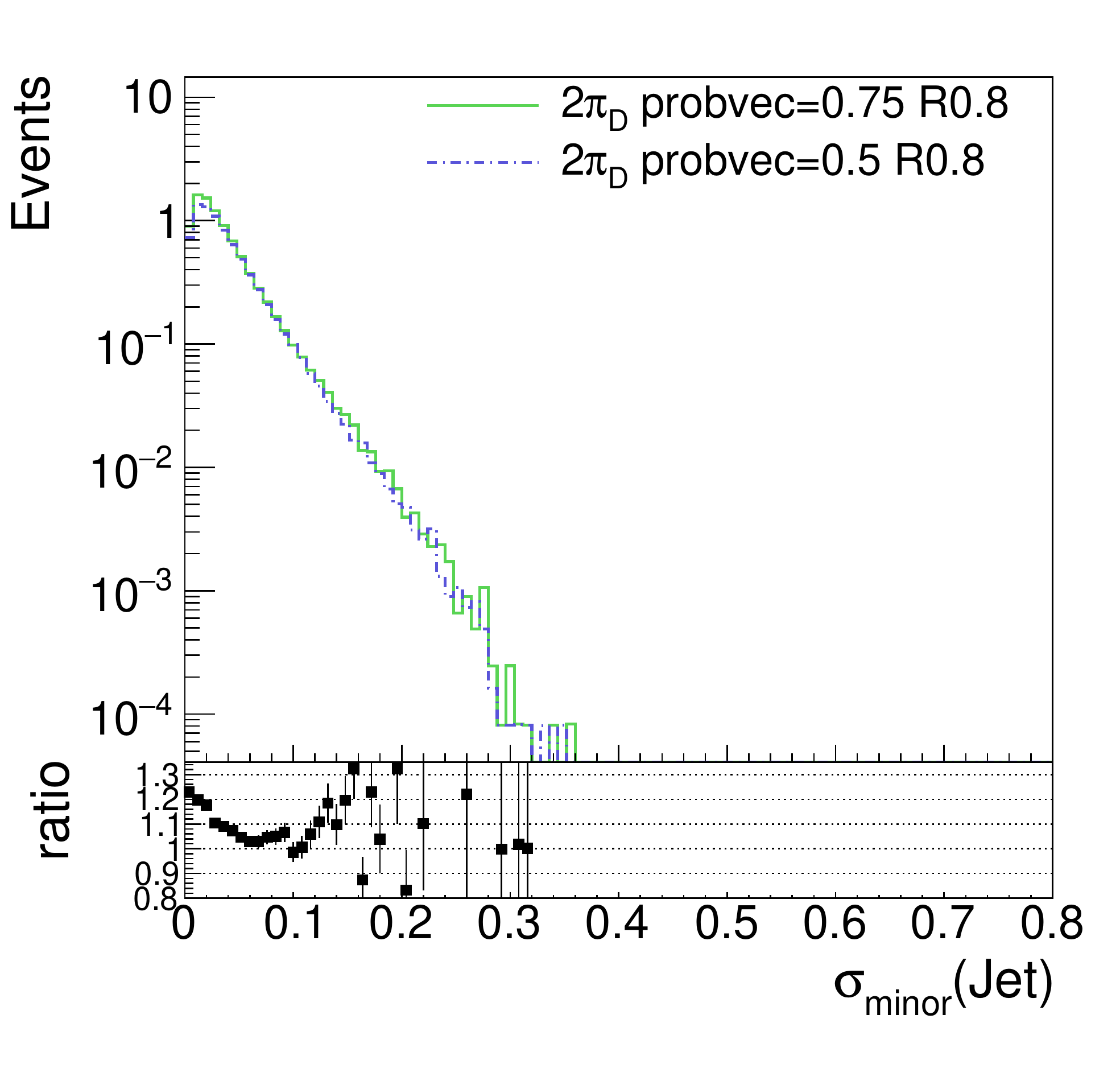}
\includegraphics[width=0.3\linewidth]{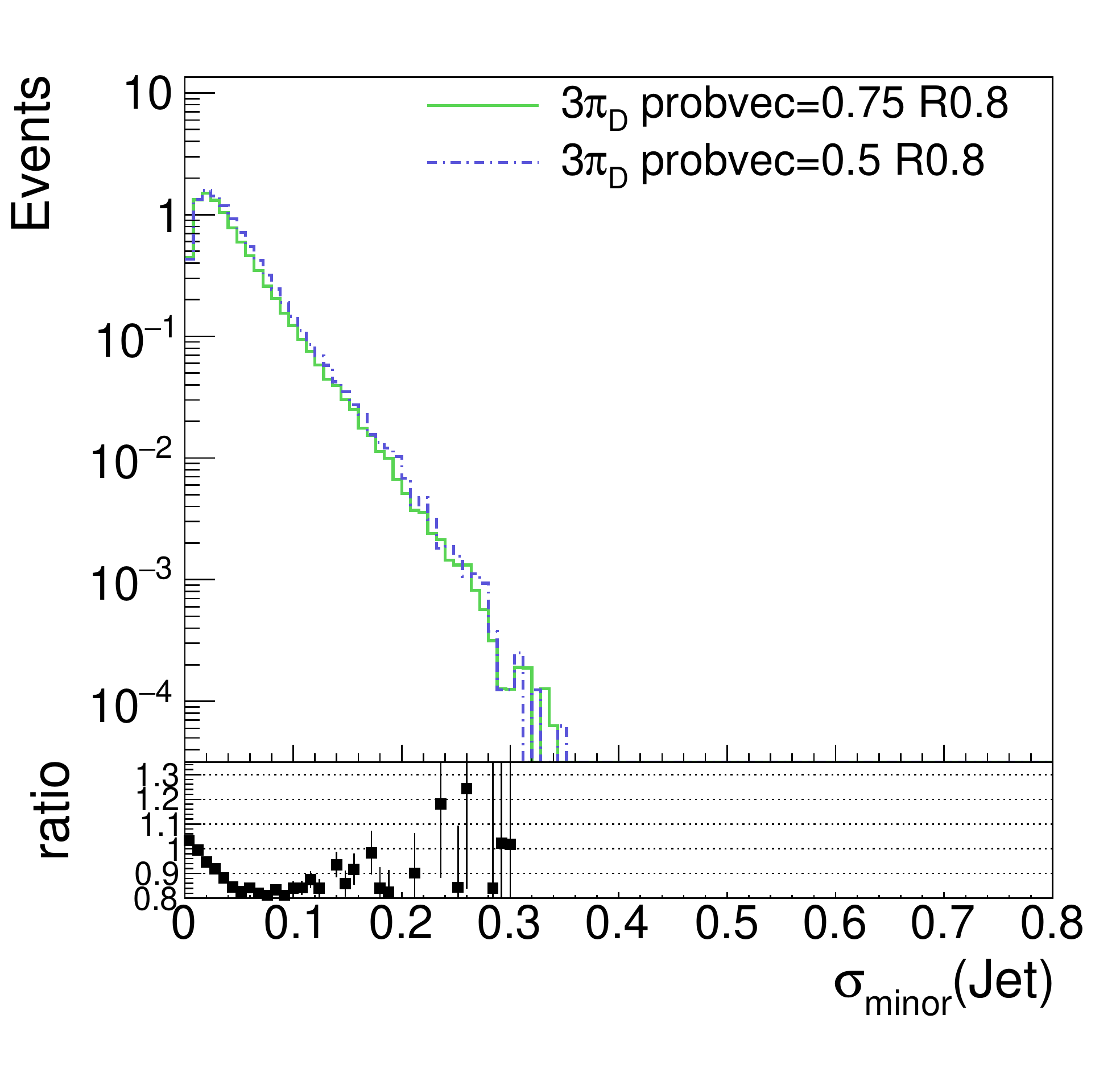}
\includegraphics[width=0.3\linewidth]{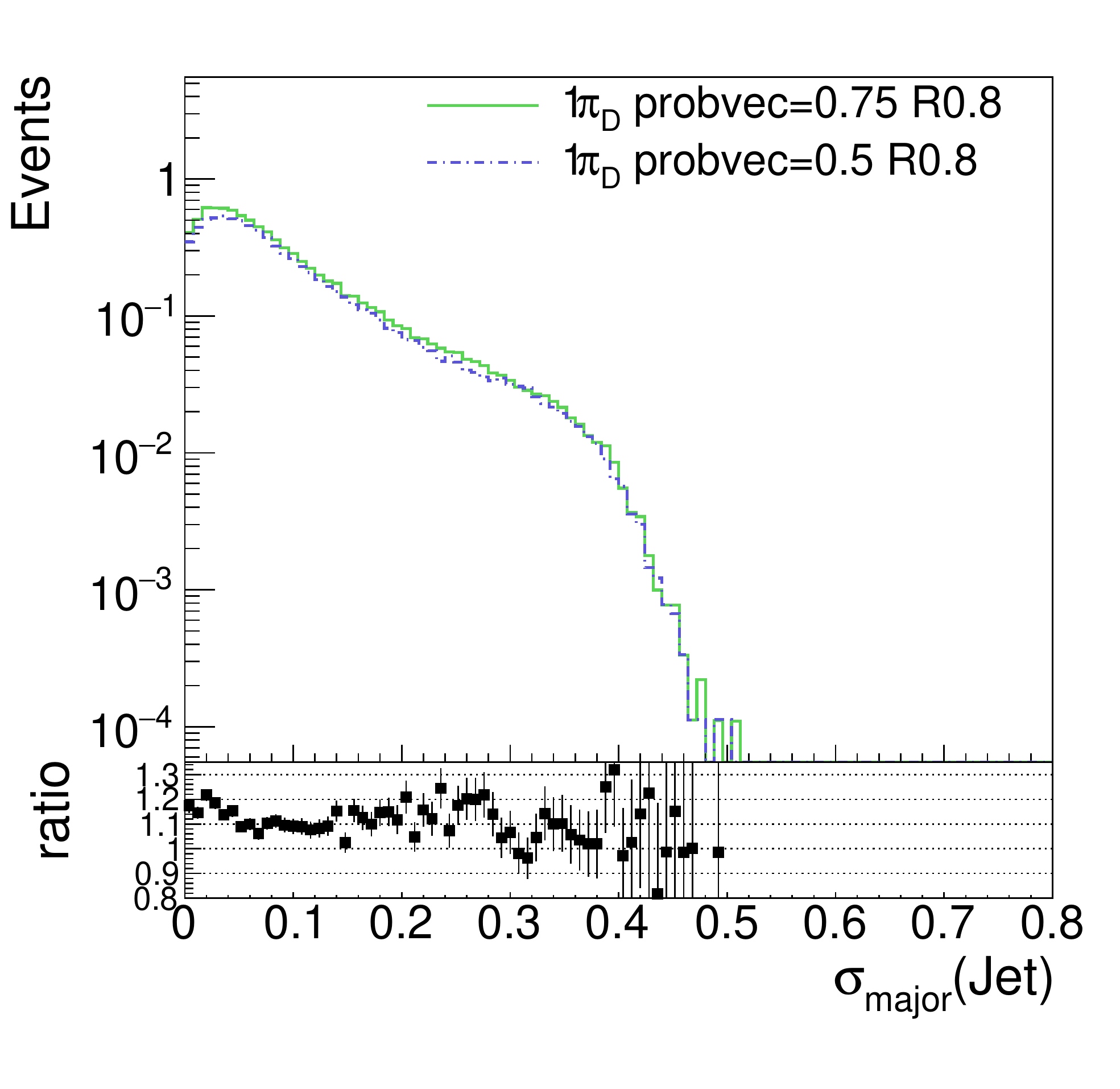}
\includegraphics[width=0.3\linewidth]{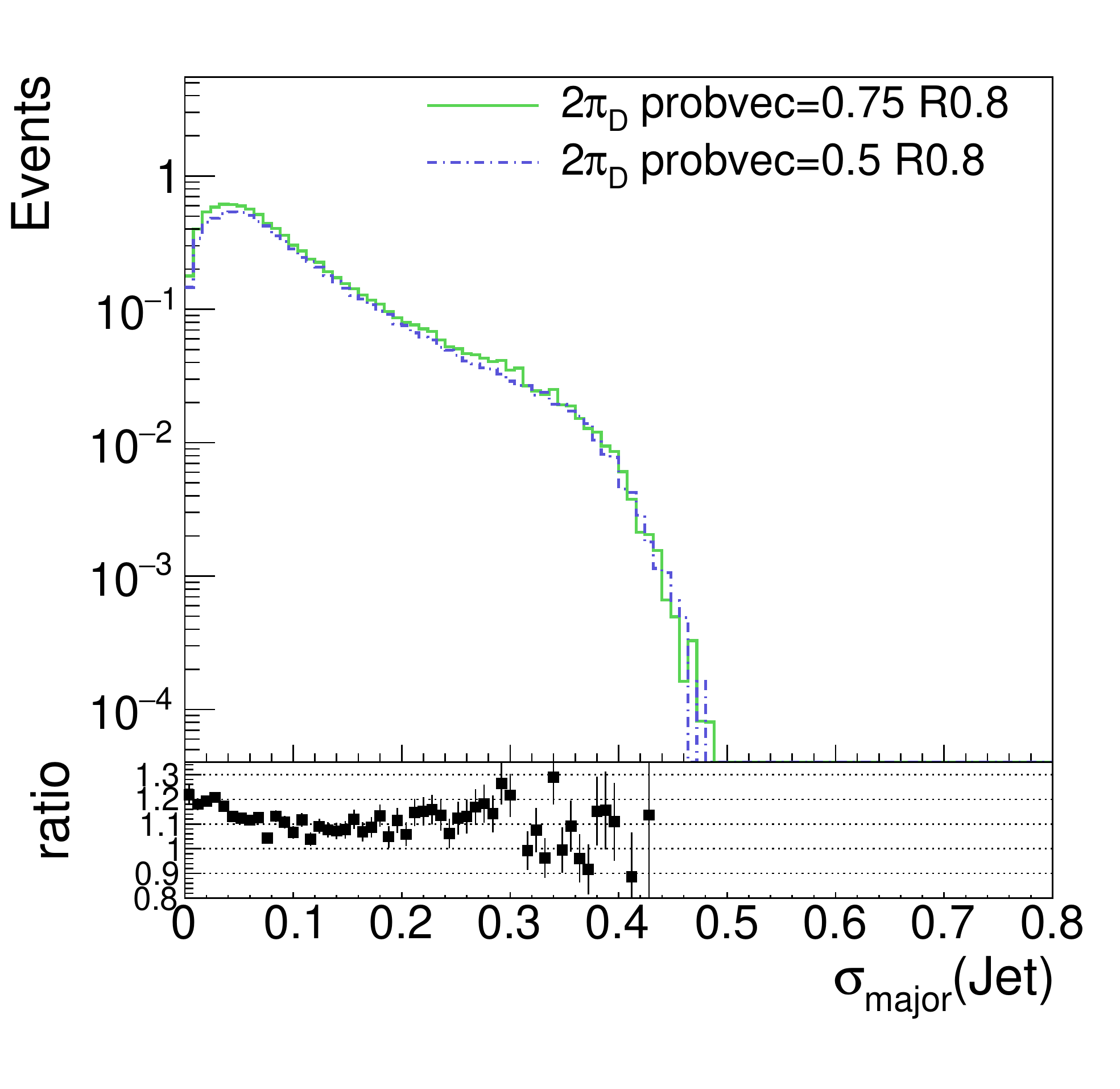}
\includegraphics[width=0.3\linewidth]{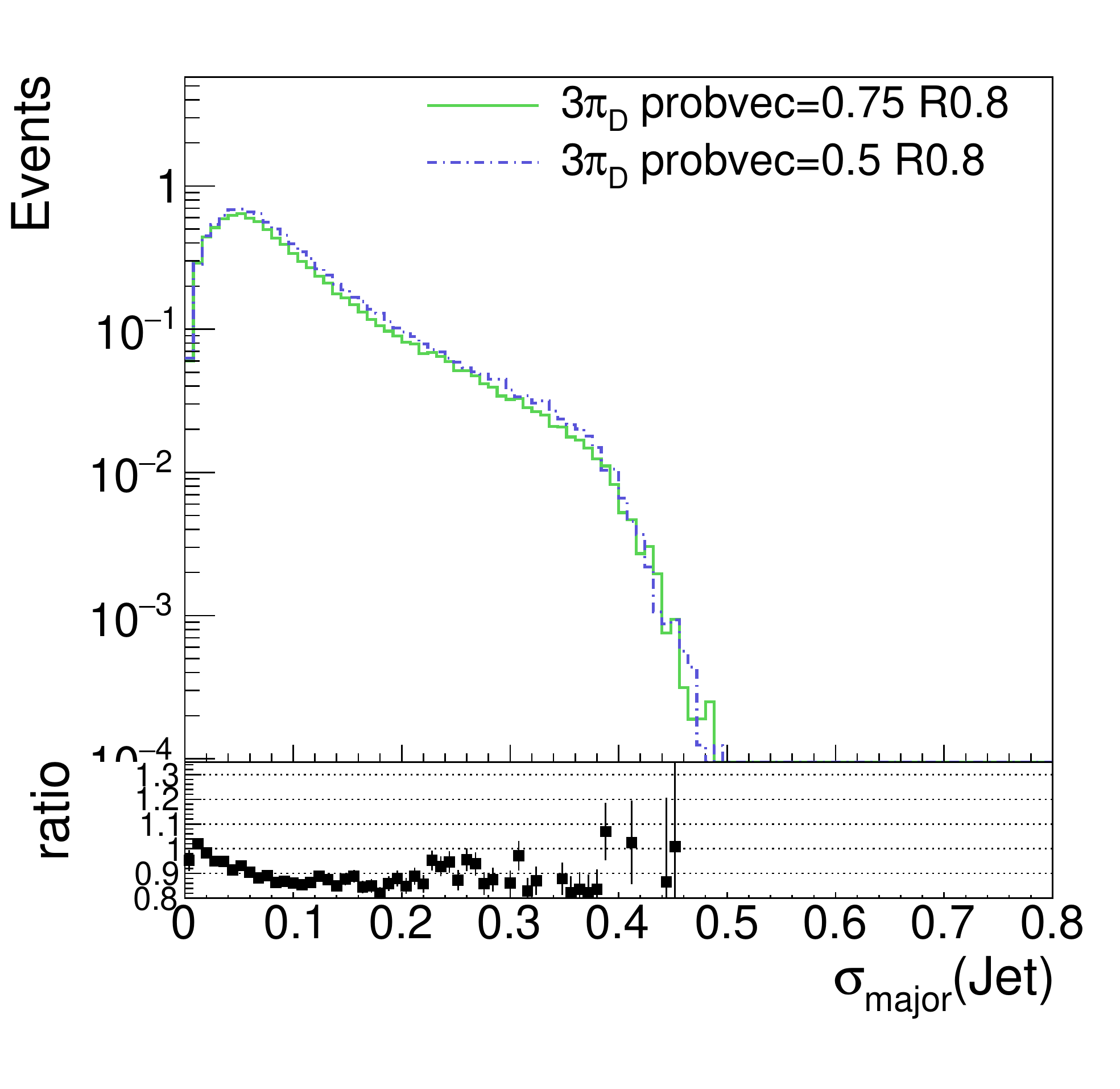}
\includegraphics[width=0.3\linewidth]{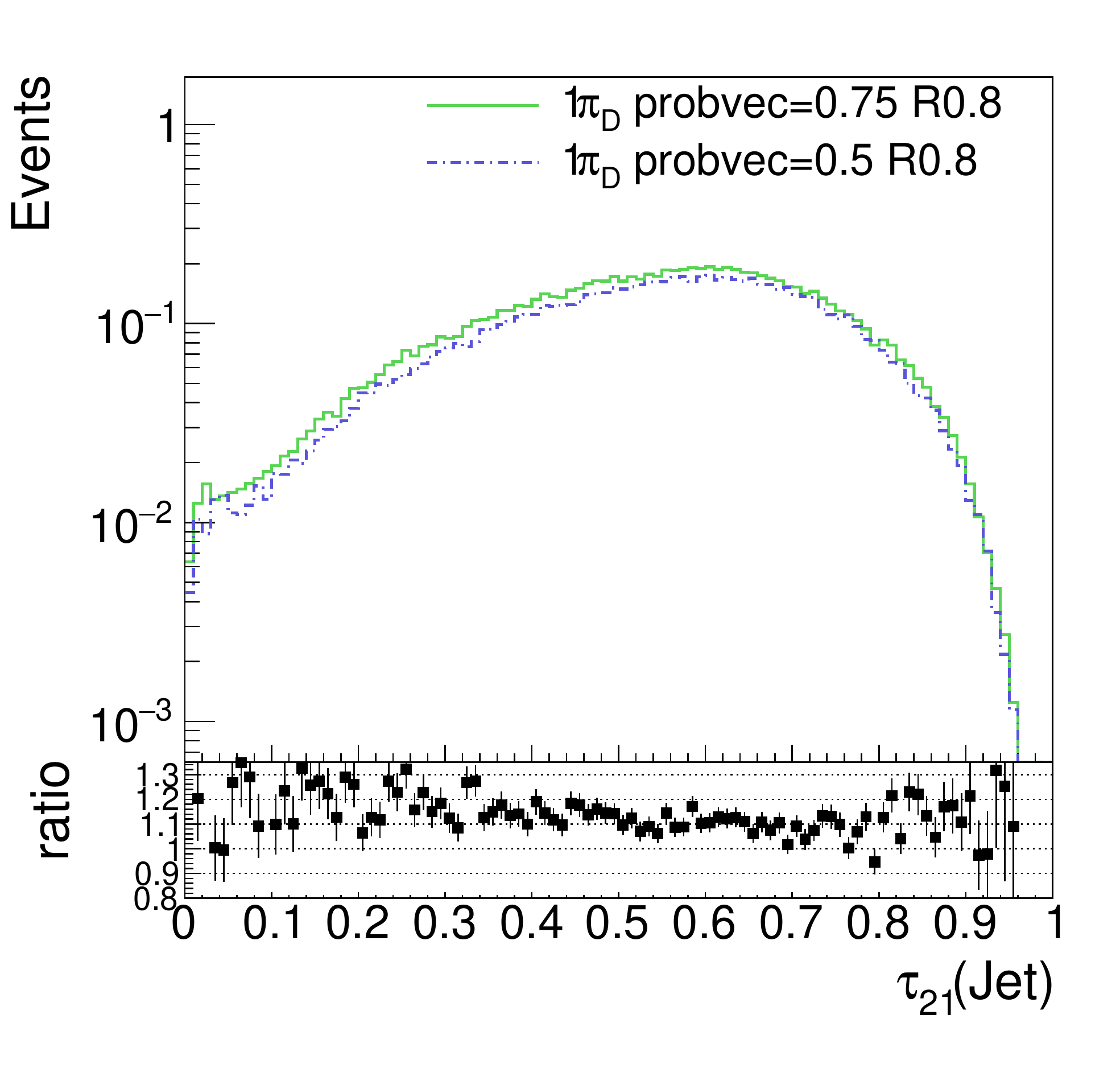}
\includegraphics[width=0.3\linewidth]{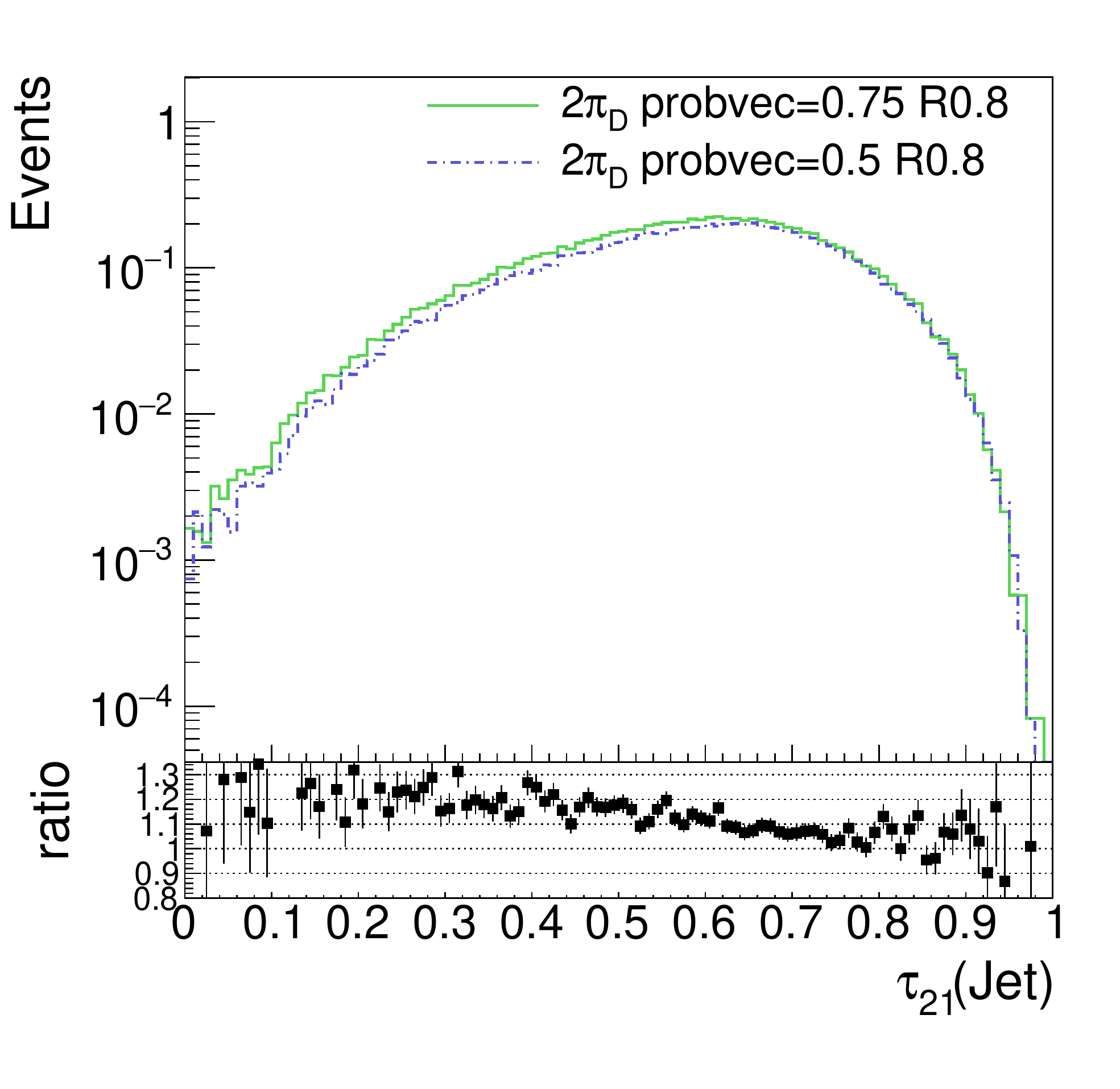}
\includegraphics[width=0.3\linewidth]{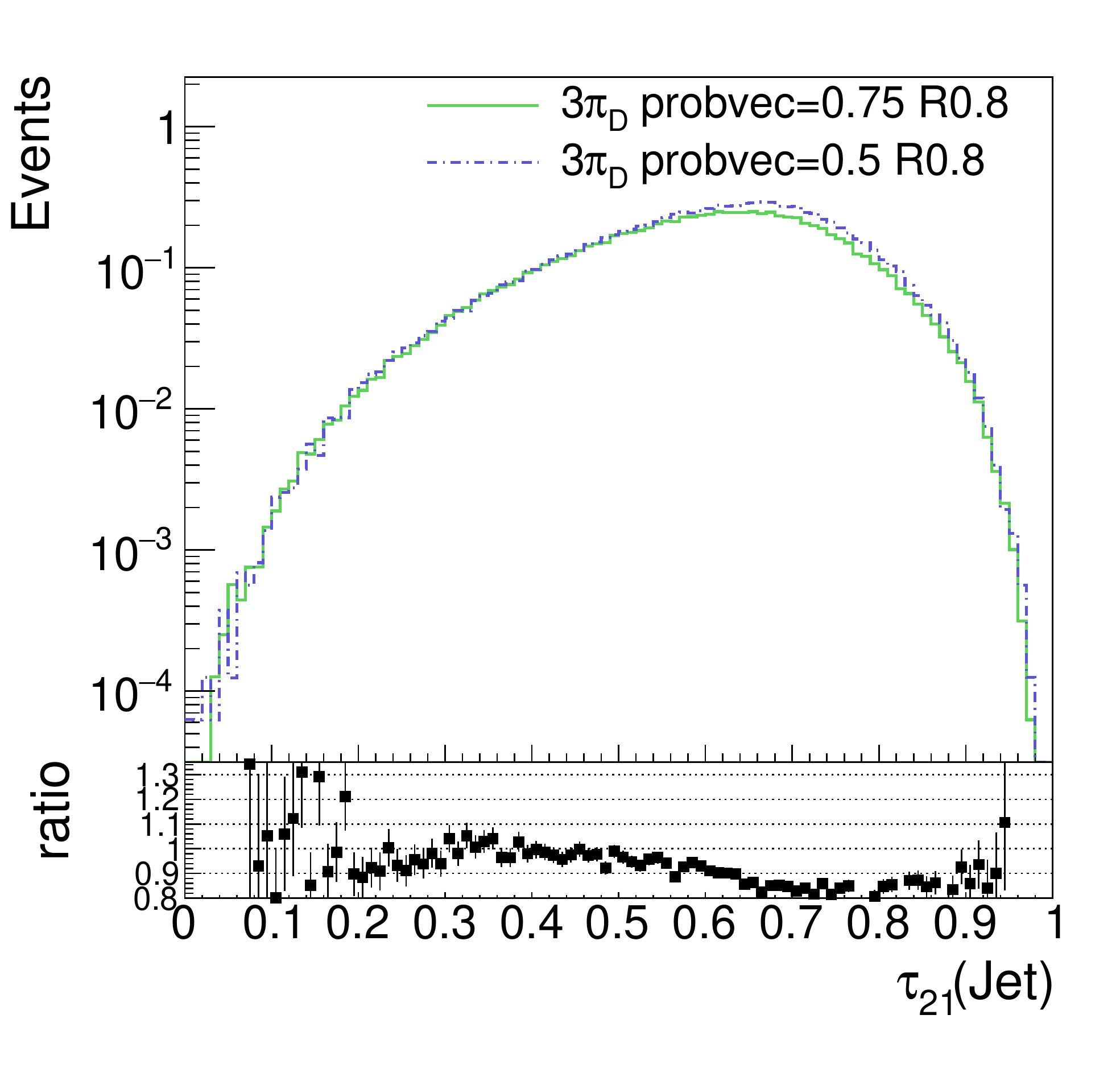}
\includegraphics[width=0.3\linewidth]{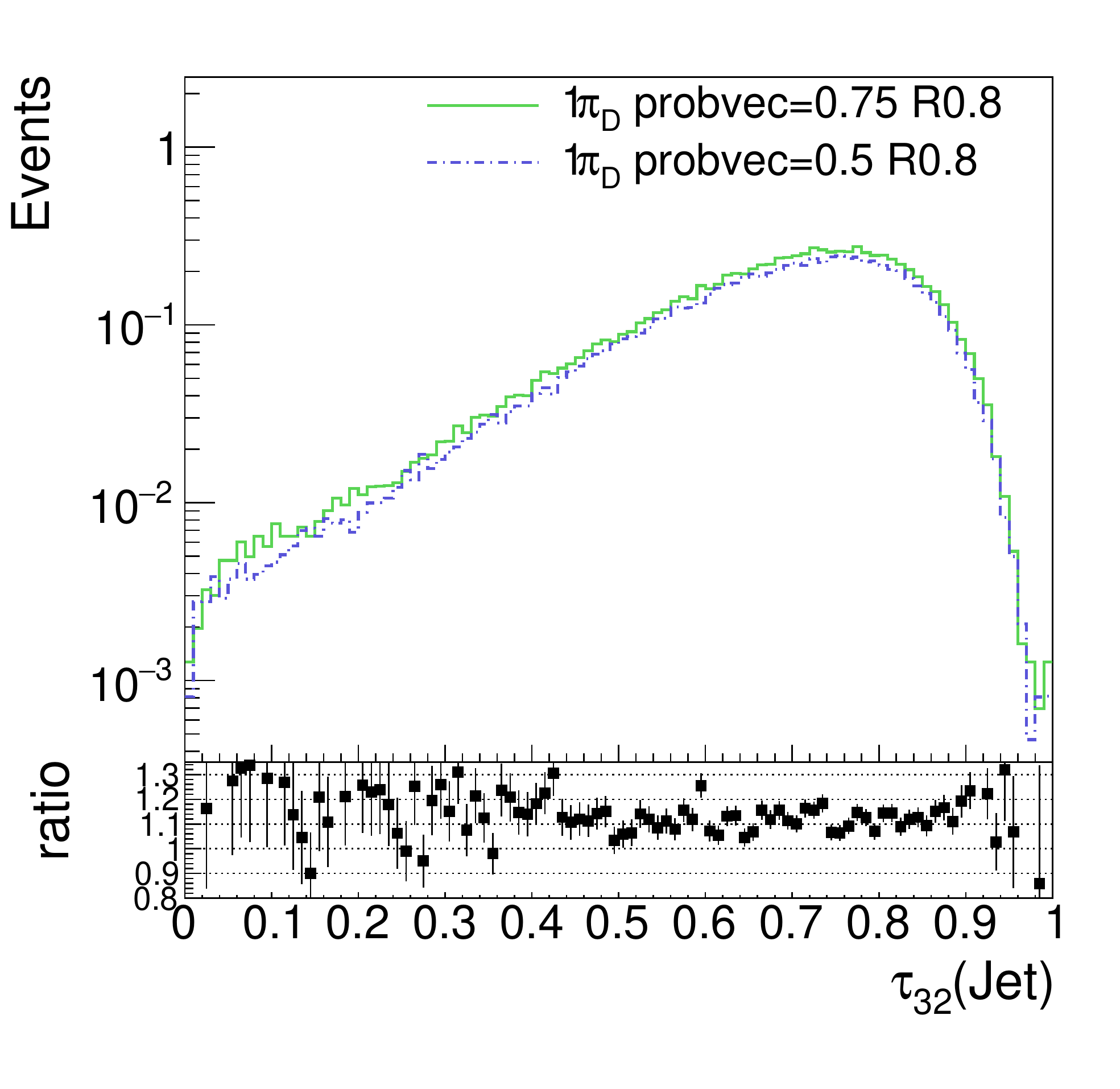}
\includegraphics[width=0.3\linewidth]{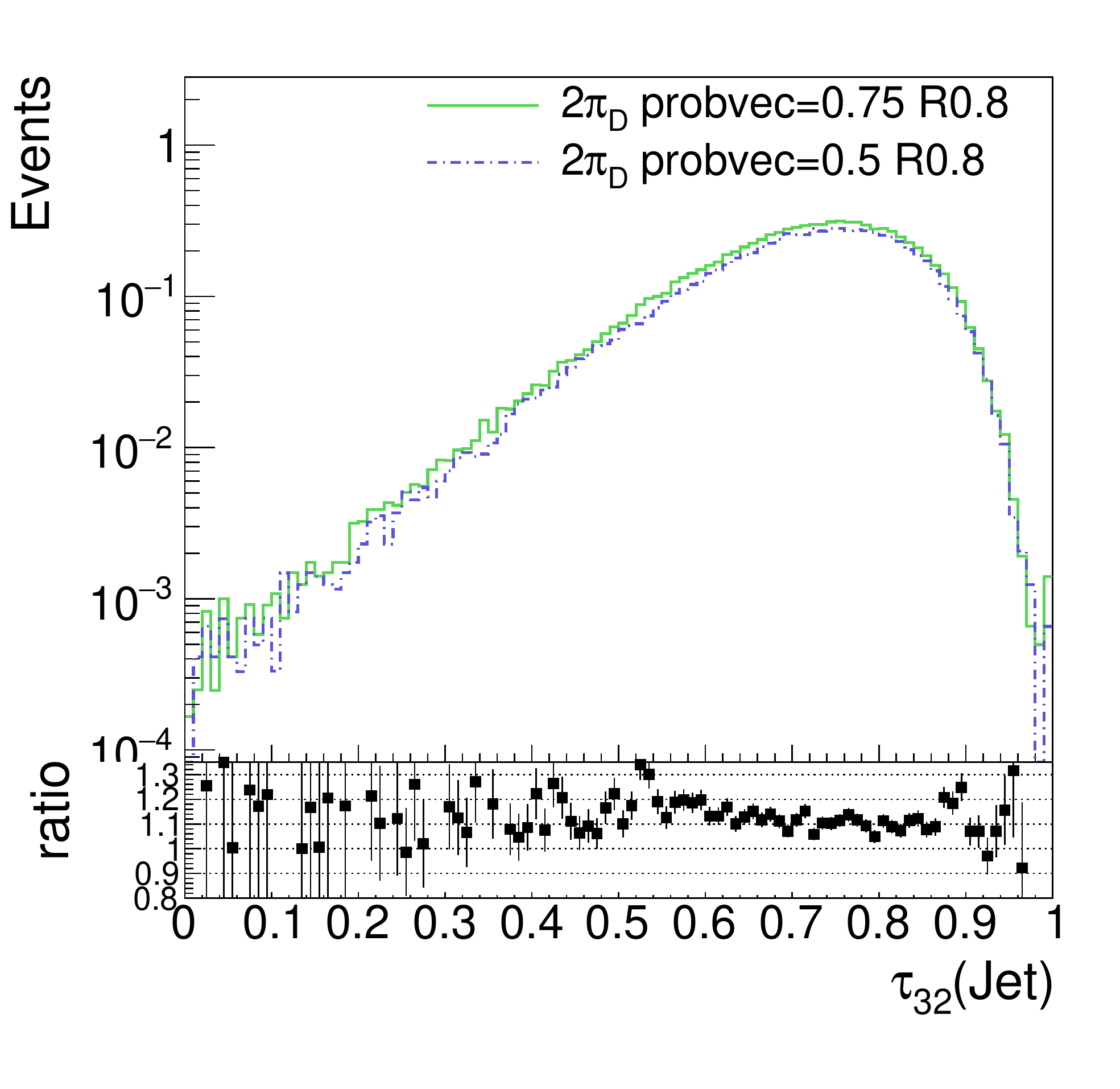}
\includegraphics[width=0.3\linewidth]{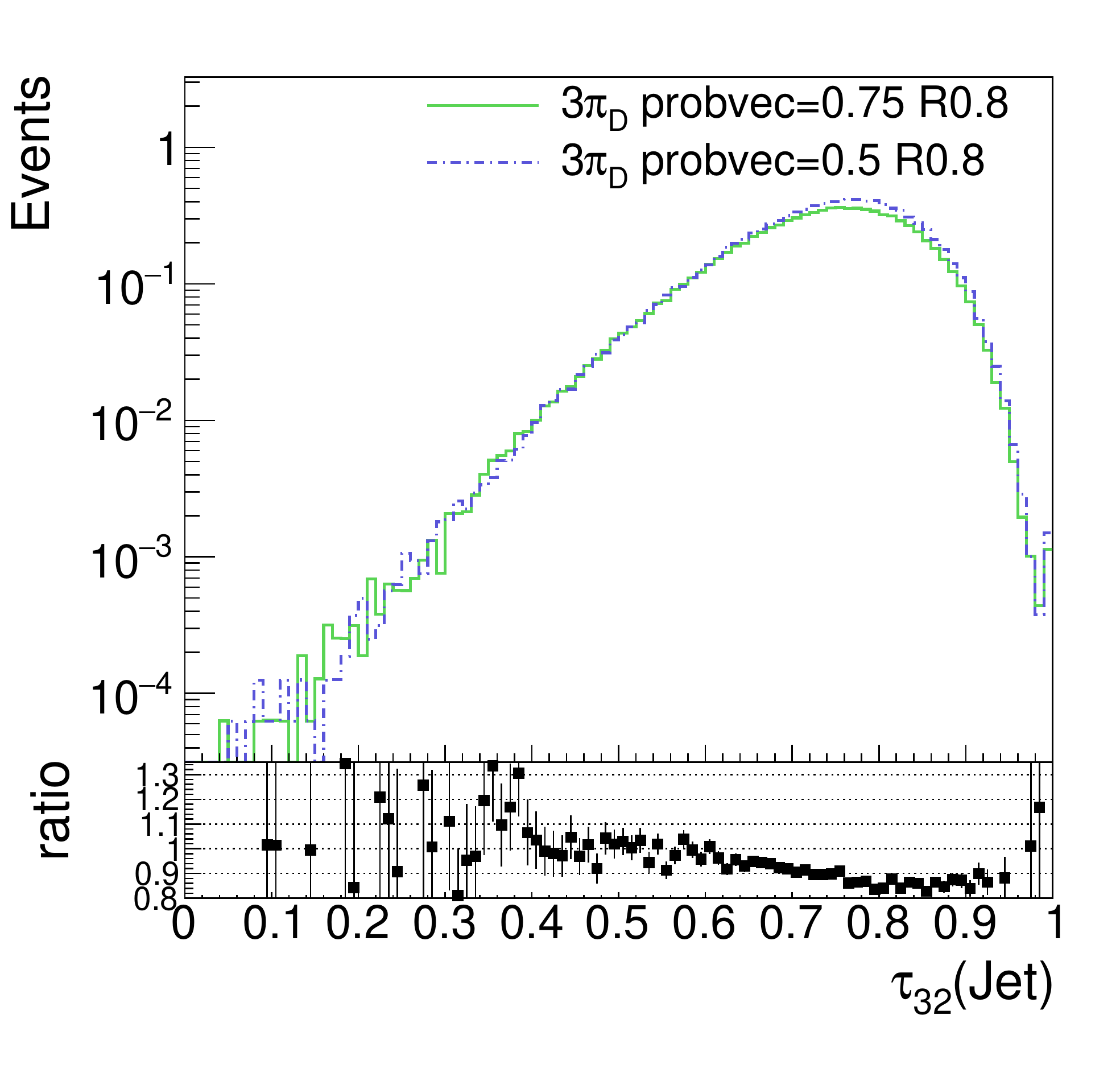}
\caption{Comparison of reco-level variables between \texttt{probVector=0.5} and \texttt{probVector=0.75} for  different number of unstable diagonal dark pions.  
The first row shows the minor axis, the second row shows the major axis, the third row shows the n-subjettiness ratio $\tau_{21}$, and the fourth row shows $\tau_{32}$. The plotted ratio is the ratio of \texttt{probVector=0.75} to \texttt{probVector=0.5}.}
\label{fig:reco_probvec_b}
\end{figure}

Figures~\ref{fig:reco_1pi2pi3pi_a} and~\ref{fig:reco_1pi2pi3pi_b} show the effect of varying the number of unstable diagonal pions on the JSS observables, for different distance parameters and \texttt{probVector=0.5}. The variables \ptd~and the $N$-subjettiness ratios show the most discrimination between the different samples.

\subsubsection{Conclusion}
  Setting the IR parameters in accordance with the UV physics in general leads to a more cohesive modelling of the signal. This modelling however necessarily suffers from uncertainties due to a lack of knowledge of the precise hadronization parameters. These parameters can be varied to understand their effect on the resulting kinematic observables. In this section we have considered several jet substructure variables. We illustrated that changes in {\tt probVector} can lead to changes in the observed jet substructure variables. It should be noted that this study concentrates only on one specific benchmark point and two values of {\tt probVector} settings. It nevertheless shows the importance of understanding the effects of hadronization uncertainties. We also discussed the importance of Infrared and Collinear (IRC) safety when using substructure variables and in particular demonstrated that $p_TD$ is not IRC safe. Our studies thus highlight the need of a more detailed analysis of widely used jet substructure techniques in the light of dark showers phenomenology.
\begin{figure}[htb!]
\centering
\includegraphics[width=0.44\linewidth]{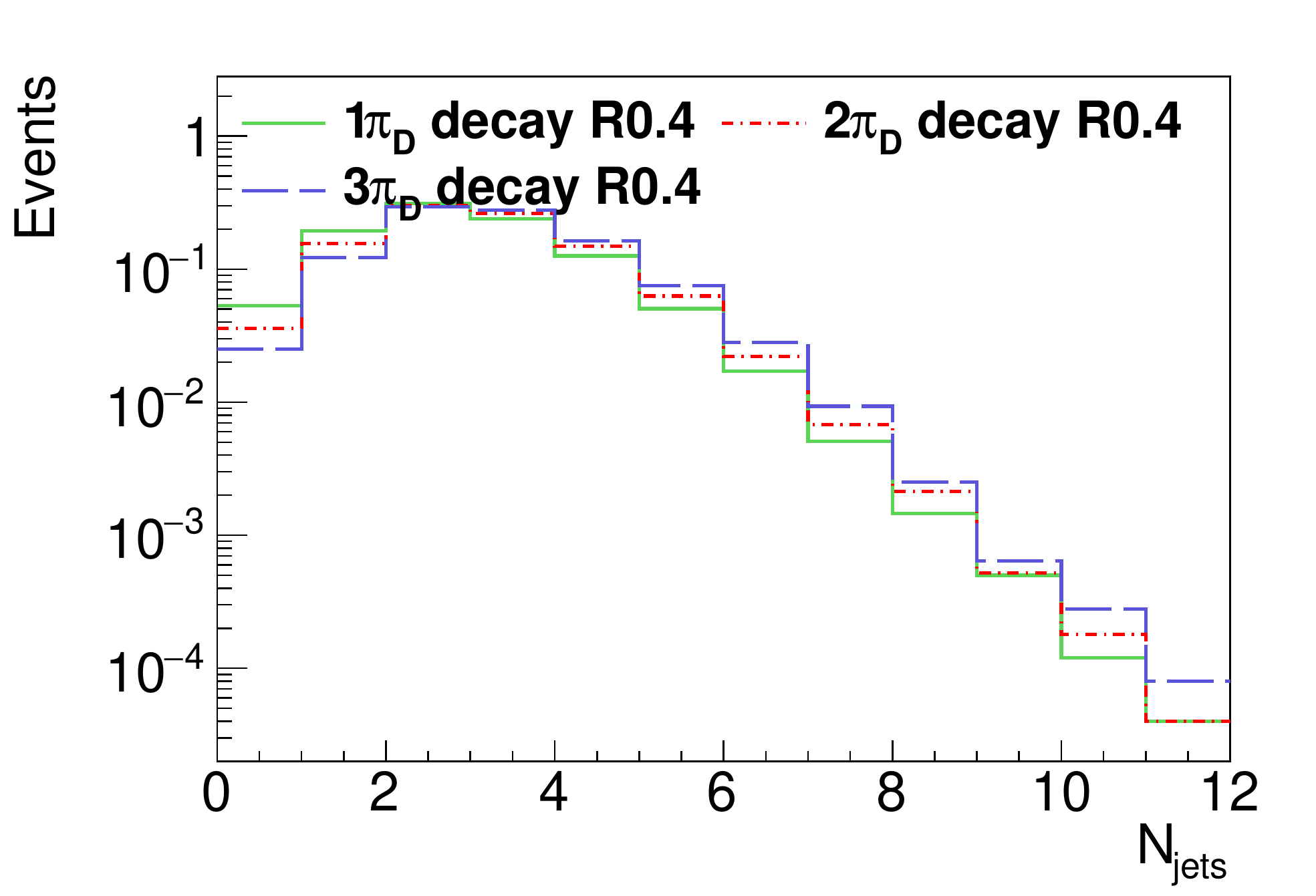}
\includegraphics[width=0.44\linewidth]{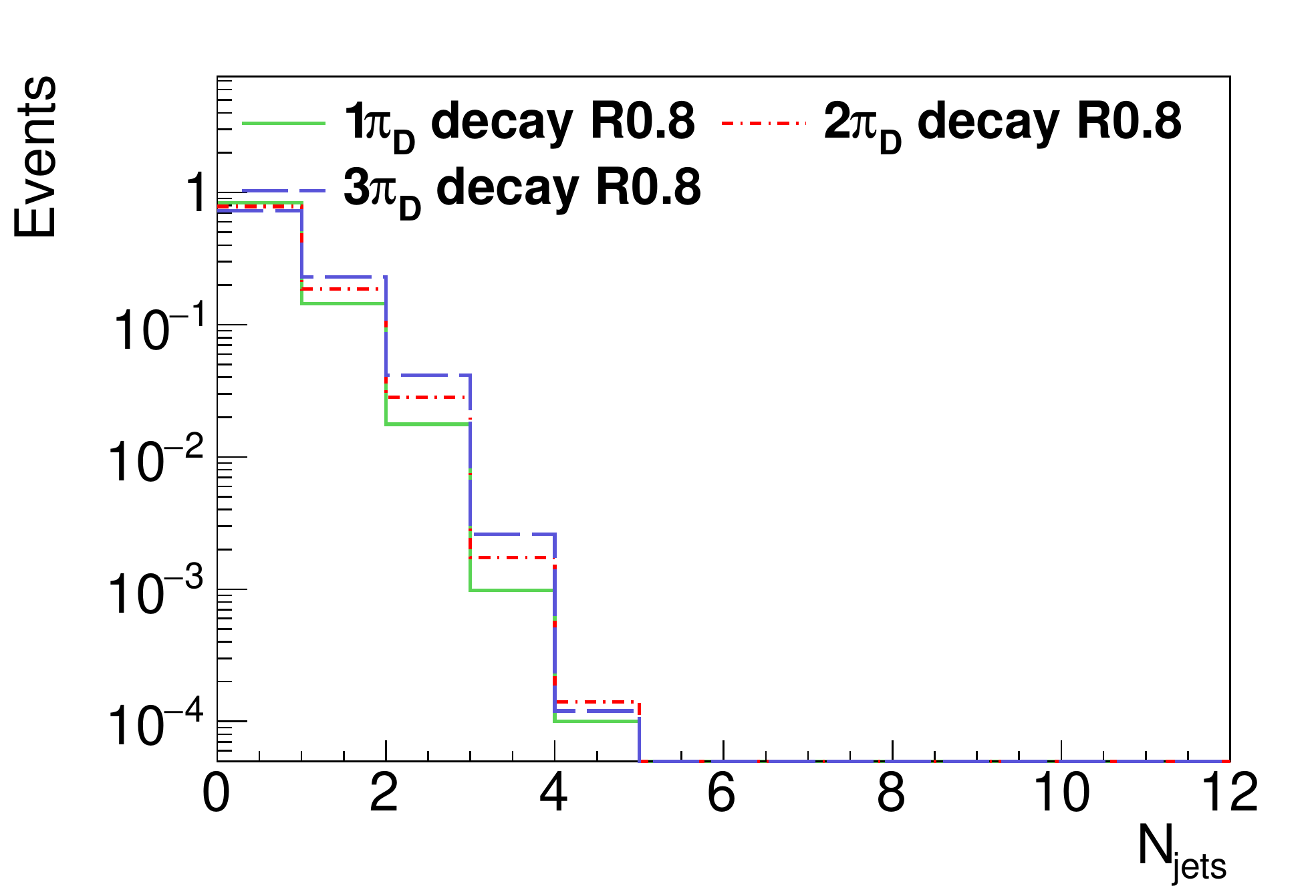}
\includegraphics[width=0.44\linewidth]{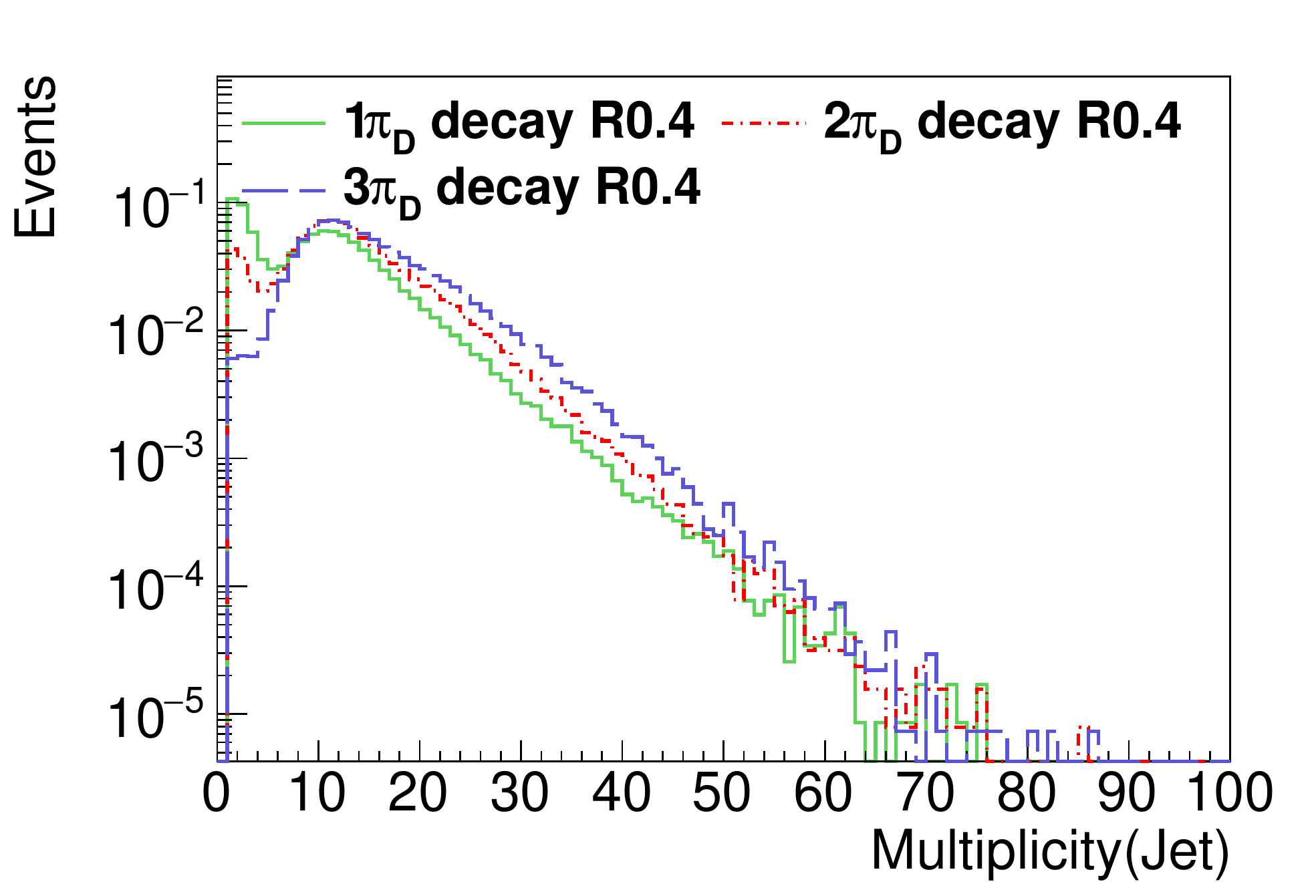}
\includegraphics[width=0.44\linewidth]{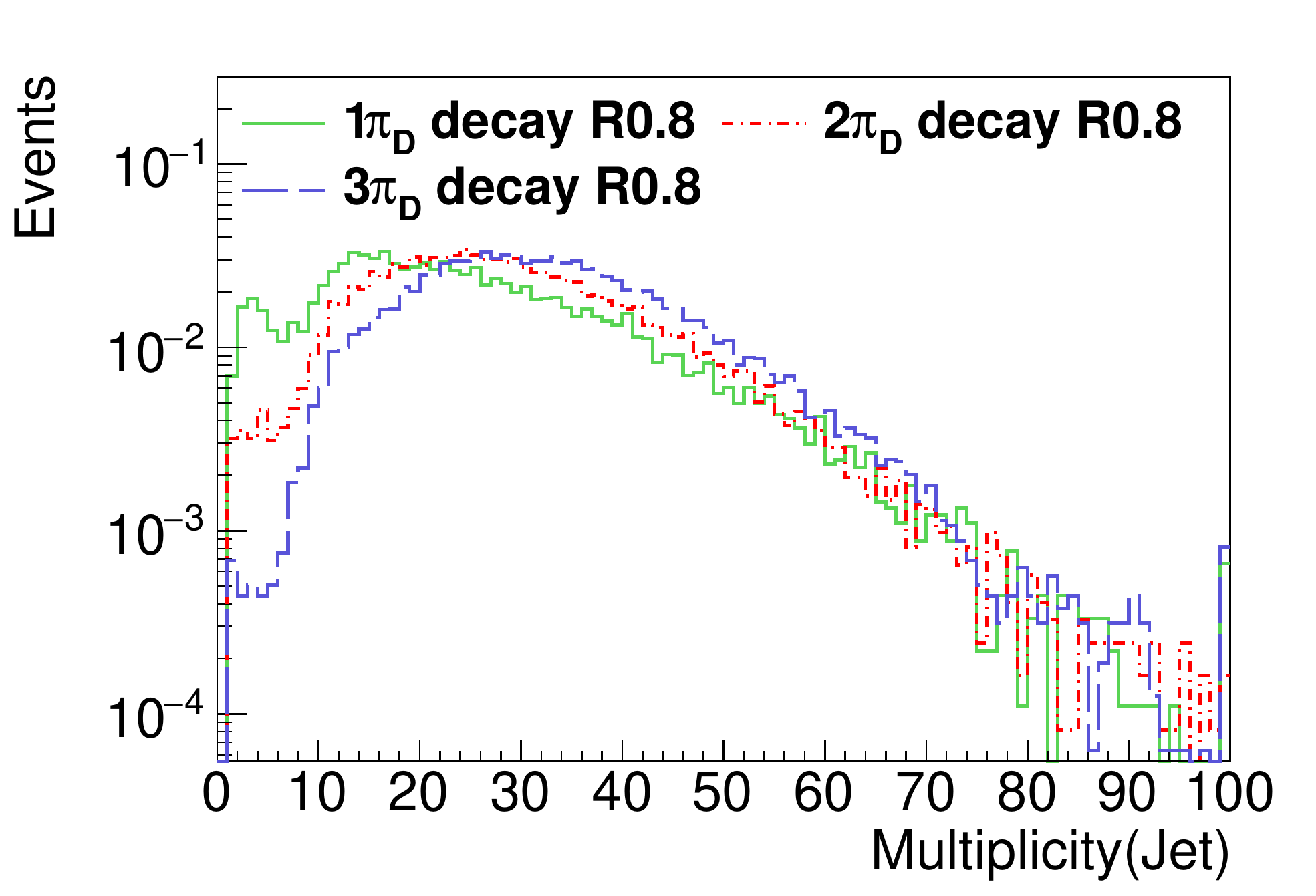}
\includegraphics[width=0.44\linewidth]{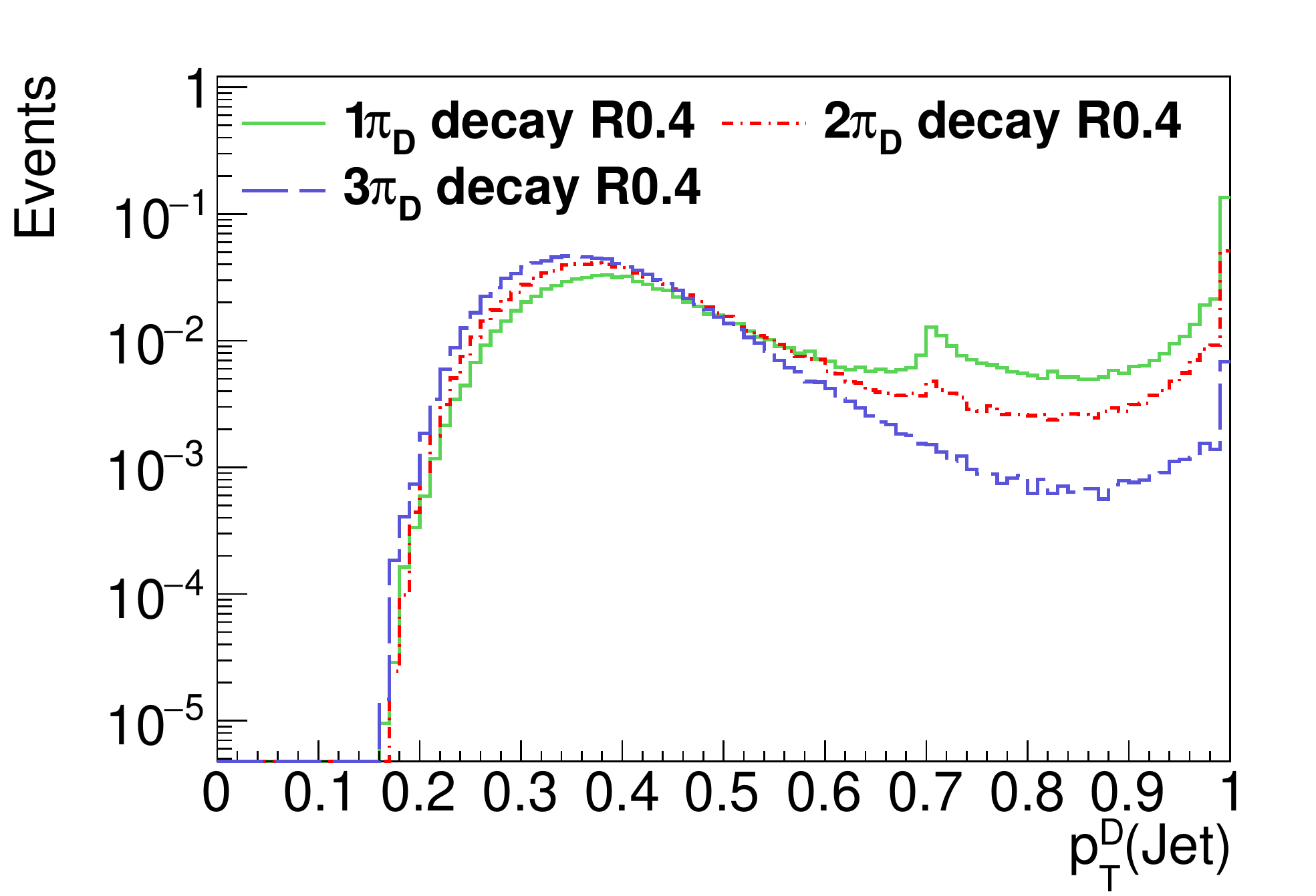}
\includegraphics[width=0.44\linewidth]{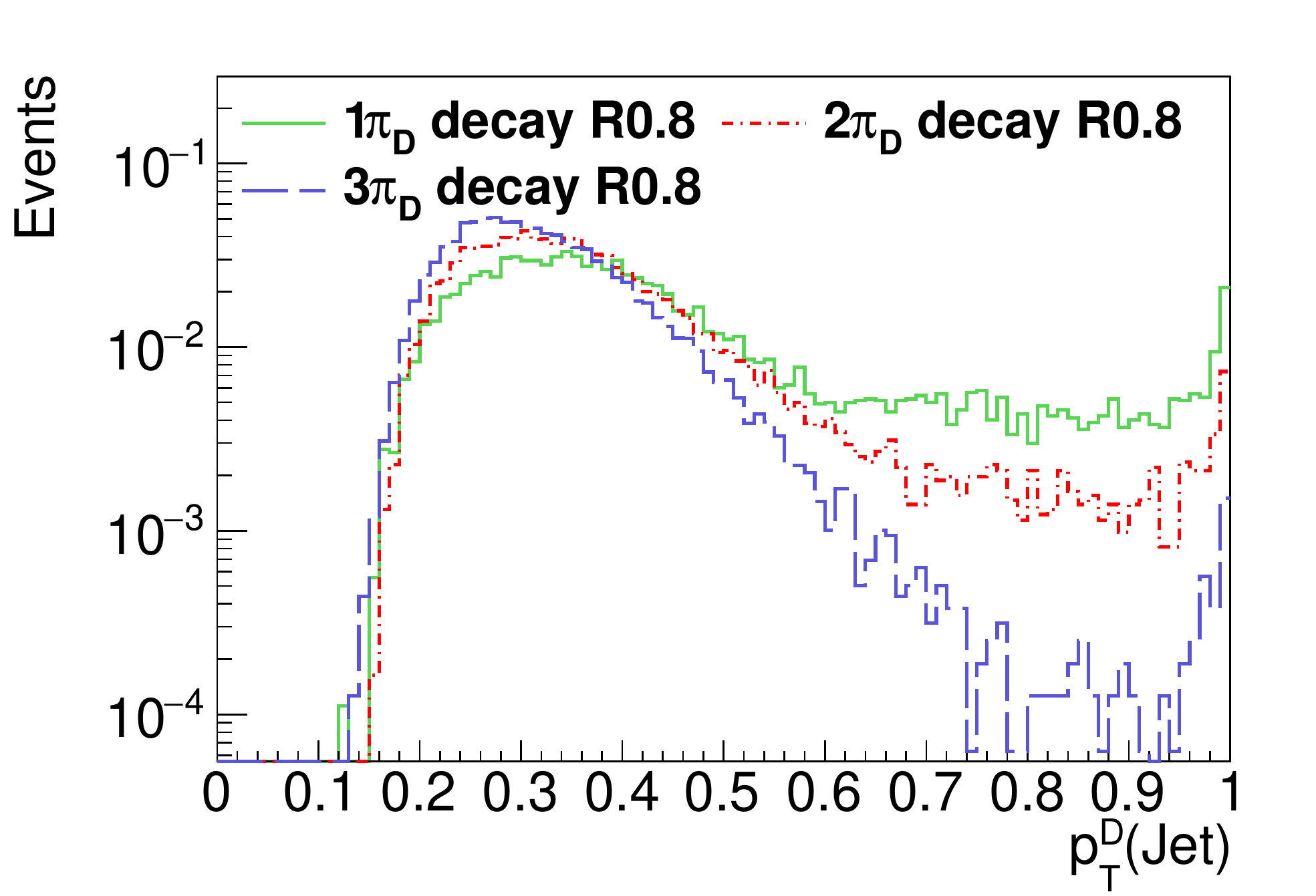}
\includegraphics[width=0.44\linewidth]{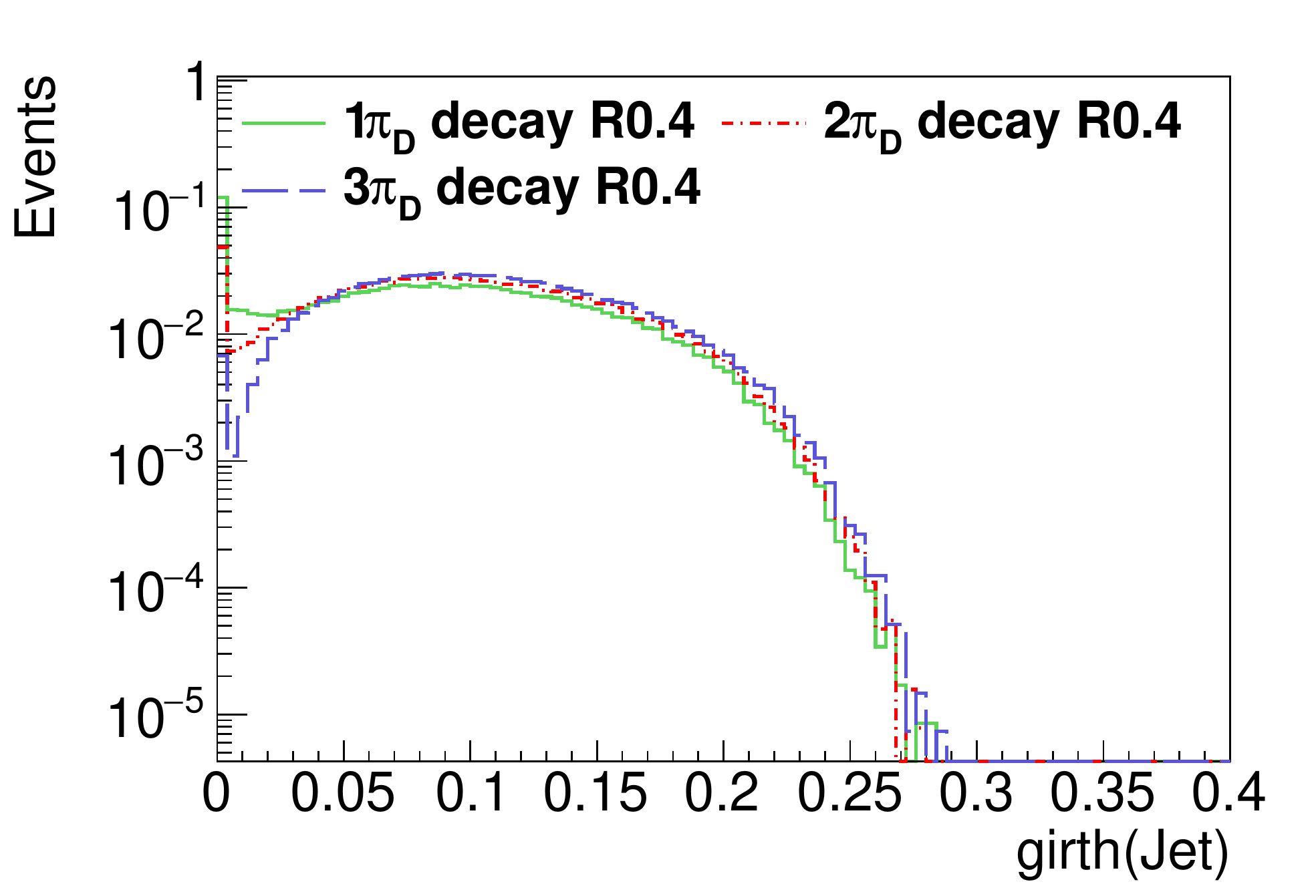}
\includegraphics[width=0.44\linewidth]{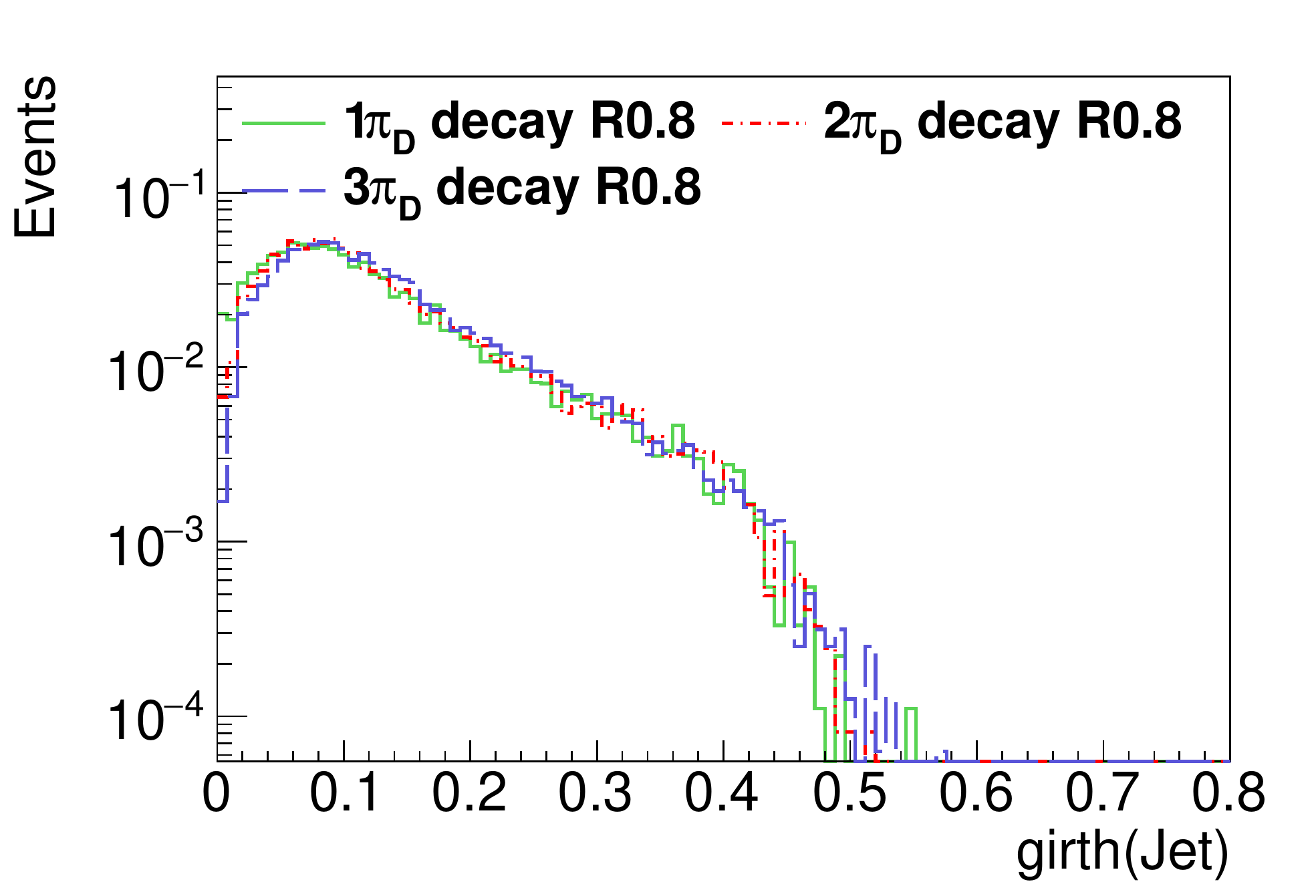}
\caption{Comparison of reco-level variables between different number of unstable diagonal dark pions for \texttt{probVector=0.5}. Left column show the $R=0.4$ jets and right column $R=0.8$ jets. First row is the number of reconstructed jets, second row is the constituent multiplicity of all jets, third row is the \ptd~of all jets, and the fourth row is the girth of all jets in the event.}
\label{fig:reco_1pi2pi3pi_a}
\end{figure}

\begin{figure}[htb!]
\centering
\includegraphics[width=0.44\linewidth]{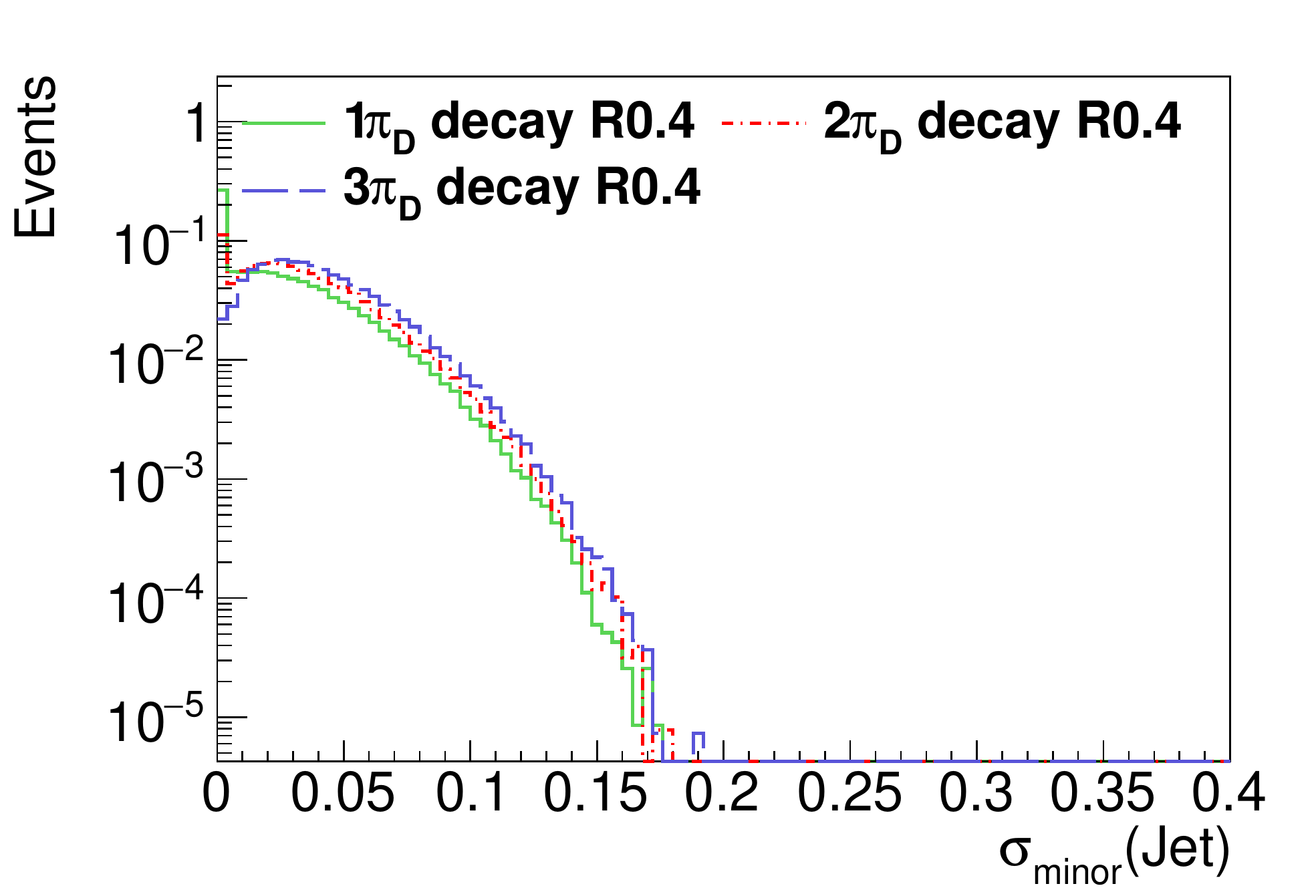}
\includegraphics[width=0.44\linewidth]{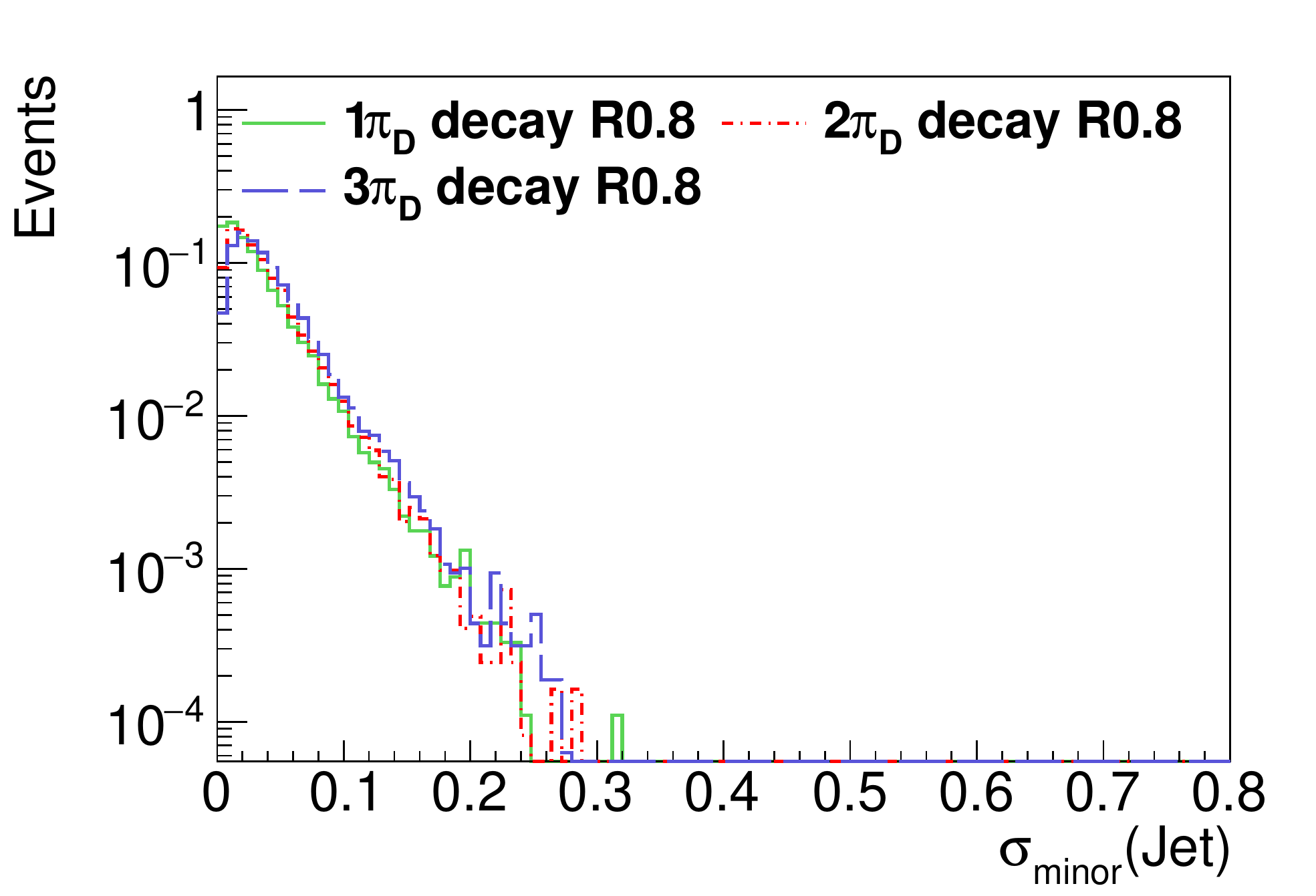}
\includegraphics[width=0.44\linewidth]{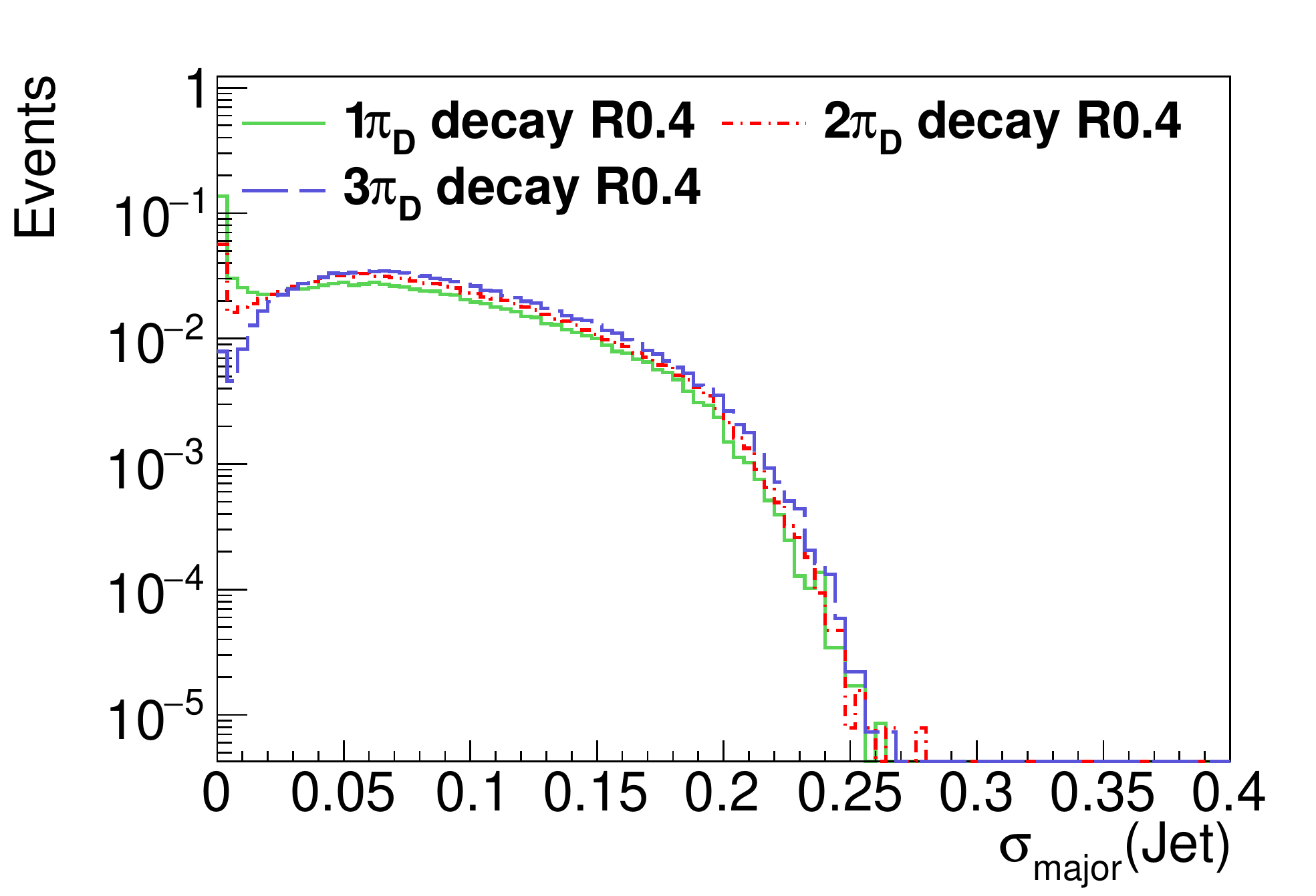}
\includegraphics[width=0.44\linewidth]{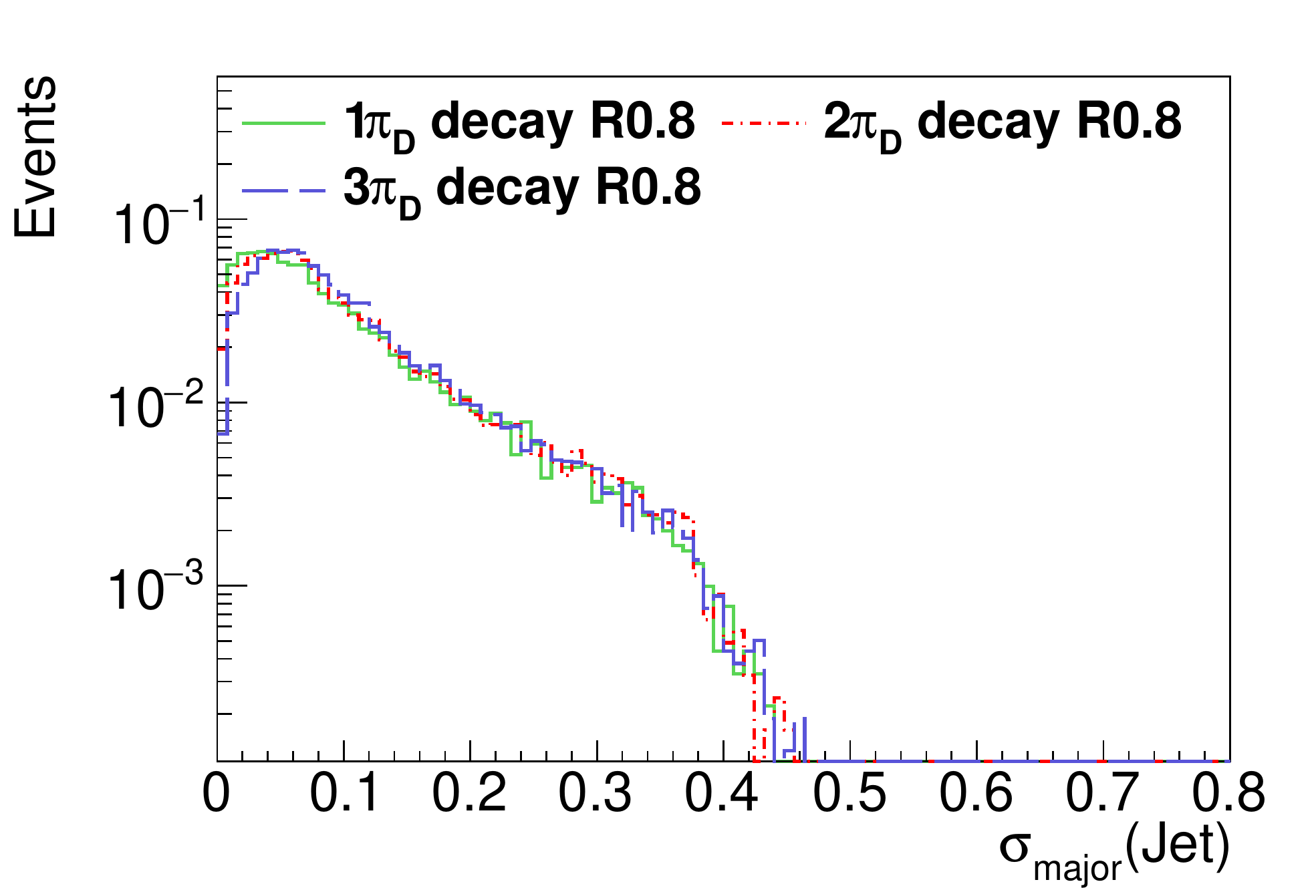}
\includegraphics[width=0.44\linewidth]{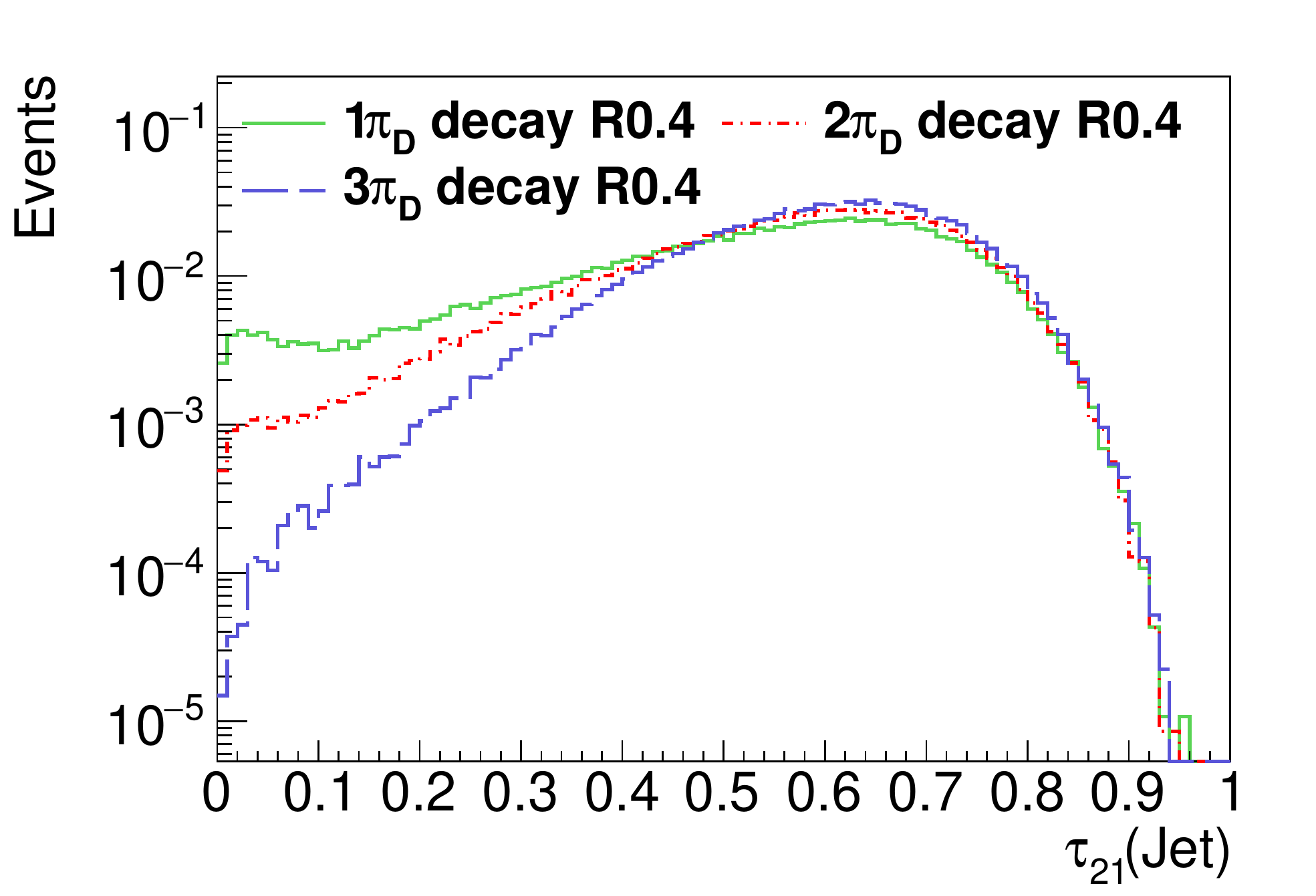}
\includegraphics[width=0.44\linewidth]{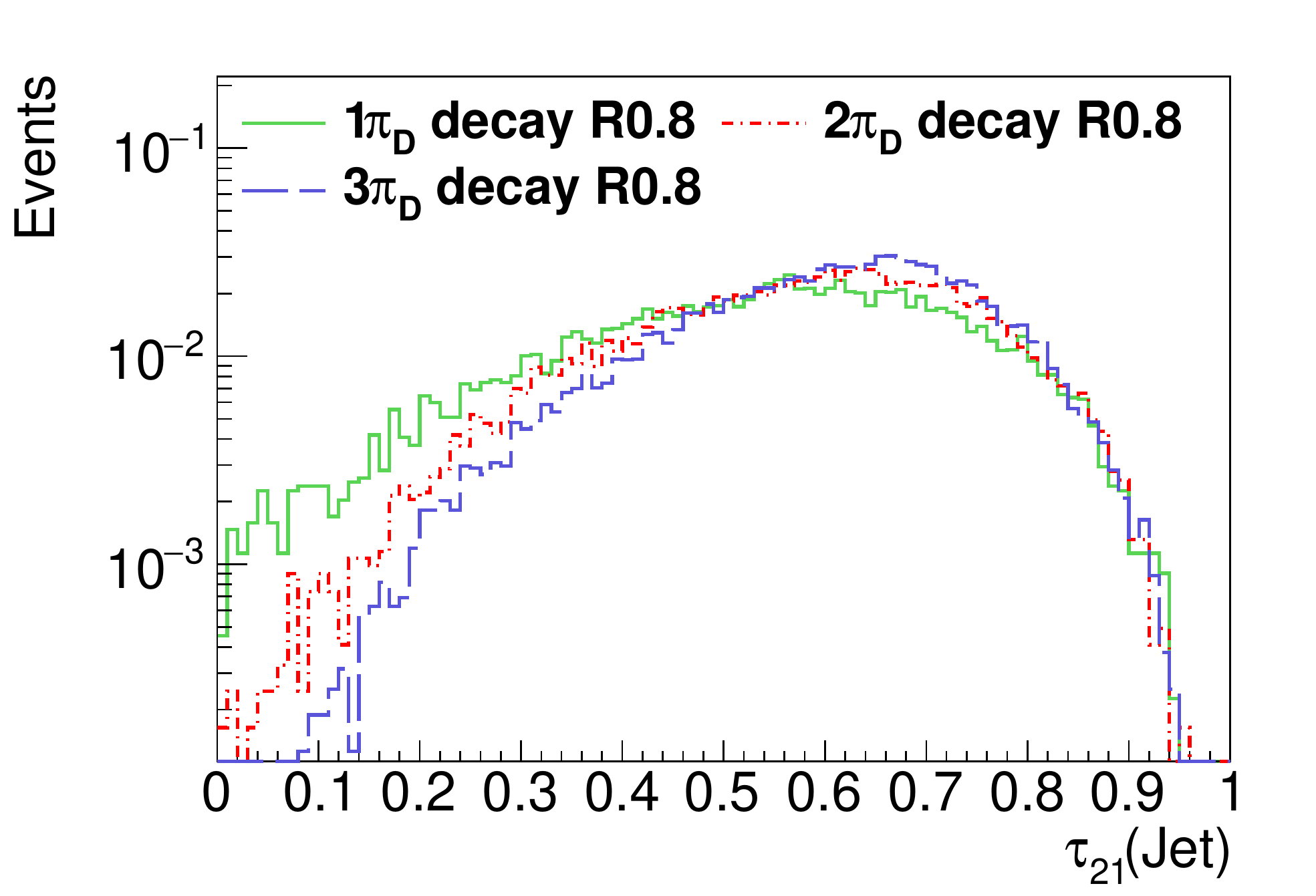}
\includegraphics[width=0.44\linewidth]{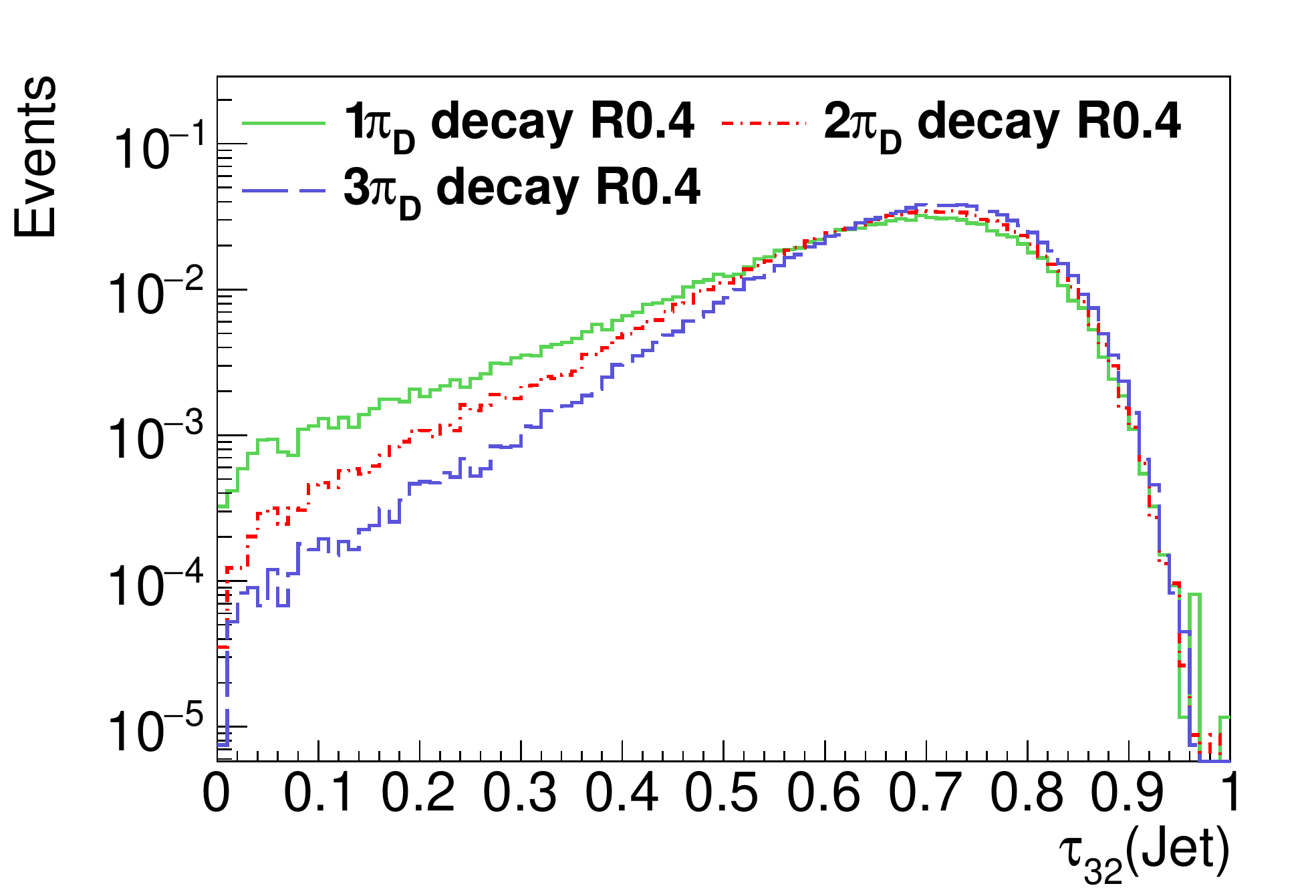}
\includegraphics[width=0.44\linewidth]{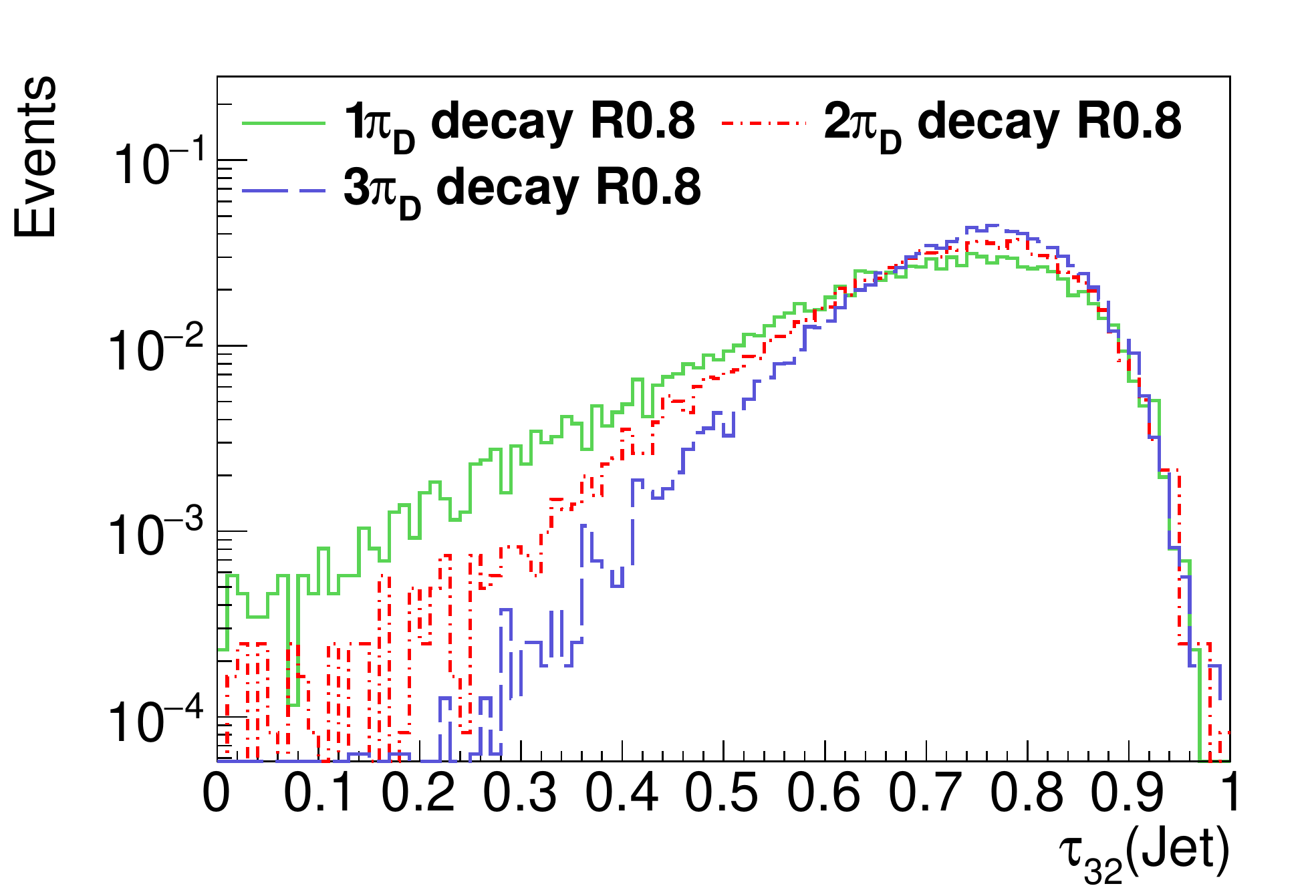}
\caption{Comparison of reco-level variables between different number of unstable diagonal dark pions for \texttt{probVector=0.5}. Left column show the $R=0.4$ jets and right column $R=0.8$ jets. First row is the axis minor, second row is the axis major, third row is the n-subjettiness ratio $\tau_{21}$ and fourth row is $\tau_{32}$.}
\label{fig:reco_1pi2pi3pi_b}
\end{figure}
\clearpage

\section{Improved search strategies}
\label{sec:improved_strategies}
The wide variety of signatures coming from the dark/hidden sector scenarios considered throughout this work also motivates advanced techniques which may enable us to distinguish between signal and background at the LHC. These techniques may involve new kinematic variables, jet substructure information (as briefly discussed in sec.~\ref{sec:phenovalid}), machine learning, or advanced triggering strategies. In this Section, we illustrate some the  avenues which have been explored in the literature, using the dark/hidden sector parametrizations presented in sec.~\ref{benchmarks}. It would be of great interest to also perform such analyses in the light of the new developments presented in  sec.~\ref{parameters}. 

\subsection{Event-level variables}\label{ELvariables}

\emph{Contributors: Hugues Beauchesne, Giovanni Grilli di Cortona }

Semi-visible jets are a characteristic signature of many confining dark sectors and consist of jets of visible hadrons intermixed with invisible stable particles. Up to now, two main search strategies have been pursued: tagging semi-visible jets (see e.g. Refs.~\cite{Aguilar-Saavedra:2017rzt, Park:2017rfb, Farina:2018fyg, Heimel:2018mkt, Cohen:2020afv, Bernreuther:2020vhm, Kar:2020bws, Canelli:2021aps}) and exploiting the special relation between the azimuthal direction of the semi-visible jets and the missing transverse momentum \met (see e.g. Refs.~\cite{Cohen:2015toa, Cohen:2017pzm, Beauchesne:2017yhh}). In Ref.~\cite{Beauchesne:2021qrw}, it was shown that these two approaches can be combined to define new event-level variables that considerably increase the sensitivity of semi-visible jet searches. The central idea is that semi-visible jets are responsible for most of \met in signals and that tagging specifies which jets are semi-visible. The tagging information then predicts the direction and magnitude of \met, which can be compared to its measurement. In this section, we present a summary of Ref.~\cite{Beauchesne:2021qrw} and refer to it for technical details.

For illustration purposes, consider the following benchmark model. Assume a new confining group $\mathcal{G}$. Introduce a dark quark $\Pqdark$ that is a fundamental of $\mathcal{G}$ and neutral under the Standard Model gauge groups. Introduce a scalar mediator $S$ that is an antifundamental of $\mathcal{G}$ and has an hypercharge of $-1$. These fields allow the Lagrangian
\begin{equation}\label{eq:LagrangianDecay}
  \mathcal{L} = \lambda_i S^\dagger \Paqdark P_R E_i + \text{h.c.},
\end{equation}
where $E_i$ are the Standard Model leptons. Assume for simplicity that the only non-negligible $\lambda_i$ is the one corresponding to the electron. If the mediators are pair-produced, they will each decay to an electron and a dark quark. The experimental signature will then be two electrons and two semi-visible jets. This is similar to the signature of leptoquark pair-production and as such preselection cuts are applied based on typical leptoquark cuts. The event is also required to contain two jets tagged as semi-visible. We focus on the \ttbar background. Events are generated using \MADGRAPH \cite{Alwall:2014hca}, \PYTHIA~8 \cite{Sjostrand:2007gs} and \texttt{Delphes}~3 \cite{deFavereau:2013fsa}. The Hidden Valley module of \PYTHIA is used with the following parameters:
  \begin{center}
	\begin{tabular}{llll}
	  \hline
		Setting         & Value  & Setting        & Value  \\
		\hline
		NGauge          & 3      & Dark pion mass & 10 Gev \\
        nFlav           & 1      & Dark rho mass  & 21 GeV \\
        FSR             & On     & pTminFSR       & 11 GeV \\
        alphaOrder      & 1      & fragment       & On     \\
        Lambda          & 10 GeV & probVec        & 0.75   \\
        Dark quark mass & 10 GeV                           \\
      \hline
	\end{tabular}
  \end{center}
All other parameters are left to their default value. Finally, we define $1 - \rinv$ as the average fraction of the dark pions that decay back to Standard Model particles.

Consider a signal event. Label the transverse momenta of the two dark quarks produced from the decay of the two $S$ as $\mathbf{p}_\mathrm{T}^{\PqdarkI}$, with $i \in \{1, 2\}$. Each $\Pqdark$ leads to a visible jet of transverse momentum $\mathbf{p}_\mathrm{T}^{D_i}\sim(1 - \rinv) \mathbf{p}_\mathrm{T}^{\PqdarkI}$ and a contribution to $\mathbf{\met}$ of $\sim \rinv \mathbf{p}_\mathrm{T}^{\PqdarkI}$. This gives
\begin{equation}\label{eq:vMETexp}
  \met \sim \frac{\rinv}{1 - \rinv}\mathbf{p}_\mathrm{T}^{D_1} + \frac{\rinv}{1 - \rinv}\mathbf{p}_\mathrm{T}^{D_2}.
\end{equation}
Consider the decomposition $\met = a_1 \mathbf{p}_\mathrm{T}^{D_1} + a_2 \mathbf{p}_\mathrm{T}^{D_2}$. The coefficients $a_1$ and $a_2$ should then peak at $\sim \rinv/(1 - \rinv)$ and can be combined in a single test statistics. This could be done in multiple ways, but a simple and powerful one is to train a fully supervised neural network on the $a_1$ and $a_2$ of both the signal and the background. Alternatively, one can encode much of the same reasoning in a single variable. Define
\begin{equation}\label{eq:DeltaPhi}
  \Delta \phi = \left|\phi_{\mathbf{p}_\mathrm{T}^{D}} - \phi_{\met}\right|,
\end{equation}
where $\phi_{\mathbf{p}_\mathrm{T}^{D}}$ ($\phi_{\met}$) is the azimuthal angle of $\mathbf{p}_\mathrm{T}^{D_1}  + \mathbf{p}_\mathrm{T}^{D_2}$ ($\met$). This quantity should peak at 0 for the signal, but unfortunately contains no information on the norm of $\met$.
We introduce two comparisons. First, the standard procedure up to now has been to compute the minimal difference in azimuthal angle between $\met$ and the leading jets \cite{Cohen:2017pzm}
\begin{equation}\label{eq:CLLM}
  \Delta \phi_{\text{CLLM}} = \min_{i \leq 4}\left\{\left|\phi_{\mathbf{p}_\mathrm{T}^{j_i}}  - \phi_{\met}\right|\right\},
\end{equation}
where in this case four jets are considered. Second, we consider a supervised neural network using $x = \{\phi_{\mathbf{p}^{D_1}}, \phi_{\mathbf{p}^{D_2}}, \eta_{\mathbf{p}^{D_1}}, \eta_{\mathbf{p}^{D_2}}, \phi_{\met}\}$. This is only meant as a comparison, as fully supervised neural network are susceptible to simulation artefacts and sculpting.

\begin{figure}[ht!]
\begin{center}
 \begin{subfigure}{0.47\textwidth}
    \centering
    \caption{$\rinv=0.2$}
    \includegraphics[width=1\textwidth]{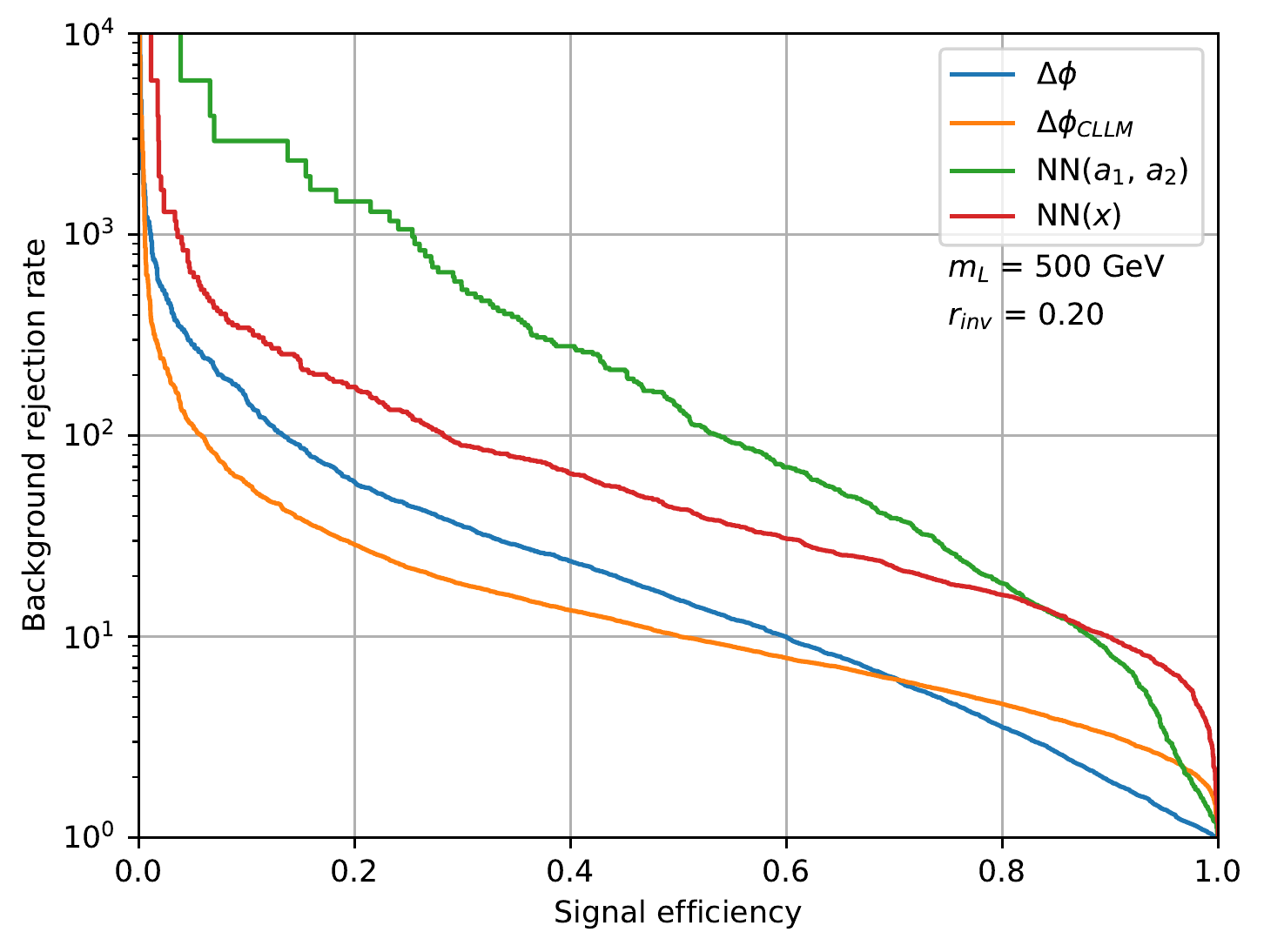}
    \label{fig:ROC_0.2_500_100_100}
  \end{subfigure}
  ~
  \begin{subfigure}{0.47\textwidth}
    \centering
    \caption{$\rinv=0.4$}
    \includegraphics[width=1\textwidth]{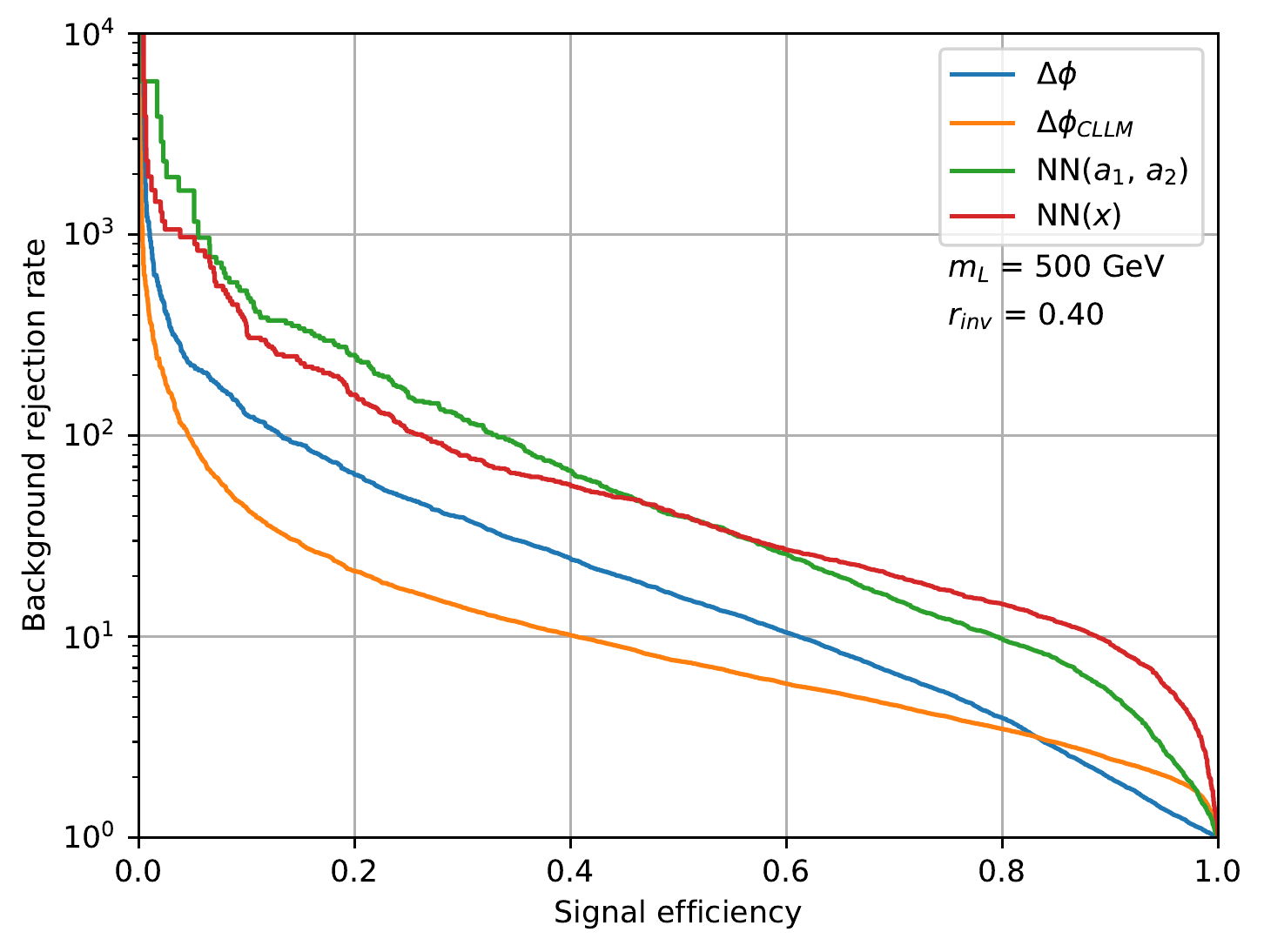}
    \label{fig:ROC_0.4_500_100_100}
  \end{subfigure}
\caption{ROC curves for $\Delta \phi$ (blue), $\Delta \phi_{CLLM}$ (orange) and for the neural networks using the input variables $a_1$ and $a_2$ (green) or the set $x$ (red) for a mediator mass (here called $L$) of 500 GeV. Taken from Ref.~\cite{Beauchesne:2021qrw}.
\label{fig:ROC}}
\end{center}
\end{figure}

Receiver Operating Characteristic (ROC) curves are shown in Fig.~\ref{fig:ROC} for different values of $\rinv$. As can be seen, the coefficients $a_1$ and $a_2$ typically provide the strongest results. They sometimes exceed the fully supervised neural network by exploiting information on the magnitude of the momenta which are not provided to the neural network. The coefficients outperform the standard approach of $\Delta \phi_{\text{CLLM}}$  by an order of magnitude for a signal rejection rate of 0.5. The variable $\Delta \phi$ also generally outperforms $\Delta \phi_{\text{CLLM}}$.

\subsection{Casting a graph net to catch dark showers}\label{findingDS}

\emph{Contributors: Elias Bernreuther }

To increase the sensitivity to dark shower signals consisting of promptly decaying dark hadrons, it is crucial to reduce the large QCD background. While backgrounds from mismeasured QCD jets mimic the signal with regards to event-level observables, such as $\Delta\phi$, differences are expected at the level of jet substructure. These can arise from differences in the shower evolution between QCD and the dark sector, the presence of visibly decaying heavy dark mesons in the jets, or invisible dark hadrons that are interspersed with visible particles. See e.g.\ Refs.~\cite{Cohen:2020afv, Kar:2020bws} for recent studies of dark shower signals in terms of classic jet substructure variables. In contrast, advances in tagging jets with modern machine learning techniques make use of low-level properties of jet constituents. Here, we summarize the results of Ref.~\cite{Bernreuther:2020vhm}, which studies the potential of deep neural networks for identifying semi-visible jets from dark showers. 

As a benchmark, dark showers of nearly mass-degenerate GeV-scale dark mesons which are produced at the LHC via a heavy $\PZprime$ vector mediator with mass on the TeV scale were considered. The underlying dark sector is the Aachen model summarized in Section~\ref{aachen} and motivated by cosmological and experimental constraints~\cite{Bernreuther:2019pfb}.
The dark quark production process $p p \to \Pqdark\Paqdark$ was simulated with \textsc{MadGraph5}~2.6.4~\cite{Alwall:2014hca} using a UFO file generated with \textsc{FeynRules}~\cite{Alloul:2013bka} and performing MLM matching with up to one additional hard jet. Showering and hadronization, both in QCD and in the hidden sector, were carried out using \PYTHIA~8.240~\cite{Sjostrand:2014zea, Carloni:2010tw, Carloni:2011kk}. The settings used in \PYTHIA's Hidden Valley module for a signal with dark meson mass $\mdark$ are summarized in table~\ref{tab:hidden_valley_settings}. The parameter probVector was set to 0.5 such that 25~\% of dark mesons are unstable, flavor-diagonal vector mesons as predicted by the benchmark model. Jet clustering is performed by \textsc{FastJet}~\cite{Cacciari:2011ma} using the anti-$k_T$ algorithm with jet radius $R=0.8$.

\begin{table}[htb]
	\caption{\label{tab:hidden_valley_settings} Settings of the \PYTHIA Hidden Valley module used for generating the dark shower signal in Ref.~\cite{Bernreuther:2020vhm}.}
	\begin{center}
		\begin{tabular}{cccccccc}
			\hline
			setting & value & setting & value & setting & value & setting & value \\
			\hline
			FSR & on & probVector & 0.5 &  Ngauge & 3 & Lambda & $\mdark$ \\
			fragment & on & dark pion mass & $\mdark$ & nFlav & 2 & pTminFSR & $1.1 \mdark$ \\
			alphaOrder & 1 & dark rho mass & $\mdark$ & spinFv & 0 & & \\
			\hline
		\end{tabular}
	\end{center}
\end{table}

A priori, it is not clear what the optimal jet representation and neural network architecture are to optimally distinguish dark shower jets from QCD jets. In Ref.~\cite{Bernreuther:2020vhm} it was shown that a dynamic graph convolutional neural network (DGCNN)~\cite{Wang:2018nkf, Qu:2019gqs} operating on particle clouds outperforms convolutional neural networks (CNNs) based on jet images~\cite{Macaluso:2018tck} and a network operating on ordered lists of Lorentz vectors~\cite{Butter:2017cot}. While a standard CNN carries out convolutions over neighboring pixels in a jet image, a DGCNN performs convolutions over edges of a graph constructed from jet constituents that are neighbors in feature space. While graph networks also represent the state of the art in tagging boosted top jets~\cite{Kasieczka:2019dbj}, their advantage over a CNN or a Lorentz Layer network is considerably larger in identifying semi-visible jets. A comparison of ROC curves showing the QCD jet background rejection $1/\epsilon_B$ as a function of the dark shower signal efficiency $\epsilon_S$ for $\mdark = 5$~GeV  is shown in figure~\ref{fig:supervisedROCcomparison}.

\begin{figure}[t]
	\begin{center}
	\includegraphics[width=0.49\textwidth]{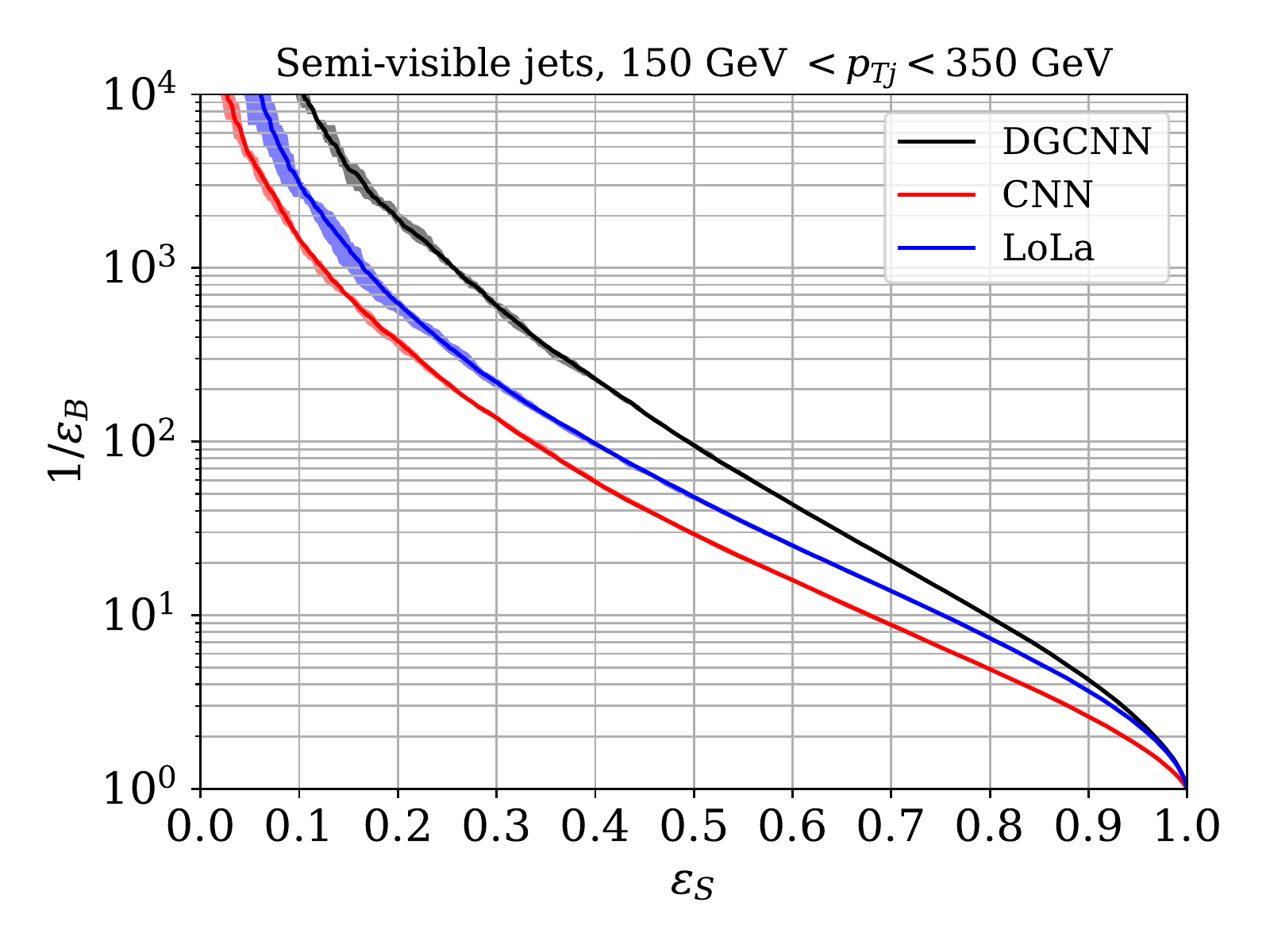}
	\includegraphics[width=0.49\textwidth]{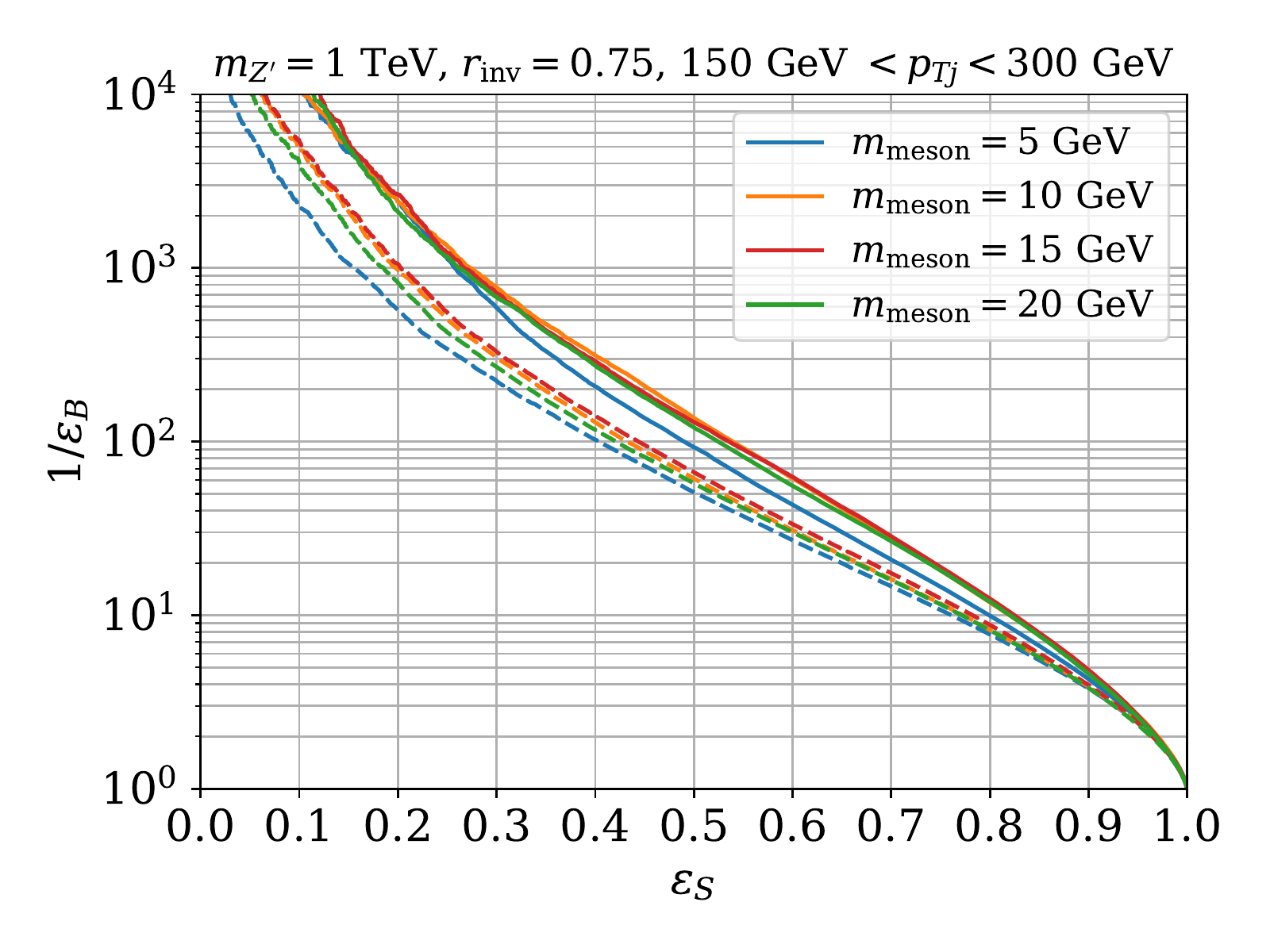}
	\caption{Left: ROC curves showing the semi-visible jet tagging efficiency $\epsilon_S$ and QCD background rejection $1/\epsilon_B$ for a DGCNN compared to a CNN and a LoLa network operating on jet images and Lorentz vectors, respectively. Right: ROC curves for a DGCNN trained on a mixed sample containing a number of different dark meson masses $\mdark$ (here called $m_\mathrm{meson}$) and tested on dark showers with $\mdark$ as stated in the legend (dashed lines), compared to a DGCNN trained and tested on dark showers with identical $\mdark$ (solid lines). Figures taken from Ref.~\cite{Bernreuther:2020vhm}.  \label{fig:supervisedROCcomparison}}
	\end{center}
\end{figure}

Since the parameters of the dark sector are a priori unknown it is a crucial question how well the classification performance of the DGCNN generalizes to dark showers with different parameter values than were used for training. Varying \rinv and $\mdark$, the performance continuously degrades the further the parameters of the dark showers in the test sample are from those in the training sample. While the effect is modest for \rinv, it is much more substantial for the dark meson mass. For example, for a network trained with $\mdark=5$~GeV, the background rejection rate for signal efficiencies between 0.1 and 0.3 is reduced by nearly an order of magnitude when tested on samples with $\mdark=20$~GeV. This suggests that the network learns to reconstruct this mass from the jet constituents. Importantly, this behavior can be mitigated by training the network on mixed samples which contain jets with a range of different dark meson masses. This yields a much more general classifier as reflected in the ROC curves in figure~\ref{fig:supervisedROCcomparison}.

Finally, it was investigated how much the sensitivity of an experimental search for dark showers can be improved by applying a DGCNN as a semi-visible jet tagger. As an example, an ATLAS search for mono-jet events with a luminosity of $36.1$~fb$^{-1}$~\cite{Aaboud:2017phn} was considered, which is sensitive to signal events where one of the two dark showers remains invisible and, thus, $\Delta\phi \approx \pi$. For an event to be accepted, it had to fulfil the original selection criteria of the search and contain at least one fat jet that is classified as a semi-visible jet by the network. The training sample consisted of jets from a dark shower signal with the benchmark parameters stated in Section~\ref{aachen} and from the dominant $\PZ$+jets background. The expected number of background events with and without the DGCNN tagger is shown in table~\ref{tab:background_signal_limit} for the signal region EM4, which is the region most sensitive to the signal when $\mZprime = 1$~TeV. In addition, the table compares the resulting expected 95~\% CL limit $S^{95}_\mathrm{exp}$ on the number of signal events in the region with and without the tagger and shows the corresponding improvement of the projected limit on the dark quark production cross section. In the benchmark scenario shown in table~\ref{tab:background_signal_limit} a DGCNN for tagging semi-visible jets can improve the sensitivity of the search to dark showers by more than one order of magnitude.

\begin{table}[t]
	\caption{\label{tab:background_signal_limit} Number of background events $B$ with systematic uncertainty in the signal region EM4 of the search with and without the dark shower tagger and corresponding expected 95~\% CL limit $S^{95}_\mathrm{exp}$ on the number of signal events. In addition, the improvement in the limit on the dark quark production cross section for the benchmark scenario described in the main text is shown relative to the search without a tagger. Table adapted from Ref.~\cite{Bernreuther:2020vhm}.}
\begin{center}
	\begin{tabular}{c c c c}
		\hline
		& $B$ & $S^{95}_\mathrm{exp}$ & $(\sigma^{95}_\mathrm{exp})^\mathrm{w/o  NN}/\sigma^{95}_\mathrm{exp}$ \\
		\hline
		without DGCNN tagger &  $27640 \pm 610$ & $1239$ & 1 \\
		with DGCNN tagger & $12.1 \pm 0.3$ & $8.2$ & $19.7$ \\
		\hline
\end{tabular}
\end{center}
\end{table}

\subsection{Autoencoders for semi-visible jets}\label{autoencoders}
\emph{Contributors: Annapaola de Cosa, Jeremi Niedziela, Kevin Pedro}

Semi-visible jets arise from Hidden Valley models of dark matter, which include strong interaction in the dark sector. They constitute a challenging experimental signature in which a fraction of jet constituents is invisible to the detector, leading to missing transverse energy \met being aligned with the jet. 

The details of the kinematics are mainly affected by the following theory parameters: \mZprime (the mass of the mediator), \mdark (the mass of the dark hadrons) and \rinv (the fraction of stable, invisible dark hadrons). However, a large total number of unknown theory parameters leads to a vast model space with a huge number of possible scenarios that can easily evade any constraints from e.g. cosmological measurements. Since it is impractical to perform dedicated searches for all possible model variations, we propose to use autoencoders (AE) as anomalous jets taggers instead \cite{Canelli:2021aps}.

The autoencoder-based anomaly detection strategy is robust against both detector effects and details of the model implementation. AEs are designed to detect objects significantly different from the training sample, without prior knowledge of signal characteristics. For reference, the AE introduced here is compared to a Boosted Decision Tree (BDT) trained on the QCD background and a mixture of different signals. For completeness, we have also studied alternative anomaly detection techniques, namely Variational Autoencoders (VAE) and Principal Component Analysis (PCA).

All architectures mentioned above were trained on high-level properties of jets: $\eta$ and $\phi$ coordinates and invariant mass \mj, as well as jet substructure variables: jet \pt dispersion \ptd, jet ellipse minor and major axes, EFP\textsubscript{1}, and ECF ratios: C\textsubscript{2} and D\textsubscript{2}. We have also considered including four-momenta of jet constituents in the training, but they were ultimately discarded since no improvement was observed.

\begin{figure}[htbp!]\centering
\includegraphics[width=1.0\textwidth]{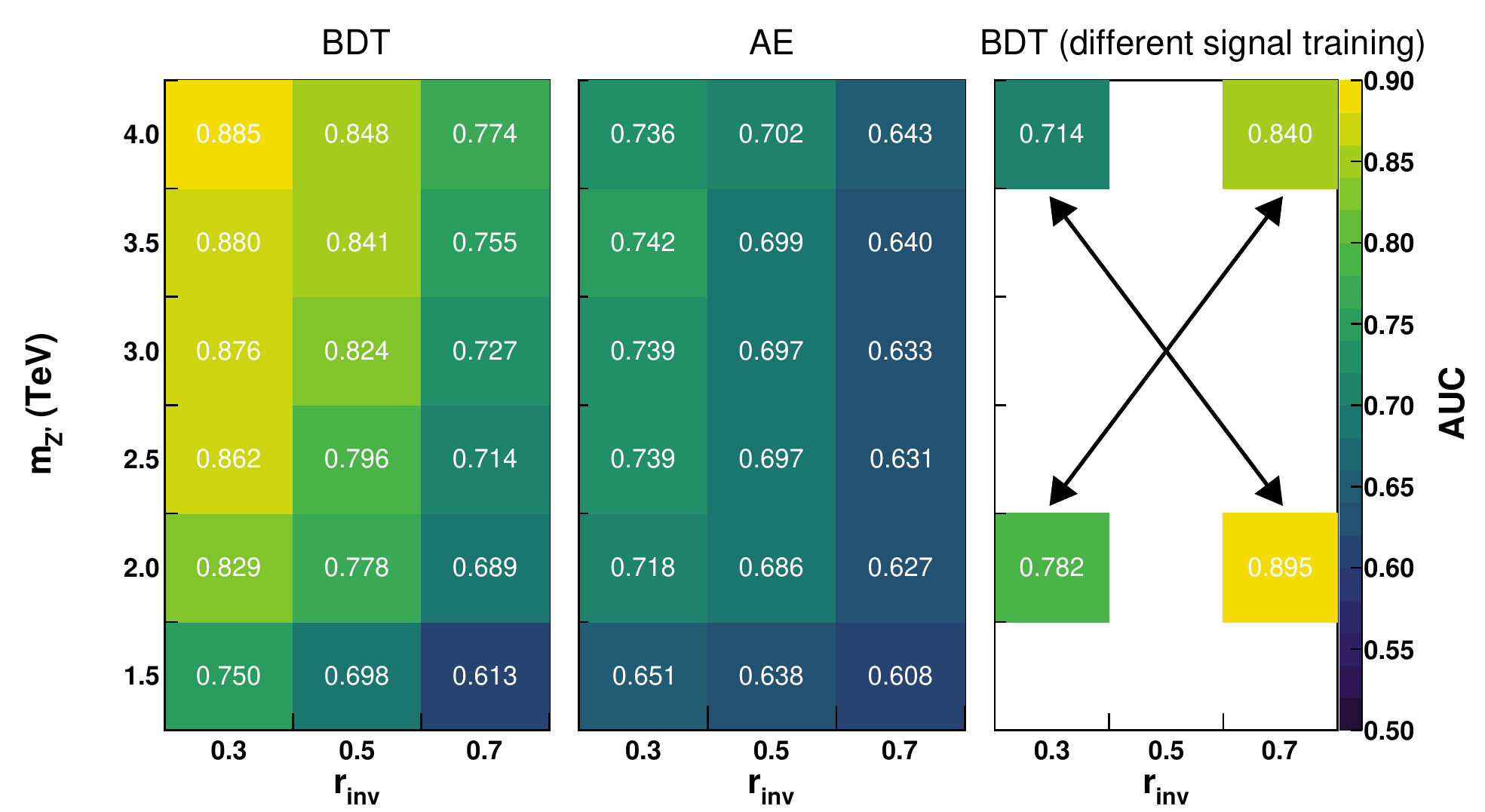}
\caption{Left and middle panels: comparison of AUC values from the autoencoder and a BDT trained on a mixture of all signal models with $\mdark = 20\,\GeV$. Right panel: AUC values for a BDT trained on a signal with parameters different from the signal used for testing, as indicated by the arrows. For example, the AUC value presented in the top left corner of the table comes from a model trained on the sample from the lower right corner. This figure is reproduced from Ref.~\cite{Canelli:2021aps}.}
\label{fig:aucs}
\end{figure}

The performance of different approaches is quantified by comparing the area under the ROC (receiver operator characteristic) curve (AUC), shown in Fig.~\ref{fig:aucs}. It was demonstrated that an AE-based jet tagger can provide satisfactory performance, compared with the fully supervised BDT approach. The PCA proved to be less efficient than other approaches. The VAE was found give the best results when trained exclusively on reconstruction loss, leading its variance to collapse to zero and therefore becoming equivalent to a regular AE.

\begin{figure}[htbp!]\centering
\includegraphics[width=0.6\textwidth]{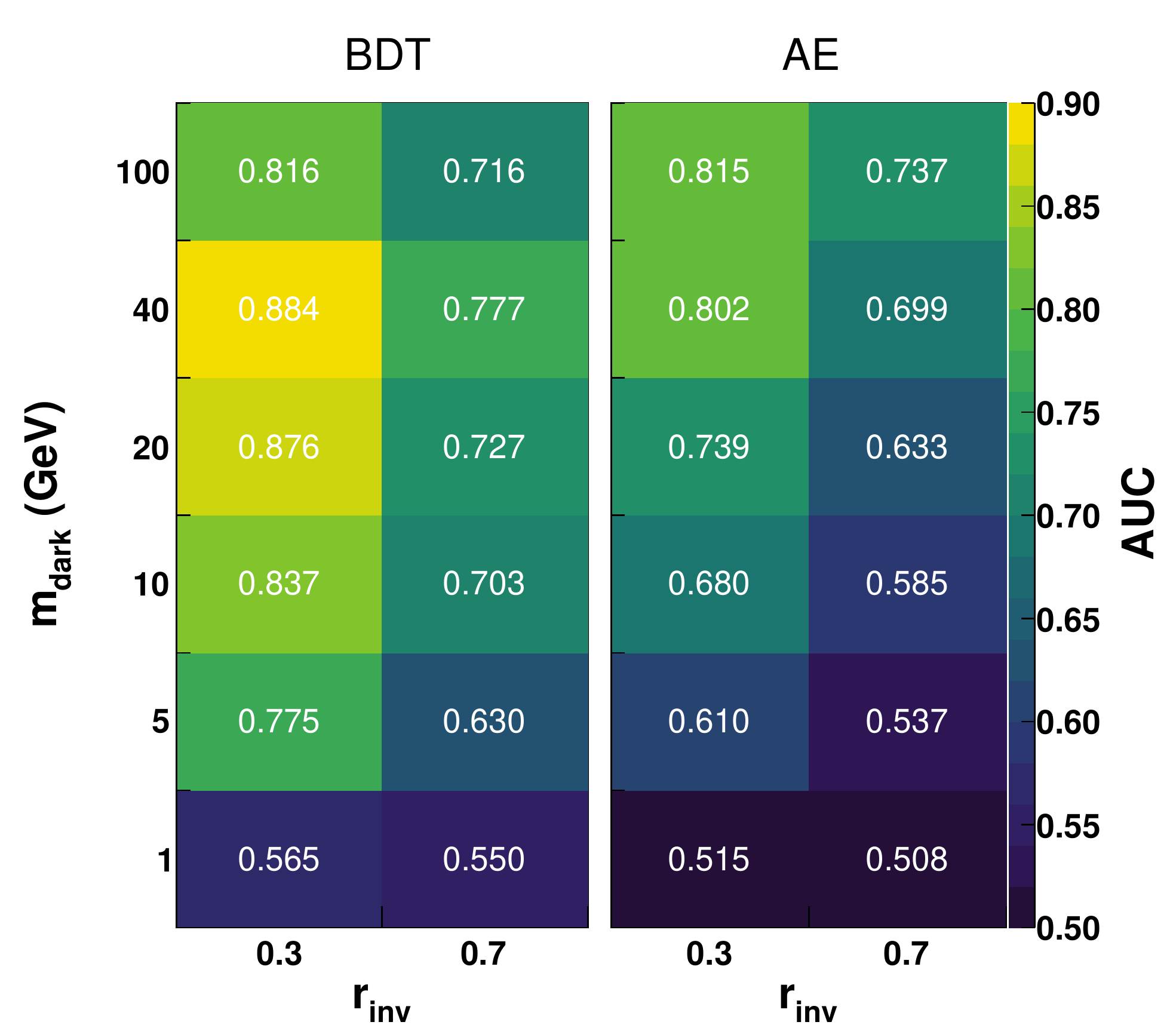}
\caption{Comparison of AUC values of the autoencoder and BDT with varying \mdark values (here called $m_{\text{dark}}$). The BDT was trained on a mixture of all signals with $\mdark = 20\,\GeV$. This figure is reproduced from Ref.~\cite{Canelli:2021aps}.}
\label{fig:aucs_mdark}
\end{figure}

Robustness against unknown model parameters was also assessed. As shown in the rightmost panel of Fig.~\ref{fig:aucs} and in Fig.~\ref{fig:aucs_mdark}, in certain cases the AE can outperform the BDT when the latter was trained on an incorrect signal hypothesis. Another interesting observation that can be made in the right panel of Fig.~\ref{fig:aucs} is that a BDT trained on \rinv = 0.3 and tested on \rinv = 0.7 performs better then the one trained on a mixture of different signals (left panel). This is caused by the fact that the low \rinv signal is more similar to the background, and therefore the BDT has to learn how to distinguish between the two more precisely. This results in a performance boost when tested on an easier case of large \rinv.

\subsection{Autoencoders for SUEP}\label{autoencodersSUEP}
\emph{Contributors: Jared Barron, David Curtin, Gregor Kasieczka, Tilman Plehn, Aris G.B. Spourdalakis}

\subsubsection{Searching for Hadronic SUEP}
The theoretical motivation and experimental phenomenology of SUEPs are described in section \ref{suep}. Strategies to overcome the experimental challenges of searches for SUEP at the LHC are still being developed. For the nightmare scenario of prompt, hadronically decaying SUEP, a search strategy was proposed in \cite{Barron:2021btf}, employing an autoencoder neural network as an anomaly detector. 

An autoencoder is an unsupervised neural network trained on background events, which attempts to minimize the difference between its output and input. Ideally the autoencoder learns to do this efficiently only for inputs that are similar to its training data, so that when evaluated on an event from outside the background distribution, a high reconstruction error flags the event as anomalous. In the case of a search for SUEP, the background events are soft, highly isotropic QCD events. The unsupervised nature of this analysis avoids the model dependence that comes from using signal simulation to develop a classifier.

\subsubsection{Signal Generation}
While the use of unsupervised machine learning techniques removes the need for signal events in the training dataset, a simulated signal dataset is still necessary to evaluate the autoencoder's performance as an anomaly detector. For this purpose, SUEP events were generated using a statistical toy model of the dark shower. 

The highly isotropic hadronic SUEP toy model simulated events were generated beginning with the production of Higgs bosons in association with a \PW or \PZ boson, simulated at center-of-mass energy 14 TeV in \PYTHIA 8 \cite{Sjostrand:2014zea}. The vector boson was then required to decay leptonically. The hard lepton(s) from the vector boson decay were used to sidestep the issue of how to trigger on SUEP for this analysis. The decay of the Higgs to a shower of dark mesons was performed with the \texttt{SUEP\_Generator} plugin in \PYTHIA 8.243, which models the dark shower as being a completely isotropic cloud with Boltzmann-distributed momenta as was presented in Eq. \ref{eq:Boltzmannsuep}, and for which 
the parameter $\Tdark$ controls the energy distribution of the dark mesons, and represents the Hagedorn temperature of the dark sector. Only one flavor of dark meson is assumed, with mass $\mdark$. Each dark meson was then forced to decay hadronically to a $u\bar{u}$ quark pair. From this point the parton showering and hadronization were performed by \PYTHIA as normal. Signal simulation was generated for $\mdark$ from 0.4 GeV to 8 GeV, and for $\Tdark/\mdark$ from 0.25 to 4. Detector simulation was performed with \Delphes3 with CMS detector settings \cite{deFavereau:2013fsa}. Due to the difficulty of disentangling the highly diffuse energy depositions of SUEP from pile-up, only charged track information was used for the analysis.

\subsubsection{Background Generation}
The simulated QCD background events necessary for the training and test datasets were created by generating di-jet plus lepton(s) events with a reduced jet $\pt$ threshold of 15 GeV in \texttt{MadGraph5\_aMC@NLO 2.6.6} with hadronization by \PYTHIA8 and detector simulation by \Delphes3 \cite{Alwall:2014hca,Frederix:2018nkq}. 

\subsubsection{Analysis}
A trigger-level selection was applied to all simulated events requiring at least one charged lepton with $\pt>40$ GeV, or two opposite-charged leptons with $\pt>30(20)$ GeV, as well as hadronic $\HT>30$ GeV. A further set of pre-selection cuts were then applied. Before feeding events into the autoencoder as training data, 98\% of the initial simulated background events were discarded by cutting on three high-level observables that encode the essential features of SUEP. 

First, the multiplicity of charged tracks was required to be $N_{charged}\geq 70$. Second, the event ring isotropy variable introduced in \cite{Cesarotti:2020hwb}, measuring the Wasserstein distance between a given event and a uniformly isotropic distribution of energy, was required to be $\mathcal{I}<0.07$. Finally, the inter-particle $\Delta R_{ij}$ distance averaged over all pairs of tracks in the event was required to be $\overline{\Delta R}>3$. Signal efficiency of these cuts varied from $1-30\%$ with $\mdark$ and $\Tdark$. 

A fully connected autoencoder with five layers was trained using QCD background events that passed the pre-selection cuts as training data. Each event was represented using a modified inter-particle distance matrix $\Delta \tilde R_{ij}$ of the 70 highest-$\pt$ charged tracks in the event. 

\begin{equation}
    \label{e.inputfeatures}
    \Delta \tilde R_{ij} \equiv \left\{
    \begin{array}{lll}
    \Delta R_{ij}=\sqrt{(\Delta \eta_{ij})^{2} + (\Delta \phi_{ij})^{2}} & & i > j\\
    p_{T,i}/\mathrm{GeV} \ \ \   & \mathrm{for} \ \ \  & i = j\\
    0 &  & i < j
    \end{array} \right.
\end{equation}

A modified mean-squared-error loss quantified the reconstruction error for each event.

\subsubsection{Results}
After training on background events, the autoencoder was fed test data including both background events and signal events across the range of simulated $\mdark$ and $\Tdark$ points. Using the reconstruction loss as an anomaly score, ROC curves were constructed for each parameter point. To estimate the physical sensitivity of the model, the minimal excludable branching ratio of Higgs to SUEP for which $S/\sqrt{B + u_{sys}B^{2}}>2$ was computed. Statistical uncertainties due to the limited size of the simulated background sample became dominant as the cut threshold was increased before the classifier's performance began to deteriorate, indicating that the sensitivity of a real search using this method could be even higher than we report here. 

\begin{figure}
    \centering
    \includegraphics[width=0.5\textwidth]{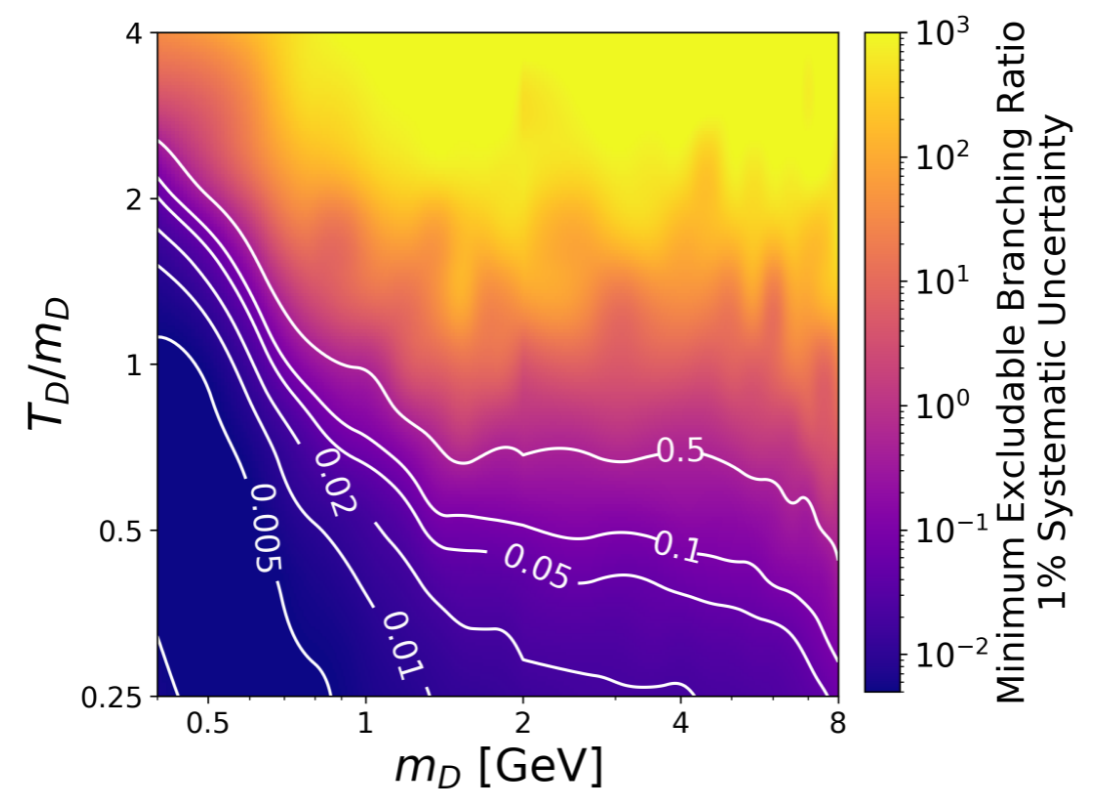}
    \\
    \caption{
    Minimum excludable $\mathrm{Br}(h \to \mathrm{SUEP})$ at the HL-LHC, assuming $1\%$ systematic uncertainty on QCD background for fully connected autoencoder.
    }
    \label{fig:minbr}
\end{figure}

As Figure \ref{fig:minbr} illustrates, this autoencoder-based analysis could exclude Higgs branching ratios to SUEP down to 1\% for dark meson mass $\mdark<~ 1$ GeV and $\Tdark/\mdark<~ 1$. If the dark shower temperature $\Tdark$ is $<~0.5 \mdark$, branching ratios down to 5\% could be probed for $\mdark$ up to $\approx 8$ GeV. 

Using a neural network architecture and event representation tailored to the essential characteristics of the SUEP signature, but without relying on the details of any signal simulation model, this study demonstrates that even the maximally challenging scenario of entirely prompt and hadronic SUEP can be probed at the HL-LHC.

\subsection{Triggering on Emerging Jets}\label{triggering}

\emph{Contributors: Daniel Stolarski  }

The original Emerging Jets (EJ) theory paper~\cite{Schwaller:2015gea} as well as the CMS search~\cite{CMS:2018bvr} 
(see also section~\ref{tchannel} of this white paper)
considers a model with a colored mediator $\Pbifun$ that is pair produced at the LHC, which leads to a final state with two QCD jets and two EJs. Those works also consider mediator masses in the regime of $\mbifun \gtrsim 600$ GeV. Given those assumptions, the vast majority of EJ events have substantial $\HT$ and thus the trigger efficiency is very high. In this section we consider relaxing both of the above assumptions and explore how one can still trigger on Emerging Jets.

In~\cite{Linthorne:2021oiz}, an $s$-channel mediator was considered 
(see also section~\ref{schannel} of this report), 
focusing on a $\PZprime$ that couples to the quark current in the SM and the dark quark current. Such a mediator produces events that typically do not include additional hard jets. That work also considered the possibility of relatively light $\PZprime$ down to masses of $\sim 50$ GeV. The typical $\HT$ of such events, particularly in the light $\PZprime$ regime, is considerably lower than typical trigger thresholds at the LHC experiments, and other techniques are needed to increase the trigger efficiency. 

The events were generated using a modified spin-1 mediator model\footnote{https://github.com/smsharma/SemivisibleJets}~\cite{Cohen:2017pzm} implemented using the \FEYNRULES~\cite{Alloul:2013bka} package. The hard process is generated with \verb!Madgraph5_aMC@NLO!~\cite{Alwall:2014hca} using a centre of mass energy of 13 TeV. This output is interfaced to the Hidden Valley~\cite{Carloni:2010tw,Carloni:2011kk} module of \PYTHIA 8~\cite{Sjostrand:2014zea}, which simulates showering and hadronization in the dark sector as well as decays of dark hadrons to either other dark hadrons or to SM states. The $\PZprime$ mass is varied and a $\PZprime$ width of $\Gamma_{\PZprime} = \mZprime/100$ is used. The remaining dark sector parameters are varied across a few benchmark models shown in Table I of~\cite{Linthorne:2021oiz}.

Initial state radiation (ISR) in QCD or EW is included at leading order in the hard processes. The resulting hadrons are clustered into jets using the Anti-$k_{t}$ algorithm~\cite{Cacciari:2008gp} implemented in \verb!FASTJET!~\cite{Cacciari:2011ma} with a jet angular parameter $R = 0.4$ and a maximum pseudorapidity of $|\eta| < 2.49$ to be compatible with the ATLAS inner tracker. MLM matching and merging procedure~\cite{Mangano:2006rw} is employed for extra QCD radiation with \verb!XQCut! of $\mZprime$/10. A crude detector volume cut is implemented at the \PYTHIA 8 stage for which particles that are outside of a cylinder of $(r = 3000$ mm$, z = 3000$ mm) are considered stable.

Two main strategies are explored to increase the trigger efficiency. The first is exploiting the possibility of SM radiation from the initial state. While electroweak ($\PW/\PZ/\gamma$) radiation was explored, the most effective strategy was to use additional QCD radiation. This radiation can increase the trigger efficiency in two complimentary ways. First, the additional hard jet(s) can be used to trigger on directly. Second, the emerging jets tend to be boosted and carry more energy. This in turn will will increase the $\HT$ ($\met$) if the dark pion states are short (long) lived.

\begin{figure*}
\centering
\begin{minipage}[c]{\textwidth}
\includegraphics[width=0.48\textwidth]{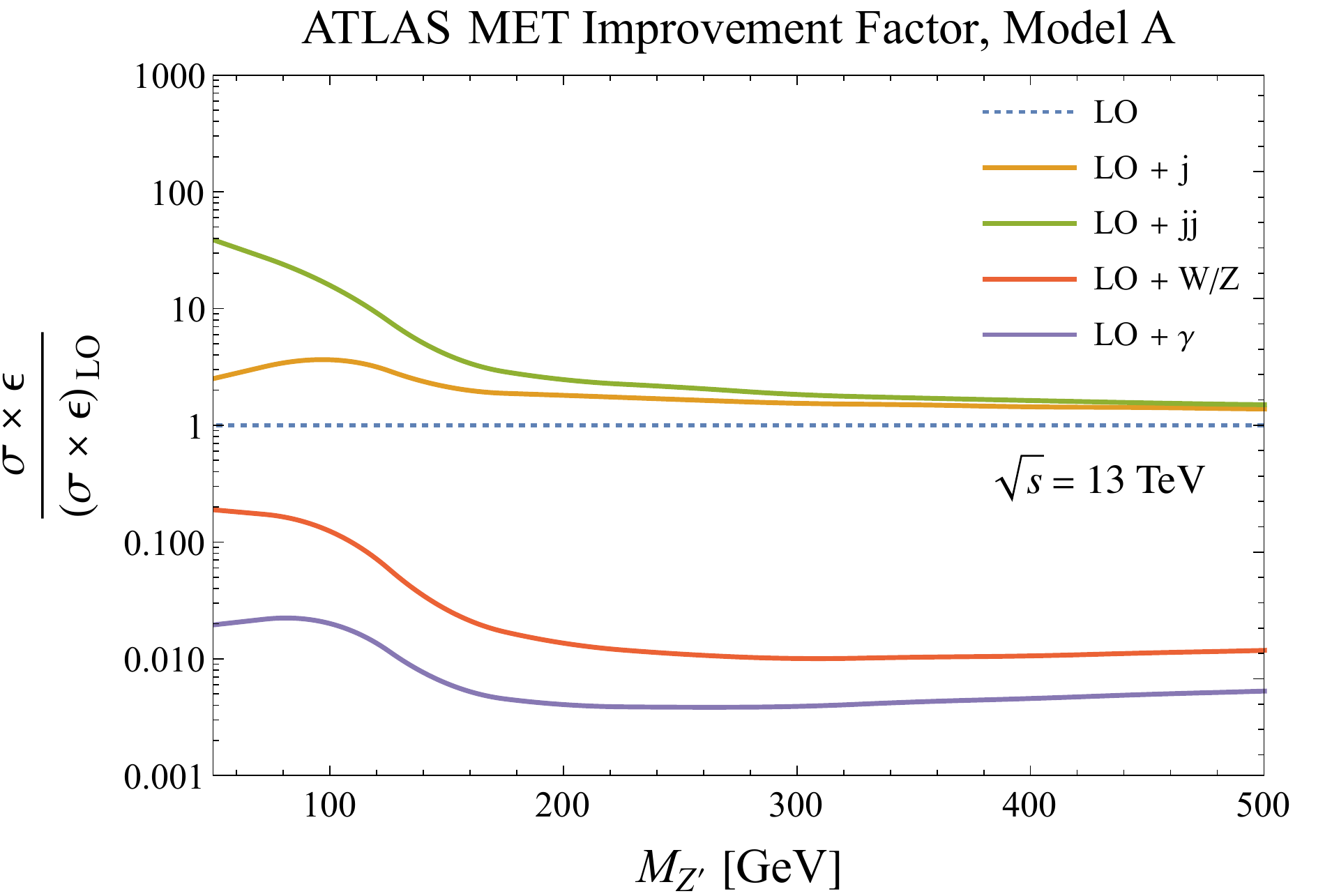}
\hfill
\includegraphics[width=0.48\textwidth]{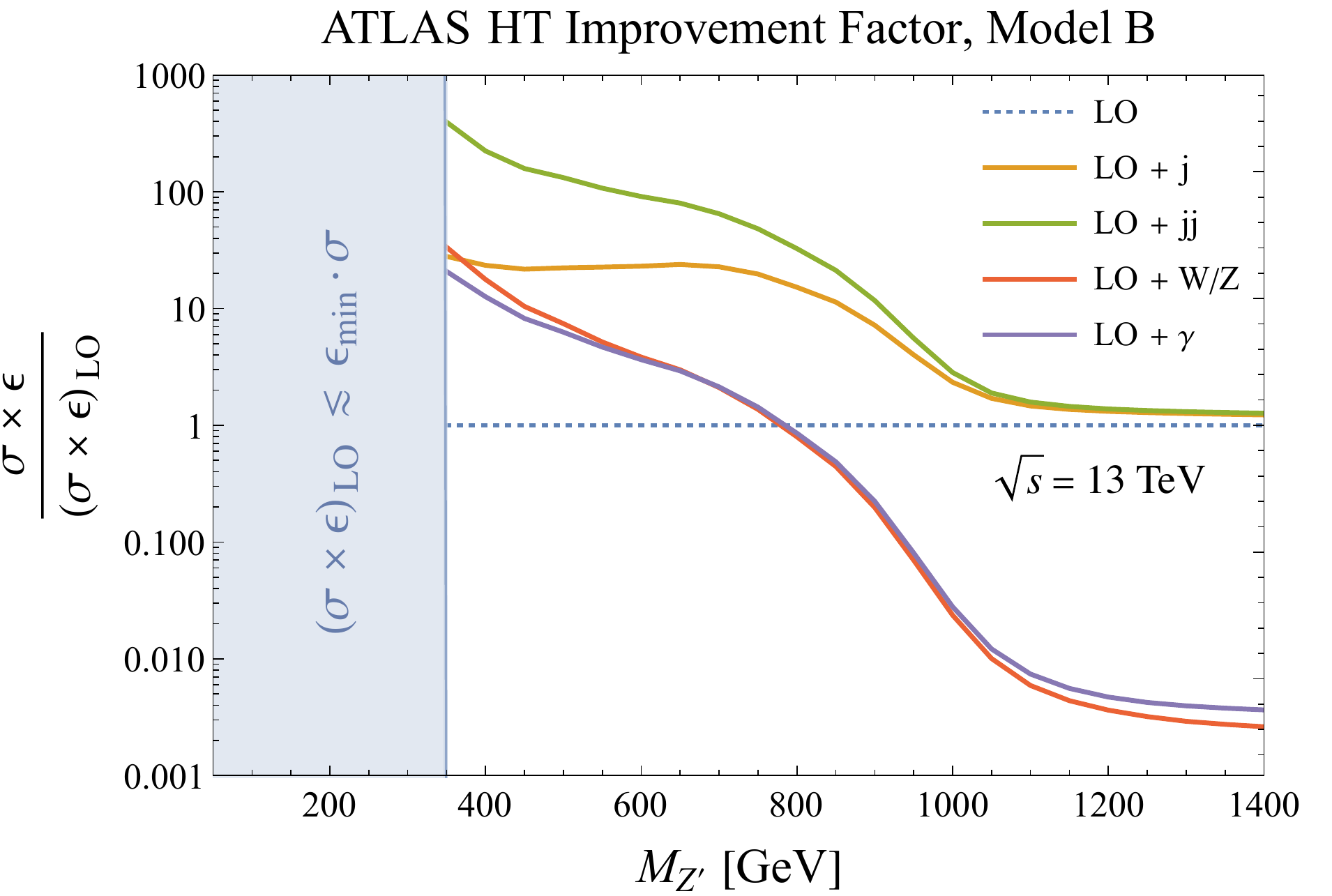}
\end{minipage}
\hfill
\caption{(Figure 4 from~\cite{Linthorne:2021oiz}) Cross section times efficiency of various processes (leading order, 1-jet ISR, 2-jet ISR, electro weak ISR) scaled by their respective leading order process. The left plot uses the \met (here called MET) trigger and Model \textbf{A} which has a dark pion lifetime of 150 mm. The right uses the $\HT$ trigger and Model \textbf{B} which has a dark pion lifetime of 5 mm. The dotted line is the leading order process. On the right plot, the blue region on the left has zero of events that we simulated pass the trigger and thus an efficiency $\epsilon \lesssim \epsilon_{\rm{ min}} = 1/ 400$ $000$. }
\label{Fig:Improvement}
\end{figure*}

Using ATLAS trigger thresholds from~\cite{ATL-DAQ-PUB-2018-002}, we estimate the improvement in rate achieved by including radiation and the results are shown in Fig.~\ref{Fig:Improvement}. In addition to increasing the trigger efficiency, events with extra radiation have reduced rates, therefore Fig.~\ref{Fig:Improvement} shows the ratio of the cross section times trigger efficiency for events with radiation to those without. We see that the largest improvement is additional radiation of two extra jets (green line). The left panel is a benchmark with a dark pion lifetime of 150 mm (Model \textbf{A}) and uses the missing energy trigger. We see that for a light $\PZprime$, more than an order of magnitude improvement in rate is possible. The right side is a model with a dark pion lifetime of 5 mm (Model \textbf{B}) and uses the $\HT$ trigger. In that benchmark, the efficiency of the leading order process is below what was simulated for $\mZprime \lesssim 350$ GeV, and the improvement is potentially even larger.

The first method considered above uses existing triggers, but~\cite{Linthorne:2021oiz} also considers implementing new triggers using modern machine learning techniques. 
As ISR is no longer relevant, \PYTHIA 8's hidden valley production process $f\bar{f} \rightarrow \PZprime$ processes is employed to generate events. Regardless of the lifetime of the dark pions, the detector subsystem with the largest number of decays is the inner tracker. Therefore, the strategy employed (which is also similar to that of~\cite{Huffman:2016wjk} proposed for $b$-tagging), is to use the tracker information but not reconstruct tracks. Rather~\cite{Linthorne:2021oiz} proposes to use hit patterns in different layers of the tracker as an input to a support vector machine\footnote{Other machine learning techniques gave similar results.} (SVM) from the TMVA toolkit~\cite{Hocker:2007ht}.
A proper detector simulation of the inner tracker is outside of the scope, but a crude detector simulation with code used in~\cite{Knapen:2016hky} which encompasses the ATLAS tracker from the Inner B-layer (IBL) to the Transition Radiation Tracker (TRT). This detector simulation assumes simple models of energy loss through each thin layer of the detector.

When proposing new triggers, backgrounds must also be considered, and the main background for this strategy is $b\bar{b}$ jets as they have a very large rate and also produce displaced hadrons. 
Simulations are performed using $ gg \rightarrow b\bar{b}$ with \PYTHIA 8's heavy flavor hard $b\bar{b}$ processes.  The inclusive background cross section is taken from the \PYTHIA 8. Pileup is added to both signal and background events with \PYTHIA 8's minimum bias events. For each signal or background event, a number of minimum bias events are added randomly sampled from a poisson distribution with mean of $\mu =  50$, mimicking the Run 2 conditions.

\begin{figure*}
\centering
\begin{minipage}[c]{\textwidth}
\includegraphics[width=.49\textwidth ]{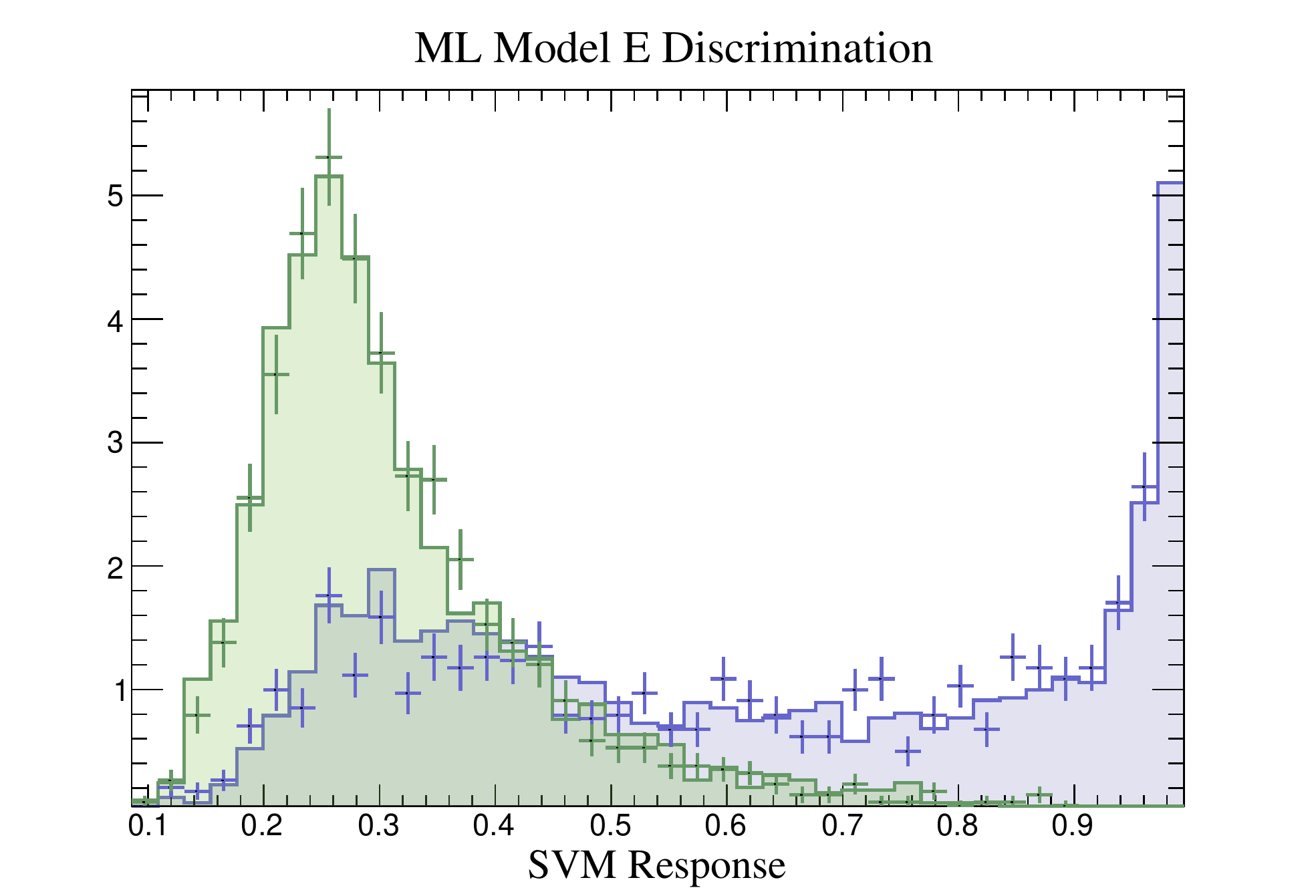}
\hfill
\includegraphics[width=.49\textwidth ]{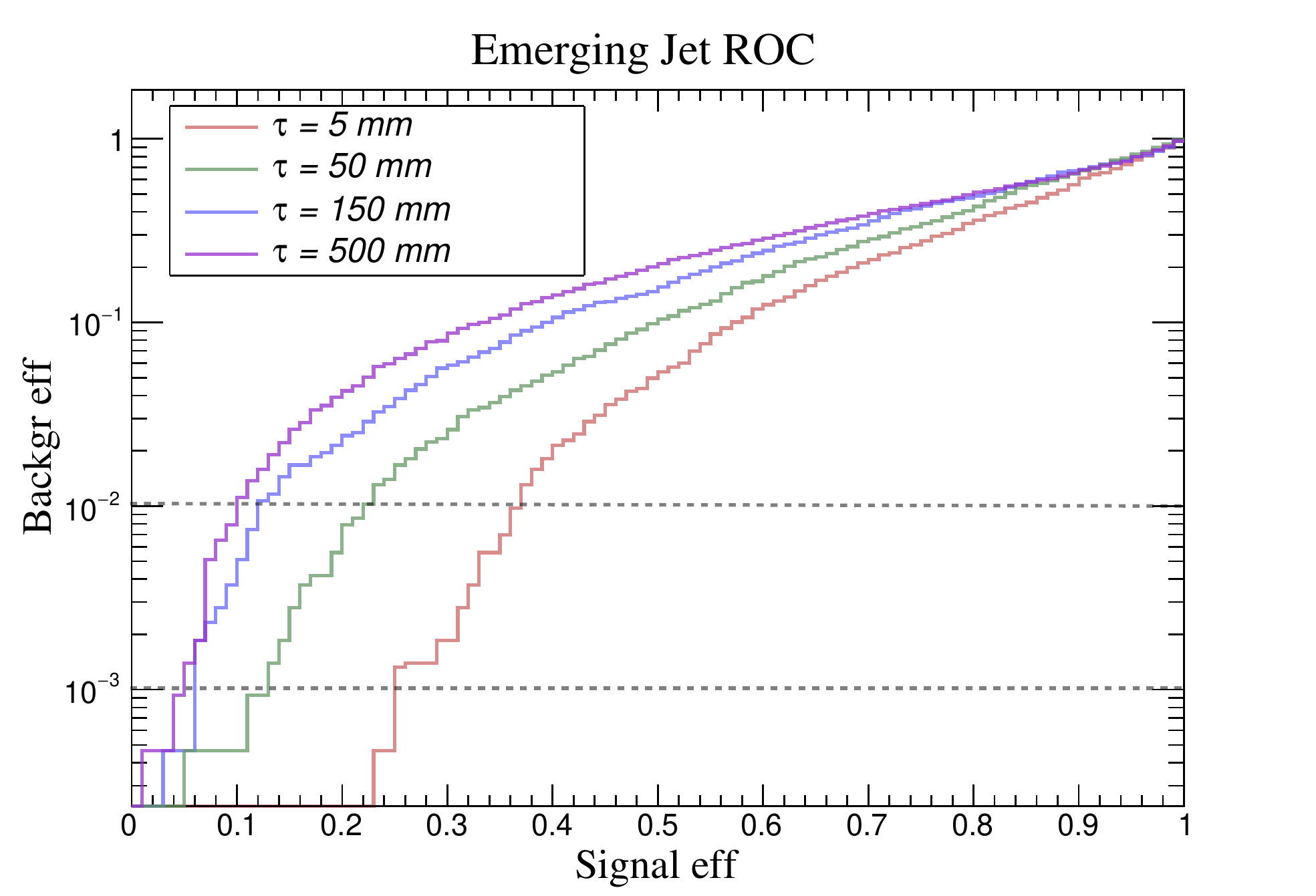} 
\end{minipage}
\hfill
\caption{(Figure 8 from~\cite{Linthorne:2021oiz}) On the left, discrimination of signal (blue) from $b\bar{b}$ background (green) using a support vector machine. The flat bars (points) correspond to the training (test) set. On the right, Receiver Operation Characteristic ROC for four different lifetimes of signal. At a given background efficiency, the expected signal efficiencies increase as the dark pion lifetimes lower. The required background rejection is estimated to lie between the horizontal dotted lines. Both figures use a mediator mass of 500 GeV.}
\label{fig:ml}
\end{figure*}

The left panel of Fig.~\ref{fig:ml} shows the ability of the SVM to distinguish signal in blue from the dominant $b\bar{b}$ background. On the right panel we show the ROC curve as a function of lifetime. A background rejection of $\sim 10^{-2}-10^{-3}$ is needed for a novel high level trigger, and we see that efficiencies of $\mathcal{O}(10\%)$ are achievable, with larger efficiencies at lower dark pions lifetimes. It is also found that using an SVM trained on one signal benchmark can also give good acceptance for other signal benchmarks, showing great promise for such a new trigger. 

\clearpage

\section{Summary and Perspectives}

In this report, we have summarised the work performed in the context of the Dark Shower Snowmass project: it is the first comprehensive effort to gather the large, pre-existing theoretical, phenomenological and experimental communities working in this field, following initial discussions in the LHC Long-lived Particles Working Group \cite{Alimena:2019zri} and also some presentations in the LHC Dark Matter Working Group. This report also concretely describes pathways for a systematic exploration of strongly interacting theories. In this context, we mainly concentrated on QCD-like scenarios leading to jetty signatures at the LHC, but we also discussed signatures such as SUEPs and glueballs which are typically associated with non-QCD like theories. 

QCD-like scenarios, which are the main focus of this report, are inherently non-trivial to analyse due their non-perturbative nature. In such theories, confinement in the IR leads to bound states whose masses and interactions are governed by the UV dynamics. While the SM QCD has been analysed in great detail in terms of UV versus IR parametrizations, little is known for arbitrary gauge groups and flavor contents. Nevertheless, due to the interesting new signatures the strongly interacting scenarios could produce at the LHC, their phenomenology is being actively explored. 

In this context, we began this report (see sec.~\ref{existing_studies}) with a review of the existing efforts and phenomenological parametrizations of QCD-like scenarios. We qualitatively illustrated the phenomenological differences obtained for various mediator mechanisms, giving rise to exotic LHC signatures such as emerging or  semi-visible jets. We also discussed some existing experimental results constraining these models and ongoing efforts to search for these signatures. 

If the dark sector is instead non QCD-like, other classes of spectacular signatures can be obtained in terms of SUEPs or glueballs. These were discussed sec.~\ref{beyond_QCDlike_theories}, in which original phenomenological SUEP studies were presented, along with recent preliminary simulation tools for these scenarios.

After this overview of existing efforts and of the signature landscape, the report also addressed in sec.~\ref{sec:model_building} possible pathways for consistent theory frameworks, especially concentrating on semi-visible jets. In that section, lattice calculations, chiral perturbation theory and an analysis of symmetry breaking due to SM-DS portals were combined to exemplify avenues in theoretical model building. Improvements to the \PYTHIA 8 Hidden Valley module, made in the context of this Dark Shower Snowmass project, were also presented along with their validation. Combining the theory developments  with the new Hidden Valley module, we then illustrated their impact on the phenomenology of semi-visible jets. 

In the final section of the report, sec.~\ref{sec:improved_strategies}, we discussed some proposed improvements to LHC search strategies. These include efforts using machine learning, trigger considerations and the definition of new event level variables. 

Strongly-interacting dark sectors are an exciting class of scenarios in which a vibrant community of theorists, phenomenologists and experimentalists is being invested. They could lead to spectacular signatures which have not yet been systematically explored at the LHC. In view of the large phenomenological interest  of such theories, a more concentrated effort in theoretical work is needed, covering  model building and classification of associated LHC signatures, a deeper understanding of hadronization physics, as well as studies of cross correlation with open problems of the SM such as the nature of dark matter. It is clear from this report that such a work involves communication among experts in SM QCD, lattice, and collider physics as well as in dark matter. We hope that our report lays down the foundations for such a wider exchange, and that this may help devising better strategies that could ultimately lead to a breakthrough in finding signals of strongly-interacting theories.

\section{Acknowledgements}

H.~Beauchesne's work was supported by the Ministry of Science and Technology, National Center for Theoretical Sciences of Taiwan.
T.~Cohen is supported by the U.S. Department of Energy, under grant number DE-SC0011640.
The research of D.~Curtin was supported in part by a Discovery Grant from the Natural Sciences and Engineering Research Council of Canada, the Canada Research Chair program, the Alfred P. Sloan Foundation, and the Ontario Early Researcher Award.
M-H.~Genest acknowledges the support of the French Agence Nationale de la Recherche (ANR), under grant ANR-21-CE31-0013 (project DMwithLLPatLHC).
G.~Grilli di Cortona is supported by the INFN Iniziativa Specifica Theoretical Astroparticle Physics (TAsP) and by the Frascati National Laboratories (LNF) through a Cabibbo Fellowship call 2019.
T.~Holmes's research is supported by U.S. Department of Energy, Office of Science, Office of Basic Energy Sciences Energy Frontier Research Centers program under Award Number DE-SC0020267. 
S.~Kulkarni is supported by Austrian Science Fund Elise Richter Fellowship  V592-N27. S.~Mee is supported by Austrian Science Fund research group funding FG1. S.~Kulkarni and S.~Mee thank Biagio Lucini, Axel Maas, Simon Pl\"atzer and Fabian Zierler for numerous discussions. 
K.~Pedro, S.~Mrenna, K.~Folan~DiPetrillo and E.~Bernreuther are supported by the Fermi National Accelerator Laboratory, managed and operated by Fermi Research Alliance, LLC under Contract No. DE-AC02-07CH11359 with the U.S. Department of Energy.
A.~Peixoto acknowledges funding from the French Programme d’investissements d’avenir through the Enigmass Labex.
The work of J.~Shelton was supported in part by the Binational Science Foundation and DOE grant DE-SC0017840.
S.~Sinha's work is based on the research supported wholly by the National Research Foundation of South Africa (Extension Support Doctoral Scholarship).
T.~Sjöstrand is supported by the  Swedish Research Council, contract number 2016-05996. 
The work of D.~Stolarski and A.~Spourdalakis is supported in part by the Natural Sciences and Engineering Research Council of Canada (NSERC).
Work in Mainz was supported by the Cluster of Excellence Precision Physics, Fundamental Interactions, and Structure of Matter (PRISMA+ EXC 2118/1) funded by the German Research Foundation(DFG) within the German Excellence Strategy (Project ID 39083149), and by grant 05H18UMCA1 of the German Federal Ministry for Education and Research (BMBF).

\bibliographystyle{jhep}
\bibliography{references_1_1, references_1_2,references_1_3,references_1_4,references_2_2,references_2_3,references_3_1,references_3_2,references_3_3,references_4_1,references_4_2,references_4_3,references_4_4,references_4_5,references_intro}
\end{document}